%% file: thesis.tex
\documentclass{report}
\usepackage{graphicx,cite,amssymb,amsbsy}
\newcommand{\bbox}[1]{\boldsymbol{#1}}

\begin{document}
\pagenumbering{roman}

\include{titlepage}

\input{signature}

\include{abstract}

\include{acknowledgments}

\include{publications}

\tableofcontents
\listoffigures

\newpage
\pagenumbering{arabic}

\include{chap1}
\include{chap2}
\include{chap3}
\include{chap4}
\include{chap5}
\include{chap6}
\include{chap7}

\include{appendix}

\bibliographystyle{prsty}
\include{references}

\end{document}

%% file: titlepage.tex
\begin{titlepage}
\begin{center}
{\Large\textbf{Nonequilibrium Phenomena in Quantum Field
Theory:}\\[1ex]\textbf{From Linear Response to Dynamical Renormalization Group}}
\end{center}
\vspace{4.99\baselineskip}

\begin{center}
{\large by}\\
\vspace{1.33\baselineskip}
{\large Shang-Yung Wang}\\
\vspace{1.33\baselineskip} {\large BS, National Taiwan University,
Taiwan, 1991\\[1ex] MS, National Tsing Hua University, Taiwan,
1993}
\end{center}

\vspace{4.99\baselineskip}

\begin{center}
{\large Submitted to the Graduate Faculty of\\[1ex]
Arts and Sciences in partial fulfillment\\[1ex]
of the requirements for the degree of\\[1ex]
Doctor of Philosophy\\}
\end{center}

\vfill

\begin{center}
{\large University of Pittsburgh\\[1ex] 2001}
\end{center}

\end{titlepage}

\setcounter{page}{2}
\newpage

%% file: signature.tex
\thispagestyle{plain}
\begin{center}
\large{UNIVERSITY OF PITTSBURGH} \\[-1.5ex] %
\makebox[1in]{\hrulefill} \\
\large{FACULTY OF ARTS AND SCIENCES} \\ %
\vfill
This dissertation was presented\\[1ex] %
by\\[2ex] %
Shang-Yung Wang\\[-3ex] %
\makebox[3in]{\hrulefill} \\ %
\vfill %
It was defended on\\[2ex] %
September 28, 2001\\[-3ex] %
\makebox[3in]{\hrulefill} \\ %
\vspace{0.5in} %
and approved by\\ %
\vspace{0.5in} %
Richard Holman \\[-3ex] %
\makebox[3in]{\hrulefill} \\[1ex] %
David Jasnow \\[-3ex] %
\makebox[3in]{\hrulefill} \\[1ex] %
Vittorio Paolone\\[-3ex] %
\makebox[3in]{\hrulefill} \\[1ex] %
David Snoke \\[-3ex] %
\makebox[3in]{\hrulefill} \\[1ex] %
Frank Tabakin \\[-3ex] %
\makebox[3in]{\hrulefill} \\[1ex] %
Daniel Boyanovsky \\[-3ex] %
\makebox[3in]{\hrulefill} \\[-1ex]%
\makebox[3in][l]{\footnotesize Committee chairperson}
\end{center}
\vspace{2\baselineskip}

%% file: abstract.tex
\thispagestyle{plain}
%
%
\vspace*{0.25in}
\begin{center}
{\large\textbf{Nonequilibrium Phenomena in Quantum Field Theory:}\\
\textbf{From Linear Response to Dynamical Renormalization Group}}\\
\vspace{2.5ex}
Shang-Yung Wang, PhD\\
\vspace{2ex}
University of Pittsburgh, 2001
\end{center}
\vspace{2ex}

\noindent This thesis is devoted to studying aspects of real-time
nonequilibrium dynamics in quantum field theory by implementing an
initial value formulation of quantum field theory. The main focus
is on the linear relaxation of mean fields and quantum kinetics in
nonequilibrium multiparticle quantum systems with potential
applications to ultrarelativistic heavy ion collisions,
cosmological phase transitions and condensed matter systems. We
first study the damping of fermion mean fields in a fermion-scalar
plasma with a view towards understanding baryon transport
phenomena during electroweak baryogenesis. We obtain a fully
renormalized, retarded and causal equation of motion that
describes the relaxation of the mean field towards equilibrium and
allows an unambiguous identification of a novel damping mechanism
and the corresponding damping rate. Secondly, we apply and extend
the renormalization group method to study nonequilibrium dynamics
in a self-interacting scalar theory, the $O(4)$ linear sigma model
and a hot QED plasma, with the goals of constructing a quantum
kinetic description that goes beyond usual Boltzmann kinetics and
understanding anomalous (nonexponential) relaxation associated
with infrared phenomena. Remarkably, within this framework the
quantum kinetic equations and equations of motion for mean fields
have the interpretation as the dynamical renormalization group
equation which describes the dynamical evolution of a
multiparticle system that is insensitive to microscopic details.
By solving these equations, we effectively integrate out fast
motion dynamics, and are left with an effective theory for slow
motion dynamics. As a by-product, the issue of pinch singularities
in nonequilibrium quantum field theory is resolved naturally in
real time from a quantum kinetic point of view. The final part of
this thesis presents a real-time kinetic analysis of direct photon
production from a quark-gluon plasma created in ultrarelativistic
heavy ion collisions. We show that the direct photon yield is
significantly enhanced by the lowest order energy-nonconserving
processes originated in the transient lifetime of the quark-gluon
plasma. In particular, transverse momentum distribution of direct
photons, which features a power law spectrum in the experimentally
relevant regime, is proposed as a nonequilibrium signature of the
quark-gluon plasma.

%% file: acknowledgments.tex
\chapter*{\centerline{Acknowledgments}}
It is my great pleasure to thank my advisor Dr.\ Daniel Boyanovsky
for introducing me to the fascinating topics of this thesis and
for his guidance, patience, and encouragement throughout the
period of this work. Daniel is not only an advisor in physics but
also a friend and a mentor in life. It is the privilege of me to
be learning from and working with him.

Collaborative work constitutes an important part in modern
scientific research. I would like to take this opportunity to
thank Drs.\ Hector J.\ de Vega, Da-Shin Lee, Kin-Wang Ng, and Yee
J.\ Ng for sharing their knowledge with me and for their
invaluable collaboration during the various stages of this work.

Special thanks to the members of my thesis committee for their
helpful comments.

Finally, I would like to express my deepest gratitude to my
parents for supporting their youngest son in choosing his own path
and to my wife Shu-Hua Kang for her care, support, understanding,
and love that accompany me throughout the good days and bad days.

I dedicate this thesis to my parents.

This work was supported in part by the Andrew Mellon Predoctoral
Fellowship and the National Science Foundation.

%% file: publications.tex
\chapter*{\centerline{List of Publications}}
Parts of this thesis are based on the following publications,
which I published with collaborators during the period of this
work.
\begin{enumerate}
\item
D.\ Boyanovsky, H.J.\ de Vega, D.-S. Lee, Y.J.\ Ng and S.-Y.\
Wang, \emph{Fermion Damping in a Fermion-Scalar Plasma}, Phys.\
Rev.\ D \textbf{59}, 105001 (1999).
\item
S.-Y. Wang, D.\ Boyanovsky, H.J.\ de Vega, D.-S. Lee, and Y.J. Ng,
\emph{Damping Rates and Mean Free Paths of Soft Fermion Collective
Excitations in a Hot Fermion-Gauge-Scalar Theory}, Phys.\ Rev.\ D
\textbf{61}, 065004 (2000).
\item
D.\ Boyanovsky, H.J.\ de Vega, and S.-Y. Wang, \emph{Dynamical
Renormalization Group Approach to Quantum Kinetics in Scalar and
Gauge Theories}, Phys.\ Rev.\ D \textbf{61}, 065006 (2000).
\item
S.-Y.\ Wang, D.\ Boyanovsky, H.J.\ de Vega, and D.-S.\ Lee,
\emph{Real-time Nonequilibrium Dynamics in Hot QED Plasmas:
Dynamical Renormalization Group Approach}, Phys. Rev. D
\textbf{62}, 105026 (2000).
\item
S.-Y.\ Wang and D.\ Boyanovsky, \emph{Enhanced Photon Production
from Quark-Gluon Plasma: Finite-Lifetime Effect}, Phys.\ Rev.\ D
\textbf{63}, 051702(R) (2001).
\item
S.-Y.\ Wang, D.\ Boyanovsky, and K.-W.\ Ng, \emph{Direct Photons:
A Nonequilibrium Signal of the Expanding Quark-Gluon Plasma},
hep-ph/0101251 (to appear in Nucl.\ Phys.\ A).
\end{enumerate}

%% file: chap1.tex
\chapter{Introduction}\label{chap:1}
\section{Motivation}\label{sec:1.1}
The theoretical study of nonequilibrium dynamics in quantum
multiparticle systems dates back to 1961 when Schwinger published
a pioneering paper, \emph{Brownian Motion of a Quantum
Oscillator}~\cite{schwinger}. Conventional applications of quantum
theory are restricted mainly to calculating transition matrices
for scattering processes. In his 1961 paper Schwinger showed, for
the first time, how quantum theory can be formulated to study
initial value problems associated with the dynamical evolution of
nonequilibrium quantum systems.

Over the past two decades, nonequilibrium dynamics in quantum
field theory has attracted a great deal of interest as new
experimental techniques in particle and condensed matter physics
continue to probe novel nonequilibrium quantum phenomena that
require field-theoretical descriptions. Specific examples of
current theoretical interest involving nonequilibrium dynamics of
quantum fields include, just to name a few, inflationary dynamics
in the early Universe, electroweak baryogenesis, the chiral phase
transition and quark-gluon plasma in ultrarelativistic heavy ion
collisions, dynamics of phase transition in Bose-Einstein
condensation and ultrafast spectroscopy of semiconductors. Such a
diversity of applications reveals the truly interdisciplinary
character of nonequilibrium dynamics in quantum field theory.

In this thesis we develop new field-theoretical techniques to
study aspects of real-time nonequilibrium dynamics in quantum
field theory with a view towards potential applications to
ultrarelativistic heavy ion collisions, cosmological phase
transitions and condensed matter systems.

\subsection*{Ultrarelativistic heavy ion collisions}
The goal of ultrarelativistic heavy ion collisions is to create
and study a new state of hot and dense matter in the laboratory.
This program began almost two decades ago with the fixed target
heavy ion experiments at BNL Alternating Gradient Synchrotron
(AGS) and CERN Super Proton Synchrotron (SPS). This new phase of
matter, the quark-gluon plasma (QGP)~\cite{qgp,book:qgp,book:hwa},
whose fundamental degrees of freedom are the deconfined quarks and
gluons is a prediction of quantum chromodynamics (QCD) and
believed to have existed in the early Universe during the first
few microseconds after the big bang when the temperature was
greater than about 200 MeV ($\sim 10^{12}$ K).

Recent assessments of the collected data from CERN SPS seem to provide positive
evidence for the creation of a new state of matter in Pb$+$Pb collisions based
on a multitude of different observations, ranging from anomalous dilepton
production, strangeness enhancement, $J/\psi$ suppression to pion
interferometry~\cite{cern}. It is fair to say, however, that while the evidence
is very suggestive it is far from conclusive. Experiments at higher collision
energies provided by the newly operating Relativistic Heavy Ion Collider (RHIC)
at BNL and the planned Large Hadron Collider (LHC) at CERN are expected to
allow for a quantitative characterization of the quark-gluon plasma and
detailed studies of its early thermalization processes and dynamical evolution.

The Relativistic Heavy Ion Collider at BNL is currently studying
Au$+$Au collisions by colliding two ultrarelativistic, highly
Lorentz contracted gold nuclei with center-of-mass energy per
nucleon pair $\sqrt{s_{NN}}\sim 200$ GeV. At RHIC energies the two
colliding nuclei penetrate one another and most of the baryons are
expected to be carried away by the receding nuclei (the
fragmentation region). The quarks and gluons composing the nuclei
collide and transfer a large amount of energy from the colliding
nuclei to the vacuum, creating a nearly baryon-free region of hot
and dense matter in the form of energetic quarks and gluons that
strongly interact with each other. The hard parton-parton
scatterings thermalize the quarks and gluons on a time scale of
about 1 fm/$c$ ($\sim 3\times 10^{-24}$ s) and produce a
deconfined and chirally restored quark-gluon plasma in local
thermal equilibrium that lives long enough to generate detectable
signals. This hot and dense quark-gluon plasma (energy density
$\epsilon\sim 4-6$ GeV/fm$^3$, corresponding to an initial
temperature $T\sim 200-300$ MeV) then expands rapidly due to
internal pressure and cools down to the deconfinement temperature
$T_\mathrm{QCD}\sim 160$ MeV, below which the quark-gluon plasma
undergoes the QCD phase transition (confinement and chiral
symmetry breaking) and condenses into a gas of hadrons. If the
transition is of first order, the quarks, gluons and hadrons
coexist in a mixed phase before the phase transition is completed.
The hadrons continue to scatter from one another, maintaining the
pressure and causing further expansion and cooling. Eventually,
the hadron gas becomes sufficiently dilute that scatterings among
the hadrons cease and the hadron gas freezes out. Estimates based
on energy deposited in the central collision region at RHIC
suggest that the quark-gluon plasma has a transient lifetime about
10 fm/$c$ and the overall freeze-out time of order about 100
fm/$c$.

Although there is some evidence that local thermal equilibrium may be achieved
at the late stages of the heavy ion collisions, colliding nuclei are inherently
nonequilibrium quantum multiparticle system as manifested in copious production
of particle at extremely high energy densities and on unprecedentedly short
time scales as well as in rapid expansion and cooling. It is therefore a
theoretical challenge to understand the full nonequilibrium dynamics of
ultrarelativistic heavy ion collisions, from the initial state of the heavy ion
reaction (i.e., the two colliding nuclei) up to the freeze-out of all initial
and produced particles after the reaction.

Over the past two decades, our theoretical picture of the quark-gluon plasma is
mainly based on thermal equilibrium models which neglect most of the
nonequilibrium dynamical effects, but make simplified assumptions on the very
initial stages of the collisions, e.g., thermalization or plasma creation. In
the last few years, microscopic (kinetic)~\cite{wang,geiger:pcm,urqmd} and
macroscopic (hydrodynamical)~\cite{hydro} transport models have been developed
to describe various stages of heavy ion collisions. On the one hand, while
these nonequilibrium approaches have provided predictions that are in
reasonable agreement with the available experimental results at lower collision
energies (mostly from AGS and SPS), they are nevertheless phenomenological
models which are crucially based on \emph{ad hoc} semiclassical assumptions and
approximations such that their validity in extreme situations expected to arise
during the early states of ultrarelativistic heave ion collisions at RHIC and
LHC is questionable at best. On the other hand, whereas a coherent picture of
the collision dynamics is emerging, a microscopic field-theoretical description
that allows extracting unambiguous signatures of quark-gluon plasma formation
remains an open question.

With the first RHIC results of Au$+$Au collisions at
$\sqrt{s_{NN}}=130$ GeV coming out
recently~\cite{PHENIX,PHOBOS,STAR,RHIC}, in order to correctly
interpret of the experimental data there is a pressing need for a
field-theoretical description of the quark-gluon plasma created in
ultrarelativistic heave ion collisions. In particular, there may
exist interesting genuine nonequilibrium phenomena which could
lead to unambiguous signatures of quark-gluon plasma formation. An
example is provided by the possible formation of disoriented
chiral condensates (DCC's), which are regions of misaligned vacuum
in the internal isospin space~\cite{DCC:1,DCC:2}. A microscopic,
first-principle description of novel nonequilibrium phenomena in
ultrarelativistic heave ion collisions will definitely require a
thorough understanding of nonequilibrium dynamics in quantum field
theory, especially QCD, at finite temperature and density.

\subsection*{Cosmological phase transitions}
The QCD phase transition is not the only phase transition expected
to occur in the early Universe. On the other hand, based on the
evidence from observational dada over the past decades it is now
widely accepted that the Universe that we observe today is a
remnant of several cosmological phase transitions that took place
at different temperature (energy) scales and with remarkable
consequences at low temperatures~\cite{book:kolb}. The idea of
phase transitions in cosmology is closely related to that of
symmetry braking in particle physics, hence the study of
cosmological phase transitions play a fundamental role in our
understanding of the interplay between cosmology and particle
physics in extreme environments.

Current theoretical ideas beyond the standard model suggest that
when the Universe was about $10^{-35}$ second old there could have
been a symmetry-breaking phase transition at the grand unified
theory (GUT) scale $T_\mathrm{GUT}\sim 10^{15}$ GeV, above which
the GUT symmetry is restored and the strong interaction is unified
with the electroweak interaction. The GUT phase transition is
usually associated with an important cosmological stage:
inflation, i.e., an epoch of accelerated expansion of the
Universe~\cite{guth,linde}. Originally introduced by
Guth~\cite{guth} in order to explain the initial conditions for
the hot big bang model, the idea of inflation has subsequently
become the cornerstone of modern cosmology. Current observations
of the temperature anisotropies in the cosmic microwave background
(CMB) seem to confirm predictions of the inflationary cosmology: a
flat Universe, an almost scale invariant spectrum of density
perturbations that are ultimately responsible for large scale
structure formation~\cite{CMB}.

The next symmetry-breaking phase transition in the early Universe
is the electroweak phase transition, which occurred at the
electroweak scale $T_\mathrm{EW}\sim 100$ GeV when the Universe
was about $10^{-12}$ second old. The most remarkable cosmological
consequence of the electroweak phase transition is the possibility
that the observed baryon asymmetry of the Universe may be
generated during this putative first-order phase transition. As
pointed out by Sakharov~\cite{sakharov} long ago, a small baryon
asymmetry may have been produced in the early Universe if three
necessary conditions are satisfied: (i) baryon number violation,
(ii) violation of $C$ (charge conjugation symmetry) and $CP$ (the
composition of parity and charge conjugation), and (iii) departure
from thermal equilibrium. The standard model and its minimal
supersymmetric extensions satisfy all three Sakharov criteria for
producing a baryon excess~\cite{trodden}. As a result, over the
last decade electroweak baryogenesis has been a popular scenario
for the generation of the baryon asymmetry of the Universe and
there has been considerable interest in nonequilibrium dynamics of
the electroweak phase transition.

The last cosmological phase transition, during which the Universe
transformed from a quark-gluon plasma to a hadron gas, took place
at the QCD scale $T_\mathrm{QCD}\sim 160$ MeV when the Universe
was about few microseconds old . Because of the relatively low QCD
scale, the QCD phase transition perhaps is the \emph{only}
cosmological phase transition that can be (and is currently)
studied directly with terrestrial accelerators in
ultrarelativistic heavy ion experiments~\cite{meyer-ortmanns}.
Based on a first-order phase transition scenario, several
potential cosmological implications of the QCD phase transition
have been proposed, e.g., (i) baryon inhomogeneities that might
have affected nucleosynthesis, (ii) solar-mass scale primordial
black holes that could be part of the cold dark matter (CDM), and
(iii) strange quark nuggets that could also be a component of the
cold dark matter~\cite{boyanovsky:1}.

Most of the theoretical investigations on cosmological phase transitions are
based on the pictures of phase transition in equilibrium that utilize the
finite-temperature effective potential, a field-theoretical analogy of the
equilibrium free energy in statistical mechanics. Although the effective
potential is useful in determining equilibrium properties of the phase
transition, e.g., the order and critical temperature of the phase transition,
it is incapable of describing the full dynamics of the phase transition, which
is crucial to our understanding of phase transition in a cosmological setting
in which the system was evolving and therefore not in thermal equilibrium.
Whereas the importance of nonequilibrium aspects of cosmological phase
transition was recognized long ago~\cite{eboli}, the self-consistent
nonequilibrium description of the inflationary dynamics has only been developed
recently~\cite{boyanovsky:2} and field-theoretical descriptions of
nonequilibrium dynamics of the electroweak and QCD phase transitions are still
in their infancy.

\subsection*{Bose-Einstein condensation and ultrafast spectroscopy}
Recent advances in modern technology have made it possible to
explore properties of condensed matter systems under unusual
laboratory conditions such as at extremely low temperatures or on
unprecedentedly short time scales. Among many novel developments,
two of them have stimulated immense theoretical interest: (i) the
realization of the Bose-Einstein condensation
(BEC)~\cite{BEC,dalfovo} in dilute atomic gases in which the atoms
were confined in magnetic traps and cooled down to temperatures of
order fractions of microkelvins, and (ii) the observation of the
ultrafast spectroscopy in semiconductors in which the hot carriers
(electrons and holes) are excited by a femtosecond laser
pulse~\cite{book:shah,book:haug}. Remarkably, the common aspects
of these two phenomena that receive intense theoretical work but
still far from clear are those of the nonequilibrium dynamics.

The theoretical description of the Bose-Einstein condensation in
weakly interacting dilute gases has a long history~\cite{griffin}
and has accounted for many experimental results~\cite{dalfovo}.
However, a full microscopic description of the nonequilibrium
dynamics in Bose-Einstein condensates such as the condensate
formation process, the damping of collective excitations and the
relevant time scales, has not yet been fully developed. As pointed
out by Stoof~\cite{stoof}, the semiclassical Boltzmann equation is
unable to treat the buildup of coherence, which is crucial for the
phase transition to occur. It is therefore evident that such a
full microscopic description goes beyond Boltzmann kinetics and
calls for a deeper understanding of nonequilibrium aspects of
phase transitions in quantum multiparticle systems.

The theoretical analysis of the relaxation dynamics in optically-excited
semiconductors is usually based on the semiclassical kinetic equation of the
Boltzmann type. A wealth of recent experimental results on the ultrafast
femtosecond spectroscopy in bulk gallium arsenide (GaAs)~\cite{ultrafast}
demonstrate that for time intervals which are short compared to the optical
lattice oscillation period ($\simeq 115$ fs in GaAs), the relaxation dynamics
cannot be described by the semiclassical Boltzmann equation in terms of
completed, energy-conserving collisions. On such extremely short time scales,
time-energy uncertainty relation comes into play and virtual (off-shell)
processes that do not conserve energy will result in important quantum coherent
effects. Instead, quantum kinetics has to be used in order to account for the
quantum coherent nature of electronic states in the band (e.g., a well-defined
phase relation between electrons and holes), which is completely neglected
within the semiclassical Boltzmann description. Furthermore, a comprehensive
theoretical framework has to be able to describe the buildup of coherence, the
dynamics of quantum decoherence and the relevant time scales. It is
indisputable that the ultimate answer can be provided only by a self-consistent
nonequilibrium quantum dynamical approach.

This brief description of timely interdisciplinary physics
highlights the necessity for a thorough understanding of
nonequilibrium dynamics in quantum field theory.

\section{New Developments in this Thesis}\label{sec:1.2}
The study of nonequilibrium dynamics in quantum field theory plays an important
role in our understanding of a variety of nonequilibrium phenomena in
ultrarelativistic heavy ion collisions, cosmological phase transitions and
condensed matter systems. In the literature, however, the conventional
theoretical framework is largely based on finite-temperature thermal field
theory and the semiclassical Boltzmann description of kinetics. On the one
hand, finite-temperature thermal field theory can only study static quantities
(of a system in thermal equilibrium) such as the time-independent damping rates
and mean free paths, which at best can provide only estimates of nonequilibrium
properties near equilibrium but certainly not the full real-time dynamics. On
the other hand, whereas Boltzmann kinetics is capable of describing real-time
nonequilibrium dynamics, it relies crucially upon the validity of the
quasiparticle approximation (well-defined ``quasiparticles'' with a long
lifetime) and Fermi's golden rule (energy-conserving processes with
time-independent transition rates).

The main theme of this thesis is to provide, from first
principles, a theoretical study of nonequilibrium dynamics in
quantum field theory \emph{directly in real time} with the
emphasis of extracting potential experimental signatures. In
particular, the following new developments distinguish the work
presented in this thesis from the existing work in the literature.
\begin{enumerate}
\item \textbf{Initial value formulation and linear response}.
We have developed an initial value formulation in quantum field
theory that allows us to obtain fully renormalized, retarded and
causal equations of motion for nonequilibrium expectation values
of quantum fields and to study nonequilibrium quantum dynamics in
linear response theory directly in real time.

\item \textbf{Dynamical renormalization group}.
The renormalization group (RG) is a powerful tool to extract the
physics of a multiparticle system that is insensitive to system
details. In the literature the renormalization group method is
usually confined to the study of static and equilibrium
properties. We have developed a renormalization group method to
study real-time nonequilibrium quantum dynamics. We have
demonstrated explicitly that the dynamical evolution of a quantum
multiparticle system corresponds to a \emph{dynamical
renormalization group flow} in real time with equilibrium being a
fixed point of the flow, and that the equation of motion for the
mean field and the quantum kinetic equation are interpreted as the
\emph{dynamical renormalization group equation} which describes
the slow motion behavior of a nonequilibrium system.

\item \textbf{Quantum kinetics directly in real time}.
We have derived from first principles the quantum kinetic equation
for the (quasi)particle distribution function in scalar and
Abelian gauge field theories that transcends the semiclassical
Boltzmann equation---in the sense that medium effects, off-shell
(energy-nonconserving) processes and infrared threshold
divergences are included consistently.

\item \textbf{Resolution of pinch singularities}.
A technical but important issue of pinch singularities in
nonequilibrium quantum field theory is clarified and resolved
directly in real time from a quantum kinetic point of view. A
real-time kinetic analysis combined with the dynamical
renormalization group reveals that pinch singularities signal the
breaking down of perturbation theory in the long-time limit and
provides a consistent resummation scheme to extract the
nonperturbative long-time dynamics.

\item \textbf{Phenomenological application and experimental
prediction}. Most importantly, based on a real-time kinetic
approach to direct photon production from the QGP, we have shown
that emission of hard direct photons from a QGP created at RHIC
and LHC energies is significantly enhanced by energy-nonconserving
effects associated with the transient QGP lifetime. In contrast to
the exponential falloff predicted by the usual equilibrium
calculations the transverse momentum distribution of direct
photons at midrapidity falls off with a power law, providing a
distinct experimental nonequilibrium signature of the QGP
formation in ultrarelativistic heavy ion collisions.
\end{enumerate}

\section{Outline of the Thesis}\label{sec:1.3}
The thesis is organized as follows. Chapter~\ref{chap:2} is a
brief introduction to the theoretical framework and most of the
techniques that we shall be using in this thesis. This includes
short reviews of the Schwinger-Keldysh closed-time-path formalism
in nonequilibrium quantum field theory and the initial value
formulation in quantum field theory for studying, in particular,
the real-time relaxation of mean fields and quantum kinetics.

In Chapter~\ref{chap:3} we study the real-time relaxation of the
fermion mean field induced by an adiabatically switched-on
external source in a fermion-scalar plasma in terms of the initial
value formulation. The emphasis is on obtaining a fully
renormalized, retarded and causal initial value problem for the
fermion mean field that allows an unambiguous identification of a
novel damping mechanism and the corresponding damping rate.

In Chapter~\ref{chap:4} we presents a novel quantum kinetic
approach that goes beyond the usual semiclassical Boltzmann
kinetics by incorporating directly in real time perturbation
theory and the dynamical renormalization group resummation.
Quantum kinetic equation describing the dynamical evolution of
(quasi)particle distribution functions is derived in a
self-interaction scalar field theory and the $O(4)$ linear sigma
model and further solved in the linearized approximation near
thermal equilibrium. We compare the dynamical renormalization
group with the familiar renormalization group in Euclidean field
theory to highlight the equivalence between the two methods. The
issue of pinch singularities in nonequilibrium quantum field
theory is discussed and a real-time resolution is provided from
the viewpoint of quantum kinetics.

In Chapter~\ref{chap:5} we study the real-time nonequilibrium
dynamics in a quantum electrodynamics (QED) plasma at high
temperature in the relaxation time and hard thermal loop
approximations. The dynamical renormalization group approach to
real-time relaxation and quantum kinetics allows us to extract
anomalous nonequilibrium dynamics of photon and fermion mean
fields associated with infrared divergences in gauge field
theories at finite temperature.

This study is phenomenologically relevant to the production of
direct photons from the QGP, which is the main subject of
Chapter~\ref{chap:6}. There we present a real-time kinetic
description of direct photon production which reveals that
production of hard photons is significantly enhanced by the
transient lifetime of the QGP, hence providing a distinct
experimental signature of the QGP formation in ultrarelativistic
heavy ion collisions.

Finally, the main results of this thesis are summarized in
Chapter~\ref{chap:7}. A pedagogical introduction to the
renormalization group (RG) method in studying asymptotic analysis
of ordinary differential equations is presented in
Appendix~\ref{chap:app}.

We would like to provide a road map for the reader on how to read
this thesis. The reader who prefers skimming this thesis without
getting into the technical details is suggested to read the
abstract, the first chapter, the thesis summary chapter and the
introduction and conclusions of Chapters~\ref{chap:3},
\ref{chap:4}, \ref{chap:5} and \ref{chap:6}. Interested reader can
continue to read Chapter~\ref{chap:2} and the main text of
Chapters~\ref{chap:3}$-$\ref{chap:6} for all the technical
details.

To conclude this introductory chapter, we emphasize that whereas
this thesis focuses mainly on nonequilibrium phenomena related to
ultrarelativistic heavy ion collisions, the fundamental work on
real-time nonequilibrium dynamics in quantum field theory
presented in this thesis is truly interdisciplinary and can be
adapted to study a variety of nonequilibrium quantum phenomena in
the realm of condensed matter physics, quantum optics,
astrophysics and cosmology.

%% file: chap2.tex
\chapter{Nonequilibrium Quantum Field Theory}\label{chap:2}
In this chapter we introduce the Green's function formalism of
nonequilibrium quantum field theory and the initial value
formulation in quantum field theory. These two techniques are
closely related to each other and constitute the basic theoretical
framework which is used to study timely nonequilibrium problems in
this thesis.

The Green's function formalism to study nonequilibrium phenomena
in quantum field theory was originally initiated by
Schwinger~\cite{schwinger}, Bakshi and Mahanthappa~\cite{kt}, and
later developed further by Keldysh~\cite{keldysh} and many
others~\cite{book:KB,craig} in the 1960s. In the literature, this
formalism is usually referred to as the Schwinger-Keldysh or
closed-time-path (CTP) formalism~\cite{chou,rammer,hu}.

The essential difference between the usual vacuum quantum field
theory and nonequilibrium quantum field theory lies in the
physical quantity that one is interested in. In the former one is
interested in obtaining the cross section in a scattering
experiment between two beams of particles with well-defined
momenta. The initial state $|\textrm{in}\rangle$ is prepared at
remote past and the final state $|\textrm{out}\rangle$ is observed
at distant future. The cross section for a given process is
related to the transition amplitude ($S$-matrix element)
$\langle\textrm{out}|\textrm{in}\rangle$ between the above
asymptotically defined \textsl{in} and \textsl{out} states. Using
the Lehmann-Symanzik-Zimmermann (LSZ) reduction formula, one can
express the $S$-matrix element in terms of the Green's function
(vacuum expectation value of time-ordered product of field
operators), which in turn has a familiar path integral
representation and diagrammatic perturbative expansion.

In nonequilibrium quantum field theory, however, we are interested in the
expectation values of physical observables. Here, the expectation value is used
in the statistical sense and is obtained by a weighted sum with the density
matrix $\rho$. Most importantly, we are want to know the time evolution of
these expectation values, which will determine the real-time dynamics of the
quantum field theoretical system under consideration. The usual $S$-matrix
formalism designed to calculate transition amplitudes between the \textsl{in}
and \textsl{out} states simply cannot fulfill this purpose, as there is no
\textit{a priori} knowledge about the asymptotic \textsl{out} state at distance
future. Hence, a physically intuitive description of nonequilibrium phenomena
is that of the initial value formulation. The closed-time-path formalism of
quantum field theory is powerful theoretical framework to describe the
expectation values of physical observables \emph{directly in real time}, thus
providing a general tool for treating initial value problems of nonequilibrium
multiparticle dynamics in quantum field theory.

\section{Closed-Time-Path Formalism}\label{sec:2.1}
The most important quantity in statistical mechanics is the equilibrium density
matrix, which contains all the information of the physical system, likewise the
most important ingredient in nonequilibrium quantum field theory is the density
matrix $\rho$ specified at an initial time $t_0$. In the Heisenberg picture the
full time dependence is contained in the fields and the density matrix $\rho$
is not time dependent as it was in the Schr\"{o}dinger picture. The expectation
value of an operator with one time argument
$\langle\mathcal{O}_\mathrm{H}(t)\rangle$ is defined by
\begin{equation}
\langle\mathcal{O}_\mathrm{H}(t)\rangle\equiv\mathrm{Tr}
[\rho\,\mathcal{O}_\mathrm{H}(t)]/\mathrm{Tr}\rho.\label{ev}
\end{equation}
The questions we would like to ask are the following: how to calculate
$\langle\mathcal{O}_\mathrm{H}(t)\rangle$ and what is its time evolution?

\begin{figure}[t]
\begin{center}
\includegraphics[width=3.5truein,keepaspectratio=true]{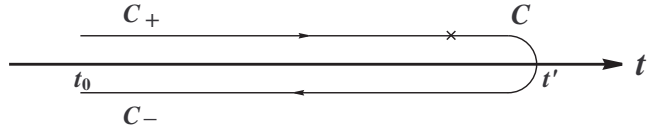}
\caption[The contour along the time axis used to evaluate the generating
functional for nonequilibrium Green's functions.]{The contour $\mathcal{C}$
along the time axis used to evaluate the generating functional for
nonequilibrium Green's functions. It consists of a forward branch
$\mathcal{C}_+$ running from $t_0$ to $t'$ and a backward branch
$\mathcal{C}_-$ from $t'$ back to $t_0$. The cross denotes insertion of an
operator.} \label{ctpcontour}
\end{center}
\end{figure}

To be specific, let us consider a system described by the time independent
Hamiltonian $H$ and the initial density matrix $\rho$. The time evolution of
$\mathcal{O}_\mathrm{H}(t)$ is determined by the Heisenberg equation of motion
\begin{equation}
i\frac{d}{dt}\mathcal{O}_\mathrm{H}(t)= [\mathcal{O}_\mathrm{H}(t),H],
\end{equation}
whose formal solution is
\begin{equation}
\mathcal{O}_\mathrm{H}(t)=U(t_0,t)\,\mathcal{O}\,U(t,t_0),\label{evol}
\end{equation}
where $\mathcal{O}$ is the operator at time $t_0$, which by definition is the
same as the time independent Schr\"{o}dinger operator and
$U(t,t_0)=e^{-iH(t-t_0)}$ is the time evolution operator in the Schr\"{o}dinger
picture. Substituting Eq.~(\ref{evol}) into Eq.~(\ref{ev}) and using the
property $U(t_0,t)U(t,t_0)=1$, we can rewrite the expectation value
$\langle\mathcal{O}_\mathrm{H}(t)\rangle$ as
\begin{equation}
\langle\mathcal{O}_\mathrm{H}(t)\rangle=\frac{\mathrm{Tr}
[\rho\,U(t_0,t)\,\mathcal{O}\,U(t,t_0)]}{\mathrm{Tr}[\rho\,U(t_0,t)\,U(t,t_0)]}.
\end{equation}
The numerator and denominator of the above expression have a simple
interpretation as the time evolution along a \emph{closed-time} contour. As
illustrated in Fig.~\ref{ctpcontour}, this closed contour starts from $t_0$ to
$t'$ along the forward branch $\mathcal{C}+$ and back to $t_0$ along the
backward branch $\mathcal{C}_-$, where any point on the forward branch is
understood at an earlier instant than any point on the backward branch. The
only difference between the numerator and the denominator is that in the
numerator the operator $\mathcal{O}$ is inserted at time $t$. The above
observation can be easily generalized to the expectation value of product of
operators with one time argument
$\langle\mathcal{O}_\mathrm{H}(t_1)\,\mathcal{O}_\mathrm{H}(t_2)
\cdots\mathcal{O}_\mathrm{H}(t_n)\rangle$, thus leading to multiple insertion
of operator $\mathcal{O}$ at times $t_n$, $\ldots$, $t_2$ and $t_1$. As usual,
the insertion of the operator $\mathcal{O}$ can be achieved by introducing
external source coupled to the operator in the time evolution operator,
constructing the generating functional and taking variational derivatives of
the generating functional with respective to the source. Note that, however,
unlike the usual situation where there is only time forward evolution, we now
have both forward and backward time evolution. Hence we need to introduce
\emph{two different sources} $J^+$ and $J^-$, respectively, for the forward and
backward time evolution.

For later convenience in the construction of the nonequilibrium Green's
functions, we will introduce sources $J^\pm$ coupled to the field operator
$\Phi$ in the time evolution operators.\footnote{Here we discuss the case for
(real) Bose fields, generalization to the case for Fermi fields is
straightforward.} The nonequilibrium generating functional is defined as
\begin{equation}
Z[J^+,J^-]\equiv\mathrm{Tr}[\rho\,U(t_0,t';J^-)\,U(t',t_0;J^+)],
\end{equation}
where $U(t,t';J)$ denotes the time evolution operator in the presence of the
external source $J$. Note that in general $J^+\ne J^-$, so that $Z[J^+,J^-]$
depends on two different sources. If these are set equal, one has
$Z[J,J]=\mathrm{Tr}\,\rho$, which is equal to unity after proper normalization,
being a statement of unitarity.

The functional $Z[J^+,J^-]$ can be represented as a path integral by imposing
boundary conditions in terms of complete sets of eigenstates of the
Schr\"{o}dinger field operator $\Phi$. Making use of the completeness of
eigenstates and the path integral representation for the time evolution
operators, one obtains the following path integral representation for
$Z[J^+,J^-]$:
\begin{eqnarray}
Z[J^+,J^-]&=&\int D\phi_1\,D\phi_2\,\langle\phi_1|\rho|\phi_2\rangle\int
D\varphi\int_{\phi_1}^\varphi\mathcal{D}\Phi^+
\int_{\phi_2}^\varphi\mathcal{D}\Phi^-\nonumber\\
&&\times\,\exp\left[i\int_{t_0}^{t'}dt\,d^3x
\Big(\mathcal{L}[\Phi^+]-\mathcal{L}[\Phi^-]+J^+ \Phi^+ - J^-
\Phi^-\Big)\right],\label{Z}
\end{eqnarray}
where $\Phi^+$ ($\Phi^-$) refers to field defined on the forward
$\mathcal{C}_+$ (backward $\mathcal{C}_-$) branch and $\mathcal{L}[\Phi]$ is
the Lagrangian density. We note that in above functional expression $\Phi^+$
and $\Phi^-$ are not independent variables, but are linked through the boundary
conditions at time $t'$ in the future. To circumvent this difficulty, we will
take $t'\rightarrow\infty$ for all practical purpose and treat $\Phi^+$ and
$\Phi^-$ as independent variables.

The generating functional $Z[J^+,J^-]$ can be written in a compact
\emph{path-ordered} form
\begin{eqnarray}
Z[J_c]&=&\int
D\phi_1\,D\phi_2\,\langle\phi_1|\rho|\phi_2\rangle\int_{\phi_1}^{\phi_2}
\mathcal{D}\Phi_c\,\exp\left[i\int_\mathcal{C}dt\,
d^3x\Big(\mathcal{L}[\Phi_c]+J_c\,\Phi_c\Big)\right],\label{Zc}
\end{eqnarray}
where
\begin{equation}
\Phi_c=\left\{
\begin{array}{l}
\Phi^+\quad \mathrm{for}\quad t\in \mathcal{C}_+\\
\Phi^-\quad \mathrm{for}\quad t\in \mathcal{C}_-
\end{array},
\right. \quad\quad
J_c=\left\{
\begin{array}{l}
J^+\quad \mathrm{for}\quad t\in \mathcal{C}_+\\
J^-\quad \mathrm{for}\quad t\in \mathcal{C}_-
\end{array},
\right.
\end{equation}
and
\begin{equation}
\int_\mathcal{C}dt=\int^\infty_{t_0\;\mathcal{C}_+}
dt-\int^\infty_{t_0\;\mathcal{C}_-}dt
\end{equation}
accounting for the opposite direction of integration along the backward time
branch.

The above results are very general and are valid for any initial density matrix
$\rho$. In fact, as can be seen from Eqs.~(\ref{Z}) and (\ref{Zc}), the only
effect of the initial density matrix $\rho$ is to specify the boundary
conditions for the path integral at $t_0$ through the matrix element
$\langle\phi_1|\rho|\phi_2\rangle$, thus in turn to impose the boundary
conditions on the nonequilibrium Green's functions. This means that the
equations of motion for the Green's functions at $t>t_0$ are not influenced by
the presence of the initial density matrix.

We note that it is always possible to express the matrix element of the initial
density matrix as an exponential of a polynomial in the
fields~\cite{hu,cooper}:
\begin{equation}
\langle\phi_1|\rho|\phi_2\rangle=e^{-K[\phi_1,\phi_2]},
\end{equation}
where
\begin{equation}
K[\phi_1,\phi_2]=K+\int d^3x\, K_a(\mathbf{x})\phi_a(\mathbf{x})+\int
d^3x\,d^3y\, \phi_a(\mathbf{x})K_{ab}(\mathbf{x},\mathbf{y})
\phi_b(\mathbf{y})+\cdots,
\end{equation}
with $a,b=1,2$. In the above expression the constant $K$ can be absorbed into
the normalization and the various coefficient functions $K_a(\mathbf{x})$,
$K_{ab}(\mathbf{x},\mathbf{y})$, etc., have the interpretations of initial
conditions on the one-point (mean field), two-point correlation functions,
etc., which contain all the information of the initial density matrix $\rho$.
For an initial density matrix which is diagonal in the basis of the
(quasi)particle number operators, it can be shown that all the initial
correlation functions higher than two point vanish.

\subsection*{Perturbation theory and Feynman rules}
The most convenient feature of the CTP formalism of nonequilibrium quantum
field theory is that it is formally analogous to standard quantum field theory,
except for the fact that the fields have contributions from both time branches.
In particular, as in usual field theory, one obtains from variations of the
generating functional $Z[J^+,J^-]$ (or, equivalently, $Z[J_c]$) the
nonequilibrium Green functions, which, however, now include all correlations
between points on either forward and backward time branches. The path-ordered
$n$-particle ($2n$-point) Green's function is defined by
\begin{equation}
G_c(x_1\ldots x_n,x'_1\ldots x'_n)=i^{\,n}\,\langle
\mathcal{T}_c\Phi(x_1)\ldots\Phi(x_n)\Phi(x'_n)\ldots\Phi(x'_1)\rangle,
\end{equation}
with \begin{equation} \langle
\mathcal{T}_c\Phi(x_1)\ldots\Phi(x_n)\Phi(x'_n)\ldots\Phi(x'_1)
\rangle=(-i)^{2n}\prod_{i=1}^n\left. \frac{\delta}{\delta
J_c(x_i)}\frac{\delta}{\delta J_c(x'_i)}\ln
Z[J_c]\right|_{J_c=0},\nonumber
\end{equation}
where $\mathcal{T}_c$ is the contour ordering operator along the
closed contour $\mathcal{C}$. Although for physical observables
the time arguments are on the forward branch, both forward and
backward branches will come into play at intermediate steps in a
self-consistent calculation.

For instance, there are four types of single-particle (two-point) Green's
functions (i.e., propagators) to be given by (here we have suppressed the space
arguments for notational simplicity)
\begin{eqnarray}
G^{++}(t,t')&=&\pm\,i\,\langle \mathcal{T}\Phi(t)\bar{\Phi}(t')\rangle\nonumber\\
&=&G^>(t,t')\theta(t-t')+G^<(t,t')\theta(t'-t),\nonumber\\
G^{--}(t,t')&=&\pm\,i\,\langle\overline{\mathcal{T}}\Phi(t)\bar{\Phi}(t')\rangle\nonumber\\
&=&G^>(t,t')\theta(t'-t)+G^<(t,t')\theta(t-t'),\nonumber\\
G^{-+}(t,t')&=&\pm\,i\,\langle\Phi(t)\bar{\Phi}(t')\rangle\;=\;G^>(t,t'),\nonumber\\
G^{+-}(t,t')&=&
i\,\langle\bar{\Phi}(t')\Phi(t)\rangle\;=\;G^<(t,t'),\label{greenfunctions}
\end{eqnarray}
where $\mathcal{T}$ ($\overline{\mathcal{T}}$) is the
(anti)time-ordering operator, the upper (lower) sign refers to
Bose (Fermi) fields, $\theta(t-t')$ is the Heaviside step function
and $\bar{\Phi}\equiv\Phi^\dagger$ for bosons but
$\bar{\Psi}\equiv\Psi^\dagger\gamma^0$ for Dirac fermions. These
nonequilibrium Green's functions are not completely independent of
each other, instead they are related by the relation
\begin{equation}
G^{++}(t,t')+G^{--}(t,t')-G^{-+}(t,t')-G^{+-}(t,t')=0,
\end{equation}
where use has been made of the identity
$\theta(t-t')+\theta(t'-t)=1$. Furthermore, we can define the
retarded and advanced Green's functions, respectively, by
\begin{eqnarray}
G_R(t,t')&=&G^{++}(t,t')-G^{+-}(t,t')\nonumber\\
&=&[G^>(t,t')-G^<(t,t')]\,\theta(t-t'),\nonumber\\
G_A(t,t')&=&G^{++}(t,t')-G^{-+}(t,t')\nonumber\\
&=&[G^<(t,t')-G^>(t,t')]\,\theta(t'-t).\label{G:retadv}
\end{eqnarray}
We note that the Wightman functions $G^\gtrless(t,t')$ determine
uniquely the nonequilibrium single-particle Green's functions,
thus playing important roles in nonequilibrium quantum field
theory. In equilibrium $G(x,x')=G(x-x')$ and the space-time
Fourier transform of the Green's functions have specific analytic
properties that allow to extract dispersion relations that are
important to our later discussion.

One can then construct a diagrammatic perturbative expansion of the Green's
functions much in the same manner as in usual field theory by decomposing the
full Lagrangian into the free field (noninteracting) part $\mathcal{L}_0$ and
the interaction part $\mathcal{L}_\mathrm{int}$ and treating the latter as
perturbation. The doubling of the sources, fields and integration contours in
the nonequilibrium generating functional results in an effective nonequilibrium
CTP Lagrangian density [see Eq.~(\ref{Z})]
\begin{equation}
\mathcal{{L}}_{\mathrm{CTP}}[\Phi^+,\Phi^-]=
\mathcal{{L}}[\Phi^+]-\mathcal{L}[\Phi^-]~,\label{CTPL}
\end{equation}
which in turn leads to the following nonequilibrium Feynman rules in
perturbation theory:
\begin{enumerate}
\item There are two types of interaction vertices, corresponding to fields
defined on the forward and backward time branch. The vertices associated with
fields on the forward branch are the usual interaction vertices, while those
associated with fields on the backward branch have the opposite sign.
\item There are four kinds of free field single-particle Green's functions
(propagators), corresponding to the possible contractions of fields among the
two branches [see Eq.~(\ref{greenfunctions})]. Besides the usual time-ordered
propagators which are associated with correlations of fields on the forward
branch, there are antitime-ordered propagators associated with correlation of
fields on the backward branch as well as the unordered Wightman functions
associated with correlation of fields on different branches.
\item The combinatoric factors, rules for loop (time and momentum) integrals, etc.,
remain the same as in usual field theory.
\end{enumerate}

In this thesis we study two different types of nonequilibrium phenomena in
quantum field theoretical systems: the relaxation of mean fields in the
linearized approximation (i.e., linear response theory) and the quantum kinetic
equation that describes the time evolution of the single-(quasi)particle
distribution function, hence we will specify below in Sec.~\ref{sec:2.2} the
initial density matrix $\rho$ used in each case.

\section{Initial Value Problems in Quantum Field Theory}\label{sec:2.2}
The initial value formulation provides a convenient description
for studying the time evolution of dynamical systems, ranging from
Hamilton's equations, kinetic equations and hydrodynamical
equations in classical physics to the time-dependent
Schr\"{o}dinger equation in nonrelativistic quantum mechanics. The
fundamental ingredients of initial value problems are the
\emph{equations of motion}, which govern the dynamical evolution,
and the \emph{initial conditions}, which contain all the dynamics
prior to some initial time $t_0$ from which on the evolution of
the state is followed. The advantage of the initial value
formulation is that the dynamics of the system for times $t>t_0$
will be determined once the equations of motions and initial
condition are prescribed.

When one studies relativistic quantum field theory, however, one
generally does not use the initial value formulation, instead a
quite different approach to relativistic quantum field theory
based on $S$-matrix elements was developed. The main reason for
this is that the majority of problems to which the theory has been
applied do not require knowledge of the detailed time evolution of
the system. Indeed, as mentioned in the beginning of this chapter,
the typical quantity of experimental interest is the cross
section, which is most easily calculated in terms of the
$S$-matrix elements. Thus, an initial value formulation in
relativistic quantum field seems superfluous.

Nevertheless, as emphasized in the Introduction, recent
development in cosmology, ultrarelativistic heavy ion collisions
and condensed matter physics reveals that there are a wide variety
of physically interesting questions which require not only a
quantum field-theoretical description but also a thorough
understanding of the time evolution of these systems. Clearly, one
must depart from the usual description in terms of the
\emph{time-independent} $S$-matrix element and treat the dynamics
with the full time evolution.

The rest of this chapter is devoted to an introduction to two important initial
value problems for studying nonequilibrium dynamics in quantum field theory:
linear relaxation of the mean fields and quantum kinetic theory.

\subsection{Linear relaxation of mean fields}
The mean fields under consideration are the expectation value of
field operators in the nonequilibrium state induced by the
external time-dependent perturbations. Our strategy to study the
irreversible relaxation (damping) of these mean fields as
\emph{initial value problems} is to prepare these mean fields for
a system via adiabatic switching-on of an external perturbation
from the remote past. Once the external perturbations are switched
off at time $t=0$, the induced mean fields must relax towards
equilibrium and we aim to study this nonequilibrium dynamics in
linear response theory~\cite{book:FW,book:kapusta,book:lebellac}
directly in real time.

To illustrate all the relevant physics and avoid the complexity associated with
the issue of thermalization, we assume the system at $t_0=-\infty$ is described
by the free (quasi)particle Hamiltonian $H_0$ and in thermodynamical
equilibrium at a temperature $T$. We further assume that the interaction
Hamiltonian $H_\mathrm{int}$ is adiabatically turned on while we prepare the
mean fields via adiabatic switching-on the external perturbation. This allows
the induced mean fields to be ``dressed'' adiabatically by the interaction and
external perturbation during the preparation time. Hence in this case the
initial density matrix $\rho$ at time $t_0=-\infty$ has the form
$\rho=\exp(-\beta H_0)/\mathrm{Tr}[\exp(-\beta H_0)]$, where $\beta=1/T$ is the
inverse temperature.

The initial value formulation for studying the linear relaxation
(damping) of the mean fields begins by introducing \emph{c-number}
external sources $J(x)$ coupled to generic (Bose or Fermi) quantum
fields $\Phi(x)$.\footnote{The reader should not confuse the
external source $J$ with the source $J^\pm$ introduced previously
in the construction of the nonequilibrium generating functional
$Z[J^+,J^-]$. The external sources $J$ introduced here, just like
those introduced in usual linear response theory, are external
perturbations to the system, hence it is set to be equal on both
forward and backward time branches.} In the presence of external
perturbation the effective CTP Lagrangian density becomes
\begin{equation}
\mathcal{L}_\mathrm{CTP}[\Phi^+(x),\Phi^-(x)]=\mathcal{L}[\Phi^+(x)]
+\left[J(x)\,\Phi^+(x)+\mathrm{H.c.}\right]-(\Phi^+\to\Phi^-).
\end{equation}
The presence of external sources will induce responses of the
system. The expectation value of $\Phi(x)$ induced by $J(x)$ in a
linear response analysis is given by\footnote{We assume here that
in the absence of the external source $J(x)$, the mean field
$\phi(x)$ vanishes identically in thermal equilibrium. This is
true for fermion and gauge mean fields and for scalar mean filed
in the unbroken symmetry phase.}
\begin{eqnarray}
\phi({\bf x},t)&\equiv&
\langle\Phi^\pm({\bf x},t)\rangle_J\nonumber\\
&=&\pm\int_{t_0}^{\infty}dt'\int d^3x'\,G_R({\mathbf x},t;{\mathbf x'},t') \,
J({\mathbf x'},t'),\label{linearreospose}
\end{eqnarray}
where the upper (lower) sign in the second equality refers to Bose (Fermi)
fields and $G_R(\mathbf{x},t;\mathbf{x}',t')$ is the retarded Green's function
defined in Eq.~(\ref{G:retadv}):
\begin{eqnarray}
G_R(\mathbf{x},t;\mathbf{x}',t')&=&\pm\,i\left[\left\langle
\Phi^\pm(\mathbf{x},t)\bar{\Phi}^+(\mathbf{x}',t')\right\rangle
-\left\langle\Phi^\pm(\mathbf{x},t)\bar{\Phi}^-(\mathbf{x}',t')
\right\rangle\right]\nonumber\\
&=&\pm\,i\left\langle\left[\Phi(\mathbf{x},t),\bar{\Phi}(\mathbf{x'},t')
\right]_\mp\right\rangle\,\theta(t-t')\nonumber\\
&=&[G^>(\mathbf{x},t;\mathbf{x}',t')-G^<(\mathbf{x},t;\mathbf{x}',t')]\,\theta(t-t').
\label{Gret}
\end{eqnarray}
In the above, $[\,\cdot\,,\,\cdot\,]_\mp$ denotes commutator ($-$)
for bosons and anticommutator ($+$) for fermions and, as before,
$\bar{\Phi}\equiv\Phi^\dagger$ for bosons but
$\bar{\Psi}\equiv\Psi^\dagger\gamma^0$ for Dirac fermions. We
emphasize that in Eq.~(\ref{Gret}) the expectation values are
calculated in the CTP formalism with full Lagrangian density
$\mathcal{L}$, but \emph{in the absence of external source} as
befits the \emph{linear response} approach.

As explained above, a practically useful initial value formulation
for the real-time relaxation of the mean field is obtained by
considering that the external source is adiabatically switched on
in time from $t_0=-\infty$ and suddenly switched off at $t=0$,
i.e.,
\begin{equation}
J({\mathbf x},t)=J({\mathbf x})\,e^{\varepsilon t}\,\theta(-t),
\quad\varepsilon\rightarrow 0^+. \label{extsource}
\end{equation}
For $t>0$, after the external perturbation has been switched off, the mean
field will relax towards its equilibrium value and our aim is to study this
relaxation directly in real time.

For the study of linear relaxation of the mean fields, we assume that the
system is in thermal equilibrium at a temperature $T$ at time $t=-\infty$ with
$\phi(\mathbf{x},t=-\infty)=0$. The retarded and equilibrium (i.e.,
translational invariant in time) nature of $G_R(\mathbf{x},t;\mathbf{x'},t')$
and the adiabatic switching on of $J({\mathbf x},t)$ entail that
\begin{equation}
\phi(\mathbf{x},t=0)=\phi_0(\mathbf{x}),\quad\dot\phi(\mathbf{x},t<0)=0,
\label{icond}
\end{equation}
where $\phi_0(\mathbf{x})$ is determined by $J(\mathbf{x})$ [and vice versa,
$\psi_0(\mathbf{x})$ can be used to fix $J(\mathbf{x})$] and, here and
henceforth, an overdot denotes time derivative. We note that
$\dot\phi(\mathbf{x},t=0)$ is unspecified even though
$\dot\phi(\mathbf{x},t<0)=0$. This is because the external source is
instantaneous switched off at $t=0$. In fact there could be initial time
singularities associated with our choice of the external source given in
Eq.~(\ref{extsource}). It has been shown in Ref.~\cite{baacke} that the
long-time behavior is insensitive to the initial time singularities, hence we
will not address this issue here as it is beyond the scope of this thesis.

In order to study the relaxation of the mean field $\phi(\mathbf{x},t)$ for
$T>0$, we need to relate the linear response problem to an initial value
problem for the equation of motion of the mean field. This can be achieved by
considering the following \emph{formal} equation of motion for the mean field
\begin{equation}
D\circ\phi(\mathbf{x},t)=\pm\,J(\mathbf{x},t),\label{EOM}
\end{equation}
where the upper (lower) sign refers to Bose (Fermi) fields and the
(integro-differential) operator $D$ is the inverse of the retarded
Green's function $G_R$. The source $J(x)$ is given by
Eq.~(\ref{extsource}) and $\phi(x)$ satisfies the condition
Eq.~(\ref{icond}) at $t=0$. For $t<0$, on the one hand,
Eq.~(\ref{EOM}) is exactly the linear response problem
Eq.~(\ref{linearreospose}) satisfying the condition
Eq.~(\ref{icond}) at $t=0$; for $t>0$, on the other hand, it
describes the time evolution of the mean field
$\phi(\mathbf{x},t)$ with the initial condition specified by
Eq.~(\ref{icond}).

It is at this stage where the nonequilibrium CTP formalism provides the most
powerful framework. The \emph{real-time} equation of motion for the mean field
$\phi(x)$ can be obtained via the tadpole method~\cite{tadpole}, which
automatically leads to a retarded and causal initial value problem for the mean
field $\phi(x)$.

The central idea of the tadpole method is to write the field into the
$c$-number expectation value plus quantum fluctuations around it, i.e., writing
\begin{equation}
\Phi^\pm({\mathbf x},t)=\phi({\mathbf x},t)+\chi^\pm({\mathbf x},t),\quad
\mbox{with}\quad\langle\chi^\pm({\mathbf x},t)\rangle=0.\label{tadpolecond}
\end{equation}
The equation of motion for the mean field $\phi({\mathbf x},t)$ can be obtained
to any order in perturbation theory by requiring the tadpole condition
$\langle\chi^\pm({\mathbf x},t)\rangle=0$ to all orders in perturbation
theory~\cite{tadpole}.

\subsection{Quantum kinetic theory}
The foundation of modern kinetic theory dates back to 1872 when
Boltzmann published his famous kinetic equation for dilute
gases~\cite{boltzmann}. The Boltzmann equation is a nonlinear
integro-differential equation for the single-particle distribution
function, a microscopic quantity introduced by Boltzmann that
describes the density of gas molecules in phase space. The success
of Boltzmann's kinetic description for dilute gasses not only
provides the first theoretical understanding of \emph{macroscopic}
physics in terms of the underlying \emph{microscopic} dynamics,
but also makes kinetic theory one of the most powerful approaches
to transport phenomena in macroscopic systems.

A necessary criterion for the validity of Boltzmann's kinetic
description is that there must be a wide separation of \emph{time
scales} in the system. A classic example is furnished by the
derivation of the classical Boltzmann equation: Boltzmann
introduced, with great intuition, ``molecular chaos'' (assumption
that particles in a dilute gas are not correlated), which
crucially depends upon the fact the duration of each individual
collision process is much shorter than the average time that
elapse between two successive collisions. Based on Boltzmann's
classical picture of completed collisions, one can easily write
down the semiclassical Boltzmann equation, also known as the
Uehling-Uhlenbeck equation~\cite{UU}, by modifying the
\emph{collision term} using Fermi's golden rule (which gives the
quantum mechanical transition probability per unit time) and
taking into account the quantum statistics of the particles
(bosons or fermions). Whereas the semiclassical Boltzmann equation
is appealing in describing a variety of transport problems,
Boltzmann kinetics is an approximation only and needs to be
justified from first principles, especially for quantum systems
under extreme nonequilibrium conditions which have been a subject
of great theoretical and experimental interest over the last
decade.

Although a wide separation of time scales is a necessary condition
for the validity of the (semi)\-classical Boltzmann kinetics, it
is far from sufficient. In particular when there is competition of
relevant time scales or quantum transient kinetics at early stages
is physically important, a full quantum kinetic equation, of which
usual Boltzmann kinetics is a semiclassical limit, must be
obtained. The most widely used method in the derivation of the
quantum kinetic equation is that of nonequilibrium Green's
functions originally developed by Kadanoff and
Baym~\cite{book:KB}. This method involves Wigner transforms of
two-point Green's functions and eventually a gradient
expansion~\cite{blaizot:be,geiger:qcd}. Such gradient expansion
assumes that the center-of-mass Wigner variables are ``slowly
varying'', but it is seldom clear at this level which are the fast
and which are the slow scales involved. A ``coarse-graining''
procedure is typically invoked that averages out microscopic
scales in the kinetic description, leading to irreversible
evolution in the resulting kinetic equation. Such an averaging
procedure is usually poorly understood and justified \textit{a
posteriori}.

In this thesis, we propose to study quantum kinetics by focusing on a
systematic, first-principle derivation of quantum kinetic equations from the
underlying quantum field theory. Specifically, we aim to provide a kinetic
description of an initially prepared nonequilibrium quantum system by studying
the time evolution of the single-particle distribution functions for the
relevant degrees of freedom directly in real time.

This novel approach begins by defining a suitable (quasi)particle number
operator whose expectation value $n_\mathbf{k}(t)$ is interpreted as the single
particle distribution function for the relevant degrees of freedom of momentum
$\mathbf{k}$ in the presence of the medium, and by separating the total
Hamiltonian into the free part that commutates with the (quasi)particle number
operator and a interaction part that describes interaction between these
(quasi)particles. As a consequence of the medium effects, the above defined
(quasi)particles are in general different from the corresponding free particles
in the vacuum. In the Heisenberg picture, the time evolution of the
distribution function $dn_\mathbf{k}(t)/dt$ is obtained by using the Heisenberg
equation of motion. Choosing the initial density matrix $\rho$ at a
\emph{finite} initial time $t=t_0$ to be diagonal in the basis of the
(quasi)particles number operators but with nonequilibrium distributions
$n_\mathbf{k}(t_0)$, one can write down a perturbative expansion for
$d\,n_\mathbf{k}(t)/dt$ as an initial value problem in terms of the
\emph{initial} distribution function. The solution of this initial value
problem contains secular terms that grow with time and invalidate the
perturbative expansion at large times as a consequence of the fact that
perturbation theory neglects the change in the distribution function as time
goes on.

We introduce a novel \emph{dynamical renormalization group} that
allows a consistent resummation of secular terms in real time and
leads to an improvement of the perturbative expansion that is
valid on kinetic time scales. This real-time initial value
formulation has the advantage that it displays the relevant
microscopic and kinetic time scales, leads to a well-defined
''coarse-graining'' procedure, allows a systematic improvement on
the kinetic description and, if necessary, consistently includes
nonexponential relaxation in a physically intuitive manner.

To conclude this chapter, we emphasize that the main theme of this thesis is
describing the dynamical development of a nonequilibrium multiparticle system,
which evolves from an initially prepared quantum state. Since the
nonequilibrium system is characterized by its dynamical evolution in time, a
physically intuitive description of nonequilibrium dynamics is that in terms of
the initial value formulation. In this thesis we propose to study real-time
relaxation of mean fields and quantum kinetics in quantum multiparticle systems
as initial value problems. The CTP formalism of nonequilibrium quantum field
theory is a powerful theoretical framework designed to describe \emph{directly
in real time} the expectation values of physical observables in the presence of
medium, thus providing a general tool for treating initial value problems of
nonequilibrium multiparticle dynamics in quantum field theory.

%% file: chap3.tex
\chapter{Fermion Damping in a Fermion-Scalar Plasma}\label{chap:3}
\section{Introduction}\label{sec:3.1}
The propagation of quarks and leptons in a medium of high temperature and/or
density is of fundamental importance in a wide variety of physically relevant
situations. In stellar astrophysics, electrons and neutrinos play a major role
in the evolution of dense stars such as white dwarfs, neutron stars and
supernovae~\cite{raffelt}. The propagation of quarks during the nonequilibrium
stages of the electroweak phase transition is conjectured to be an essential
ingredient for baryogenesis at the electroweak scale both in non-supersymmetric
and supersymmetric extensions of the standard model~\cite{trodden}.
Furthermore, medium effects can enhance neutrino oscillations as envisaged in
the Mikheyev-Smirnov-Wolfenstein (MSW) effect~\cite{MSW} and dramatically
modify the neutrino electromagnetic couplings~\cite{adams}.

In-medium propagation of particles is dramatically different from that in the
vacuum. The medium modifies the dispersion relation of the excitations and
introduces a width to the propagating
excitation~\cite{weldon:1,weldon:2,book:kapusta,book:lebellac} that results in
damping of the amplitude of the propagating mode. Whereas the propagation of
quarks and leptons in a QED or QCD plasma has been studied
thoroughly~\cite{lebedev,book:lebellac}, a similar study for a scalar plasma
has not been carried out to the same level of detail. Recently some attention
has been given to understanding the thermalization time scales of boson and
fermion excitations in a plasma of gauge~\cite{vilja:1} and scalar
bosons~\cite{vilja:2}. It has also been shown that fermion thermalization is an
important ingredient in models of baryogenesis mediated by
scalars~\cite{krishna}. Most of the studies of fermion thermalization focus on
the mechanism of fermion scattering off the gauge quanta in the heat bath
(Landau damping). Although the scalar contribution to the fermion self-energy
to one-loop order has been obtained a long time ago~\cite{weldon:2}, scant
attention has been paid to a more detailed understanding of the contribution
from the scalar degrees of freedom to the fermion relaxation and
thermalization. As mentioned above, this issue becomes of pressing importance
in models of baryogenesis and more so in models in which the scalars carry
baryon number~\cite{krishna}.

In this chapter we focus on several aspects of propagation of fermion
excitations in a fermion-scalar plasma. Specifically, we offer a detailed and
general study of fermion relaxation and thermalization through the interactions
with the scalars in the plasma directly in real time for arbitrary scalar and
fermion masses, temperature and fermion momentum. More importantly, we focus on
a novel mechanism of damping for fermion excitations that occurs whenever the
effective mass of the scalar particle allows its kinematic \emph{decay} into
fermion pairs. This phenomenon only occurs in the medium and is interpreted as
an induced damping due to the presence of scalars in the medium. It is a
process different from collisional damping and Landau damping which are the
most common processes that lead to relaxation and thermalization. This process
results in new thermal cuts in the fermion self-energy and, for heavy scalars,
a quasiparticle pole structure in the fermion propagator that provides a finite
width to the fermion excitation. The remarkable and perhaps nonintuitive aspect
of our analysis is that the decay of the scalar in the medium leads to
\emph{damping} of the fermion excitations and their propagation as
quasiparticle resonances.

The effective real-time Dirac equation in the medium allows a direct
interpretation of the damping of the fermion excitation and leads to a clear
definition of the damping rate. By analyzing the quasiparticle wave functions
we obtain an \emph{all-order} expression for the damping rate that confirms and
generalizes recent results for the massless chiral case~\cite{dolivo:1,nieves}.

In order to provide a complementary understanding of the process of induced
decay of the heavy scalars in the medium and the resulting fermion damping, we
study relaxation of the fermion distribution function by using a semiclassical
Boltzmann equation. Linearizing the Boltzmann equation near the equilibrium
distribution, we obtain the relation between the thermalization rate for the
distribution function in the relaxation time approximation and the damping rate
for the amplitude of the fermion mean fields to lowest order in the Yukawa
coupling. This analysis provides a real-time confirmation of the oft quoted
relation between the interaction rate (obtained from the Boltzmann kinetic
equation in the relaxation time approximation) and the damping rate for the
mean field~\cite{weldon:2,book:lebellac}. More importantly, this analysis
reveals directly, via a kinetic approach in real time how the process of
induced decay of a heavy scalar in the medium results in damping and
thermalization of the fermion excitations. A study of the relation between the
interaction rate and the damping rate has been presented recently for gauge
theories within the context of the imaginary time formulation~\cite{ayala}. Our
results provide a real-time confirmation for the scalar case.

This chapter is organized as follows. In Sec.~\ref{sec:3.2} we obtain the
renormalized effective in-medium Dirac equation for the fermion mean field in
real time starting from the linear response to an external (Grassmann) source
that induces the mean field. The renormalization aspects are addressed in
detail. We then study in detail the structure of the renormalized fermion
self-energy and establish the presence of new cuts of thermal origin. In
Sec.~\ref{sec:3.3} we present a real-time analysis of the evolution of the
fermion mean field and clarify the difference between complex poles and
resonances (often misunderstood). An analysis of the structure of the
self-energy and an interpretation of the exact quasiparticle spinor wave
functions allows us to provide an all-order expression for the damping rate of
the fermion mean fields. In Sec.~\ref{sec:3.4} we present an analysis of the
evolution of the fermion distribution function in real time by obtaining a
(Boltzmann) kinetic equation in the relaxation time approximation. We clarify
to lowest order the relation between the damping rate of the fermion
excitations and the interaction rate of the distribution function. In
Sec.~\ref{sec:3.5} we summarize our results.

\section{Effective Dirac Equation in the Medium}\label{sec:3.2}
As mentioned in the Introduction, whereas the damping of collective and
quasiparticle excitations via the interactions with gauge bosons in the medium
has been the focus of most attention, understanding of the influence of scalars
has not been pursued so vigorously.

Although we are ultimately interested in studying the damping of fermion
excitations in a plasma with scalars and gauge fields within the realm of
electroweak baryogenesis in either the Standard Model or generalizations
thereof, we will begin by considering only the coupling of a massive Dirac
fermion to a scalar via a simple Yukawa interaction. The model dependent
generalizations of the Yukawa couplings to particular cases will differ
quantitatively in the details of the group structure but the qualitative
features of the effective Dirac equation in the medium as well as the
kinematics of the thermal cuts that lead to damping of the fermion excitations
will be rather general.

We consider a Dirac fermion $\Psi$ with the bare mass $M_0$ coupled to a scalar
$\phi$ with the bare mass $m_0$ via a bare Yukawa coupling $y_0$. The bare
fermion mass could be the result of spontaneous symmetry breaking in the scalar
sector, but for the purposes of our studies we need not specify its origin. The
Lagrangian density is given by
\begin{equation}
\mathcal{L}[\Psi,\phi]=\bar{\Psi}(i{\not\!{\partial}}-M_0)\Psi + \frac{1}{2}
\partial_\mu\phi\,\partial^\mu\phi-\frac{1}{2} m_0^2 \phi^2 - {\cal
L}_I[\phi]-y_0\bar{\Psi}\phi\Psi+\bar{\eta}\Psi+\bar{\Psi}\eta+j\phi.
\label{yukalan}
\end{equation}
The self-interaction of the scalar field accounted for by the term ${\cal
L}_I[\phi]$ need not be specified to lowest order. In the above expression,
$\eta$ and $j$ are the respective external fermionic and scalar sources that
are introduced in order to provide an initial value problem for fermion
relaxation. We now write the bare fields and sources in terms of the
renormalized quantities (referred to with a subscript $r$) by introducing the
renormalization constants and counterterms:
\begin{eqnarray}
&&\Psi=Z_\psi^{1/2}\,\Psi_r,\quad\phi=Z_\phi^{1/2}\,\phi_r,\quad
\eta=Z_{\psi}^{-1/2}\eta_r,\quad j=Z_\phi^{-1/2}j_r,\nonumber\\
&& y=y_0 Z_\phi^{1/2} Z_\psi / Z_y,\quad m^2_0=\left(\delta_m
+m^2\right)/Z_{\phi},\quad M_0=\left(\delta_M +M\right)/Z_{\psi}.\label{wfren}
\end{eqnarray}
With the above definitions, the Lagrangian can be rewritten as (here and
henceforth, we have suppressed the subscript $r$ for notational simplicity)
\begin{eqnarray}
{\cal{L}}&=&\bar{\Psi}(i{\not\!{\partial}}-M)\Psi+\frac{1}{2}
\partial_\mu\phi\,\partial^\mu\phi-\frac{1}{2}m^2
\phi^2 -\mathcal{L}_I[\phi]-y \bar{\Psi}\phi\Psi +\bar{\eta}\Psi+
\bar{\Psi}\eta+j\phi\nonumber\\
&&+\,\frac{1}{2} \delta_\phi \partial_\mu\phi\,\partial^\mu\phi -\frac{1}{2}
\delta_m\phi^2+\bar{\Psi}(i\delta_\psi{\not\!{\partial}}-\delta_M)\Psi-
y\delta_y\bar{\Psi}\phi\Psi+\delta\mathcal{L}^c_{I}[\phi],\label{renyukalan}
\end{eqnarray}
where $m$ and $M$ are the renormalized masses, and $y$ is the renormalized
Yukawa coupling. The terms with the coefficients
$$
\delta_\psi=Z_\psi-1,\quad\delta_\phi=Z_\phi-1,\quad \delta_M=Z_\psi
M_0-M,\quad\delta_m=Z_\phi\,m^2_0 -m^2,\quad\delta_y=Z_y-1,
$$
and $\delta\mathcal{L}^c_I$ are the counterterms to be determined consistently
in the perturbative expansion by choosing a renormalization prescription. As it
will become clear below this is the most natural manner for obtaining a fully
renormalized Dirac equation in a perturbative expansion.

Our goal is to understand the relaxation of inhomogeneous fermion mean field
$\psi(\mathbf{x},t)=\langle \Psi(\mathbf{x},t)\rangle$ induced by external
source that is adiabatically switched-on at $t=-\infty$. As explained in
Sec.~\ref{sec:2.2}, the equation of motion for the mean field can be obtained
to any order in perturbative theory via the tadpole method~\cite{tadpole} by
writing
\begin{equation}
\Psi^\pm(\mathbf{x},t)=\psi(\mathbf{x},t)+\chi^\pm(\mathbf{x},t),\quad
\mbox{with}\quad\langle \chi^\pm(\mathbf{x},t)\rangle=0.
\end{equation}

The essential ingredients for perturbative calculations are the
following free real-time Green's functions (in momentum space):

(i) Scalar Propagators
\begin{eqnarray}
G_0^{++}(\mathbf{k},t,t')&=& G_0^{>}(\mathbf{k},t,t')\,\theta(t-
t')+G_0^{<}(\mathbf{k},t,t')\,\theta(t'-t),\nonumber \\
G_0^{--}(\mathbf{k},t,t') &=&
G_0^{>}(\mathbf{k},t,t')\,\theta(t'-
t)+G_0^{<}(\mathbf{k},t,t')\,\theta(t-t'),\nonumber \\
G_0^{-+}(\mathbf{k},t,t')&=& G_0^{>}(\mathbf{k},t,t'),\quad
G_0^{+-}(\mathbf{k},t,t')= G_0^{<}(\mathbf{k},t,t'), \nonumber \\
G_0^{>}(\mathbf{k},t,t')&=& i \int d^3x
\,e^{-i\mathbf{k}\cdot\mathbf{x}} \,
\langle \phi(\mathbf{x},t) \phi(\mathbf{0},t') \rangle \nonumber \\
&=&\frac{i}{2\omega_\mathbf{k}}\left[[1+n_B(\omega_\mathbf{k})]\,
e^{-i\omega_\mathbf{k}(t-t')}+n_B(\omega_\mathbf{k})\,
e^{i\omega_\mathbf{k}(t-t')}\right], \nonumber  \\
G_0^{<}(\mathbf{k},t,t')&=& i \int d^3x \,
e^{-i\mathbf{k}\cdot\mathbf{x}} \,
\langle \phi(\mathbf{0},t') \phi(\mathbf{x},t) \rangle \nonumber \\
&=&\frac{i}{2\omega_\mathbf{k}}\left[n_B(\omega_\mathbf{k})\,
e^{-i\omega_\mathbf{k}(t-t')}+[1+n_B(\omega_\mathbf{k})]\,
e^{i\omega_\mathbf{k}(t-t')}\right],\label{scalarprops}
\end{eqnarray}
where $\omega_\mathbf{k}=\sqrt{\mathbf{k}^2+m^2}$ and $n_B(\omega)=
1/(e^{\beta\omega}-1)$ is the Bose-Einstein distribution.

(ii) Fermion Propagators (zero fermion chemical potential)
\begin{eqnarray}
S_0^{++}(\mathbf{k},t,t')&=& S_0^{>}(\mathbf{k},t,t')\,\theta(t-
t')+S_0^{<}(\mathbf{k},t,t')\,\theta(t'-t),\nonumber \\
S_0^{--}(\mathbf{k},t,t') &=& S_0^{>}(\mathbf{k},t,t')\,\theta(t'-
t)+S_0^{<}(\mathbf{k},t,t')\,\theta(t-t'),\nonumber \\
S_0^{-+}(\mathbf{k},t,t')&=& S_0^{>}(\mathbf{k},t,t'),\quad
S_0^{+-}(\mathbf{k},t,t')= S_0^{<}(\mathbf{k},t,t'), \nonumber \\
S_0^{>}(\mathbf{k},t,t')&=&-i\int
d^3x\,e^{-i\mathbf{k}\cdot\mathbf{x}}\,
\langle\Psi(\mathbf{x},t) \bar{\Psi}(\mathbf{0},t')\rangle\nonumber\\
&=&-\frac{i}{2\bar{\omega}_\mathbf{k}}
\Big[(\gamma^0\bar{\omega}_\mathbf{k}-\bbox{\gamma}\cdot\mathbf{k}+M)
[1-n_F(\bar{\omega}_\mathbf{k})]\,e^{-i\bar{\omega}_\mathbf{k}(t-t')}\nonumber\\
&&+\,(\gamma^0\bar{\omega}_\mathbf{k}+\bbox{\gamma}\cdot\mathbf{k}-M)\,
n_F(\bar{\omega}_\mathbf{k})\,e^{i\bar{\omega}_\mathbf{k}(t-t')}\Big], \nonumber \\
S_0^{<}(\mathbf{k},t,t')&=& i\int d^3x \,e^{-i\mathbf{k}\cdot
\mathbf{x}} \,
\langle\bar{\Psi}(\mathbf{0},t')\Psi(\mathbf{x},t)\rangle \nonumber \\
&=&\frac{i}{2\bar{\omega}_\mathbf{k}}
\Big[(\gamma^0\bar{\omega}_\mathbf{k}+\bbox{\gamma}\cdot\mathbf{k}-M)
\,n_F(\bar{\omega}_\mathbf{k})\,e^{-i\bar{\omega}_\mathbf{k}(t-t')}\nonumber\\
&&+\,(\gamma^0\bar{\omega}_\mathbf{k}+\bbox{\gamma}\cdot\mathbf{k}-M)\,
[1-n_F(\bar{\omega}_\mathbf{k})]\,e^{i\bar{\omega}_\mathbf{k}(t-t')}\Big],
\label{fermionprops}
\end{eqnarray}
where $\bar{\omega}_\mathbf{k}=\sqrt{\mathbf{k}^2+M^2}$ and
$n_F(\omega)=1/(e^{\beta\omega}+1)$ is the Fermi-Dirac distribution. We note
that these \emph{free} propagators given in Eqs.~(\ref{scalarprops}) and
(\ref{fermionprops}) are thermal because, as discussed in Sec.~\ref{sec:2.2},
the initial state is chosen to be in thermal equilibrium and the interaction is
assumed to be turned on adiabatically.

In momentum space, we find the effective \emph{real-time} Dirac equation for
the mean field of momentum $\mathbf{k}$ reads
\begin{eqnarray}
&&\left[\left(i\gamma^0\partial_t - \bbox{\gamma}\cdot\mathbf{k}-M\right)+
\delta_\psi\left(i\gamma^0\partial_t -
\bbox{\gamma}\cdot\mathbf{k}\right)-\delta_M\right]
\psi(\mathbf{k},t)\hspace*{1truein}\nonumber\\
&&\hspace*{2.0truein}-\int_{-\infty}^t dt'\,\Sigma(\mathbf{k},t-t')\,
\psi(\mathbf{k},t')= - \eta(\mathbf{k},t),
\end{eqnarray}
where $\partial_t\equiv\partial/\partial t$, $\Sigma(\mathbf{k},t-t')$ is the
retarded fermion self-energy and
$$
\psi(\mathbf{k},t)\equiv \int d^3 x\,e^{-i \mathbf{k}\cdot\mathbf{x}}\,
\psi(\mathbf{x},t).
$$
Using the real-time free scalar and fermion propagators given by
Eqs.~(\ref{scalarprops}) and (\ref{fermionprops}), respectively, we find to
one-loop order, that $\Sigma(\mathbf{k},t-t')$ is given by
\begin{equation}
\Sigma(\mathbf{k},t-t')=i\gamma^0\,\Sigma^{(0)}(\mathbf{k},t-t')+
\bbox{\gamma}\cdot\mathbf{k}\,\Sigma^{(1)}(\mathbf{k},t-t')+
\Sigma^{(2)}(\mathbf{k},t-t'),
\end{equation}
with
\begin{eqnarray}
\Sigma^{(0)}(\mathbf{k},t-t')&=&-y^2 \int \frac{d^3 q}{(2\pi)^3
}\frac{1}{2\omega_\mathbf{p}}
\Big[\cos[(\omega_\mathbf{p}+\bar{\omega}_\mathbf{q})(t-t')]\,
[1+n_B(\omega_\mathbf{p})-n_F(\bar{\omega}_\mathbf{q})]\nonumber\\
&&+\,\cos[(\omega_\mathbf{p}-\bar{\omega}_\mathbf{q})(t-t')]\,
[n_B(\omega_\mathbf{p})+n_F(\bar{\omega}_\mathbf{q})]\Big],\nonumber\\
\Sigma^{(1)}(\mathbf{k},t-t')&=&  -y^2 \int \frac{d^3 q}{(2\pi)^3}
\frac{\mathbf{k}\cdot\mathbf{q}}{2k^2 \omega_\mathbf{p}\bar{\omega}_\mathbf{q}}
\Big[\sin[(\omega_\mathbf{p}+\bar{\omega}_\mathbf{q})(t-t')]\,
[1+n_B(\omega_\mathbf{p})\nonumber\\
&&-\,n_F(\bar{\omega}_\mathbf{q})]-\sin[(\omega_\mathbf{p}-\bar{\omega}_\mathbf{q})(t-t')]\,
[n_B(\omega_\mathbf{p})+n_F(\bar{\omega}_\mathbf{q})]\Big],\nonumber\\
\Sigma^{(2)}(\mathbf{k},t-t')&=& -y^2 \int \frac{d^3 q}{(2\pi)^3}
\frac{M}{2\omega_\mathbf{p}\bar{\omega}_\mathbf{q}}
\Big[\sin[(\omega_\mathbf{p}+\bar{\omega}_\mathbf{q})(t-t')]\,
[1+n_B(\omega_\mathbf{p})-n_F(\bar{\omega}_\mathbf{q})]\nonumber\\
&&-\,\sin[(\omega_\mathbf{p}-\bar{\omega}_\mathbf{q})(t-t')]\,
[n_B(\omega_\mathbf{p})+n_F(\bar{\omega}_\mathbf{q})]\Big],
\end{eqnarray}
where $\mathbf{p=q+k}$ and $k=|\mathbf{k}|$.

As mentioned before, the source is taken to be switched on adiabatically from
$t=-\infty$ and switched off at $t=0$ to provide the initial conditions [see
Eq.~(\ref{icond})]
\begin{eqnarray}
&&\psi(\mathbf{k},t=0)=\psi(\mathbf{k},0),\qquad\dot{\psi}(\mathbf{k},t < 0)=0.
\end{eqnarray}
Introducing an auxiliary quantity $\sigma(\mathbf{k},t-t')$ defined as
\begin{equation}
\Sigma(\mathbf{k},t-t')=\partial_{t'}\sigma(\mathbf{k},t-t') \label{sigma}
\end{equation}
and imposing $\eta(\mathbf{k},t>0)=0$, we obtain the equation of motion for
$t>0$ to be given by
\begin{eqnarray}
\lefteqn{\left[\left(i\gamma^0\partial_t
-\bbox{\gamma}\cdot\mathbf{k}-M\right)+ \delta_\psi\left(i\gamma^0\partial_t
-\bbox{\gamma}\cdot\mathbf{k}\right) -\sigma(\mathbf{k},0) -\delta_M \right]
\psi_\mathbf{k}(t)}\hspace{2.5truein}\nonumber\\
&&+\int_0^t dt'\,\sigma_{\mathbf{k}}(t-t')\,
\dot{\psi}_{\mathbf{k}}(t')=0.\label{fermion:eom}
\end{eqnarray}
This equation of motion can be solved by Laplace transform as befits an initial
value problem. The Laplace transformed equation of motion is given by
\begin{eqnarray}
\lefteqn{\left[i\gamma^0
s-\bbox{\gamma}\cdot\mathbf{k}-M+\delta_\psi\left(i\gamma^0 s
-\bbox{\gamma}\cdot\mathbf{k}\right)-\delta_M -\sigma(\mathbf{k},0)+s\,
\widetilde{\sigma}(s,\mathbf{k})\right]\widetilde{\psi}(s,\mathbf{k})}
\hspace{2.5truein}\nonumber\\
&&=\left[i\gamma^0 + i\delta_\psi \gamma^0+
\widetilde{\sigma}(s,\mathbf{k})\right]\psi(\mathbf{k},0),
\end{eqnarray}
where $\widetilde{\psi}(s,\mathbf{k})$ and $\widetilde{\sigma}(s,\mathbf{k})$
are, respectively, the Laplace transforms of $\psi(\mathbf{k},t)$ and
$\sigma(\mathbf{k},t)$
$$
\widetilde{\psi}(s,\mathbf{k})\equiv \int_0^{\infty} dt \,e^{-st}\,
\psi(\mathbf{k},t), \quad\widetilde{\sigma}(s,\mathbf{k})\equiv \int_0^{\infty}
dt \, e^{-st}\, \sigma(\mathbf{k},t),
$$
with $\mathrm{Re}\,s>0$.

\subsection{Renormalization}
Before proceeding further, we address the issue of renormalization by analyzing
the ultraviolet divergences of the self-energy. As usual the ultraviolet
divergences are those of zero temperature field theory, since the finite
temperature distribution functions are exponentially suppressed at large
momenta~\cite{book:lebellac}. Therefore the ultraviolet divergences are
obtained by setting to zero the distribution functions for the scalar and
fermion.

With Eq.~(\ref{sigma}), one can write $\sigma(\mathbf{k},t-t')$ as
\begin{equation}
\sigma(\mathbf{k},t-t')=i\gamma^0 \sigma^{(0)}(\mathbf{k},t-t')+
\bbox{\gamma}\cdot\mathbf{k}\,\sigma^{(1)}(\mathbf{k},t-t')+
\sigma^{(2)}(\mathbf{k},t-t').
\end{equation}
A straightforward calculation leads to the following ultraviolet divergences
\begin{equation}
\sigma^{(1)}(\mathbf{k},0)\simeq\frac{y^2}{16\pi^2}\ln\frac{\Lambda}{K},\quad
\sigma^{(2)}(\mathbf{k},0)\simeq -\frac{y^2
M}{8\pi^2}\ln\frac{\Lambda}{K},\quad
\widetilde{\sigma}^{(0)}(s,\mathbf{k})\simeq
\frac{y^2}{16\pi^2}\ln\frac{\Lambda}{K},
\end{equation}
where, $\widetilde{\sigma}^{(i)}(s,\mathbf{k})$ ($i=0,1,2$) are the Laplace
transform of $\sigma^{(i)}(\mathbf{k},t)$, $\Lambda$ is an ultraviolet momentum
cutoff, $K$ is an arbitrary renormalization scale, and ``$\simeq$'' denotes
only the divergent contribution is included. Therefore, the counterterms
$\delta_\psi$ and $\delta_M$ are chosen to be given by
\begin{equation}
\delta_\psi = -\frac{y^2}{16\pi^2}\ln\frac{\Lambda}{K}+\mbox{finite},\quad
\delta_M =\frac{y^2 M}{8\pi^2} \ln\frac{\Lambda}{K}+\mbox{finite},
\label{counterterms}
\end{equation}
and the components of the self-energy are rendered finite
\begin{equation}
\widetilde{\sigma}(s,\mathbf{k})+i\gamma^0 \delta_\psi = \mathrm{finite},\quad
\sigma(\mathbf{k},0) +\bbox{\gamma}\cdot\mathbf{k}\,\delta_\psi+\delta_M =
\mathrm{finite}, \label{renker}
\end{equation}
The finite parts of the counterterms in Eq.~(\ref{counterterms}) are fixed by
prescribing a renormalization scheme. There are two important choices of
counterterms: (i) determining the counterterms from an on-shell condition,
including finite temperature effects and (ii) determining the counterterms from
a zero temperature on-shell condition. Obviously these choices only differ by
finite quantities, but the second choice allows us to separate the dressing
effects of the medium from those in the vacuum. For example, by choosing to
renormalize the self-energy with the zero-temperature counterterms on the mass
shell, the poles in the particle propagator will have unit residue at zero
temperature. Then, in the medium the residues at the finite temperature poles
(or the position of the resonances) are \emph{finite}, smaller than one and
determined solely by the properties of the medium. Thus the formulation of the
initial value problem as presented here yields an unambiguous separation of the
vacuum and in-medium renormalization effects.

Hence we obtain the \emph{renormalized} effective Dirac equation in the medium
and the corresponding initial value problem for the fermion mean field
\begin{equation}
\left[i\gamma^0 s - \bbox{\gamma}\cdot\mathbf{k}-M-
\widetilde{\Sigma}(s,\mathbf{k})\right] \widetilde{\psi}(s,\mathbf{k}) =
\left[i\gamma^0 +
\widetilde{\sigma}(s,\mathbf{k})\right]\psi(\mathbf{k},0),\label{reneom}
\end{equation}
where $\widetilde{\Sigma}(s,\mathbf{k})=
\sigma(\mathbf{k},0)-s\,\widetilde{\sigma}(s,\mathbf{k})$ is the Laplace
transform of the \emph{renormalized} retarded fermion self-energy [see
Eq.~(\ref{renker})], which can be written in its most general form
\begin{equation}
\widetilde{\Sigma}(s,\mathbf{k})= i\gamma^0 s
\,\widetilde{\mathcal{E}}^{(0)}(s,\mathbf{k})+ \bbox{\gamma}\cdot
\mathbf{k}\,\widetilde{\mathcal{E}}^{(1)}(s,\mathbf{k})+M\,
\widetilde{\mathcal{E}}^{(2)}(s,\mathbf{k}).\label{epsilons}
\end{equation}
The solution to Eq.~(\ref{reneom}) is given by
\begin{equation}
\widetilde{\psi}(s,\mathbf{k})= \frac{1}{s}\left\{ 1 + S(s,\mathbf{k})
\Big[\bbox{\gamma}\cdot\mathbf{k}+M
+\widetilde{\Sigma}(0,\mathbf{k})\Big]\right\}\psi(\mathbf{k},0),
\label{laplakern}
\end{equation}
where $S(s,\mathbf{k})$ is the fermion propagator in terms of the Laplace
variable $s$
\begin{eqnarray}
\hspace*{-.4in}S(s,\mathbf{k})& = & \left[i\gamma^0 s -
\bbox{\gamma}\cdot\mathbf{k}-M
-\widetilde{\Sigma}(s,\mathbf{k})\right]^{-1} \nonumber \\
&=& -\,\frac{i\gamma^0 s[1-\widetilde{\mathcal{E}}^{(0)}(s,\mathbf{k})] -
\bbox{\gamma}\cdot\mathbf{k}[1+\widetilde{\mathcal{E}}^{(1)}(s,\mathbf{k})]
+M[1+\widetilde{\mathcal{E}}^{(2)}(s,\mathbf{k})]}{
s^2[1-\widetilde{\mathcal{E}}^{(0)}(s,\mathbf{k})]^2
+k^2[1+\widetilde{\mathcal{E}}^{(1)}(s,\mathbf{k})]^2
+M^2[1+\widetilde{\mathcal{E}}^{(2)}(s,\mathbf{k})]^2}.\label{fermpropofs}
\end{eqnarray}
We note that the square of the denominator in Eq.~(\ref{fermpropofs}) is
recognized as~\cite{book:lebellac}
\begin{equation}
\det \left[i\gamma^0 s -\bbox{\gamma}\cdot\mathbf{k}-M
-\widetilde\Sigma(s,\mathbf{k})\right].
\end{equation}

The real-time evolution of $\psi(\mathbf{k},t)$ is obtained by performing the
inverse Laplace transform along the Bromwich contour in the complex $s$-plane
parallel to the imaginary axis and to the right of all singularities of
$\widetilde{\psi}(s,\mathbf{k})$. Therefore to obtain the real time evolution
we must first understand the singularities of the Laplace transform in the
complex $s$-plane.

\subsection{Structure of the self-energy and damping processes}
To one-loop order, the Laplace transform of the components
$\widetilde{\mathcal{E}}^{(i)}(s,\mathbf{k})$ ($i=0,1,2$) of the fermion
self-energy $\widetilde{\Sigma}(s,\mathbf{k})$ [see Eq.~(\ref{epsilons})] can
be written as dispersion integrals in terms of spectral functions
$\rho^{(i)}(k_0,\mathbf{k})$
\begin{equation}
\left\{\begin{array}{c}
\widetilde{\mathcal{E}}^{(0)}(s,\mathbf{k})  \\
\widetilde{\mathcal{E}}^{(1)}(s,\mathbf{k})  \\
\widetilde{\mathcal{E}}^{(2)}(s,\mathbf{k})
\end{array} \right\} =
 \int_{-\infty}^{+\infty}  \frac{dk_0}{s^2+k_0^2}
\left\{\begin{array}{c}
\rho^{(0)}(k_0,\mathbf{k})  \\
k_0\, \rho^{(1)}(k_0,\mathbf{k})  \\
k_0\, \rho^{(2)}(k_0,\mathbf{k})
\end{array}
\right\}+
\left\{\begin{array}{c}
-\delta_{\psi}  \\
\delta_{\psi} \\
\frac{\delta_M}{M}
\end{array}
\right\},
\end{equation}
with the one-loop spectral functions given by the expressions
\begin{eqnarray}
\rho^{(0)}(k_0,\mathbf{k})&=& -y^2 \int \frac{d^3 q}{(2\pi)^3} \frac{1}{2
\omega_\mathbf{p}}\Big[\delta(k_0-\omega_\mathbf{p}-\bar{\omega}_\mathbf{q})\,
[1+n_B(\omega_\mathbf{p})-n_F(\bar{\omega}_\mathbf{q})]\nonumber\\
&&+\,\delta(k_0-\omega_\mathbf{p}+\bar{\omega}_\mathbf{q})\,[n_B(\omega_\mathbf{p})
+n_F(\bar{\omega}_\mathbf{q})]\Big],
\nonumber\\
\rho^{(1)}(k_0,\mathbf{k})&=& -y^2 \int \frac{d^3 q}{(2\pi)^3}
\frac{\mathbf{k}\cdot\mathbf{q}}{2k^2 \omega_\mathbf{p}\bar{\omega}_\mathbf{q}}
\Big[\delta(k_0-\omega_\mathbf{p}-\bar{\omega}_\mathbf{q})\,
[1+n_B(\omega_\mathbf{p})-n_F(\bar{\omega}_\mathbf{q})]\nonumber\\
&&-\,\delta(k_0-\omega_\mathbf{p}+\bar{\omega}_\mathbf{q})\,[n_B(\omega_\mathbf{p})
+n_F(\bar{\omega}_\mathbf{q})]\Big],\nonumber \\
\rho^{(2)}(k_0,\mathbf{k})&=& -y^2 \int \frac{d^3 q}{(2\pi)^3 } \frac{1}{ 2
\omega_\mathbf{p}
\bar{\omega}_\mathbf{q}}\Big[\delta(k_0-\omega_\mathbf{p}-\bar{\omega}_\mathbf{q})\,
[1+n_B(\omega_\mathbf{p})-n_F(\bar{\omega}_\mathbf{q})]\nonumber\\
&&-\,\delta(k_0-\omega_\mathbf{p}+\bar{\omega}_\mathbf{q})\,[n_B(\omega_\mathbf{p})
+n_F(\bar{\omega}_\mathbf{q})]\Big].\label{rhos}
\end{eqnarray}

The analytic continuation of the retarded self-energy
$\widetilde{\Sigma}(s,\mathbf{k})$ and its components
$\widetilde{\mathcal{E}}^{(i)}(s,\mathbf{k})$ in the complex $s$-plane
(physical sheet) are defined by
\begin{eqnarray}
\Sigma(\omega,\mathbf{k})&\equiv&\widetilde{\Sigma}(s=-i\omega+0^+,\mathbf{k})
= \mathrm{Re}\Sigma(\omega,\mathbf{k})+
i\,\mathrm{Im}\Sigma(\omega,\mathbf{k}), \nonumber \\
\mathcal{E}^{(i)}(\omega,\mathbf{k})&\equiv&
\widetilde{\mathcal{E}}^{(i)}(s=-i\omega+0^+,\mathbf{k}) =
\mathrm{Re}\mathcal{E}^{(i)}(\omega,\mathbf{k})+i\,
\mathrm{Im}\mathcal{E}^{(i)}(\omega,\mathbf{k}).
\label{realandimaginary}
\end{eqnarray}
Using the relation
\begin{equation}
\frac{1}{x+i0^+}=\frac{\wp}{x}-i\,\pi\,\delta(x),
\end{equation}
where $\wp$ denotes the principal value, one finds
\begin{equation}
\left\{\begin{array}{c}
\mathrm{Re}\widetilde{\mathcal{E}}^{(0)}(s,\mathbf{k})  \\
\mathrm{Re}\widetilde{\mathcal{E}}^{(1)}(s,\mathbf{k})  \\
\mathrm{Re}\widetilde{\mathcal{E}}^{(2)}(s,\mathbf{k})
\end{array} \right\} =
 \int_{-\infty}^{+\infty}dk_0\,\frac{\wp}{k_0^2-\omega^2}
\left\{\begin{array}{c}
\rho^{(0)}(k_0,\mathbf{k})  \\
k_0\, \rho^{(1)}(k_0,\mathbf{k})  \\
k_0\, \rho^{(2)}(k_0,\mathbf{k})
\end{array}
\right\}+ \left\{\begin{array}{c}
-\delta_{\psi}  \\
\delta_{\psi} \\
\frac{\delta_M}{M}
\end{array}
\right\},
\end{equation}
and
\begin{eqnarray}
\mathrm{Im}\mathcal{E}^{(0)}(\omega,\mathbf{k}) & = &
\frac{\pi}{2|\omega|}\,\mbox{sgn}(\omega)\left[
\rho^{(0)}(|\omega|,\mathbf{k})+\rho^{(0)}(-|\omega|,\mathbf{k})\right],
\nonumber \\
\mathrm{Im}\mathcal{E}^{(1)}(\omega,\mathbf{k}) & = &
\frac{\pi}{2}\,\mbox{sgn}(\omega)\left[\rho^{(1)}(|\omega|,\mathbf{k})-
\rho^{(1)}(-|\omega|,\mathbf{k})\right], \nonumber   \\
\mathrm{Im}\mathcal{E}^{(2)}(\omega,\mathbf{k}) & = &
\frac{\pi}{2}\,\mbox{sgn}(\omega)\left[ \rho^{(2)}(|\omega|,\mathbf{k})-
\rho^{(2)}(-|\omega|,\mathbf{k})\right]. \label{imagipart}
\end{eqnarray}
We note that the real (imaginary) parts of the retarded self-energy are even
(odd) functions of $\omega$.

The denominator of the analytically continued fermion propagator in
Eq.~(\ref{fermpropofs}) can be written in a compact form $\omega^2 -
\bar{\omega}^2_\mathbf{k}-\Pi(\omega,\mathbf{k})$ where
\begin{eqnarray}
\Pi(\omega,\mathbf{k})&=&-2\left[\omega^2
\mathcal{E}^{(0)}(\omega,\mathbf{k})+k^2\mathcal{E}^{(1)}(\omega,\mathbf{k})+M^2
\mathcal{E}^{(2)}(\omega,\mathbf{k})\right] \nonumber\\
&&-\,\omega^2 \, [\mathcal{E}^{(0)}(\omega,\mathbf{k})]^2 + k^2
\,[\mathcal{E}^{(1)}(\omega,\mathbf{k})]^2 + M^2 \,
[\mathcal{E}^{(2)}(\omega,\mathbf{k})]^2. \label{totalpi}
\end{eqnarray}
We recognize that the \emph{lowest order} term (but \emph{certainly not the
higher order terms}) of this effective self-energy can be written in the
familiar form~\cite{weldon:2,book:kapusta,book:lebellac}
\begin{eqnarray}
\Pi(\omega,\mathbf{k}) &=& - 2\left[\omega^2
\mathcal{E}^{(0)}(\omega,\mathbf{k})+k^2\mathcal{E}^{(1)}(\omega,\mathbf{k})+M^2
\mathcal{E}^{(2)}(\omega,\mathbf{k})\right]\nonumber\\
&=& -\frac{1}{2}\mathrm{Tr}\left[
(\gamma^0\omega-\bbox{\gamma}\cdot\mathbf{k}+M)\,
{\Sigma}(\omega,\mathbf{k})\right].\label{Pi}
\end{eqnarray}

The imaginary part of $\Pi(\omega,\mathbf{k})$ evaluated on the fermion mass
shell will be identified with the fermion damping rate (see below). The
expression given by Eq.~(\ref{Pi}) leads to the familiar form of the damping
rate, but we point out that Eq.~(\ref{Pi}) is a \emph{lowest order} result. The
full imaginary part must be obtained from the full function
$\Pi(\omega,\mathbf{k})$ in Eq.~(\ref{totalpi}) and the generalization to
\emph{all orders} will be given in a later section below.

For fixed $M$ and $m$, the delta function constraints in the spectral functions
$\rho^{(i)}(\omega,\mathbf{k})$ can only be satisfied for certain ranges of
$\omega$. Since the imaginary parts are odd functions of $\omega$ we only
consider the case of positive $\omega$. The delta function
$\delta(|\omega|-\omega_\mathbf{p}-\bar{\omega}_\mathbf{q})$ has support only
for $|\omega|
> \sqrt{k^2+(m+M)^2}$ and corresponds to the usual two-particle cuts that are
present at zero temperature corresponding to the process $f \rightarrow s + f$.
Hence, the two-particle cuts (both for positive and negative $\omega$) do not
give a contribution to $\mathrm{Im}\Pi(\omega,\mathbf{k})$ on the fermion mass
shell and the only contributions are from terms proportional to
$n_B(\omega_\mathbf{p})+n_F(\bar{\omega}_\mathbf{q})$ in Eq.~(\ref{imagipart}).
We find
\begin{eqnarray}
\mathrm{Im}\Pi(\omega,\mathbf{k}) &=& -\pi\,y^2\,\mathrm{sgn}(\omega) \int
\frac{d^3q}{(2\pi)^3}
\frac{n_B(\omega_\mathbf{p})+n_F(\bar{\omega}_\mathbf{q})}{2
\omega_\mathbf{p}\bar{\omega}_\mathbf{q}}\,
\Big[(|\omega|\bar{\omega}_\mathbf{q}-\mathbf{k}\cdot\mathbf{q}-M^2)\nonumber\\
&&\times\,\delta(|\omega|-\omega_\mathbf{p}+\bar{\omega}_\mathbf{q})
+(|\omega|\bar{\omega}_\mathbf{q}+\mathbf{k}\cdot\mathbf{q}+M^2)\,
\delta(|\omega|+\omega_\mathbf{p}-\bar{\omega}_\mathbf{q})\Big].
\label{Piimag}
\end{eqnarray}
In the above equation the first delta function
$\delta(|\omega|-\omega_\mathbf{p}+\bar{\omega}_\mathbf{q})$ determines a cut
in the region $|\omega|< \sqrt{k^2+(m-M)^2}$ and originates in the process $s
\rightarrow f + \bar{f}$, whereas the second delta function
$\delta(|\omega|+\omega_\mathbf{p}-\bar{\omega}_\mathbf{q})$ determines a cut
in the region $0<|\omega|<k$ and originates in the process $f + s \rightarrow
f$. We note that the first cut originates in the process of decay of the scalar
into fermion pairs and the second cut ($\omega^2<k^2$) is associated with
Landau damping. As to be shown below, both delta functions will result in
restriction in the range of the momentum integration in
$\mathrm{Im}\Pi(\omega,\mathbf{k})$.

For $m>2M$ the scalar can kinematically decay into a fermion pair,
and in this case the fermion pole is embedded in the cut
$|\omega|< \sqrt{k^2+(m-M)^2}$ becoming a quasiparticle pole and
only the first cut contributes to the quasiparticle width. This is
a remarkable result, \emph{the fermions acquire a width through
the induced decay of the scalar in the medium}. This process only
occurs in the medium (obviously vanishing at $T=0$) and its origin
is very different from either collisional broadening or Landau
damping. A complementary interpretation of the origin of this
process as a medium induced decay of the scalars into fermions and
the resulting quasiparticle width for the fermion excitation will
be highlighted in Sec.~\ref{sec:3.4} using the kinetic approach to
relaxation.

The width is obtained to lowest order from
$\mathrm{Im}\Pi(\bar{\omega}_\mathbf{k},\mathbf{k})$, and the on-shell delta
function
$\delta(\bar{\omega}_\mathbf{k}-\omega_\mathbf{p}+\bar{\omega}_\mathbf{q})$ is
recognized as the energy conservation condition for the process $s \rightarrow
f +\bar{f}$. To lowest order we find the following expression for
$\mathrm{Im}\Pi(\bar{\omega}_\mathbf{k},\mathbf{k})$ for arbitrary scalar and
fermion masses with $m>2M$ and arbitrary fermion momentum and temperature:
\begin{eqnarray}
\mathrm{Im}\Pi(\bar{\omega}_\mathbf{k},\mathbf{k})&=&-\pi y^2\int \frac{d^3q}{
(2\pi)^3 } \frac{\bar{\omega}_\mathbf{k}\bar{\omega}_\mathbf{q} -
\mathbf{k}\cdot\mathbf{q}-M^2}{2\bar{\omega}_\mathbf{q}\omega_\mathbf{p}}\,
[n_B(\omega_\mathbf{p})+n_F(\bar{\omega}_\mathbf{q})]\,
\delta(\bar{\omega}_\mathbf{k}-\omega_\mathbf{p}+
\bar{\omega}_\mathbf{q})\nonumber\\
&=&-\frac{y^2 m^2 T}{16 \pi k }\left(1-\frac{4M^2}{m^2}\right)
\ln\left.\left[\frac{1-e^{-\beta(\bar{\omega}_\mathbf{q}+
\bar{\omega}_\mathbf{k})}}{1+e^{-\beta\bar{\omega}_\mathbf{q}}}\right]\right|
^{q=q^+}_{q=q^-}, \label{dampingrate}
\end{eqnarray}
where
\begin{eqnarray}
q^{\pm}&=& \frac{m^2}{2 M^2} \left|\, k\left(1-\frac{2M^2}{m^2}\right)\pm
\sqrt{(k^2+M^2)\left(1-\frac{4M^2}{m^2}\right)}\,\right|,\label{q2star}
\end{eqnarray}
with $q\in (q^-,q^+)$ being the support of
$\delta(\bar{\omega}_\mathbf{k}-\omega_\mathbf{p}+\bar{\omega}_\mathbf{q})$ for
fixed $\mathbf{k}$.

\section{Real-Time Evolution of Mean Fields}\label{sec:3.3}
\begin{figure}[t]
\begin{center}
\includegraphics[width=4.0truein,keepaspectratio=true]{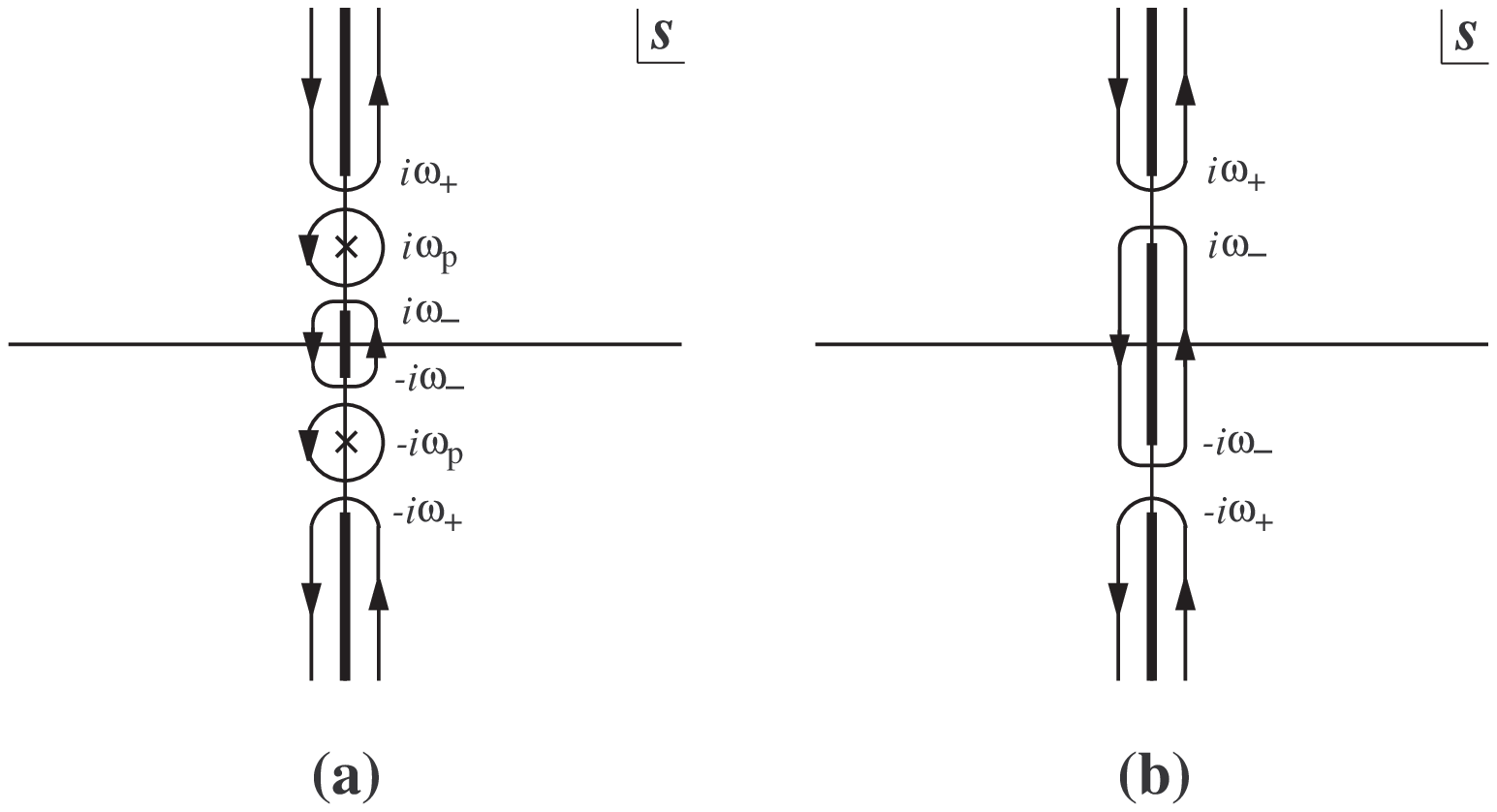}
\caption{The analytic structure of the fermion propagator $S(s,\mathbf{k})$ and
the complex contour used for evaluating the mean field $\psi(\mathbf{k},t)$ for
(a) $m<2M$ and (b) $m>2M$.}\label{laplace}
\end{center}
\end{figure}

The real time evolution is obtained by performing the inverse Laplace transform
as explained above. This requires analyzing the singularities of
$\widetilde{\psi}(s,\mathbf{k})$ given in Eq.~(\ref{fermpropofs}) in the
complex $s$-plane. It is straightforward to see that the putative pole at $s=0$
has vanishing residue, therefore the singularities are those arising from the
fermion propagator $S(s,\mathbf{k})$. For $m < 2M$, the fermion poles at
$\omega =\omega_\mathrm{p}(k)$ are real and isolated away from the
multiparticle cuts along the imaginary axis $s= -i\omega$ with $|\omega|<
\omega_-(k)$ and $|\omega|>\omega_+(k)$, where $\omega_{\pm}(k)=\sqrt{k^2+(m\pm
M)^2}$. In this case, the inverse Laplace transform can be performed by
deforming the Bromwich contour, circling the isolated poles, and wrapping
around the cuts as depicted in Fig.~\ref{laplace}~(a). One finds the inverse
Laplace transformation is dominated by the pole contribution and, to lowest
order, the fermion mean field $\psi(\mathbf{k},t)$ oscillates at late times
with a constant amplitude and frequency $\omega_\mathrm{p}(k)$.

On the other hand, for $m>2M$ (i.e., when the scalar particle can decay into
fermion pairs) the fermion poles become complex and are embedded in the lower
cut, and one must find out if they are complex poles in the physical sheet (the
domain of integration) or moves off to the unphysical (second) sheet.

\subsection[Complex poles or resonances]{Complex poles or resonances}
For $m>2M$ the fermion poles become complex and are embedded in the cut $s=
-i\omega$ with $|\omega|<\omega_-(k)$. The position of the complex poles are
determined from the zeros of $\omega^2
-\bar{\omega}^2_\mathbf{k}-\Pi(\omega,\mathbf{k})$ in the denominator of the
analytically continued fermion propagator for $\omega=\omega_\mathrm{p}(k)
-i\gamma(k)$, where $\omega_\mathrm{p}(k)$ [$\gamma(k)$] is the real
(imaginary) part of the complex pole. In the following discussion, we consider
the narrow width approximation, $|\gamma(k)/\omega_\mathrm{p}(k)|\ll 1$, where
the physical concept of a propagating mode is still meaningful.

With the expressions for the discontinuities in the physical sheet given by
Eq.~(\ref{realandimaginary}), the position of the complex pole is determined by
\begin{equation}
[\omega_\mathrm{p}(k) -i\gamma(k)]^2 -\bar{\omega}^2_\mathbf{k} -
\mathrm{Re}\Pi(\omega_\mathrm{p}(k),\mathbf{k})+i \,
\mathrm{sgn}[\gamma(k)]\,\mathrm{Im}\Pi(\omega_\mathrm{p}(k),\mathbf{k})=0.
\label{complexpole}
\end{equation}
Retaining terms at most linear in $\gamma(k)$, one finds the real and imaginary
parts of Eq.~(\ref{complexpole}), respectively, become
\begin{equation}
\omega^2_\mathrm{p}(k)=
\bar{\omega}^2_\mathbf{k}+\mathrm{Re}\Pi(\omega_\mathrm{p}(k),\mathbf{k}),
\label{realpole}
\end{equation}
and
\begin{equation}
\gamma(k)=\mathrm{sgn}[\gamma(k)]\,
\frac{\mathrm{Im}\Pi(\omega_\mathrm{p}(k),\mathbf{k})}{2\omega_\mathrm{p}(k)}.
\label{imagipole}
\end{equation}
Since $\mathrm{Re}\Pi(\omega,\mathbf{k})$ is an even function of $\omega$, here
and henceforth, we choose $\omega_\mathrm{p}(k)$ to be the positive solution of
Eq.~(\ref{realpole}). To lowest order, one finds
\begin{eqnarray}
\omega_\mathrm{p}(k) &=&\bar{\omega}_\mathbf{k}+\frac{\mathrm{Re}
\Pi(\bar{\omega}_\mathbf{k},\mathbf{k})}{2\bar{\omega}_\mathbf{k}}\nonumber\\
&=&\bar{\omega}_\mathbf{k} +\frac{1}{4\bar{\omega}_\mathbf{k}}
\mathrm{Tr}[(\gamma^0\omega_\mathbf{k}-\bbox{\gamma}\cdot\mathbf{k}+M)\,
\mathrm{Re}\Sigma(\bar{\omega}_\mathbf{k},\mathbf{k})]. \label{lowestorderpole}
\end{eqnarray}
To this order, the solution for $\gamma(k)$ is obtained by approximating
$\omega_\mathrm{p}(k)\approx\bar{\omega}_\mathbf{k}$ in Eq.~(\ref{imagipole}).

A close inspection, however, shows that Eq.~(\ref{imagipole}) \emph{cannot}
have a solution because $\mathrm{Im}\Pi(\omega,\mathbf{k})$ is an odd function
of $\omega$ and $\mathrm{Im}\Pi(\bar{\omega}_\mathbf{k},\mathbf{k})<0$.
Therefore, there is \emph{no} complex pole in the physical sheet. Indeed, this
is a fairly well-known (but seldom noticed) result: if the imaginary part of
the self-energy on the mass shell is negative, then there is no complex pole in
the physical sheet and the pole has moved off into the unphysical (second)
sheet.

In the case that $\mathrm{Im}\Pi(\bar{\omega}_\mathbf{k},\mathbf{k})>0$,
complex poles appear in the physical sheet, but in such case there are
\emph{two poles} with both signs for $\gamma(k)$ (one corresponds to a decaying
exponential in time and the other a growing exponential in time), which is the
signal of an instability, not of damping. However since we confirm that
$\mathrm{Im}\Pi(\bar{\omega}_\mathbf{k},\mathbf{k})$ is negative in the case
under consideration, the complex poles are in the unphysical sheet and
correspond to resonances.

\subsection{Scalar decay implies fermion damping}
We have shown that there are \emph{no} complex poles in the physical sheet in
the complex $s$-plane (the integration region), hence the only singularities
are the cuts along the imaginary axis $s= -i\omega$ with $|\omega|<
\omega_-(k)$ and $|\omega|>\omega_+(k)$. The contour of integration can be
deformed to wrap around these cuts as depicted in Fig.~\ref{laplace}~(b). Since
the resonance is below (and away from) the multiparticle cut
$|\omega|>\omega_+(k)$, in the narrow width approximation and consistent with
perturbation theory the contribution from the cut $|\omega|< \omega_-(k)$
becomes the dominant one while that from the multiparticle cut is always
perturbatively small.

It proves convenient to write the product in Eq.~(\ref{laplakern}) in the
compact form
\begin{equation}
S(s,\mathbf{k}) \left[\bbox{\gamma}\cdot\mathbf{k}+M
+\widetilde{\Sigma}(0,\mathbf{k})\right] = \frac{\mathcal{N}(s,k)}{-s^2
-\bar{\omega}^2_\mathbf{k}-\Pi(s,k)}, \label{numerator}
\end{equation}
which defines $\mathcal{N}(s,k)$, and to change variables to $s=-i\omega\pm
0^+$ on the right-hand ($+$) and left-hand ($-$) side of the cut. After some
algebra, one finds the following contribution to the real time evolution of the
mean field
\begin{eqnarray}
\psi(\mathbf{k},t) &=& \frac{1}{\pi}\int_{-\omega_-}^{+\omega_-}
\frac{d\omega}{\omega}\,e^{-i\omega t} \left[
\frac{\mathrm{Re}\mathcal{N}(\omega,\mathbf{k})\,
\mathrm{Im}\Pi(\omega,\mathbf{k})}{[\omega^2-\bar{\omega}^2_\mathbf{k}-
\mathrm{Re}\Pi(\omega,\mathbf{k})]^2+\mathrm{Im}\Pi(\omega,\mathbf{k})^2}\right.
\nonumber \\
&&+\left.\frac{\mathrm{Im}\mathcal{N}(\omega,\mathbf{k})
[\omega^2-\bar{\omega}^2_\mathbf{k}-\mathrm{Re}\Pi(\omega,\mathbf{k})] }
{[\omega^2-\bar{\omega}^2_\mathbf{k}-\mathrm{Re}\Pi(\omega,\mathbf{k})]^2+
\mathrm{Im}\Pi(\omega,\mathbf{k})^2}\right] \psi(\mathbf{k},0).
\end{eqnarray}
The term proportional to $\mathrm{Re}\mathcal{N}(\omega,\mathbf{k})$ features a
typical Breit-Wigner resonance near the real part of the complex pole, where
$\omega^2_{\mathrm{p}}-\bar{\omega}^2_\mathbf{k}-
\Pi_R(\omega_{\mathrm{p}},\mathbf{k})=0$, since, for $m>2M$, the imaginary part
of the self-energy at this value of $\omega$ (perturbatively close to
$\pm\omega_\mathbf{k}$) is nonvanishing. On the other hand, the term
proportional to $\mathrm{Im}\mathcal{N}(\omega,\mathbf{k})$ is a representation
of the principal part in the limit of small $\mathrm{Re}\Pi(\omega,\mathbf{k})$
and is therefore subleading.

The sharply peaked resonances at $\omega = \pm\omega_\mathrm{p}(k)$ dominate
the integral and give the largest contribution to the real-time evolution of
$\psi(\mathbf{k},t)$. In the limit of a narrow resonance, the $\omega$ integral
is performed by taking the integration limits to infinity and approximating
near the resonances at $\omega =\pm\omega_\mathrm{p}(k)$
\begin{equation}
-\frac{1}{\omega}\frac{\mathrm{Im}\Pi(\omega,\mathbf{k})}
{[\omega^2-\bar{\omega}^2_\mathbf{k}-\mathrm{Re}\Pi(\omega,\mathbf{k})]^2+
\mathrm{Im}\Pi(\omega,\mathbf{k})^2} \approx
\frac{Z(k)}{2\,\omega^2_\mathrm{p}(k)}\,\frac{\gamma(k)}{[\omega\mp
\omega_\mathrm{p}(k)]^2+ \gamma^2(k)},\label{breit}
\end{equation}
where
\begin{equation}
Z(k)=\left[1-\frac{\partial \mathrm{Re}\Pi(\omega,\mathbf{k})}{\partial
\omega^2}\right]_{\omega=\omega_\mathrm{p}(k)}^{-1},\qquad \gamma(k)
=-\frac{Z(k)}{2\omega_\mathrm{p}(k)}\mathrm{Im}
\Pi(\omega_\mathrm{p}(k),\mathbf{k}).
\end{equation}
The wave function renormalization constant $Z(k)$ is \emph{finite}, since the
effective self-energy $\Pi(\omega,\mathbf{k})$ has been rendered finite by an
appropriate choice of counterterms. Only when the counterterms are chosen to
provide a subtraction of the self-energy at the position of the resonance
$\omega\approx\pm\omega_\mathrm{p}(k)$ will result in $Z(k)=1$. On the other
hand, as noted above, if the counterterms are chosen to renormalize the theory
on the fermion mass shell at \emph{zero temperature} then $Z(k)$ describes the
dressing of the medium and is smaller than one. As will be shown shortly, the
width of the Breit-Wigner resonance $\gamma(k)$ is the damping rate of the
fermion mean field.

The integration in the variable $\omega$ can be performed under these
approximations (justified for narrow width), leading to the real-time evolution
\begin{equation}
\psi(\mathbf{k},t) \approx
\frac{Z(k)}{2\omega^2_\mathrm{p}(k)}\left[\mathrm{Re}\mathcal{N}
(\omega_\mathrm{p}(k),\mathbf{k})\,e^{-i\omega_\mathrm{p}(k)t}+
\mathrm{Re}\mathcal{N}(-\omega_\mathrm{p}(k),\mathbf{k})\,e^{i\omega_\mathrm{p}(k)
t}\right]\,e^{-\gamma(k) t}\,\psi(\mathbf{k},0). \label{finalrealtime}
\end{equation}
Gathering the results for
$\mathrm{Im}\Pi(\bar{\omega}_\mathrm{p}(k),\mathbf{k})$ given in
Eq.~(\ref{dampingrate}), we find the damping rate for the fermion mean fields
at one-loop order to be given by (for $m>2M$)
\begin{equation}
\gamma(k)=\frac{y^2 m^2 T}{32 \pi k \bar{\omega}_\mathbf{k}
}\left(1-\frac{4M^2}{m^2}\right)
\ln\left.\left[\frac{1-e^{-\beta(\bar{\omega}_\mathbf{q}+
\bar{\omega}_\mathbf{k})}}{1+e^{-\beta\bar{\omega}_\mathbf{q}}}\right]\right
|^{q=q^+}_{q=q^-}, \label{dampingratefin}
\end{equation}
where $q^\pm$ are given in Eq.~(\ref{q2star}).

\begin{figure}[t]
\begin{center}
\includegraphics[width=3.5truein,keepaspectratio=true]{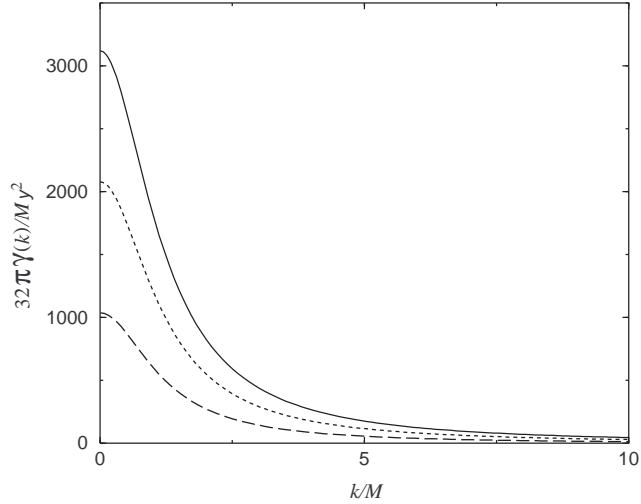}
\caption[The fermion damping rate plotted as a function of the momentum for
$m/M=4$ and $T/M=150$, 100, and 50.]{The fermion damping rate plotted as a
function of the momentum for $m/M=4$ and $T/M=150$ (solid), 100 (dotted), and
50 (dashed).}\label{damping:1}
\end{center}
\end{figure}

\begin{figure}[t]
\begin{center}
\includegraphics[width=3.5truein,keepaspectratio=true]{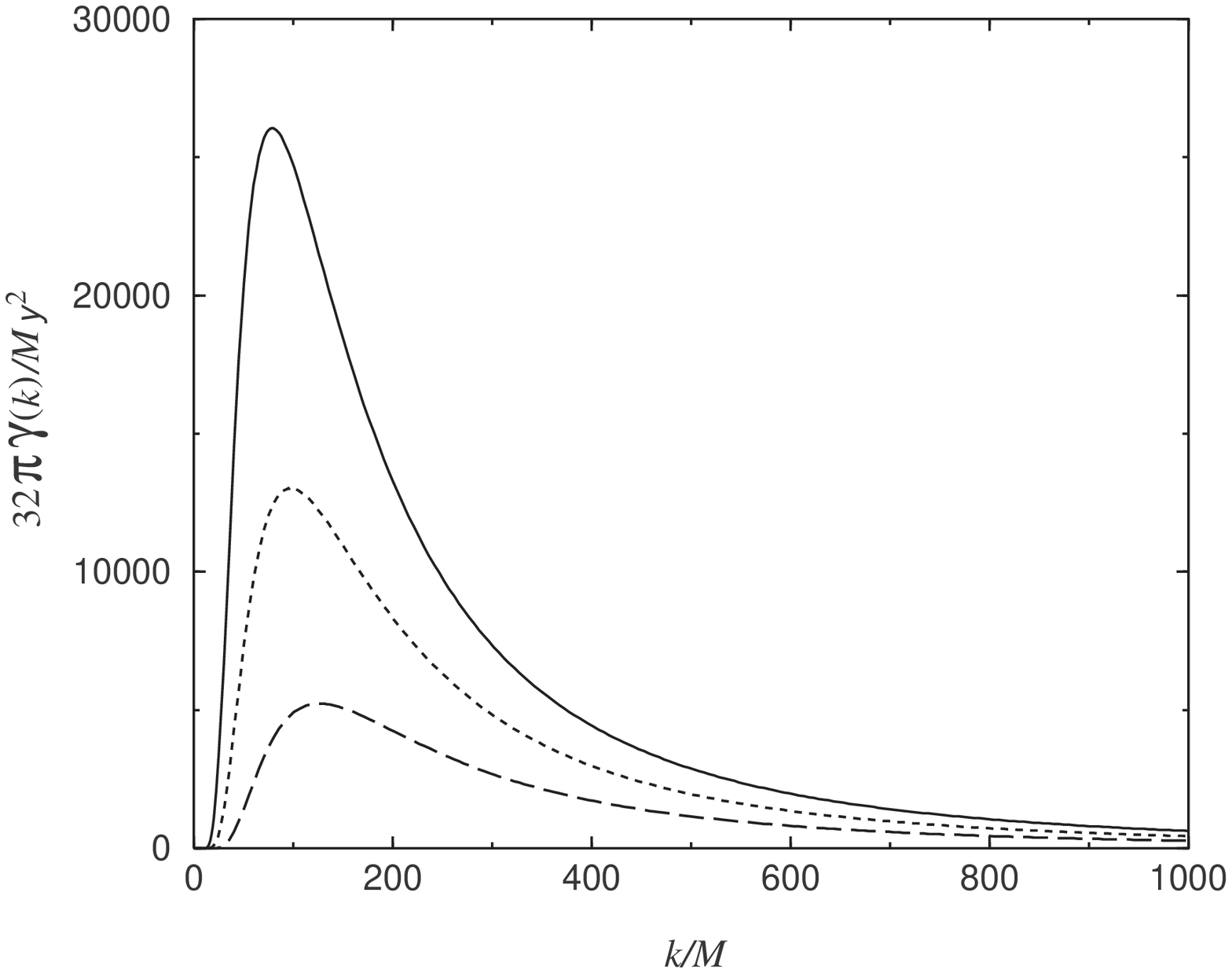}
\caption[The fermion damping rate plotted as a function of the momentum for
$m/M=800$ and $T/M= 1000$, 800, and 600.]{The fermion damping rate plotted as a
function of the momentum for $m/M=800$ and $T/M= 1000$ (solid), 800 (dotted),
and 600 (dashed).}\label{damping:2}
\end{center}
\end{figure}

Figures~\ref{damping:1} and \ref{damping:2} display the behavior of $\gamma(k)$
for several ranges of the parameters. We have chosen a wide range of parameters
for the ratios of the scalar  to fermion masses ($m/M$) and the temperature to
fermion mass ($T/M$) to illustrate in detail the important differences.  The
damping rate features a strong peak as a function of the ratio $k/M$. This peak
is at very small momentum when the ratio of scalar to fermion mass is not much
larger than 2, but moves to larger values of the fermion momentum when this
ratio is very large. Figure~\ref{damping:2} displays this feature in an extreme
case ($m/M=800$) to highlight this behavior. The height of the peak is a
monotonically increasing function of temperature as expected. This is one of
the important results of this work: the decay of the heavy scalar into fermion
pairs results in a \emph{induced damping} of the amplitude of fermionic
excitations.

\subsection{All-order expression for the damping rate}
Since there is no complex pole solution in the physical sheet for $m>2M$, there
are no solutions of the effective in-medium Dirac equation. However, we can
define the spinor wave function of the \emph{resonance} by considering the
solutions of the in-medium Dirac equation with only the \emph{real part} of the
self-energy at $\omega =\pm\omega_\mathrm{p}(k)$~\cite{dolivo:1,nieves}. The
form given by Eq.~(\ref{epsilons}) for the self-energy is general and not
restricted to perturbation theory, hence our analysis below is valid to
\emph{all orders} in perturbation theory.

Using the expression for the self-energy given by Eq.~(\ref{epsilons}) for
$s=\mp i\omega_\mathrm{p}(k)+0^+$, and introducing the following variables
\begin{eqnarray}
\mathcal{W}&=&\omega_\mathrm{p}(k)\,
[1-\mathrm{Re}\mathcal{E}^{(0)}(\omega_\mathrm{p}(k),\mathbf{k})], \nonumber\\
\bbox{\mathcal{K}}&=&
\mathbf{k}\,[1+\mathrm{Re}\mathcal{E}^{(1)}(\omega_\mathrm{p}(k),\mathbf{k})],\nonumber\\
\mathcal{M}&=&M\,[1+\mathrm{Re}\mathcal{E}^{(2)}(\omega_\mathrm{p}(k),\mathbf{k})],
\label{newvariables}
\end{eqnarray}
we find the resonance wave functions
$\Psi_s(\pm\omega_\mathrm{p}(k),\mathbf{k})$ obey the following effective
in-medium Dirac equation
\begin{equation}
\left(\pm\gamma^0 \mathcal{W} -\bbox{\gamma}\cdot
\bbox{\mathcal{K}}-\mathcal{M}\right)\Psi_s(\pm\omega_\mathrm{p}(k),\mathbf{k})=0,
\label{diracsolution}
\end{equation}
where $s=1,2$ labels the spin. Furthermore, using the properties that
$\mathrm{Re}\,\mathcal{E}^{(i)}(\omega,\mathbf{k})$ are even functions of
$\omega$ and depend on $k$, we can define two in-medium \emph{particle} (i.e.,
positive energy) solutions
$U_s(\mathbf{k})=\Psi_s(\omega_\mathrm{p}(k),\mathbf{k})$ and two in-medium
\emph{antiparticle} (i.e., negative energy) solutions $V_s(\mathbf{k})=
\Psi_s(-\omega_\mathrm{p}(k),-\mathbf{k})$. These in-medium Dirac spinors
satisfy
\begin{eqnarray}
&&\left(\gamma^0 \mathcal{W}-\bbox{\gamma}\cdot
\bbox{\mathcal{K}}-\mathcal{M}\right)\,U_s(\mathbf{k})=0,\nonumber\\
&&\left(\gamma^0\mathcal{W}-\bbox{\gamma}\cdot
\bbox{\mathcal{K}}+\mathcal{M}\right)\,V_s(\mathbf{k})=0, \label{parantipar}
\end{eqnarray}
and can be constructed from the usual free Dirac spinors through the
replacement $\bar{\omega}_\mathbf{k} \rightarrow \mathcal{W}$,
$\mathbf{k}\rightarrow \bbox{\mathcal{K}}$, and $M \rightarrow \mathcal{M}$.

Combining Eqs.~(\ref{epsilons}), (\ref{realandimaginary}) and (\ref{totalpi}),
we obtain, from the definition of the damping rate given by Eq.~(\ref{breit}),
the following expression for the width of the resonance to \emph{all orders} in
perturbation theory:
\begin{eqnarray}
\gamma(k)&=& -\frac{Z(k)}{4\,\omega_\mathrm{p}(k)} \mathrm{Tr}\left[(\gamma^0
\mathcal{W}-\bbox{\gamma}\cdot \bbox{\mathcal{K}}+\mathcal{M})\,
\mathrm{Im}\Sigma(\omega_\mathrm{p}(k),\mathbf{k})\right] \nonumber\\
&=&-\frac{\mathcal{M}Z(k)}{2\,\omega_\mathrm{p}(k)}
\sum_{s=1,2}\left[\bar{U}_s(\mathbf{k})\,
\mathrm{Im}\Sigma(\omega_\mathrm{p}(k),\mathbf{k})\,
U_s(\mathbf{k})\right],\label{exactwidth}
\end{eqnarray}
where we have alternatively written the expression for the width in terms of
the \emph{exact} resonance spinors in the medium. This result confirms those
found in Refs.~\cite{dolivo:1,nieves} and leads to the often quoted expression
for the width in lowest order~\cite{weldon:2,book:lebellac}.

\section{Kinetics of Fermion Relaxation}\label{sec:3.4}
To clarify and confirm independently the result of the previous section that
decay of the scalar leads to a quasiparticle width of the fermionic
excitations, we now provide an analysis of the relaxation of the fermion
distribution function via a kinetic Boltzmann equation. This analysis will also
provide, directly in real time, a firm relationship between the damping rate
and the interaction rate in the relaxation time approximation (i.e.,
linearization near equilibrium).

Because here we are only interested in the kinetics of fermions, we assume that
the scalars are in thermal equilibrium. Let us denote the distribution function
for fermions of momentum $\mathbf{k}$ and spin $s$ by
$\bar{n}_{s,\mathbf{k}}(t)$. In the kinetic approach, the time derivative of
this distribution function is obtained from the Boltzmann equation. Since for a
fixed spin component the matrix elements for the transition probabilities are
rather cumbersome, we study the spin-averaged fermion distribution function
defined as $ \bar{n}_{\mathbf{k}}(t) = \frac{1}{2} \sum_{s}
\bar{n}_{\mathbf{k},s}(t) $.

For $m>2M$, two processes are responsible for the change in the fermion
populations in a fermion-scalar plasma: $s\rightarrow f+\bar{f}$, which
provides the ``gain'' term in the balance equation, and $f+\bar{f} \rightarrow
s$, which provides the ``loss'' term. Using the standard approach to obtain the
semiclassical Boltzmann equation, we find
\begin{eqnarray}
\frac{d}{dt}\bar{n}_{\mathbf{k}}(t)& = &\pi y^2\int \frac{d^3q}{(2 \pi)^3}
\frac{\bar{\omega}_\mathbf{k} \bar{\omega}_\mathbf{q} -
\mathbf{k}\cdot\mathbf{q}-M^2}{2\bar{\omega}_\mathbf{q}\bar{\omega}_\mathbf{k}
\omega_\mathbf{p}}\,
\delta(\bar{\omega}_\mathbf{k}+\bar{\omega}_\mathbf{q}-\omega_\mathbf{p})\nonumber\\
&&\times\left[n_B(\omega_\mathbf{p})[1-\bar{n}_{\mathbf{k}}(t)][1-\bar{n}_{\mathbf{q}}(t)]-
[1+n_B(\omega_\mathbf{p})]\bar{n}_{\mathbf{k}}(t)
\bar{n}_{\mathbf{q}}(t)\right], \label{fermion:BE}
\end{eqnarray}
where $\mathbf{p=k+q}$. It is easy to check that the above equation has an
equilibrium solution given by $\bar{n}_\mathbf{k}(t)=
n_F(\bar{\omega}_\mathbf{k})$ for all momentum $\mathbf{k}$.

Let us now consider small departure for fermions from this equilibrium solution
and study the relaxation to equilibrium in the relaxation time approximation.
This is achieved by introducing a disturbance in the distribution function for
fermion of (fixed) momentum $\mathbf{k}$ so that
\begin{equation}
\bar{n}_{\mathbf{k}}(t) = n_F(\bar{\omega}_\mathbf{k}) +
\delta\bar{n}_{\mathbf{k}}(t),
\end{equation}
where $\delta\bar{n}_{\mathbf{k}}(t)/n_F(\bar{\omega}_\mathbf{k})\ll 1$.
Retaining terms at most linear in $\delta\bar{n}_{\mathbf{k}}(t)$, one obtains
from Eq.~(\ref{fermion:BE}) an equation for $\delta\bar{n}_{\mathbf{k}}(t)$
\begin{equation}
\frac{d}{dt}\,\delta\bar{n}_{\mathbf{k}}(t) =-\Gamma(k)\,
\delta\bar{n}_\mathbf{k}(t),
\end{equation}
where $\Gamma(k)$ is the interaction rate, whose inverse characterizes the time
scale for the fermion distribution to approach equilibrium~\cite{weldon:2}
\begin{eqnarray}
\Gamma(k) &=& \pi y^2\int \frac{d^3q}{ (2\pi)^3 }
\frac{\bar{\omega}_\mathbf{k}\bar{\omega}_\mathbf{q} -
\mathbf{k}\cdot\mathbf{q}-M^2}{2\bar{\omega}_\mathbf{k}
\bar{\omega}_\mathbf{q}\omega_\mathbf{p}}\,
[n_B(\omega_\mathbf{p})+n_F(\bar{\omega}_\mathbf{q})]\,
\delta(\bar{\omega}_\mathbf{k}-\omega_\mathbf{p}+
\bar{\omega}_\mathbf{q})\nonumber\\
&=&\frac{y^2 m^2 T}{16 \pi k
\bar{\omega}_\mathbf{k}}\left(1-\frac{4M^2}{m^2}\right)
\ln\left.\left[\frac{1-e^{-\beta(\bar{\omega}_\mathbf{q}
+\bar{\omega}_\mathbf{k})}}{1+e^{-\beta\bar{\omega}_\mathbf{q}}}
\right]\right|^{q=q^+}_{q=q^-},
\label{relaxtime}
\end{eqnarray}
with $q^\pm$ given in Eq.~(\ref{q2star}).

Upon comparing the interaction rate with the damping rate found in the previous
section [see Eq.~(\ref{dampingratefin})], we provide a real-time confirmation
of the result
\begin{equation}
\Gamma(k) = 2\,\gamma(k). \label{relation}
\end{equation}
The kinetic analysis confirms that the damping of the fermionic quasiparticle
excitations in the medium is a consequence of the decay of the heavy scalar.
Furthermore, this analysis is carried out \emph{directly in real time} and
clearly establishes the relation between the interaction rate in the relaxation
time approximation and the damping rate of the mean field at least to lowest
order.

Recently a detailed investigation between the damping and interaction rates for
chiral fermions in gauge theories at finite temperature has been reported in
Ref.~\cite{ayala} within the framework of the imaginary-time (Matsubara)
formalism of finite-temperature field theory. Our analysis provides a
complementary confirmation of this result in real time both for the relaxation
of the mean field and that of the spin-averaged fermion distribution function.

\section{Conclusions}\label{sec:3.5}
In this chapter we have focused on studying the propagation of fermion
excitations in a fermion-scalar plasma, as a complement to the more studied
issue of fermion propagation in a hot (Abelian or non-Abelian) plasma. Our
motivation was to provide a real-time analysis of the propagation of fermions
that could eventually be used in other problems such as neutrino oscillations
in the medium and nonequilibrium processes in electroweak baryogenesis.

The first step of the program is to obtain the effective in-medium Dirac
equation for the fermion mean field directly in real time. This is achieved by
relating the problem of linear response to an initial value problem for the
fermion mean field induced by an external Grassmann source that is
adiabatically switched on from remote past and switched off at time $t=0$. The
resulting effective Dirac equation for the mean field is fully renormalized,
retarded and causal, thus allowing a direct study of real-time relaxation of
the mean field. We study in detail the structure of the renormalized fermion
self-energy and establish the presence of new cuts of thermal origin in the
fermion propagator. We found that when the scalar mass is large enough that
decay of the scalar into fermion pairs is kinematically allowed, this decay
process results in an induced damping of the fermion excitations and their
propagation in the medium as quasiparticle resonances. Solving the effective
Dirac equation by Laplace transform, we obtained the time evolution of the
fermion mean field, which leads to a clear identification of the damping rate.
We calculate the damping rate in the narrow width approximation to one loop
order for arbitrary values of the fermion and scalar masses (provided the
scalar is heavy enough to decay), temperature and fermion momentum.

An all-order expression for the fermion damping rate (in the narrow width
approximation) is obtained from an analysis of the structure of the self-energy
and the exact quasiparticle spinor wave functions. A kinetic approach based on
a semiclassical Boltzmann kinetic equation for the spin-averaged fermion
distribution function reveals that the interaction rate (obtained in the
relaxation time approximation) is twice the damping rate of the mean fields at
least to lowest order in the Yukawa coupling. We emphasize that this relation
is established from a real-time analysis both for the mean field and the
distribution function.

Although the expression for the damping rate was obtained for arbitrary values
of the scalar and fermion masses, temperature and fermion momentum, a deeper
analysis is required if the theory undergoes a second-order (or weakly
first-order) phase transition. The reason being that if the fermion masses are
a result of spontaneous symmetry breaking in the scalar sector, near a
second-order (or weak first-order) phase transition both the scalar mass and
the chiral breaking fermion mass vanish. In this case the kinematic region in
momentum space for which the energy conservation delta functions are fulfilled
shrinks and one must understand if, for soft fermion momentum $k\sim yT$, a
resummation akin to the hard thermal loops (HTL's)~\cite{htl} in hot gauge
field theory is required~\cite{thoma:fermion}.

%% file: chap4.tex
\chapter{Dynamical Renormalization Group Approach to Quantum Kinetics}
\label{chap:4}
\section{Introduction}
The ultrarelativistic heavy ion experiments at the BNL Relativistic Heavy Ion
Collider (RHIC) and the forthcoming CERN Large Hadron Collider (LHC) have the
potential of providing clear evidence for the formation of the quark-gluon
plasma (QGP)~\cite{qgp,book:qgp} and hopefully to study the chiral phase
transition~\cite{DCC:1}. Since this is perhaps the only opportunity to study
phase transitions that are conjectured to occur in particle physics with
earth-bound accelerators, a great theoretical effort parallel to the
experimental work has been devoted to understanding the signatures of the QGP
and the chiral phase transition.

A essential part of this theoretical program is to establish, from first
principles, a consistent quantum kinetic description of transport phenomena in
a hot nonequilibrium multiparton system. Such a kinetic description has the
potential of providing a detailed understanding of experimental signatures such
as photon and dilepton production, charmonium suppression, strangeness
enhancement, freeze-out of hadrons, and collective flow~\cite{signals,flow}
that will lead to an unambiguous determination of the formation of the QGP and
the observables of chiral phase transitions in ultrarelativistic heavy ion
collisions. The kinetic description is also of fundamental importance in the
understanding of the emergence of hydrodynamics in the long-wavelength limit of
quantum field theories~\cite{jeon}. During the last few years, there have been
significant advances ranging from first-principle derivations of kinetic and
transport equations in QCD~\cite{heinz,mrowczynski,geiger:qcd,blaizot:be} and
scalar field theories~\cite{frenkel,lawrie,boyanrelkin,rau,niegawa:1},
numerical codes that describe the space-time evolution of heavy ion collisions
such as the screened perturbative QCD (pQCD) parton model~\cite{wang,eskola},
the parton cascade model~\cite{geiger:pcm} and the hydrodynamical transport
model~\cite{urqmd} to lattice simulations of nonequilibrium dynamics in quantum
field theories~\cite{smit,rajantie} and classical gauge
theories~\cite{venugopalan,poeschl,hindmarsh}. Furthermore, an effective
quantum kinetic equation has been advocated as a description of the dynamics of
soft collective excitations in hot non-Abelian
plasmas~\cite{bodeker,arnold,basagoiti,litim,blaizot:be}.

The typical approach to derive quantum kinetic equation involves a Wigner
transform of the two-point Green's functions and a gradient
expansion~\cite{geiger:qcd,blaizot:be,heinz,mrowczynski,danielewicz} (a gauge
covariant Wigner transform in the case of gauge theories) and often requires a
quasiparticle
approximation~\cite{blaizot:be,mrowczynski,niegawa:1,danielewicz}. The gradient
expansion assumes that the center-of-mass Wigner variables are slowly varying,
but it is seldom clear at this level which are the fast and which are the slow
scales involved. A ``coarse-graining'' procedure is typically invoked that
averages out microscopic scales in the kinetic description, leading to
irreversible evolution in the resulting kinetic equation. Such an averaging
procedure is usually poorly understood and justified \textit{a posteriori}. The
quasiparticle approximation consists in neglecting the broadening of the
single-particle states when computing the collision terms. A derivation of
quantum kinetic equation for a hot QCD plasma along these lines has recently
been reported recently in Ref.~\cite{blaizot:be}, however the collision terms
obtained in the quasiparticle approximation turn out to be infrared divergent.

The goal of this chapter is to provide a novel and alternative
derivation of quantum kinetic equations directly in real time from
the underlying quantum field theory by implementing a dynamical
renormalization group (DRG) resummation. The strategy to be
followed is a generalization of the dynamical renormalization
group method introduced in Ref.~\cite{boyanrgir} to study
anomalous relaxation in Abelian gauge theories, but adapted to the
description of quantum kinetics~\cite{boyanrelkin,boyanhtl}. The
starting point is the identification of the distribution function
of the quasiparticles which could require a resummation of medium
effects (the equivalent of hard thermal loops~\cite{htl}). The
equation of motion for this distribution function is solved in a
perturbative expansion in terms of nonequilibrium Feynman
diagrams. The perturbative solution in real time gives rise
\emph{secular terms} that grow rapidly with time and invalidate
the perturbative expansion beyond a particular time scale
(recognized \textit{a posteriori} as the kinetic time scale). The
dynamical renormalization group implements a systematic
resummation of these secular terms and the resulting
renormalization group equation is the quantum kinetic equation.

The validity of this approach hinges upon the basic assumption of a wide
separation between the microscopic and the kinetic time scales. Such an
assumption underlies every approach to a kinetic description and is generally
justified in weakly coupled theories. Unlike other approaches in terms of a
truncation of the equations of motion for the Wigner distribution function, the
main ingredient in the approach presented here is a perturbative diagrammatic
evaluation of the time evolution of the distribution function in real
time~\cite{boyanrelkin} improved via a renormalization group resummation of the
secular divergences.

An important bonus of this approach is that it illuminates the origin and
provides a natural resolution of pinch singularities~\cite{landsmann,altherr}
found in perturbation theory out of equilibrium. The perturbative real-time
approach combined with the renormalization group resummation reveal clearly
that these are indicative of the nonequilibrium evolution of the distribution
functions. In this framework, pinch singularities are the manifestation of
secular terms.

This chapter is organized as follows. In Sec.~\ref{sec:4.2} we study in detail
the dynamical renormalization group approach to quantum kinetics in the
familiar case of a self-interacting scalar theory, including in addition a
nonequilibrium resummation akin to the hard thermal loops to account for the
effective scalar masses in the medium and therefore the relevant microscopic
time scales. In Sec.~\ref{sec:4.3} we discuss the main features of the
dynamical renormalization group method and compare it to the more familiar
renormalization group in Euclidean field theory.

In Sec.~\ref{sec:4.4} we apply these techniques to obtain the kinetic equations
for pions and sigma mesons in the $O(4)$ linear sigma model in the chiral
limit. In relaxation time approximation we obtain the relaxation rates for
pions and sigma mesons. This case allows us to highlight the power of this the
dynamical renormalization group approach to study threshold effects on the
relaxation of resonances, in particular the crossover between two different
relaxation regimes as a function of the momentum of the resonance. This aspect
becomes phenomenological important in view of the recent studies by Hatsuda
\textit{et al}.~\cite{hatsuda} that reveal a dropping of the sigma mass near
the chiral phase transition and an enhancement of threshold effects with
potential observational consequences in ultrarelativistic heavy ion collisions.
In Sec.~\ref{sec:4.5} we discuss the issue of pinch singularities found in
calculations in the so-called real-time formalism of nonequilibrium field
theory and provide a resolution via the dynamical renormalization group. In
Sec.~\ref{sec:4.6} we summarize our results and discuss further implications.

\section{Self-Interacting Scalar Theory}\label{sec:4.2}
The starting point is the following Lagrangian for a hot scalar theory:
\begin{equation}
{\cal{L}}[\phi] = \frac{1}{2} \left(\partial_{\mu}\phi\right)^2 - \frac{1}{2}
m^2\phi^2-\frac{\lambda}{4!}\phi^4, \label{scalar:barelagrangian}
\end{equation}
where $m$ and $\lambda$ are the physical parameters defined in the
vacuum (or, at zero temperature). Since we focus on medium
(finite-temperature) effects, in the following discussion we
ignore the vacuum (zero-temperature) ultraviolet divergences that
can be absorbed into the renormalization of $m$ and $\lambda$.

The first step towards understanding the kinetic regime is the identification
of the \emph{microscopic} time scales in the problem. In the medium, the bare
particles are dressed by the interactions becoming ``quasiparticles''. One is
interested in describing the relaxation of these quasiparticles. Thus the
important microscopic time scales are those associated with the quasiparticles
and not the bare particles. If a kinetic equation is obtained in some
perturbative scheme, such a scheme should be in terms of the quasiparticles,
which already implies a resummation of the perturbative expansion accounting
for the medium effects. This resummed perturbative expansion is precisely the
rationale behind the hard thermal loop (HTL) program developed by Braaten and
Pisarski~\cite{htl} in hot gauge field theory and also behind the
self-consistent treatment in nonequilibrium scalar field theory proposed by
Lawrie~\cite{lawrie}.

Since we are interested in obtaining a kinetic equation for the quasiparticle
distribution function, the most natural initial state corresponds to an initial
density matrix $\rho$ prepared at time $t=t_0$ that is diagonal in the basis of
free quasiparticles number operator (defined below) but with nonequilibrium
quasiparticle distribution functions $n_\mathbf{k}(t_0)$ (see
Sec.~\ref{sec:2.2}). In this picture the width of the quasiparticles arises
from their interaction and is related to the relaxation rate of the
distribution function in the relaxation time approximation. This point will
become more clear below.

With such a choice of initial density matrix, perturbative expansions are
carried out with the following nonequilibrium free quasiparticle Wightman
functions (in momentum space):
\begin{eqnarray}
G_0^{>}(\mathbf{k},t,t')&=&\frac{i}{2\omega_\mathbf{k}}\left[[1+n_\mathbf{k}(t_0)]\,
e^{-i\omega_\mathbf{k}(t-t')}+n_\mathbf{k}(t_0)\,
e^{i\omega_\mathbf{k}(t- t')}
\right],\nonumber\\
G_0^{<}(\mathbf{k},t,t')&=&\frac{i}{2\omega_\mathbf{k}}\left[
n_\mathbf{k}(t_0)\, e^{-i\omega_\mathbf{k}(t- t')}
+[1+n_\mathbf{k}(t_0)]\,e^{i\omega_\mathbf{k}(t-t')}\right],
\label{fqp:gsmall}
\end{eqnarray}
where $\omega_\mathbf{k}=\sqrt{\mathbf{k}^2+m^2_\mathrm{eff}}$ is the energy of
the free quasiparticle and $m_\mathrm{eff}$ is the quasiparticle effective mass
to be determined self-consistently below. We note that it is easy to check the
relation
\begin{equation}
G_0^{>}(\mathbf{k},t,t') = G_0^{<}(\mathbf{k},t',t),
\label{gfrelation}
\end{equation}
which will be useful in following calculations.

In the self-interacting scalar field theory the HTL resummation can be
implemented by writing $m$ in the Lagrangian as
\begin{equation}
m^2= m^2_\mathrm{eff}+\delta m^2,
\label{counterterm}
\end{equation}
where $m_\mathrm{eff}$ is the effective quasiparticle mass which enters in the
effective propagators for free quasiparticles [see Eq.~(\ref{fqp:gsmall})], and
$\delta m^2$ is a counter term which will cancel a subset of Feynman diagrams
in the perturbative expansion and is considered part of the interaction
Lagrangian. The effective mass $m_\mathrm{eff}$ will in general depend on the
state of the system (e.g., distribution functions of the scalars) through the
medium effects. In particular, in thermal equilibrium $m_\mathrm{eff}$ depends
on the temperature and is called the thermal mass~\cite{book:lebellac}.

We note that one could also introduce counter term for the coupling constant
and proceed with a resummed perturbative expansion in terms of the in-medium
effective propagators and vertices, akin to those in the HTL program for hot
gauge theory~\cite{htl,book:lebellac}. However, to lowest order that we are
interested in only the mass counter term is relevant.

The implementation of this resummation scheme to lowest in perturbation theory
is achieved by requiring that the mass counter term $\delta m^2$ cancel the
one-loop tadpole contribution to the scalar self-energy~\cite{parwani}. This
leads to the following self-consistent gap equation for
$m^2_\mathrm{eff}$~\cite{boyanrelkin}
\begin{equation}
m^2_\mathrm{eff}= m^2 +\frac{\lambda}{2} \int \frac{d^3 k}{(2\pi)^3}
\frac{n_\mathbf{k}(t_0)}{\sqrt{k^2+m^2_\mathrm{eff}}}, \label{scalar:gapeq}
\end{equation}
where $n_\mathbf{k}(t_0)$ is the initial (nonequilibrium) distribution of the
quasiparticles at time $t=t_0$. We emphasize that $m_\mathrm{eff}$ depends on
the initial quasiparticle distribution functions. This observation will become
important later when we discuss the time evolution of the quasiparticle
distribution functions and therefore of the effective mass.

Assuming that the system is in thermal equilibrium, one can make an assessment
of the (thermal) effective mass $m_\mathrm{eff}(T)$. Upon substituting the
replacement of equilibrium distribution functions $n_\mathbf{k}(t_0)\rightarrow
n_B(\sqrt{k^2+m^2_\mathrm{eff}(T)})$ in Eq.~(\ref{scalar:gapeq}), one finds
that for $T \gg m_\mathrm{eff}(T)$ the solution of the gap equation is given
by~\cite{parwani,dolan}
\begin{equation}
m^2_\mathrm{eff}(T)= m^2+\frac{\lambda}{2}\left[\frac{T^2}{12}-
\frac{T}{4\pi}m_\mathrm{eff}(T)+ \mathcal{O}\left(m^2_\mathrm{eff}(T)\,\ln
\frac{m_\mathrm{eff}(T)}{T}\right)\right]. \label{scalar:thermalmass1}
\end{equation}
In particular, for $T\gg\sqrt{\lambda}\,T\gg m$ (justified in the weak coupling
limit), one can neglect the mass $m$ and obtains
\begin{equation}
m^2_\mathrm{eff}(T)=\frac{\lambda\,T^2}{24}\left[1+
\mathcal{O}(\lambda^{1/2})\right]. \label{scalar:thermalmass2}
\end{equation}
We note that this effective mass determines the important time scales in the
medium but is \emph{not} the position of the quasiparticle pole (or, strictly
speaking, resonance).

In the massless case, $m_\mathrm{eff}(T)$ serves as an infrared cutoff for the
loop integrals~\cite{parwani,elmfors}, hence the leading term of
Eq.~(\ref{scalar:thermalmass2}) provides the correct microscopic time scale at
high temperature. Furthermore, when the temperature is much larger than the
vacuum (zero temperature) mass $m$, the self-consistent resummation is needed
to incorporate the physically relevant microscopic time and length scales in
the perturbative expansion, which, in the weakly coupling limit, is given by
$t_\mathrm{mic} \sim 1/m_\mathrm{eff}(T)\sim 1/\sqrt{\lambda}T$.

\subsection{Quantum kinetic equation for quasiparticles}
Having explained the notion of quasiparticles in the medium and the microscopic
time scales associated with them, we now proceed to obtain the kinetic equation
for the quasiparticle distribution functions. Here we present an alternative to
the usual derivation in which correlation functions are written in terms of
relative and center-of-mass space-time coordinates and a Wigner transform in
the relative coordinates is performed. In this approach the kinetic equation is
obtained in a gradient expansion assuming that the dependence on center-of-mass
coordinates is weak. Here, however, we will not assume such a situation, but
instead analyze which are the relevant time scales over which a coarse graining
procedure must be implemented.

In the presence of the medium the quasiparticles will have an effective mass
$m_\mathrm{eff}$ resulting from medium effects, much in the same manner as the
temperature dependent thermal mass in the equilibrium situation discussed
above. This effective mass $m_\mathrm{eff}$ will be very different from the
vacuum mass $m$ and must be taken into account for the correct assessment of
the microscopic time scales. In the kinetic approach, this is achieved by
writing the Hamiltonian of the theory as $H=H_0 + H_\mathrm{int}$, where
\begin{equation}
H_0=\frac{1}{2} \int d^3x \left[\Pi^2 +({\bf \nabla}\phi)^2 + m^2_\mathrm{eff}
\phi^2\right],\quad H_\mathrm{int} =\int d^3x
\left[\frac{\lambda}{4!}\,\phi^4+\frac{\delta m^2}{2}\phi^2\right],
\label{scalar:hint}
\end{equation}
with $\Pi(\mathbf{x},t)=\partial\phi(\mathbf{x},t)/\partial t$ being the
conjugate momentum of $\phi(\mathbf{x},t)$. In the above definition, the free
Hamiltonian $H_0$ describes free quasiparticles of effective mass
$m_\mathrm{eff}$ and is diagonal in terms of free quasiparticle creation and
annihilation of operators $ a^{\dagger}(\mathbf{k})$ and $a(\mathbf{k})$,
respectively, and the mass counter term has been absorbed in the interaction
Hamiltonian $H_\mathrm{int}$. The pole mass of the quasiparticles will acquire
corrections in perturbation theory, but these will remain perturbatively. With
this definition, the lifetime of the quasiparticles will be a consequence of
interactions. In this manner, the nonequilibrium equivalent of the hard thermal
loops (in the sense that the distribution functions are non-thermal), which in
this theory amount to local (momentum independent) terms, have been absorbed in
the definition of the effective mass. This guarantees that the microscopic time
scales are explicit in the quasiparticle Hamiltonian.

As discussed above, we consider that the initial density matrix $\rho$ at time
$t=t_0$ is diagonal in the basis of free quasiparticles number operator, but
with nonequilibrium initial distribution functions $n_\mathbf{k}(t_0)$. Hence,
the Heisenberg quasiparticle field operators at time $t$ (in momentum space)
can be written as
\begin{equation}
\phi(\mathbf{k},t)=\frac{1}{\sqrt{2\omega_\mathbf{k}}}\left[a(\mathbf{k},t)
+a^{\dagger}(-\mathbf{k},t) \right],\quad \Pi(\mathbf{k},t)=i
\sqrt{\frac{\omega_\mathbf{k}}{2}}
\left[a^{\dagger}(-\mathbf{k},t)-a(\mathbf{k},t)\right],
\label{initialexpansion}
\end{equation}
where $ a^{\dagger}(\mathbf{k},t) $ and $ a(\mathbf{k},t) $, respectively, are
the creation and annihilation operators in the Heisenberg picture and
$\omega_\mathbf{k}=\sqrt{\mathbf{k}^2+m^2_\mathrm{eff}}$ as defined in
Eq.~(\ref{fqp:gsmall}) above. The expectation value of Heisenberg quasiparticle
number operators $n_\mathbf{k}(t)$ is interpreted as the time dependent
quasiparticle distribution function and can be expressed in terms of the field
$\phi(\mathbf{k},t)$ and the conjugate momentum $\Pi(\mathbf{k},t)$ as follows
\begin{eqnarray}
n_\mathbf{k}(t) &=&
\langle a^{\dagger}(\mathbf{k},t)\,a(\mathbf{k},t)\rangle \nonumber\\
&=& {1 \over 2\omega_\mathbf{k} }\left[\left\langle
\Pi(\mathbf{k},t)\,\Pi(-\mathbf{k},t)\right\rangle + \omega^2_\mathbf{k}
\left\langle\phi(\mathbf{k},t)\,\phi(-\mathbf{k},t)\right\rangle\right]
-\frac{1}{2}, \label{ocupation}
\end{eqnarray}
where the bracket $\langle\,\cdot\,\rangle$ denotes the nonequilibrium
expectation value with the initial (Gaussian) density matrix specified by the
initial distribution functions $n_\mathbf{k}(t_0)$.

The interaction Hamiltonian $H_\mathrm{int}$ in momentum space is given by
\begin{eqnarray}
H_\mathrm{int}&=&\frac{\lambda}{4!}\int\prod_{i=1}^{4}\frac{d^3
q_i}{(2\pi)^3}\,\phi({\mathbf{q}_i},t)\,(2\pi)^3\,
\delta^3(\mathbf{q}_1 +\mathbf{q}_2 +\mathbf{q}_3 +\mathbf{q}_4) \nonumber\\
&&+\,\frac{\delta m^2}{2}\int\frac{d^3
q}{(2\pi)^3}\,\phi(\mathbf{q},t)\phi(-\mathbf{q},t).\label{intTlarge}
\end{eqnarray}
Taking the derivative of $n_\mathbf{k} (t)$ with respect to time and using the
Heisenberg field equations, one finds
\begin{eqnarray}
\frac{d}{dt}n_\mathbf{k}(t) &=&-\frac{\lambda}{12\omega_\mathbf{k}}\,\Big[
\left\langle[\phi^3(\mathbf{k},t)]\,\Pi(-\mathbf{k},t)\right\rangle+
\left\langle\Pi(\mathbf{k},t)\,[\phi^3(\mathbf{k},t)]
\right\rangle\Big]\nonumber \\
&&+\,\frac{\delta m^2}{2\omega_\mathbf{k}}\,\Big[\left\langle\phi(\mathbf{k},t)
\,\Pi(-\mathbf{k},t)\right\rangle +\left\langle\Pi(\mathbf{k},t)
\,\phi(-\mathbf{k},t)\right\rangle\Big],\label{nkdot}
\end{eqnarray}
where we use the compact notation
\begin{equation}
[\phi^3(\mathbf{k},t)]\equiv\int \prod_{i=1}^{3} \frac{d^3q_i}{(2\pi)^3}\,
\phi({\mathbf{q}_i},t)\,\delta^3
(\mathbf{k}-\mathbf{q}_1-\mathbf{q}_2-\mathbf{q}_3).
\end{equation}
In a perturbative expansion care is needed to handle the conjugate momentum and
the scalar field at the same time because of the Schwinger terms. This
ambiguity is avoided by noticing that
\begin{eqnarray}
\left\langle \Pi(\mathbf{k},t)\,[\phi^3(-\mathbf{k},t)] \right\rangle &=&
\lim_{t' \rightarrow t} {\partial \over \partial t'} {\rm
Tr}\left[[\phi^3(-\mathbf{k},t)]^+ \,\rho\,\phi^-(\mathbf{k},t')
\right]\nonumber\\
&=&\lim_{t' \rightarrow t} {\partial \over \partial t'}\left\langle
[\phi^3(-\mathbf{k},t)]^+ \,\phi^-(\mathbf{k},t')\right\rangle,
\end{eqnarray}
where we used the cyclic property of the trace and, as usual, the $\pm$
superscripts for the fields refer to field defined in the forward ($+$) and
backward ($-$) time branch in the CTP formalism.

We now use the canonical commutation relation between $\Pi$ and $\phi$ and
define the mass counterterm $\delta m^2=\lambda\Delta/6$ to write the above
expression as
\begin{eqnarray}
\frac{d}{dt}n_\mathbf{k}(t)  &=& -\frac{\lambda}{12\omega_\mathbf{k}}
\bigg[\lim_{t'\to t}\frac{\partial}{\partial t'}
\bigg(2\left\langle[\phi^3(\mathbf{k},t)]^+\,
\phi^-(-\mathbf{k},t')\right\rangle+\Delta
\left[\left\langle\phi^+(\mathbf{k},t)\phi^-(-\mathbf{k},t')
\right\rangle\right. \nonumber\\
&&+\left.\left\langle\phi^+(\mathbf{k},t')\phi^-(-\mathbf{k},t)
\right\rangle\right]\bigg) +3i\int \frac{d^3
q}{(2\pi)^3}\left\langle\phi^+(\mathbf{q},t)
\phi^-(-\mathbf{q},t)\right\rangle\bigg].\label{nkdot1}
\end{eqnarray}
The expectation values on the right-hand side of Eq.~(\ref{nkdot1}) can be
obtained perturbatively in weak coupling expansion in $\lambda$. Such a
perturbative expansion is carried out in terms of the nonequilibrium Feynman
rules with the free quasiparticle Wightman functions given by
Eqs.~(\ref{fqp:gsmall}).

Before proceeding further, an important point to notice is that the free
propagators entering in the perturbation expansion are the resummed
propagators, which include the proper microscopic scales as the contribution of
the hard thermal loops have been incorporated consistently by summing the
tadpole diagrams in the scalar self-energy. As a result, the terms with
$\Delta$ are required to cancel the tadpole contribution in Eq.~(\ref{nkdot1})
to all orders in perturbation theory. Thus after $\Delta$ is properly chosen in
order to cancel the tadpole diagrams in the scalar self-energy, from the
formidable expression given in Eq.~(\ref{nkdot1}) only the first term remains
(with the understanding that no tadpole diagrams being included). Physically,
this requirement guarantees that the mass in the propagators is the effective
mass that includes the microscopic time scales.

To lowest order the condition that the tadpole contributions are canceled leads
to the following condition on $\Delta$
\begin{equation}
\Delta = -3\int\frac{d^3 k}{(2\pi)^3 \omega_\mathbf{k}}\,n_\mathbf{k}(t_0),
\end{equation}
which in turn entails that the effective mass $m_\mathrm{eff}$ is determined
self-consistently by the gap equation given in Eq.~(\ref{scalar:gapeq}). We
highlight that the requirement that the term proportional to $\Delta$ in
Eq.~(\ref{nkdot1}) cancels the tadpole contributions is equivalent to the HTL
resummation in the equilibrium case~\cite{parwani,dolan} and makes explicit
that $m_\mathrm{eff}$ is a functional of the \emph{initial} nonequilibrium
quasiparticle distribution functions $n_\mathbf{k}(t_0)$.

\begin{figure}[t]
\begin{center}
\includegraphics[width=3.5truein,keepaspectratio=true]{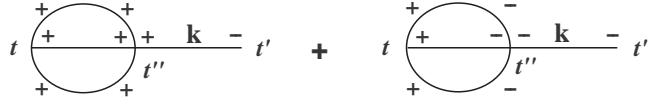}
\caption{The Feynman diagrams that contribute to the quantum kinetic equation
for a self-interacting scalar theory up to two-loop order.}
\label{fig:scalarloops}
\end{center}
\end{figure}

At order $\lambda$, the right-hand side of Eq.~(\ref{nkdot1}) vanishes
identically. This is a consequence of the fact that both the free quasiparticle
Hamiltonian $H_0$ and the initial density matrix $\rho$ are diagonal in the
basis of free quasiparticles number operator. Hence, the relaxation of
quasiparticle distribution functions is solely due to the interaction between
quasiparticles.

To order $\lambda^2$, we find that the final form of the kinetic equation is
given by
\begin{equation}
\frac{d}{dt}n_\mathbf{k}(t)= -\frac{\lambda}{6\omega_\mathbf{k}}\,\lim_{t'\to
t}\frac{\partial} {\partial t'} \left\langle
[\phi^3(\mathbf{k},t)]^+\,\phi^-(-\mathbf{k},t') \right\rangle,\label{ndotequa}
\end{equation}
with the understanding that \emph{no tadpole diagrams} being included as they
are automatically canceled by the terms containing $\Delta$ and the last term
in Eq.~(\ref{nkdot1}). To two-loop order, the Feynman diagrams that contribute
to Eq.~(\ref{ndotequa}) is displayed in Figure~\ref{fig:scalarloops} and the
resultant time evolution of the quasiparticle distribution function is given by
\begin{eqnarray}
\frac{d}{dt}n_\mathbf{k}(t) &=& \frac{\lambda^2}{6\omega_\mathbf{k}}\int
\prod_{i=1}^{3}\frac{d^3 q_i}{(2\pi)^3
2\omega_{\mathbf{q}_i}}\int^t_{t_0}dt''\,(2\pi)^3
\delta^3(\mathbf{k}-\mathbf{q}_1-\mathbf{q}_2-\mathbf{q}_3)\nonumber\\
&& \times \Big[\mathcal{N}_1(t_0)\cos[\Delta\omega_1(t-t'')]+
3\mathcal{N}_2(t_0)\cos[\Delta\omega_2(t-t'')] \nonumber \\
&& +\,3\mathcal{N}_3(t_0) \cos[\Delta\omega_3(t-t'')]+\mathcal{N}_4(t_0)
\cos[\Delta\omega_4(t-t'')]\Big], \label{scalar:perturbkineq}
\end{eqnarray}
where
\begin{eqnarray}
&&\Delta\omega_1=\omega_\mathbf{k}+
\omega_{\mathbf{q}_1}+\omega_{\mathbf{q}_2}+ \omega_{\mathbf{q}_3},\quad
\Delta\omega_2=\omega_\mathbf{k}+\omega_{\mathbf{q}_1}+
\omega_{\mathbf{q}_2}-\omega_{\mathbf{q}_3},\nonumber\\
&&\Delta\omega_3=\omega_\mathbf{k}-\omega_{\mathbf{q}_1}-
\omega_{\mathbf{q}_2}+\omega_{\mathbf{q}_3},\quad
\Delta\omega_4=\omega_\mathbf{k}-\omega_{\mathbf{q}_1}-
\omega_{\mathbf{q}_2}-\omega_{\mathbf{q}_3},
\end{eqnarray}
and
\begin{eqnarray}
\mathcal{N}_1(t) &=& [1+ n_\mathbf{k}(t)][1+ n_{\mathbf{q}_1}(t)] [1+
n_{\mathbf{q}_2}(t)][1+ n_{\mathbf{q}_3}(t)]- n_\mathbf{k}(t) \,
n_{\mathbf{q}_1}(t) \, n_{\mathbf{q}_2}(t) \, n_{\mathbf{q}_3}(t),\nonumber\\
\mathcal{N}_2(t) &=& [1+ n_\mathbf{k}(t)][1+ n_{\mathbf{q}_1} (t)] [1+
n_{\mathbf{q}_2}(t)] \, n_{\mathbf{q}_3} (t)- n_\mathbf{k}(t) \,
n_{\mathbf{q}_1}(t) \, n_{\mathbf{q}_2}(t) \,
[1+n_{\mathbf{q}_3}(t)],\nonumber\\
\mathcal{N}_3(t) &=& [1+ n_\mathbf{k} (t)] \, n_{\mathbf{q}_1}(t)  \,
n_{\mathbf{q}_2}(t) \, [1+ n_{\mathbf{q}_3} (t)] - n_\mathbf{k}(t)
\,[1+n_{\mathbf{q}_1}(t)][1+ n_{\mathbf{q}_2}(t)]
\,n_{\mathbf{q}_3}(t),\nonumber\\
\mathcal{N}_4(t) &=& [1+ n_\mathbf{k} (t) ] \, n_{\mathbf{q}_1}(t) \,
n_{\mathbf{q}_2}(t)\,n_{\mathbf{q}_3}(t)- n_\mathbf{k}(t) \, [1+
n_{\mathbf{q}_1}(t)] [1+
n_{\mathbf{q}_2}(t)][1+n_{\mathbf{q}_3}(t)].\nonumber\\
\label{scalar:ocupanumber}
\end{eqnarray}
The kinetic equation (\ref{scalar:perturbkineq}) is retarded and causal. The
different contributions have a physical interpretation in terms of the `gain
minus loss' processes in the plasma. The first term describes the creation of
four particles minus the destruction of four particles in the plasma, the
second and fourth terms describe the creation of three particles and
destruction of one minus destruction of three and creation of one, the third
term is the \emph{scattering} of two particles off two particles and is the
usual Boltzmann term.

As will be discussed in detail below, such a perturbative expansion will be
meaningful for times $t\ll t_\mathrm{kin}$, where $t_\mathrm{kin}$ is the
kinetic time scale for the nonequilibrium distribution functions. For small
enough coupling we expect that $t_\mathrm{kin}$ will be large enough such that
there is a wide separation between the microscopic and the kinetic time scales
that will warrant such an approximation (see below).

Since the propagators entering in the perturbative expansion of the evolution
equation are in terms of the quasiparticle distribution functions at the
initial time $t_0$, the time integration in Eq.~(\ref{scalar:ocupanumber}) can
be done straightforwardly leading to the following equation:
\begin{equation}
\frac{d}{dt}n_\mathbf{k}(t)  = \frac{\lambda^2}{3}\int_{-\infty}^{+\infty}
d\omega\,\mathcal{R}[\omega,\mathbf{k};\mathcal{N}_j(t_0)]\,
\frac{\sin[(\omega-\omega_\mathbf{k})(t-t_0)]}{\pi(\omega-\omega_\mathbf{k})},
\label{ke:scalar}
\end{equation}
where
\begin{eqnarray}
\mathcal{R}[\omega,\mathbf{k};\mathcal{N}_j(t_0)]&=&
\frac{\pi}{2\omega_\mathbf{k}}\,\int \prod_{i=1}^{3}\frac{d^3 q_i}{(2\pi)^3
2\omega_{\mathbf{q}_i}}\,(2\pi)^3
\delta^3(\mathbf{k}-\mathbf{q}_1-\mathbf{q}_2-\mathbf{q}_3)
\Big[\mathcal{N}_1(t_0)\,\delta(\Delta\omega_1)\nonumber\\
&&+\,3\,\mathcal{N}_2(t_0)\,\delta(\Delta\omega_2)+
3\,\mathcal{N}_3(t_0)\,\delta(\Delta\omega_3)+
\mathcal{N}_4(t_0)\,\delta(\Delta\omega_4)\Big], \label{bigR}
\end{eqnarray}
with $j=1,\ldots,4$.

We are now ready to solve the kinetic equation derived above. Since $
\mathcal{R}[\omega,\mathbf{k};\mathcal{N}_j(t_0)] $ is fixed at initial time $
t_0 $, Eq.~(\ref{ke:scalar}) can be solved by direct integration over $ t $,
thus leading to
\begin{equation}
n_\mathbf{k}(t)=n_\mathbf{k}(t_0)+\frac{\lambda^2}{3}\int_{-\infty}^{+\infty}
d\omega\,\mathcal{R}[\omega,\mathbf{k};\mathcal{N}_j(t_0)]
\frac{1-\cos[(\omega-\omega_\mathbf{k})(t-t_0)]}
{\pi(\omega-\omega_\mathbf{k})^2}. \label{integrated}
\end{equation}
It is noted that the above expression gives the time evolution of the
quasiparticle distribution function to lowest order in perturbation theory, but
only for early times. To make this statement more precise, let us consider the
following approximation
\begin{equation}
\lim_{t-t_0\to\infty}\frac{1-\cos[(\omega-\omega_\mathbf{k})(t-t_0)]}
{\pi(\omega-\omega_\mathbf{k})^2} = (t-t_0) \,
\delta(\omega-\omega_\mathbf{k}). \label{FGR}
\end{equation}
Thus, provided that $\mathcal{R}[\omega,\mathbf{k};\mathcal{N}_j(t_0)]$ is
finite at $\omega=\omega_\mathbf{k}$, one finds that $n_\mathbf{k}(t)$ at large
times is given by
\begin{eqnarray}
n_\mathbf{k}(t)&=& n_\mathbf{k}(t_0)+\frac{\lambda^2}{3}\,
\mathcal{R}[\omega_\mathbf{k},\mathbf{k};\mathcal{N}_j(t_0)]\,(t-t_0) +
\mbox{nonsecular terms}, \label{scalar:perturb}
\end{eqnarray}
where the \emph{nonsecular terms} denotes terms that are bound at all times. In
the above expression, the term that grows (linearly) with time is referred to
as a \emph{secular term}. The approximation above, replacing the oscillatory
terms with resonant denominators by $t\,\delta(\omega-\omega_\mathbf{k})$ is
the same as that invoked in ordinary time-dependent perturbation theory leading
to Fermi's golden rule in elementary time-dependent perturbation theory .

Clearly, the presence of secular terms in time restricts the validity of the
perturbative expansion to a time interval $ t-t_0\ll t_\mathrm{kin}$ with
\begin{equation}
t_\mathrm{kin}(k) \approx \frac{3\,n_\mathbf{k}(t_0)}{\lambda^2 \,
\mathcal{R}[\omega_\mathbf{k},\mathbf{k};\mathcal{N}_j(t_0)]}.
\label{timeinter}
\end{equation}
Since the time scales in the integral in Eq.~(\ref{integrated}) are of the
order of or shorter than $t_\mathrm{mic} \sim 1/m_\mathrm{eff}$ the asymptotic
form given by Eq.~(\ref{scalar:perturb}) is valid for $t-t_0 \gg
t_\mathrm{mic}$. Therefore for weak coupling there is a regime of
\emph{intermediate asymptotics} in time defined by
\begin{equation}
t_\mathrm{mic} \ll t-t_0 \ll t_\mathrm{kin}(k), \label{interasym}
\end{equation}
such that perturbation theory is \emph{valid} and the corrections to the
quasiparticle distribution function is dominated by the secular term.

We note two important features of this analysis:

(i) In the intermediate asymptotic regime (\ref{interasym}) the only
\emph{explicit} dependence on the initial time $t_0$ is in the secular term,
since $\mathcal{R}[\omega_\mathbf{k},\mathbf{k};\mathcal{N}_j(t_0)]$ depends on
$t_0$ only implicitly through the initial distribution functions. These
observations will become important for the analysis that follows below.

(ii) The function(al)
$\mathcal{R}[\omega_\mathbf{k},\mathbf{k};\mathcal{N}_j(t_0)]$ given by
Eq.~(\ref{bigR}) \emph{vanishes} if the initial distribution functions are the
equilibrium ones as a result of the energy conservation delta functions and the
equilibrium relation $1+n_B(\omega) =e^{\beta\omega} n_B(\omega)$. In this case
there are no secular terms in the perturbative expansion.

To highlight the significance of the second point above in a manner that will
allow us to establish contact with the issue of pinch singularities in a later
section, we note that the secular term in Eq.~(\ref{scalar:perturb})
corresponds to the net change of quasiparticles distribution function in the
time interval $t-t_0$. To see this more explicitly, we note that
$\mathcal{R}[\omega_\mathbf{k},\mathbf{k};\mathcal{N}_j(t_0)]$ can be rewritten
as
\begin{equation}
\frac{\lambda^2}{3}\mathcal{R}[\omega_\mathbf{k},\mathbf{k};\mathcal{N}_j(t_0)]=
\frac{i}{2\omega_\mathbf{k}}
\Big[[1+n_\mathbf{k}(t_0)]\,\Sigma^{<}(\omega_\mathbf{k},\mathbf{k})-
n_\mathbf{k}(t_0)\,\Sigma^{>}(\omega_\mathbf{k},\mathbf{k})\Big],
\label{secularpart}
\end{equation}
where $\Sigma^{>}(\omega_\mathbf{k},\mathbf{k})-
\Sigma^{<}(\omega_\mathbf{k},\mathbf{k})=
2\,i\,\mathrm{Im}\Sigma_R(\omega_\mathbf{k},\mathbf{k})$ with
$\Sigma_R(\omega_\mathbf{k},\mathbf{k})$ the on-shell retarded
self-energy calculated to two-loop order in terms of the initial
distribution functions $n_\mathbf{k}(t_0)$~\cite{boyanrelkin} .
Thus, the absence of secular term for a system in thermal
equilibrium [for which $n_\mathbf{k}(t_0)=
n_B(\omega_\mathbf{k})$] is a consequence of the
Kubo-Martin-Schwinger (KMS) condition for the equilibrium
self-energy
\begin{equation}
\Sigma^{>}(\omega,\mathbf{k})=e^{\beta\omega} \Sigma^{<}(\omega,\mathbf{k}).
\end{equation}
Furthermore, it can be recognized that the first (second) term in
Eq.~(\ref{secularpart}), corresponds to the ``gain'' (``loss'') part in the
usual Boltzmann type collision term. Hence, one can readily interpret
$\lambda^2\mathcal{R}[\omega_\mathbf{k},\mathbf{k};\mathcal{N}_j(t_0)]/3$ as
the \emph{net production rate of quasiparticles per unit time}.

\subsection{Dynamical renormalization group equation}
Secular divergences are an ubiquitous feature in the perturbative
solution of differential equations with oscillatory behavior. A
resummation method based on the idea of the renormalization group
(RG) that improves the perturbative solutions of differential
equations was introduced by Goldenfeld and
collaborators~\cite{goldenfeld} to study pattern formation in
condensed matter systems.\footnote{For a pedagogical introduction
to the renormalization group method, see Appendix~\ref{chap:app}.}
As explained in details in Refs.~\cite{goldenfeld,kunihiro}, this
powerful method not only allows a consistent resummation of the
secular terms in the perturbative expansion but also provides a
consistent reduction of the dynamics to the slow degrees of
freedom. In a recent development, this method has been generalized
to the realm of nonequilibrium quantum field theory as a dynamical
renormalization group to resum the perturbative series for the
real time evolution of nonequilibrium expectation
values~\cite{boyanrgir,salgado}. This generalization is a major
step since in nonequilibrium systems the equations of motion for
expectation values are nonlocal and, as will be shown below,
require a resummation of Feynman diagrams.

Having recognized that a time scale emerges in the perturbative expansion
(\ref{scalar:perturb}) due to the presence of the secular term, we now
implement the dynamical renormalization group resummation of secular
divergences to improve the perturbative expansion following the formulation
presented in Ref.~\cite{boyanrgir}.

To lowest order, this resummation is achieved by introducing the
``renormalized'' initial distribution functions $n_\mathbf{k}(\tau)$ which are
related to the ``bare'' initial distribution function $n_\mathbf{k}(t_0)$ via a
renormalization constant $\mathcal{Z}_\mathbf{k}(\tau,t_0)$ given by
\begin{equation}
n_\mathbf{k}(t_0)= \mathcal{Z}_\mathbf{k}(\tau,t_0)\,n_\mathbf{k}(\tau),\qquad
\mathcal{Z}_\mathbf{k}(\tau,t_0)=1 + \frac{\lambda^2}{3}\,
z_\mathbf{k}(\tau,t_0)+\mathcal{O}(\lambda^4), \label{scalar:renormal}
\end{equation}
where $\tau$ is an arbitrary ``renormalization scale'' and
$z_\mathbf{k}(\tau,t_0)$ will be chosen to cancel the secular term on a time
scale $\tau$. Substitute Eq.~(\ref{scalar:renormal}) into
Eq.~(\ref{scalar:perturb}), to order $\lambda^2$ we obtain
\begin{equation}
n_\mathbf{k}(t)=n_\mathbf{k}(\tau)+\frac{\lambda^2}{3} \,
\left[z_\mathbf{k}(\tau,t_0)n_\mathbf{k}(\tau)+(t-t_0)\,
\mathcal{R}[\omega_\mathbf{k},\mathbf{k};\mathcal{N}_j(\tau)]\,\right]
+\mathcal{O}(\lambda^4).
\end{equation}
Thus, to this order, the choice
\begin{equation}
z_\mathbf{k}(\tau,t_0)=-(\tau-t_0)
\frac{\mathcal{R}[\omega_\mathbf{k},\mathbf{k};\mathcal{N}_j(\tau)]}
{n_\mathbf{k}(\tau)}
\end{equation}
leads to
\begin{equation}
n_\mathbf{k}(t)=n_\mathbf{k}(\tau)+\frac{\lambda^2}{3}\,(t-\tau)\,
\mathcal{R}[\omega_\mathbf{k},\mathbf{k};\mathcal{N}_j(\tau)]
+\mathcal{O}(\lambda^4). \label{scalar:rensol}
\end{equation}
Whereas the original perturbative solution was only valid for times such that
the contribution from the secular term remains very small compared to the
initial distribution function at time $t_0$, the renormalized solution
Eq.~(\ref{scalar:rensol}) is valid for time intervals $t-\tau$ such that the
secular term remains small, thus by choosing $\tau$ arbitrarily close to $t$ we
have improved the perturbative expansion~\cite{goldenfeld,kunihiro}.

To find the dependence of $n_\mathbf{k}(\tau)$ on $\tau$, we make use of the
fact that the ``physical'' distribution function $n_\mathbf{k}(t)$ does not
depend on the arbitrary scale $\tau$ as a change in the renormalization point
$\tau$ is compensated by a change in the ``renormalized'' initial distribution
functions $n_\mathbf{k}(\tau)$. This $\tau$-independence leads to the
\emph{dynamical renormalization group equation}, which to lowest order is given
by
\begin{equation}
\frac{d}{d\tau} n_\mathbf{k}(\tau)-\frac{\lambda^2}{3}\,
\mathcal{R}[\omega_\mathbf{k},\mathbf{k};\mathcal{N}_j(\tau)]=0.
\label{scalar:drge}
\end{equation}

This renormalization of the distribution function  also affects the effective
mass of the quasiparticles since $m^2_\mathrm{eff}$ is determined from the
self-consistent equation (\ref{scalar:gapeq}), which is a consequence of the
tadpole cancelation consistently in perturbation theory. Since the effective
mass is a functional of the distribution function it will be renormalized
consistently. This is physically correct since the in-medium effective masses
will change under the time evolution of the distribution functions. Choosing
the arbitrary scale $\tau$ to coincide with the time $t$ in
Eq.~(\ref{scalar:drge}), we obtain the \emph{resummed} quantum kinetic
equation:
\begin{eqnarray}
\frac{d}{dt}n_\mathbf{k}(t) &=& \frac{\pi\lambda^2}{6\omega_\mathbf{k}}
\int\frac{d^3 q_1}{(2 \pi)^3 2\omega_{\mathbf{q}_1}} \frac{d^3 q_2}{(2 \pi)^3
2\omega_{\mathbf{q}_2}} \frac{d^3 q_3}{(2 \pi)^3 2\omega_{\mathbf{q}_3}}
(2\pi)^3 \delta^3 (\mathbf{k}-\mathbf{q}_1 -\mathbf{q}_2-
\mathbf{q}_3)\nonumber \\
&& \times \Big[ \delta( \omega_\mathbf{k} + \omega_{\mathbf{q}_1} +
\omega_{\mathbf{q}_2}+ \omega_{\mathbf{q}_3})\,\mathcal{N}_1(t)+
3\,\delta(\omega_\mathbf{k} +\omega_{\mathbf{q}_1}+
\omega_{\mathbf{q}_2}-\omega_{\mathbf{q}_3})\,\mathcal{N}_2(t)\nonumber \\
&&+\,3\,\delta(\omega_\mathbf{k} -\omega_{\mathbf{q}_1}-\omega_{\mathbf{q}_2}
+\omega_{\mathbf{q}_3})\,\mathcal{N}_3(t) +\delta(\omega_\mathbf{k}
-\omega_{\mathbf{q}_1}-\omega_{\mathbf{q}_2}-\omega_{\mathbf{q}_3})\,
\mathcal{N}_4(t)\Big],\nonumber\\
\label{resucinetica}
\end{eqnarray}
where the $\mathcal{N}_j(t)$ are given in Eq.~(\ref{scalar:ocupanumber}).

To avoid cluttering of notation in the above expression, we have not made
explicit the fact that the frequencies
$\omega_\mathbf{q}=\sqrt{\mathbf{q}^2+m_\mathrm{eff}^2}$ \emph{depend on time}
through the time dependence of $m_\mathrm{eff}$, which is in turn determined by
the time dependence of the quasiparticle distribution function. Indeed, the
dynamical renormalization group resummation leads at once to the conclusion
that the cancelation of tadpole terms by a proper choice of $\Delta$ requires
that at every time $t$ the effective mass is the solution of the
\emph{time-dependent} gap equation
\begin{equation}
m^2_\mathrm{eff}(t) = m^2 + \frac{\lambda}{2}\int\frac{d^3q}{(2\pi)^3}
\frac{n_\mathbf{q}(t)}{\sqrt{q^2+m^2_\mathrm{eff}(t)}}, \label{selfconmassoft}
\end{equation}
where $n_\mathbf{q}(t)$ is the solution of the kinetic equation
(\ref{resucinetica}). Thus, the \emph{full} quantum kinetic equation that
includes a nonequilibrium generalization of the hard thermal loop resummation
in this scalar theory is given by Eq.~(\ref{resucinetica}) with the replacement
$\omega_\mathbf{q}\rightarrow\omega_\mathbf{q}(t)$, where
$\omega_\mathbf{q}(t)$ are the solutions of the self-consistent time-dependent
gap equation (\ref{selfconmassoft}).

The quantum kinetic equation (\ref{resucinetica}) is therefore \emph{more
general} than the familiar semiclassical Boltzmann equation for a scalar field
theory in that it includes the proper in-medium modifications of the
quasiparticle masses. This approach provides an alternative derivation of the
self-consistent method proposed in Ref.~\cite{lawrie}.

It is now evident that the dynamical renormalization group systematically
resums the secular terms and the corresponding dynamical renormalization group
equation extracts the \emph{slow evolution} of the nonequilibrium system.

For small departures from equilibrium the time scales for relaxation can be
obtained by linearizing the kinetic equation (\ref{resucinetica}) around the
equilibrium solution at $ t=t_0 $. This is the relaxation time approximation
which assumes that the quasiparticle distribution function for a fixed mode of
momentum $\mathbf{k}$ is perturbed slightly off equilibrium such that
$n_\mathbf{k}(t_0)=n_B(\omega_\mathbf{k})+\delta n_\mathbf{k}(t_0)$, while all
the other modes remain in equilibrium.

Recognizing that only the delta function $\delta(\omega_\mathbf{k}
-\omega_{\mathbf{q}_1}-\omega_{\mathbf{q}_2} +\omega_{\mathbf{q}_3})$ that
multiplies the scattering term $\mathcal{N}_3(t)$ in Eq.~(\ref{resucinetica})
is fulfilled, one finds that the linearized kinetic equation reads
\begin{equation}
\frac{d}{dt}\delta n_\mathbf{k}(t)  = -\Gamma(k)\, \delta
n_\mathbf{k}(t),\label{linearboltzt}
\end{equation}
where $\Gamma(k)$ is the scalar interaction rate
\begin{eqnarray}
\Gamma(k)&=&\frac{\lambda^2\pi}{2\omega_\mathbf{k}} \int \frac{d^3 q_1}{(2
\pi)^3 2\omega_{\mathbf{q}_1}}\frac{d^3 q_2}{(2 \pi)^3 2\omega_{\mathbf{q}_2}}
\frac{d^3 q_3}{(2 \pi)^3 2\omega_{\mathbf{q}_3}}
(2\pi)^3 \delta^3 (\mathbf{k}-\mathbf{q}_1
-\mathbf{q}_2- \mathbf{q}_3)\nonumber\\
&&\times\,\delta(\omega_\mathbf{k}-\omega_{\mathbf{q}_1}-\omega_{\mathbf{q}_2}+
\omega_{\mathbf{q}_3}) \Big[ [1+n_B(\omega_{\mathbf{q}_1})]\,[1+
n_B(\omega_{\mathbf{q}_2})]\, n_B(\omega_{\mathbf{q}_3})\nonumber \\
&&-\,n_B(\omega_{\mathbf{q}_1})\,n_B(\omega_{\mathbf{q}_2})
\,[1+n_B(\omega_{\mathbf{q}_3})]\Big],
\end{eqnarray}
where $\omega_\mathbf{k}=\sqrt{\mathbf{k}^2+m_\mathrm{eff}^2(T)}$. Solving
Eq.~(\ref{linearboltzt}) with the initial distribution $\delta
n_\mathbf{k}(t_0)$, one finds that the quasiparticle distribution function in
the relaxation time approximation evolves in time in the following manner
\begin{equation}
\delta n_\mathbf{k}(t)=\delta n_\mathbf{k}(t_0)\,e^{-\Gamma(k)(t-t_0)},
\label{expsol}
\end{equation}
which gives the time scales for relaxation close to thermal equilibrium.

In the case of soft momentum ($ k\ll m_\mathrm{eff}(T)\ll T$) and high
temperature ($\lambda T^2\gg m^2$), one obtains~\cite{boyanrelkin}
\begin{equation}
t_\mathrm{kin}(k\approx 0)\equiv \frac{1}{\Gamma({k\approx 0})}\approx
\frac{32\sqrt{24\pi}}{\lambda^{3/2}~T}. \label{trelax}
\end{equation}
For very weak coupling (as we have assumed), the kinetic time scale is much
larger that the microscopic one $t_\mathrm{mic}\sim 1/m_\mathrm{eff} \sim
1/\sqrt{\lambda}\,T$, since $t_\mathrm{kin}/t_\mathrm{mic}\sim 1/\lambda\gg 1$.
This verifies the assumption of the separation of microscopic and kinetic time
scales in the weak coupling limit.

\section{Comparison to the Euclidean Renormalization Group}\label{sec:4.3}
In the section we establish a very close relationship between the
dynamical renormalization group discussed above and the usual
renormalization group in Euclidean field theory by demonstrating
that the rationale behind the two methods are identical and that
there exists an one-to-one correspondence between them.

First we review the idea of the Euclidean renormalization group in
zero-temperature field theory. Let us consider a (massless) scalar
field theory in four dimensions with an upper momentum cutoff
$\Lambda$ described by the Lagrangian density
\begin{equation}
{\cal{L}}[\phi] = \frac{1}{2}
\left(\partial_{\mu}\phi\right)^2-\frac{\lambda_0}{4!}\phi^4,
\label{barelagrangian}
\end{equation}
where $\lambda_0$ is the bare coupling. As an example, we calculate the two
particles to two particles scattering amplitude in perturbation theory. The
one-particle-irreducible (1PI) four-point function (i.e., two particles to two
particles scattering amplitude) at an off-shell symmetric point to one-loop
order in Euclidean space is given by
\begin{equation}
\Gamma^{(4)}(p,p,p,p)= \lambda_0-\frac{3}{2}\lambda_0^2\,\ln\frac{\Lambda}{p}+
\mathcal{O}(\lambda_0^3),\label{gamma4}
\end{equation}
where $p$ is the Euclidean four-momentum. Clearly perturbation theory breaks
down for $\Lambda/p \gtrsim e^{1/\lambda_0^2} $.

Let us introduce the renormalized coupling constant at a momentum scale
$\kappa$ as usual as
\begin{equation}
\lambda_0 = \mathcal{Z}_{\lambda}(\kappa)\,\lambda(\kappa),\quad
\mathcal{Z}_{\lambda}(\kappa) = 1 + z_1(\kappa)\,\lambda(\kappa)
+\mathcal{O}(\lambda^3),
\end{equation}
and choose $z_1(\kappa)$ to cancel the logarithmic divergence at an arbitrary
renormalization scale $\kappa$. Then in terms of $\lambda(\kappa)$ the
scattering amplitude $\Gamma^{(4)}(p,p,p,p)$ becomes
\begin{equation}
\Gamma^{(4)}(p,p,p,p)=\lambda(\kappa)+\frac{3}{2}\lambda^2(\kappa)\,
\ln\frac{p}{\kappa}+\mathcal{O}(\lambda^3),\label{RGFT}
\end{equation}
with $\Gamma^{(4)}(\kappa,\kappa,\kappa,\kappa)=\lambda(\kappa)$. The
scattering amplitude does not depend on the arbitrary renormalization scale $
\kappa $ and this independence implies $ \kappa\,\partial\Gamma^{(4)}(p,p,p,p)
/\partial \kappa =0 $, which in turn to lowest order leads to the
\emph{renormalization group equation}
\begin{equation}
\kappa\,\frac{d\lambda(\kappa)}{d\kappa} = \beta(\lambda),\qquad \beta(\lambda)
= \frac{3}{2}\lambda^2(\kappa) + \mathcal{O}(\lambda^3), \label{RGEFT}
\end{equation}
where $\beta(\lambda)$ is referred to as the renormalization group beta
function. Solving this renormalization group equation with an initial condition
$\lambda(\bar{p})=\bar{\lambda}$ that determines the scattering amplitude at
some value of the momentum and choosing $ \kappa=p $ in Eq.~(\ref{RGFT}), one
obtains the renormalization group improved scattering amplitude (at an
off-shell symmetric point)
\begin{equation}
\Gamma^{(4)}(p,p,p,p;\bar{\lambda}(\bar{p}))= \lambda(p),\label{RGimproved}
\end{equation}
with $\lambda(p)$ the solution of the renormalization group equation
\begin{equation}
\lambda(p)=\frac{\bar{\lambda}}{\raisebox{-.5ex}{$1 - (3\bar{\lambda}/2)
\ln(p/\bar{p})$}}.
\end{equation}

The connection between the renormalization group in momentum space and the
dynamical renormalization group in real time (resummation of secular terms)
used in the previous section is immediate through the identification
$$
t_0 \Leftrightarrow \ln(\Lambda/\bar{p}),\qquad
t\Leftrightarrow\ln(p/\bar{p}),\qquad \tau\Leftrightarrow\ln(\kappa/\bar{p}),
$$
which when replaced into Eq.~(\ref{gamma4}) illuminates the formal equivalence
with secular terms.

This simple analysis highlights how the \emph{dynamical
renormalization group} does precisely the same in the real time as
the renormalization group in zero-temperature Euclidean field
theory. Much in the same manner that the renormalization group
improved scattering amplitude given by Eq.~(\ref{RGimproved}) is a
\emph{resummation} of the perturbative expansion, the kinetic
equation obtained from the dynamical renormalization group
improvement represents a resummation of the perturbative expansion
in real time. The lowest order renormalization group equation
(\ref{RGEFT}) resums the leading logarithms, while the lowest
order \emph{dynamical renormalization group equation} resums the
leading secular terms.

\subsubsection{Dynamical renormalization group: fixed points and coarse-graining}
The similarity between the renormalization of distribution
functions and the renormalization of couplings suggests that the
collision terms of the quantum kinetic equation can be interpreted
as beta functions of the dynamical renormalization group and that
the space of distribution functions can be interpreted as the
space of coupling constants. The dynamical renormalization group
trajectories determine the flow in this space, therefore fixed
points of the dynamical renormalization group describe stationary
solutions with given distribution functions. Thermal equilibrium
distributions are thus fixed points of the dynamical
renormalization group. Furthermore, there can be other stationary
solutions with non-thermal distribution functions, e.g., those
describing turbulent behavior~\cite{zakharov}. Linearizing the
renormalization group equation around the fixed points corresponds
to linearizing the kinetic equation near equilibrium and the
linear eigenvalues are related to the relaxation rates in the
relaxation time approximation.

One can establish a closer relationship to the usual renormalization program of
field theory in its momentum shell version with the following alternative
interpretation of the secular terms and their resummation.

The initial distribution at a time $t_0$ is evolved in time perturbatively up
to a time $t_0+\Delta t$ such that the perturbative expansion is still valid,
i.e., $t_\mathrm{kin}\gg \Delta t$ with $t_\mathrm{kin}$ the kinetic time
scale. Secular terms begin to dominate the perturbative expansion on a time
scale $\Delta t\gg t_\mathrm{mic}$ with $t_\mathrm{mic}$ the microscopic time
scale. Thus if there is a separation of time scales such that
$t_\mathrm{kin}\gg \Delta t\gg t_\mathrm{mic}$, then in this intermediate
asymptotic regime perturbation theory is reliable but secular terms appear and
can be isolated.

A renormalization of the distribution function absorbs the contribution from
the secular terms. The ``renormalized'' distribution function is used as an
initial condition at $t_0+\Delta t$ to iterate forward in time to $t_0+2 \Delta
t$ using the perturbative expansion, but with \emph{the propagators in terms of
the distribution function at the time $t_0+\Delta t$}. This procedure can be
carried out ``infinitesimally'' (in the sense compared with the kinetic time
scale) and the differential equation that describes the changes of the
distribution function under the intermediate asymptotic time evolution is the
dynamical renormalization group equation.

This has an obvious similarity to the renormalization group in terms of
integrating in momentum shells. The result of integrating out degrees of
freedom in a momentum shell are absorbed into a renormalization of the
couplings and an effective theory at a lower scale but in terms of the
effective couplings. This procedure is carried out infinitesimally and the
differential equation that describes the changes of the couplings under
integrating out degrees of freedom in these momentum shells is the
renormalization group equation.

An important aspect of this procedure of evolving in time and ``resetting'' the
distribution functions is that in this process it is implicitly assumed that
the density matrix \emph{at later times} remains diagonal in the basis of free
quasiparticle number operators. Clearly, if at the initial time the density
matrix was diagonal in this basis, because the interaction Hamiltonian does not
commute with the density matrix, off-diagonal density matrix elements will be
generated upon time evolution. In resetting the distribution functions and
using the propagators in terms of these updated distribution functions we have
neglected off-diagonal correlations, e.g., correlations of the form $\langle
a(\mathbf{k},t)a(\mathbf{k},t)\rangle$ and its complex conjugate will be
generated upon time evolution. In neglecting these off-diagonal terms we are
introducing \emph{coarse-graining}. Indeed, two stages of coarse-graining had
been introduced: (i) integrating in time up to an intermediate asymptotics and
resumming the secular terms neglect transient phenomena, i.e., averages over
the microscopic time scales, and (ii) off-diagonal matrix elements (in the
basis of free quasiparticle number operators) had been neglected. The coarse
graining also has an counterpart in the language of the renormalization group
in Euclidean field theory, i.e., neglecting the \emph{irrelevant couplings}
that are generated upon integrating out shorter scales. Keeping all of the
correlations in the density matrix would be equivalent to Wilson's approach to
renormalization group, in which all possible couplings are included in the
Lagrangian and all of them are maintained in the renormalization on the same
footing.

\section[Application: $O(4)$ Linear Sigma Model]{Application: $\bbox{O(4)}$ Linear Sigma Model}
\label{sec:4.4}

In this section we apply the dynamical renormalization group
method to derive kinetic equations for pions and sigma mesons in
the $O(4)$ linear sigma model, which is an effective theory of low
energy QCD describing the hadronic world. We recall that the sigma
condensate $\langle\sigma\rangle$ represents the quark condensate
$\langle\bar{q}q\rangle$ in the sense that they both have the same
transformation properties, corresponding to two flavors of
massless quarks. At high temperatures, quarks and gluons exist in
a deconfined, chirally symmetric phase. At some critical
temperature of order 160 MeV a transition to a hadronic phase
occurs. In this confined and chiral symmetry broken phase the
quark condensate is nonzero.

The Lagrangian density of the $O(4)$ linear sigma model in the
limit of exact chiral symmetry, i.e., without an explicit
symmetry-breaking term, is given by
\begin{equation}
{\cal L}[{\bbox\pi},\sigma] = \frac{1}{2}
\left(\partial_{\mu}{\bbox\pi}\right)^2+ \frac{1}{2}
\left(\partial_{\mu}\sigma\right)^2
-\frac{\lambda}{4}\left({\bbox\pi}^2+\sigma^2-f_{\pi}^2\right)^2,
\end{equation}
where $\bbox{\pi}=(\pi^1,\pi^2,\pi^3)$ and $f_\pi\sim 93$ MeV is the pion decay
constant. At high temperature $T>T_c$, where $T_c\sim f_{\pi}$ is the critical
temperature, the $O(4)$ symmetry is restored by a second order phase
transition~\cite{bochkarev}.

In the symmetric phase pions and sigma mesons are degenerate, and the linear
sigma model reduces to a self-interacting scalar theory, analogous to that
discussed in Sec.~\ref{sec:4.2}. Thus, we limit our discussion here to the low
temperature broken symmetry phase in which the temperature $ T\ll f_{\pi} $.
Since at low temperature the $O(4)$ symmetry is spontaneously broken via the
sigma condensate $\langle\sigma\rangle$, we shift the sigma field
$\sigma(\mathbf{x},t)=\bar{\sigma}(\mathbf{x},t)+v $, where $v$ is temperature
dependent and yet to be determined.

In thermal equilibrium $v$ is fixed by requiring that $
\left\langle\bar{\sigma}(\mathbf{x},t)\right\rangle=0 $ to all orders in
perturbation theory for temperature $ T<T_c $. In nonequilibrium this split
must be performed on both forward $(+)$ and backward $(-)$ time branches,
writing the sigma field $\sigma^{\pm}(\mathbf{x},t)$ as
\begin{equation}
\sigma^{\pm}(\mathbf{x},t)=\bar{\sigma}^{\pm}(\mathbf{x},t)+v,\qquad
\langle\bar{\sigma}^{\pm}(\mathbf{x},t)\rangle=0,
\end{equation}
where $\langle\,\cdot\,\rangle$ denotes the nonequilibrium expectation value.
The expectation value $v$ is determined by requiring that the expectation value
of $\bar{\sigma}(\mathbf{x},t)$ vanishes to all orders in perturbation theory
via the tadpole method~\cite{tadpole}. One finds to one-loop order the equation
that determines $v$ is given by
\begin{equation}
v\left[v^2-f_{\pi}^2+\langle{\bbox\pi}^2\rangle+
3\langle\bar{\sigma}^2\rangle\right]=0.
\label{vev}
\end{equation}
We emphasize that in perturbation theory $v$ depends on the initial
nonequilibrium distribution functions of the pions and sigma mesons, which in
turn implies that after dynamical renormalization group resummation $v$
acquires implicit time dependence through the time dependence of the
distribution functions (see Sec.~\ref{sec:3.2}). Once the solution of this
equation for $v$ is used in the perturbative expansion to one loop order, the
tadpole diagrams that arise from the shift in the field cancel automatically to
this order. This feature of cancelation of tadpole diagrams that would result
in an expectation value of $\bar{\sigma}$ by the consistent use of the tadpole
equation persists to all orders in perturbation theory.

A solution of Eq.~(\ref{vev}) with $v\neq 0$ signals broken
symmetry and massless pions (in the strict chiral limit).
Therefore once the correct expectation value $v$ is used, the
one-particle-reducible (1PR) tadpole diagrams do not contribute in
the perturbative expansion of the kinetic equation. To this order
the inverse pion propagator for zero energy and momentum reads
\begin{equation}
G^{-1}_{\pi}(\omega=0,\mathbf{k}=\mathbf{0})=-\lambda\left[
 v^2-f_{\pi}^2 +\langle{\bbox\pi}^2\rangle+3\langle\bar{\sigma}^2
\rangle\right],
\end{equation}
which vanishes whenever $v\neq 0$ by the tadpole condition
Eq.~(\ref{vev}), hence the Goldstone theorem is satisfied and the
pions are the Goldstone bosons.

The study of the relaxation of sigma mesons (resonance) and pions near and
below the chiral phase transition is an important phenomenological aspect of
low energy chiral phenomenology with relevance to heavy ion collisions.
Furthermore, recent studies have revealed interesting features associated with
the dropping of the sigma mass near the chiral transition and the enhancement
of threshold effects with potential experimental consequences~\cite{hatsuda}.
The kinetic approach described here could prove useful to further assess the
contributions to the width of the sigma meson near the chiral phase transition,
this is an important study on its own right and we expect to report on these
issues in the near future.

With the purpose of comparing to recent results, we now focus on the situation
at low temperatures under the assumption that the distribution functions of
sigma mesons and pions are not too far from equilibrium, i.e., \emph{cool}
pions and sigma mesons. At low temperatures the relaxation of pions and sigma
mesons will be dominated by the one-loop contributions, and the scattering
contributions will be subleading. The scattering contributions are of the same
form as those discussed for the scalar theory and involve at least two
distribution functions and are subdominant in the low temperature limit as
compared to the one-loop contributions described below.

Since the linear sigma model is renormalizable and we focus on the medium
(finite-temperature) effects, we ignore the vacuum (zero-temperature)
ultraviolet divergences which can be absorbed into a renormalization of
$f_\pi$. For a small departure from thermal equilibrium, one can approximate
$\langle{\bbox\pi}^2\rangle$ and $\langle\bar{\sigma}^2\rangle$ by their
thermal equilibrium values
\begin{equation}
\langle{\bbox\pi}^2\rangle=\frac{T^2}{4},\qquad \langle\bar{\sigma}^2\rangle =
\int\frac{d^3q}{(2\pi)^3\omega_\mathbf{q}}\,n_B(\omega_\mathbf{q}),
\end{equation}
respectively, where $\omega_\mathbf{q}=\sqrt{\mathbf{q}^2+m_{\sigma}^2}$ with
the sigma mass $m_{\sigma}^2=2\lambda v^2$ being determined self-consistently.
In the low temperature limit $T \ll f_{\pi}$, one finds
$v^2=f_{\pi}^2\,[1-\mathcal{O}(T^2/f_{\pi}^2)]$ and
$m_{\sigma}=\sqrt{2\lambda}\,f_{\pi}[1-\mathcal{O}(T^2/f_{\pi}^2)]$. Thus in
our discussion below where $T\ll f_{\pi}$, we set $v=f_{\pi}$ and
$m_{\sigma}=\sqrt{2\lambda}f_{\pi}$. The main reason behind this analysis is to
display the microscopic time scales for the mesons: $t^\sigma_\mathrm{mic}\leq
1/m_{\sigma}$ and $t^\pi_\mathrm{mic} = 1/k$ with $k$ being the momentum of the
pion. The validity of a kinetic description will hinge upon the kinetic time
scales being much longer than these microscopic scales.

Upon the above replacement, one can rewrite the Lagrangian for the cool linear
sigma model to lowest order as
\begin{equation}
{\cal L}[{\bbox\pi},\sigma]= \frac{1}{2}
\left(\partial_{\mu}{\bbox\pi}\right)^2+\frac{1}{2}
\left(\partial_{\mu}\sigma\right)^2 -\frac{1}{2} m_{\sigma}^2 \sigma^2 -
\lambda f_{\pi} \left(\sigma{\bbox\pi}^2+\sigma^3\right)
-\frac{\lambda}{4}\left({\bbox\pi}^2+\sigma^2\right)^2,
\end{equation}
where the bar over the shifted sigma field has been omitted for simplicity of
notation.

Our goal in this section is to derive the kinetic equations describing pion and
sigma meson relaxation to lowest order. The unbroken $O(3)$ isospin symmetry
ensures that all the pions have the same relaxation rate, and sigma meson
relaxation rate is proportional to the number of pion species. Hence for
notational simplicity the pion index will be suppressed. We now begin to study
the quantum kinetics of cool pions and sigma mesons.

\subsection{Quantum kinetics of cool pions and sigma mesons}
Without loss of generality, in what follows we discuss the relaxation for one
isospin component, say $\pi^3$, but we suppress the indices for simplicity of
notation. As before, we consider the case in which at an initial time $t=t_0$,
the density matrix is diagonal in the basis of free quasiparticles number
operators, but with nonequilibrium initial distribution functions
$n^{\pi}_\mathbf{k}(t_0)$ and $n^{\sigma}_\mathbf{k}(t_0)$.

\subsubsection{Pions}
\begin{figure}[t]
\begin{center}
\includegraphics[width=3.5truein,keepaspectratio=true]{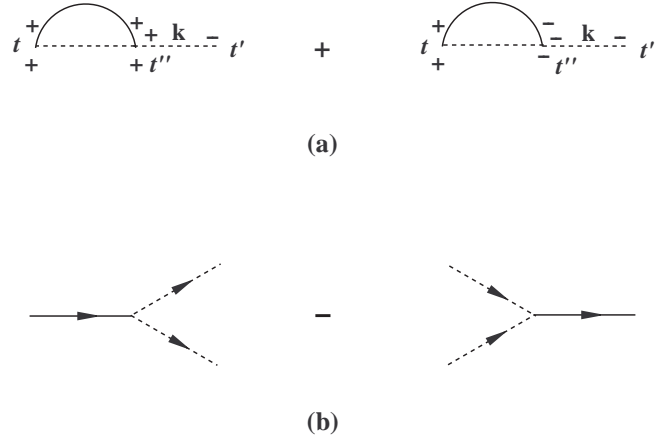}
\caption[The Feynman diagrams that contribute to the quantum kinetic equation
for the pion distribution function at one-loop order.]{(a) The Feynman diagrams
that contribute to the quantum kinetic equation for the pion distribution
function at one loop order. The solid line is the sigma meson propagator and
the dashed line is the pion propagator. (b) The only contribution on-shell is
the decay of a sigma meson into two pions minus the reverse
process.}\label{fig:pionsigma}
\end{center}
\end{figure}

The pion field operator and the corresponding conjugate momentum in the
Heisenberg picture can be written as (in momentum space)
\begin{equation}
\pi(\mathbf{k},t)=\frac{1}{\sqrt{2k}} \left[a_{\pi}(\mathbf{k},t)+
 a^{\dagger}_{\pi}(-\mathbf{k},t)\right],\quad
\Pi_{\pi}(\mathbf{k},t)=-i\sqrt{\frac{k}{2}} \left[a_{\pi}(\mathbf{k},t)-
 a^{\dagger}_{\pi}(-\mathbf{k},t)\right],
\end{equation}
The expectation value of pion number operator can be expressed in terms of
$\pi(\mathbf{k},t)$ and $\Pi_{\pi}(\mathbf{k},t)$ as
\begin{eqnarray}
n^{\pi}_\mathbf{k}(t)&=&
\langle a^{\dagger}_{\pi}(\mathbf{k},t)a_{\pi}(\mathbf{k},t)\rangle\nonumber\\
&=&\frac{1}{2k}\left[\left\langle \Pi_{\pi}(\mathbf{k},t)
\Pi_{\pi}(-\mathbf{k},t)\right\rangle+k^2\left\langle\pi(\mathbf{k},t)
\pi(-\mathbf{k},t)\right\rangle\right]-\frac{1}{2}.
\end{eqnarray}
Use the Heisenberg equations of motion, to leading order in $\lambda$, we
obtain (no tadpole diagrams are included since these are canceled by the choice
of $v$)
\begin{equation}
\frac{d}{dt}n^{\pi}_\mathbf{k}(t)= -\frac{2\lambda f_{\pi}}{k}
\lim_{t'\rightarrow t}\frac{\partial}{\partial t'}\int\frac{d^3q}{(2\pi)^{3/2}}
\left\langle\sigma^{+}(\mathbf{k}-\mathbf{q},t)\pi^{+}(\mathbf{q},t)
\pi^{-}(-\mathbf{k},t')\right\rangle. \label{pion:ndot}
\end{equation}
The expectation values can be calculated perturbatively in terms of
nonequilibrium vertices and Green's functions. To $\mathcal{O}(\lambda)$ the
right-hand side of Eq.~(\ref{pion:ndot}) vanishes identically.
Fig.~\ref{fig:pionsigma}a shows the Feynman diagrams that contribute to order $
\lambda^2 $. After some algebra, one obtains
\begin{eqnarray}
\frac{d}{dt}n^{\pi}_\mathbf{k}(t)\!\!&=&\!\!\frac{\lambda^2 f_{\pi}^2}{k} \int
\frac{d^3 q}{(2\pi)^3}\frac{1}{q\,\omega_\mathbf{p}}\int^t_{t_0}dt''
\Bigl[\mathcal{N}_1(t_0)\cos[\Delta\omega_1(t-t'')]
+\mathcal{N}_2(t_0)\cos[\Delta\omega_2(t-t'')]\nonumber\\
&&+\,\mathcal{N}_3(t_0)\cos[\Delta\omega_3(t-t'')]
+\mathcal{N}_4(t_0)\cos[\Delta\omega_4(t-t'')]\Big],
\end{eqnarray}
where $\mathbf{p=k+q}$,
\begin{eqnarray}
&&\Delta\omega_1=k+q+\omega_\mathbf{p},
\quad\Delta\omega_2=k-q-\omega_\mathbf{p},\nonumber\\
&&\Delta\omega_3=k-q+\omega_\mathbf{p},
\quad\Delta\omega_4=k+q-\omega_\mathbf{p},
\end{eqnarray}
and
\begin{eqnarray}
\mathcal{N}_1(t) &=&[1+ n^{\pi}_\mathbf{k}(t)][1+ n^{\pi}_\mathbf{q}(t)] [1+
n^{\sigma}_\mathbf{p}(t)]-n^{\pi}_\mathbf{k}(t) \,
n^{\pi}_\mathbf{q}(t)\, n^{\sigma}_\mathbf{p}(t),\nonumber\\
\mathcal{N}_2(t) &=&[1+ n^{\pi}_\mathbf{k}(t)]\, n^{\pi}_{\mathbf{q}}(t) \,
n^{\sigma}_\mathbf{p}(t) -n^{\pi}_\mathbf{k}(t)\, [1+n^{\pi}_\mathbf{q}(t)]
[1+n^{\sigma}_\mathbf{p}(t)], \nonumber\\
\mathcal{N}_3(t) &=&[1+ n^{\pi}_\mathbf{k}(t)] \,n^{\pi}_{\mathbf{q}}(t) [1+
n^{\sigma}_\mathbf{p}(t)]-n^{\pi}_\mathbf{k}(t)\,
[1+n^{\pi}_\mathbf{q}(t)] \, n^{\sigma}_\mathbf{p}(t),\nonumber\\
\mathcal{N}_4(t) &=&[1+ n^{\pi}_\mathbf{k}(t)] [1+n^{\pi}_{\mathbf{q}}(t)]
\,n^{\sigma}_\mathbf{p}(t) -n^{\pi}_\mathbf{k}(t)\,
n^{\pi}_\mathbf{q}(t)\,[1+n^{\sigma}_\mathbf{p}(t)]. \label{pion:ocupanumber}
\end{eqnarray}
The different contributions have a very natural interpretation in terms of
``gain minus loss'' processes. The first term in brackets corresponds to the
process $0\rightarrow \sigma+\pi+\pi$ minus the process
$\sigma+\pi+\pi\rightarrow 0$, the second and third terms correspond to the
scattering $\pi+\sigma\rightarrow\pi$ minus $\pi\rightarrow\pi+\sigma$, and the
last term corresponds to the decay of the sigma meson
$\sigma\rightarrow\pi+\pi$ minus the inverse process
$\pi+\pi\rightarrow\sigma$.

Just as in the scalar case, since the propagators entering in the perturbative
expansion of the kinetic equation are in terms of the distribution functions at
the initial time, the time integration can be done straightforwardly leading to
the following equation:
\begin{equation}
\frac{d}{dt}n^{\pi}_\mathbf{k}(t)= \lambda^2\int_{-\infty}^{+\infty} d\omega\,
\mathcal{R}_{\pi}[\omega,\mathbf{k};\mathcal{N}_j(t_0)] \,
\frac{\sin[(\omega-k)(t-t_0)]}{\pi(\omega-k)},\label{pion:ke}
\end{equation}
where $\mathcal{R}_{\pi}[\omega,\mathbf{k};\mathcal{N}_j(t_0)]$ is given by
\begin{eqnarray}
\mathcal{R}_{\pi}[\omega,\mathbf{k};\mathcal{N}_j(t_0)]&=&\frac{f_{\pi}^2}{k}
\,\int\frac{d^3 q}{(2 \pi)^3} \frac{1}{q\,\omega_\mathbf{p}}
\Bigl[\delta(\Delta\omega_1)\,\mathcal{N}_1(t_0)
+\delta(\Delta\omega_2)\,\mathcal{N}_2(t_0)\nonumber\\
&&+\,\delta(\Delta\omega_3)\,\mathcal{N}_3(t_0)
+\delta(\Delta\omega_4)\,\mathcal{N}_4(t_0)\Bigr].
\end{eqnarray}
Eq.(\ref{pion:ke}) can be solved by direct integration over $t$ with the given
initial condition at $t_0$, leading to
\begin{equation}
n^{\pi}_\mathbf{k}(t)=n^{\pi}_\mathbf{k}(t_0)+\lambda^2
\int_{-\infty}^{+\infty} d\omega\,
\mathcal{R}_{\pi}[\omega,\mathbf{k};\mathcal{N}_j(t_0)] \,
\frac{1-\cos[(\omega-k)(t-t_0)]}{\pi(\omega-k)^2}. \label{ndot:pion}
\end{equation}
Potential secular term arises at large times when the resonant denominator in
(\ref{ndot:pion}) vanishes, i.e., $\omega\approx k$. A detailed analysis
reveals that $\mathcal{R}_{\pi}[\omega,\mathbf{k};\mathcal{N}_j(t_0)]$ is
regular at $\omega=k$, hence upon using (\ref{FGR}) one finds that at
intermediate asymptotic time $k(t-t_0)\gg 1$, the time evolution of the pion
distribution function reads
\begin{equation}
n^{\pi}_\mathbf{k}(t)= n^{\pi}_\mathbf{k}(t_0)+\lambda^2\, \mathcal{R}_{\pi}[k
,\mathbf{k};\mathcal{N}_j(t_0)]\,(t-t_0) + \mbox{nonsecular terms},
\label{pion:perturb}
\end{equation}
where $\mathcal{R}_{\pi}[k ,\mathbf{k};\mathcal{N}_j(t_0)]$ does not depend on
$t_0$ explicitly.

At this point we would be tempted to follow the same steps as in the scalar
case and introduce the dynamical renormalization of the pion distribution
function. However, much in the same manner as the renormalization program in a
theory with several coupling constants, in the case under consideration the
$\bbox\pi$ field and the $\sigma$ field are coupled. Therefore one must
renormalize \emph{all} of the distribution functions on the same footing. Hence
our next task is to obtain the kinetic equations for the sigma meson
distribution functions.

\subsubsection{Sigma mesons}
The field operator for the sigma meson and the corresponding conjugate momentum
in the Heisenberg picture can be written as (in momentum space)
\begin{equation}
\sigma(\mathbf{k},t)=\frac{1}{\sqrt{2\omega_\mathbf{k}}}
\left[a_{\sigma}(\mathbf{k},t)+
 a^{\dagger}_{\sigma}(-\mathbf{k},t)\right],\quad
\Pi_{\sigma}(\mathbf{k},t)=-i\sqrt{\frac{\omega_\mathbf{k}}{2}}
\left[a_{\sigma}(\mathbf{k},t)-a^{\dagger}_{\sigma}(-\mathbf{k},t)\right],
\end{equation}
with $\omega_\mathbf{k}=\sqrt{k^2+m_{\sigma}^2}$. Again, for notational
simplicity we suppress the pion isospin index. The expectation value of sigma
meson number operator can be expressed in terms of $\sigma(\mathbf{k},t)$ and
$\Pi_{\sigma}(\mathbf{k},t)$ as
\begin{eqnarray}
n^{\sigma}_\mathbf{k}(t)&=& \langle
a^{\dagger}_{\sigma}(\mathbf{k},t)a_{\sigma}(\mathbf{k},t)\rangle\nonumber\\
&=& \frac{1}{2\omega_\mathbf{k}}\left[\left\langle \Pi_{\sigma}
(\mathbf{k},t)\Pi_{\sigma}(-\mathbf{k},t) \right\rangle + \omega_\mathbf{k}^2
\left\langle\sigma(\mathbf{k},t)\sigma(-\mathbf{k},t)\right\rangle\right]
-\frac{1}{2}.
\end{eqnarray}
Using the Heisenberg equations of motion to leading order in $\lambda$, and
requiring again that the tadpole diagrams are canceled by the proper choice of
$v$, we obtain
\begin{eqnarray}
\frac{d}{dt}n^{\sigma}_\mathbf{k}(t)&=& -\frac{3\lambda
f_{\pi}}{\omega_\mathbf{k}}\lim_{t'\rightarrow
t}\frac{\partial}{\partial t'} \int\frac{d^3q}{(2\pi)^{3/2}}
\Bigl[\big\langle\pi^{+}(\mathbf{k}-\mathbf{q},t)\pi^{+}(\mathbf{q},t)
\sigma^{-}(-\mathbf{k},t')\big\rangle\nonumber\\
&&+\,3\big\langle\sigma^{+}(\mathbf{k}-\mathbf{q},t)\sigma^{+}(\mathbf{q},t)
\sigma^{-}(-\mathbf{k},t')\big\rangle\Bigr] ,\label{sigma:ndot}
\end{eqnarray}
where the factor three for the first term accounts for three isospin components
of the pion field.

\begin{figure}[t]
\begin{center}
\includegraphics[width=3.5truein,keepaspectratio=true]{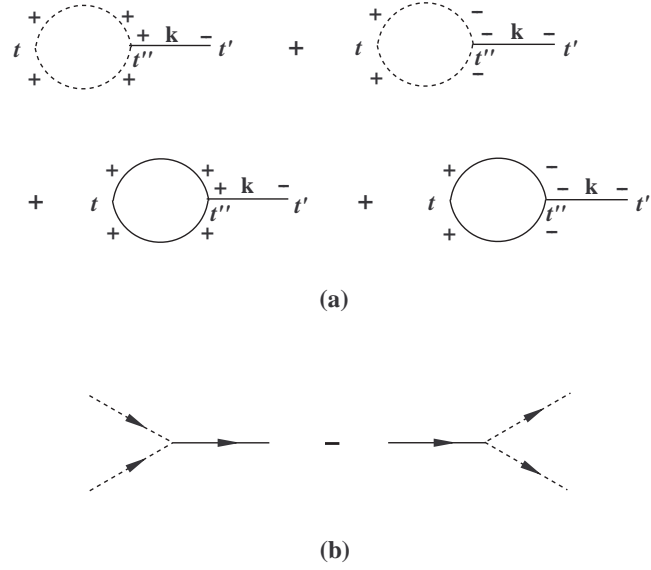}
\caption[The Feynman diagrams that contribute to the quantum kinetic equation
for the sigma meson distribution function at one-loop order.]{(a) The Feynman
diagrams that contribute to the quantum kinetic equation for the sigma meson
distribution function at one-loop order. The solid line is the sigma meson
propagator and the dashed line is the pion propagator. (b) The only
contribution on-shell is recombination of two pions into a sigma meson minus
the decay of a sigma meson into two pions.}\label{fig:sigmapion}
\end{center}
\end{figure}

To $ \mathcal{O}(\lambda) $ the right hand side of Eq.~(\ref{sigma:ndot})
vanishes identically. Fig.~\ref{fig:sigmapion}a depicts the one-loop Feynman
diagrams that enter in the kinetic equation for the sigma meson to order $
\lambda^2 $. To the same order there will be the same type of two loops
diagrams as in the self-interacting scalar theory studied in the previous
section, but in the low temperature limit the two-loop diagrams will be
suppressed with respect to the one-loop diagrams. Furthermore, in the low
temperature limit, the focus of our attention here, only the pion loops will be
important in the relaxation of the sigma mesons. A straightforward calculation
leads to the following expression
\begin{eqnarray}
\frac{d}{dt}n^{\sigma}_\mathbf{k}(t)&=& \frac{3\lambda^2
f_{\pi}^2}{2\omega_\mathbf{k}}\, \int \frac{d^3 q}{(2 \pi)^3}
\int^t_{t_0}dt''\biggl[\frac{1}{p\,q}\Bigl(\mathcal{N}^{\pi}_1(t_0)
\cos[\Delta\omega^\pi_1(t-t'')] \nonumber \\
&&+\,\mathcal{N}^{\pi}_2(t_0) \cos[\Delta\omega^\pi_2(t-t'')] +
\mathcal{N}^{\pi}_3(t_0) \cos[\Delta\omega^\pi_3(t-t'')] \nonumber \\
&&+\, \mathcal{N}^{\pi}_4(t_0)\cos[\Delta\omega^\pi_4(t-t'')]\Bigr)+
\frac{9}{\omega_\mathbf{q}\,\omega_\mathbf{p}}
\Bigl(\pi\rightarrow\sigma\Bigr)\biggr], \label{sigmadot0}
\end{eqnarray}
where $\mathbf{p=k+q}$,
\begin{eqnarray}
&&\Delta\omega^{\pi(\sigma)}_1=\omega_\mathbf{k}+q~(\omega_\mathbf{q})
+p~(\omega_\mathbf{p}),
\quad\Delta\omega^{\pi(\sigma)}_2=\omega_\mathbf{k}+q~(\omega_\mathbf{q})
-p~(\omega_\mathbf{p}),\nonumber\\
&&\Delta\omega^{\pi(\sigma)}_3=\omega_\mathbf{k}-q~(\omega_\mathbf{q})
+p~(\omega_\mathbf{p}),
\quad\Delta\omega^{\pi(\sigma)}_4=\omega_\mathbf{k}-q~(\omega_\mathbf{q})
-p~(\omega_\mathbf{p}),
\end{eqnarray}
and
\begin{eqnarray}
\mathcal{N}^{\pi(\sigma)}_1(t) &=&[1+ n^{\sigma}_\mathbf{k}(t)]
[1+n^{\pi(\sigma)}_\mathbf{q}(t)] [1+ n^{\pi(\sigma)}_\mathbf{p}(t)]
-n^{\sigma}_\mathbf{k}(t)\, n^{\pi(\sigma)}_\mathbf{q}(t)\,
n^{\pi(\sigma)}_\mathbf{p}(t),\nonumber \\
\mathcal{N}^{\pi(\sigma)}_2(t) &=&[1+ n^{\sigma}_\mathbf{k}(t)]
[1+n^{\pi(\sigma)}_{\mathbf{q}}(t)]
n^{\pi(\sigma)}_\mathbf{p}(t)-n^{\sigma}_\mathbf{k}(t)\,
n^{\pi(\sigma)}_\mathbf{q}(t)\,[1+n^{\pi(\sigma)}_\mathbf{p}(t)],\nonumber\\
\mathcal{N}^{\pi(\sigma)}_3(t) &=&[1+ n^{\sigma}_\mathbf{k}(t)] \,
 n^{\pi(\sigma)}_{\mathbf{q}}(t)\, [1+ n^{\pi(\sigma)}_\mathbf{p}(t)]
-n^{\sigma}_\mathbf{k}(t)\, [1+n^{\pi(\sigma)}_\mathbf{q}(t)]\,
n^{\pi(\sigma)}_\mathbf{p}(t),\nonumber\\
\mathcal{N}^{\pi(\sigma)}_4(t) &=&[1+ n^{\sigma}_\mathbf{k}(t)]\,
n^{\pi(\sigma)}_{\mathbf{q}}(t) \,n^{\pi(\sigma)}_\mathbf{p}(t)
-n^{\sigma}_\mathbf{k}(t)\, [1+n^{\pi(\sigma)}_\mathbf{q}(t)]
[1+n^{\pi(\sigma)}_\mathbf{p}(t)]. \label{sigma:ocupanumber1}
\end{eqnarray}

Although the above expression is somewhat unwieldy, the different contributions
have a very natural interpretation in terms of the ``gain minus loss''
processes. In the first parentheses (the pion contribution) the first term
corresponds to the process $0\rightarrow \sigma+\pi+\pi$ minus the process
$\sigma+\pi+\pi\rightarrow 0$, the second and third terms correspond to the
scattering $\pi\rightarrow\pi+\sigma$ minus $\pi+\sigma\rightarrow\pi$, and the
last term corresponds to the decay of the sigma meson
$\sigma\rightarrow\pi+\pi$ minus the inverse process
$\pi+\pi\rightarrow\sigma$. Likewise, in the second parentheses (the sigma
meson contribution) the first term corresponds to the process
$0\rightarrow\sigma+\sigma+\sigma$ minus the process
$\sigma+\sigma+\sigma\rightarrow 0$, the second and third terms correspond to
annihilation of two sigma mesons and creation of one sigma meson minus the
inverse process, and the last term corresponds to annihilation of a sigma meson
and creation of two sigma mesons minus the inverse process.

Since the propagators entering in the perturbative expansion of the kinetic
equation are in terms of the distribution functions at the initial time, the
time integration can be done straightforwardly leading to the following
equation:
\begin{equation}
\frac{d}{dt}n^{\sigma}_\mathbf{k}(t)= \lambda^2\int_{-\infty}^{+\infty}
d\omega\, \mathcal{R}_{\sigma}[\omega,\mathbf{k};\mathcal{N}_j(t_0)] \,
\frac{\sin[(\omega-\omega_\mathbf{k})(t-t_0)]}
{\pi(\omega-\omega_\mathbf{k})},\label{sigma:ke}
\end{equation}
where
\begin{eqnarray}
\mathcal{R}_{\sigma}[\omega,\mathbf{k};\mathcal{N}_j(t_0)]&=&
\frac{3f_{\pi}^2}{2\omega_\mathbf{k}}\,\int \frac{d^3 q}{(2 \pi)^3}
\biggl[\frac{1}{p\,q}\Bigl(\delta(\Delta\omega^\pi_1)\,\mathcal{N}^{\pi}_1(t_0)
+\delta(\Delta\omega^\pi_2)\,\mathcal{N}^{\pi}_2(t_0)+
\delta(\Delta\omega^\pi_3)\nonumber\\&&\times\,\mathcal{N}^{\pi}_3(t_0)+
\delta(\Delta\omega^\pi_4)\,\mathcal{N}^{\pi}_4(t_0)\Bigr)+
\frac{9}{\omega_\mathbf{q}\,\omega_\mathbf{p}}
\Bigl(\pi\rightarrow\sigma\Bigr)\biggr]. \label{Rsigma}
\end{eqnarray}

Just as before $\mathcal{R}_{\sigma}[\omega,\mathbf{k};\mathcal{N}_j(t_0)]$ is
fixed at initial time $t_0$, Eq.~(\ref{sigma:ke}) can be integrated over $t$
with the given initial condition at $t_0$, thus leading to
\begin{equation}
n^{\sigma}_\mathbf{k}(t)=n^{\sigma}_\mathbf{k}(t_0)+\lambda^2
\int_{-\infty}^{+\infty} d\omega\,
\mathcal{R}_{\sigma}[\omega,\mathbf{k};\mathcal{N}_j(t_0)]
\frac{1-\cos[(\omega-\omega_\mathbf{k})(t-t_0)]}
{\pi(\omega-\omega_\mathbf{k})^2}.\label{sigma:solofke}
\end{equation}
At intermediate asymptotic times $m_{\sigma}(t-t_0)\gg 1$, potential secular
term arises when $\omega\approx\omega_\mathbf{k}$ in Eq.~(\ref{sigma:solofke}).
We notice that, although
$\mathcal{R}_{\sigma}[\omega,\mathbf{k};\mathcal{N}_j(t_0)]$ has (infrared)
threshold singularities at $\omega=\pm k$, it is regular on the sigma meson
mass shell. This observation will allow us to explore a crossover behavior for
very large momentum later.

Since $\mathcal{R}_{\sigma}[\omega,\mathbf{k};\mathcal{N}_j(t_0)]$ is regular
near the resonance region $\omega=\pm \omega_\mathbf{k}$, the behavior at
intermediate asymptotic times is given by
\begin{equation}
n^{\sigma}_\mathbf{k}(t)=n^{\sigma}_\mathbf{k}(t_0)+\lambda^2\,
\mathcal{R}_{\sigma}[\omega_\mathbf{k},\mathbf{k};\mathcal{N}_j(t_0)](t-t_0)
+\mbox{nonsecular terms}.\label{sigmasecular}
\end{equation}
We note that the perturbative expansions for the pion and sigma meson
distribution functions contain secular terms that grow linearly in time,
\emph{unless} the system is initially prepared in thermal equilibrium. To resum
the secular terms, we must now renormalize both Eqs.~(\ref{pion:perturb}) and
(\ref{sigmasecular}) \emph{simultaneously}, because, as discussed in
Sec.~\ref{sec:4.3}, this situation corresponds to that with two \emph{relevant}
couplings in Euclidean field theory. The renormalization is achieved by
introducing the renormalized initial distribution functions
$n^{\pi(\sigma)}_\mathbf{k}(\tau)$ for pions (sigma mesons) that are related to
the bare initial distribution functions $n^{\pi(\sigma)}_\mathbf{k}(t_0)$ via
the renormalization constants $\mathcal{Z}^{\pi(\sigma)}_\mathbf{k}(\tau,t_0)$
\begin{equation}
n^{\pi(\sigma)}_\mathbf{k}(t_0)=\mathcal{Z}^{\pi(\sigma)}_\mathbf{k}(\tau,t_0)\,
n^{\pi(\sigma)}_\mathbf{k}(\tau),\quad
\mathcal{Z}^{\pi(\sigma)}_\mathbf{k}(\tau,t_0)= 1 + \lambda^2 \,
z^{\pi(\sigma)}_\mathbf{k}(\tau,t_0)+\mathcal{O}(\lambda^4),\label{pion:renormal}
\end{equation}
where $\tau$ is an arbitrary renormalization scale at which the secular terms
will be canceled. As in the case of the scalar theory, the renormalization
constants $z^{\pi(\sigma)}_\mathbf{k}(\tau,t_0)$ are chosen so as to cancel the
secular terms at the arbitrary scale $\tau$. Upon substitute
Eq.~(\ref{pion:renormal}) into Eq.~(\ref{pion:perturb}) and choosing
\begin{equation}
z^{\pi(\sigma)}_\mathbf{k}(\tau,t_0)=
-(\tau-t_0)\frac{\mathcal{R}_{\pi(\sigma)}
[k,\mathbf{k};\mathcal{N}_j(\tau)]}{n^{\pi(\sigma)}_\mathbf{k}(\tau)},
\end{equation}
consistently up to $\mathcal{O}(\lambda^2)$, one obtains
\begin{equation}
n^{\pi(\sigma)}_\mathbf{k}(t)=n^{\pi(\sigma)}_\mathbf{k}(\tau)
+\lambda^2\,(t-\tau)\,\mathcal{R}_{\pi(\sigma)}[k\,(\omega_\mathbf{k})
,\mathbf{k};\mathcal{N}_j(\tau)]+\mathcal{O}(\lambda^4).
\end{equation}
The independence of $n^{\pi(\sigma)}_\mathbf{k}(t)$ on the arbitrary
renormalization scale $\tau$ leads to the simultaneous set of dynamical
renormalization group equations to lowest order:
\begin{eqnarray}
&&\frac{d}{d\tau}n^{\pi}_\mathbf{k}(\tau)=\lambda^2\,
\mathcal{R}_{\pi}[k,\mathbf{k};\mathcal{N}_j(\tau)],\nonumber\\
&&\frac{d}{d\tau}n^{\sigma}_\mathbf{k}(\tau)=\lambda^2\, \mathcal{R}_{\sigma}
[\omega_\mathbf{k},\mathbf{k};\mathcal{N}_j(\tau)].
\end{eqnarray}
These equations have an obvious resemblance to a set of usual renormalization
group equations for ``couplings'' $ n^{\pi}_\mathbf{k}$ and
$n^{\sigma}_\mathbf{k}$, where the right-hand sides are the corresponding
``beta functions''.

As before choosing the arbitrary scale $\tau$ to coincide with the time $t$ and
keeping only the terms whose delta functions have support on the mass shells,
we obtain the quantum kinetic equations for pions and sigma mesons:
\begin{eqnarray}
\frac{d}{dt}n^{\pi}_\mathbf{k}(t)\!\!&=&\!\!\frac{\pi\lambda^2 f_{\pi}^2}{k} \,
\int\frac{d^3 q}{(2\pi)^3}
\frac{\delta(k+q-\omega_\mathbf{p})}{q\,\omega_\mathbf{p}} \Big\{
[1+n^{\pi}_\mathbf{k}(t)] [1+ n^{\pi}_{\mathbf{q}}(t)] \,
n^{\sigma}_\mathbf{p}(t)\nonumber\\
&&-\,n^{\pi}_\mathbf{k}(t) \, n^{\pi}_\mathbf{q}(t)\,
[1+n^{\sigma}_\mathbf{p}(t)]\Big\},\label{kin:pion}\\
\frac{d}{dt}n^{\sigma}_\mathbf{k}(t)\!\!&=&\!\!\frac{3\pi\lambda^2
f_{\pi}^2}{2\omega_\mathbf{k}} \, \int \frac{d^3
q}{(2\pi)^3}\frac{\delta(\omega_\mathbf{k}-q-p)}{p\,q}
\Big\{[1+n^{\sigma}_\mathbf{k}(t)]\,n^{\pi}_{\mathbf{q}}(t)\,
n^{\pi}_\mathbf{p}(t)\nonumber\\
&&-\,n^{\sigma}_\mathbf{k}(t)\,[1+n^{\pi}_\mathbf{q}(t)]
[1+n^{\pi}_\mathbf{p}(t)]\Big\}. \label{kin:sigma}
\end{eqnarray}
The processes that contribute to Eq.~(\ref{kin:pion}) are depicted in Fig.~2b
and those that contribute to Eq.~(\ref{kin:sigma}) are depicted in Fig.~3b.

\subsection{Relaxation rate for pions and sigma mesons}
Thermal equilibrium is a {\em fixed point} of the dynamical renormalization
group equations, i.e., a stationary solution of the quantum kinetic equations
(\ref{kin:pion}) and (\ref{kin:sigma}). Near equilibrium a Linearized kinetic
equation for pions (sigma mesons) can be obtained in the relaxation time
approximation, in which only distribution function for pions (sigma mesons) of
momentum $\mathbf{k}$ are slightly perturbed off equilibrium whereas all the
other modes are in equilibrium. In the relaxation time approximation for pions
(sigma mesons), one obtains
\begin{equation}
\frac{d}{dt}\delta n^{\pi(\sigma)}_\mathbf{k}(t) =
-\Gamma_{\pi(\sigma)}(k)\,\delta n^{\pi(\sigma)}_\mathbf{k}(t),
\end{equation}
where $\Gamma_{\pi(\sigma)}(k)$ is the pion (sigma meson) relaxation rate,
which is identified with twice the damping rate of the corresponding mean
field.

Linearizing Eq.~(\ref{kin:pion}) in the relaxation time approximation, we
obtain the pion relaxation rate to be given by
\begin{eqnarray}
\Gamma_{\pi}(k)&=&\frac{\pi\lambda^2 f_{\pi}^2}{k} \, \int \frac{d^3
q}{(2\pi)^3}\frac{n_B(q)-n_B(\omega_\mathbf{p})}{q\,\omega_\mathbf{p}}\,
\delta(k+q-\omega_\mathbf{p})\nonumber\\
&=&\frac{\lambda^2 f_{\pi}^2 T}{4\pi k^2}\,
\ln\frac{1-e^{-\beta(m_{\sigma}^2/4k+k)}}{1-e^{-\beta
m_{\sigma}^2/4k}}.\label{relrate:pion}
\end{eqnarray}
This is a remarkable expression because it reveals that the physical processes
that contribute to pion relaxation are the \emph{decay} of a sigma meson to two
pions and its inverse process, i.e., $\sigma\rightleftarrows\pi+\pi$. The sigma
mesons present in the medium can decay into pions and this increases the number
of pions, but at the same time pions recombine into sigma mesons, and since
there are more pions in the medium because they are lighter the loss part of
the process prevails producing a non-zero relaxation rate for the pion
distribution function. This is an induced phenomenon in the medium in the very
definitive sense that the decay of the heavier sigma meson induces the
relaxation of the pion distribution function, it is a \emph{collisionless}
process. We note that such induced relaxation of pions is analogous to the
relaxation of fermions in a fermion-scalar plasma discusses in
Chap.~\ref{chap:3}, the latter is induced by the decay of a massive scalar into
light fermion pairs.

For soft, cool pions $(k\ll T\ll f_{\pi})$, the relaxation rate reads
\begin{equation}
\gamma_{\pi}(k\ll T)\approx\frac{\lambda^2 f_{\pi}^2}{4\pi k}\exp\left(
-\frac{m_{\sigma}^2}{4kT}\right).
\end{equation}
The exponential suppression in the above expression is a consequence of the
heavy sigma mass. Our results of the pion relaxation rate are in agreement with
the pion damping rate found in Ref.~\cite{pisarski:coolpion}. These results
(accounting for the factor two necessary to relate the relaxation rate to the
damping rate) also agree with those reported recently in Ref.~\cite{patkos},
wherein a related and clear analysis of pion and sigma meson damping rates was
presented.

Likewise, the sigma meson relaxation rate in the relaxation time approximation
is found to be given by
\begin{eqnarray}
\Gamma_{\sigma}(k)&=& \frac{3\pi\lambda^2 f_{\pi}^2}{2\omega_\mathbf{k}}\, \int
\frac{d^3 q}{(2\pi)^3} \frac{\left[1+n_B(q)+n_B(p)\right]}{p\,q}\,
\delta(\omega_\mathbf{k}-q-p)\nonumber\\
&=&\frac{3\lambda^2 f_{\pi}^2}{8\pi\omega_\mathbf{k}}\left[1 +\frac{2T}{k}\,
\ln\frac{1-e^{-\beta(\omega_\mathbf{k}+k)/2}}
{1-e^{-\beta(\omega_\mathbf{k}-k)/2}}\right]. \label{relrate:sigma}
\end{eqnarray}
We note that the first temperature-independent term in $\Gamma_{\sigma}(k)$ is
the usual zero-temperature sigma meson decay rate~\cite{serot}, whereas the
finite-temperature terms result from the \emph{same} processes $\sigma
\rightleftarrows \pi+\pi$ that also determine $\Gamma_{\pi}(k)$. For soft, cool
sigma meson ($k\ll T\ll f_{\pi}$), one obtains
\begin{equation}
\Gamma_{\sigma}({k\approx 0})\approx \frac{3\lambda^2 f_{\pi}^2}{8\pi
m_{\sigma}} \coth\left(\frac{m_{\sigma}}{4 T}\right),
\end{equation}
which agrees with the decay rate for a sigma meson at
rest~\cite{patkos,sigmadecay,csernai}.

On the other hand, let us consider the \emph{theoretical} high temperature and
large momentum limit $k\gg m_{\sigma}\gtrsim T$ and $\omega_\mathbf{k}-k \ll
T$. In this limit $\Gamma_{\sigma}(k)$ becomes logarithmic (infrared)
divergent. The reason for this divergence is that, as was mentioned above,
$\mathcal{R}_{\sigma}[\omega,\mathbf{k};\mathcal{N}_j(t_0)]$ has an infrared
threshold singularity at $\omega=k$ arising from the terms proportional to
$\mathcal{N}^{\pi}_4(t_0)$, which accounts for the emission and absorption of
collinear massless pions. In the presence of this threshold singularity, we can
no longer apply Fermi's golden rule (\ref{FGR}). Instead, we must analyze the
long-time behavior of Eq.~(\ref{sigmadot0}) more carefully.

Understanding the influence of threshold behavior of the sigma meson on its
relaxation could be important in view of the recent proposal by Hatsuda and
collaborators~\cite{hatsuda} that near the chiral phase transition the mass of
the sigma meson drops and threshold effects become enhanced with distinct
phenomenological consequences.

\subsection{Threshold singularities and crossover in relaxation}
In order to understand how the (infrared) threshold divergence modifies the
long-time behavior, let us focus on the sigma mesons with large momentum $k\gg
m_{\sigma}\gtrsim T$. This situation is not relevant to the phenomenology of
the cool pion-sigma meson system for which relevant temperatures are $T\ll
m_{\sigma}$. However studying this theoretical limiting case will reveal
important insight on how threshold divergences invalidate the usual Fermi's
golden rule analysis, which leads to energy conservation delta functions in the
intermediate asymptotic regimes. This issue will become physical relevant in
the case of gauge field theories studied below (see Chap.~\ref{chap:5}).

To present this case in the simplest and clearest manner, we will proceed in
the relaxation time approximation. Keeping only the infrared divergent
contribution in $\mathcal{R}_{\sigma}[\omega,\mathbf{k};\mathcal{N}_j(t_0)]$,
i.e., the term proportional to $\mathcal{N}^{\pi}_4(t_0)$, one finds that
Eq.~(\ref{sigma:solofke}) simplifies to
\begin{equation}
\delta n^{\sigma}_\mathbf{k}(t)=\delta n^{\sigma}_\mathbf{k}(t_0)\left[
1-\int_{-\infty}^{+\infty} d\omega\, \mathcal{R}(\omega,\mathbf{k}) \,
\frac{1-\cos[(\omega-\omega_\mathbf{k})(t-t_0)]}
{\pi(\omega-\omega_\mathbf{k})^2}\right],\label{linearsigma}
\end{equation}
where
\begin{eqnarray}
\mathcal{R}(\omega,\mathbf{k})&=& \frac{3\pi\lambda^2
f_{\pi}^2}{2\omega_\mathbf{k}}\,\int\frac{d^3 q}{(2\pi)^3}\,
\frac{1+n_B(q)+n_B(p)}{p\,q}\,\delta(\omega-q-p)\nonumber\\
&=&\frac{3\lambda^2 f_{\pi}^2}{8\pi\omega_\mathbf{k}}\left[1+\frac{2T}{k}\,
\ln\frac{1-e^{-\beta(\omega+k)/2}}
{1-e^{-\beta(\omega-k)/2}}\right],\label{gammasigma}
\end{eqnarray}
which has a infrared threshold singularity at $\omega =k$. At intermediate
asymptotic times $m_{\sigma}(t-t_0)\gg 1$, the integral over $\omega$ in
Eq.(\ref{linearsigma}) is dominated by the region near $\omega\approx k$. In
the limit $k\gg m_{\sigma}\gtrsim T$, by approximating
\begin{equation}
\mathcal{R}(\omega,\mathbf{k})\,\buildrel{\omega\to k}\over=\,\frac{3\lambda^2
f_\pi^2 T}{4\pi k^2}\,\ln\left[\frac{2T}{\omega-k}\right] +
\mathcal{O}(\omega-k), \label{gammathreshold}
\end{equation}
the $\omega$ integral (denoted as $\mathcal{I}$) for $m_{\sigma}(t-t_0)\gg 1$
can be evaluated in a closed form
\begin{equation}
\mathcal{I}\approx \frac{3\lambda^2 f_{\pi}^2 \bar{T}}{4\pi
k^2}\,\mathcal{F}(\bar{t}),
\end{equation}
where $\bar{T}=T/(\omega_\mathbf{k}-k)\approx 2kT/m_\sigma^2$ and
$\bar{t}=(\omega_\mathbf{k}-k)(t-t_0)\approx m_\sigma^2(t-t_0)/2k$ are
dimensionless variables and
\begin{equation}
\mathcal{F}(x)=x\left(\ln 2\bar{T} +\mathrm{ci}(x)- \frac{\sin x}{x}\right),
\end{equation}
with $\mathrm{ci}(x)$ being the cosine integral function
\begin{equation}
\mathrm{ci}(x)=-\int_x^{+\infty}dt\,\frac{\cos t}{t}.
\end{equation}

The function $\mathcal{F}(x)$ has the following asymptotic behaviors
\begin{equation}
\mathcal{F}(x)=\left\{
\begin{array}{ll}
x\left[\ln 2x\bar{T}+\gamma-1 +\mathcal{O}\left(x^2\right)
\right] &\quad \mathrm{for}\quad x \ll 1,\\
x\left[\ln 2\bar{T}+ \mathcal{O}\left(x^{-2}\right)\right]
&\quad\mathrm{for}\quad x\gg 1,\end{array}\right. \label{Fx}
\end{equation}
where $\gamma = 0.577\ldots$ is the Euler-Mascheroni constant. Thus, for fixed
and large $k$ one can see that there is a \emph{crossover} time scale
$t_c\approx 2k/m_\sigma^2$ on which the time dependence of
$\mathcal{F}(\bar{t})$ changes from the $t\ln t$ dependence for $t-t_0\ll t_c$
to the linear dependence for $t-t_0\gg t_c$. Consequently, in the large
momentum limit as the sigma meson mass shell approaches the threshold, this
crossover time scale becomes very long such that an ``anomalous'' (nonlinear)
secular term of the form $t\ln t$ dominates during most of the time whereas the
usual linear secular term ensues at very large times.

The secular terms can be resumed by implementing the dynamical renormalization
group method [see Eq.~(\ref{sigmasecular})] with the following choice of
renormalization constant (in terms of dimensionless time variable)
\begin{equation}
z^{\sigma}_{\mathbf{k}}(\tau,\bar{t}_0)= \frac{3f_\pi^2\bar{T}}{4\pi
k^2}\,\mathcal{F}(\tau-\bar{t}_0).
\end{equation}
The resultant dynamical renormalization group equation reads
\begin{equation}
\frac{d}{d\bar{t}}\,\delta n^{\sigma}_\mathbf{k}(\bar{t})+ \frac{3\lambda^2
f_\pi^2 \bar{T}}{4\pi k^2}\,\frac{d}{d\bar{t}}
\,\mathcal{F}(\bar{t}-\bar{t}_0)= 0, \label{RGlogi}
\end{equation}
whose solution is given by
\begin{equation}
\delta n^{\sigma}_\mathbf{k}(\bar{t})= \delta n^{\sigma}_\mathbf{k}(\bar{t}_0)
\exp\left[-\frac{3\lambda^2 f_\pi^2 \bar{T}}{4\pi
k^2}\mathcal{F}(\bar{t}-\bar{t}_0)\right].
\end{equation}
In the large momentum limit, using Eq.~(\ref{Fx}) we find that the crossover in
the form of the secular terms results in a crossover in the sigma meson
relaxation: an ``anomalous'' (nonexponential) relaxation will dominate during
most of the time and the usual exponential relaxation ensues at very large
times.

This above discussion has revealed several important features highlighted by a
consistent resummation via the dynamical renormalization group:

(i) Threshold infrared divergences result in a breakdown of Fermi's golden
rule. The secular terms of the perturbative expansion are no longer linear in
time but include logarithmic contributions arising from these infrared
divergences.

(ii) The concept of the damping rate is directly tied to exponential
relaxation. The infrared divergences of the damping rate reflect the breakdown
of Fermi's golden rule and signal a very different relaxation from a simple
exponential.

(iii) Whereas the usual calculation of damping rates will lead to a divergent
result arising from the infrared threshold divergences, the dynamical
renormalization group approach recognizes that these threshold divergences
result in secular terms that are non-linear in time as discussed above. While
in relaxation time approximation linear secular terms lead to exponential
relaxation and therefore to an unambiguous definition of the damping rate,
non-linear secular terms lead to novel nonexponential relaxation phenomena for
which the concept of a damping rate may not be appropriate.

This discussion of threshold singularities and anomalous relaxation has paved
the way to studying the case of gauge field theories in next chapter, wherein
the emission and absorption of magnetic photons that are only dynamically
screened lead to a similar anomalous relaxation~\cite{boyanrgir}.

\section{Pinch Singularities and their Resolution}\label{sec:4.5}
An important difference between the approach to nonequilibrium evolution
described by quantum kinetic equations advocated in this work and that often
presented in the literature is that we work directly in \emph{real time} not
taking Fourier transforms in time. This has to be contrasted with the so-called
real-time formalism (RTF) of (non)equilibrium quantum field theory, in which
there are also a closed-time-path contour and four propagators but the
propagators and quantities computed therefrom are all in terms of temporal
Fourier transforms.

In thermal equilibrium the Fourier representation of these four propagators for
a scalar field are given by~\cite{book:lebellac,chou,landsmann}
\begin{eqnarray}
&&G_0^{++}(K)=-G_0^{--}(K)^{\ast}=-\frac{1}{K^2-m^2_\mathrm{eff}+i\varepsilon}
+2\pi i n_B(|k_0|)\,\delta(K^2-m^2_\mathrm{eff}),\nonumber\\
&&G_0^{-+}(K)= 2\pi i\,[\theta(k_0)+n_B(|k_0|)]\,
\delta(K^2-m^2_\mathrm{eff}),\nonumber\\
&&G_0^{+-}(K)= 2\pi i\,[\theta(-k_0)+n_B(|k_0|)]\,
\delta(K^2-m^2_\mathrm{eff}),\label{propagator1-ft}
\end{eqnarray}
where $K=(k_0,\mathbf{k})$ is the four-momentum with $K^2=k_0^2-k^2$, whereas
out of equilibrium the distribution functions are simply replaced by
non-thermal ones, i.e., $n_B(|k_0|)\rightarrow n_\mathbf{k}(t_0)$. Using the
integral representation of the Heaviside step function
\begin{equation}
\theta(t)=\frac{i}{2\pi}\int_{-\infty}^{+\infty}
\frac{d\omega}{\omega+i\varepsilon}
\,e^{-i\omega t},\label{thetat}
\end{equation}
one can easily show that the propagators given in Eq.~(\ref{propagator1-ft})
and the ones obtained by replacing the thermal equilibrium distributions by the
nonequilibrium ones are, respectively, the \emph{temporal} Fourier transforms
of those given in Eqs.~(\ref{scalarprops}) and (\ref{fqp:gsmall}). The Fourier
transforms of the free retarded and advanced propagators are obtained similarly
and read
\begin{equation}
G_0^{R/A}(K)=-\frac{1}{K^2-m^2_\mathrm{eff}\pm
i\,\textrm{sgn}(k_0)\,\varepsilon}.
\end{equation}

Several authors have pointed out that the calculations using the CTP formalism
in terms of the standard form of free propagators in Eq.~(\ref{propagator1-ft})
or those obtained by the replacement of the distribution functions by the
nonequilibrium ones, lead to \emph{pinch
singularities}~\cite{landsmann,altherr,bedaque,niegawa:2,greiner,dadic,carrington}.

In a consistent perturbative expansion both the retarded and advanced
propagators contribute and pinch singularities arise from the product of these,
for example for  a scalar field this product is of the form
\begin{equation}
G_0^R(K)\,G_0^A(K)=\frac{1}{[K^2-m^2_\mathrm{eff}+i\,
\textrm{sgn}(k_0)\,\varepsilon]
[K^2-m^2_\mathrm{eff}-i\,\textrm{sgn}(k_0)
\,\varepsilon]}.\label{pinchterm}
\end{equation}
For finite $\varepsilon$ this expression is regular, whereas when
$\varepsilon\rightarrow 0^+$ it gives rise to singular products such as
$[\delta(K^2-m^2_\mathrm{eff})]^2$ as discussed in
Refs.~\cite{landsmann,altherr,bedaque,niegawa:2,greiner,dadic,carrington}.
Singularities of this type are ubiquitous in nonequilibrium and are not
particular to scalar field theories.

A detailed analysis of these pinch terms reveals that they do not cancel each
other in perturbation theory unless the system is in thermal
equilibrium~\cite{landsmann,altherr,bedaque,niegawa:2,greiner}. Indeed, this
severe problem has cast doubt on the validity or usefulness of the CTP
formulation to describe nonequilibrium phenomena~\cite{altherr}. Although this
singularities have been found in many circumstances and analyzed and discussed
in the literature often, a systematic and satisfactory treatment of these
singularities is still lacking. In Ref.~\cite{carrington} it was suggested that
including an in-medium width of the quasiparticles to replace Feynman's
``$i\varepsilon$-prescription'' does provide a physically reasonable solution,
however this clearly casts doubt on the consistency of any perturbative
approach to describe even weakly out of equilibrium phenomena.

Recently some authors have conjectured that pinch singularities in perturbation
theory might be attributed to a misuse of Fourier transforms (for a detailed
discussion see Refs.~\cite{bedaque,niegawa:2,greiner}). As an illustrative and
simple example of these type of pinch singularities, these authors discussed
the elementary derivation of Fermi's golden rule in time-dependent perturbation
theory in quantum mechanics. In calculating total transition probabilities
there appears  the square of energy conservation delta function, which arises
due to taking the infinite time limit of scattering probabilities. In this
setting, such terms are interpreted as the elapsed scattering time multiplied
by the energy conservation constraint rather than a pathological singularity.

A close inspection of Eq.~(\ref{pinchterm}) reveals that the pinch term is the
square of on-shell condition for the free quasiparticle, which implies a
temporal Fourier transform in the infinite time limit and of the same form as
the square of the energy conservation constraint for the transition probability
obtained in time-dependent perturbation theory.

By assuming that the interaction duration time is large but finite,
Ni\'egawa~\cite{niegawa:2} and Greiner and Leupold~\cite{greiner} showed that
for a self-interacting scalar field the pinch part of the distribution function
can be regularized by the interaction duration time as~\cite{niegawa:2,greiner}
\begin{equation}
n^\mathrm{pinch}_\mathbf{k}(t)\simeq
(t-t_0)\,\Gamma^\mathrm{net}_\mathbf{k}(t_0), \label{pinchpart}
\end{equation}
where ``$\simeq$'' denotes that only the pinch singularity contribution is
included, $ t-t_0 $ is the interaction duration time, and
$\Gamma^\mathrm{net}_\mathbf{k}(t_0)$ is the net gain rate of the quasiparticle
distribution function per unit time
\begin{equation}
\Gamma^\mathrm{net}_\mathbf{k}(t_0)=
\frac{i}{2\omega_\mathbf{k}}\Big[[1+n_\mathbf{k}(t_0)]
\Sigma^{<}(\omega_\mathbf{k},\mathbf{k})-n_\mathbf{k}(t_0)
\Sigma^{>}(\omega_\mathbf{k},\mathbf{k})\Big]. \label{gammapinch}
\end{equation}
In the above expression, the on-shell self-energies
$\Sigma^\gtrless(\omega_\mathbf{k},\mathbf{k})$ are calculated in terms of the
initial distribution functions $n_\mathbf{k}(t_0)$~\cite{niegawa:2,greiner}.

Upon comparing Eqs.~(\ref{pinchpart}) and (\ref{gammapinch}) with
Eqs.~(\ref{scalar:perturb}) and (\ref{secularpart}), respectively, we clearly
see the \emph{equivalence} between the linear secular terms in the perturbative
expansion and the presence of pinch singularities in the usual CTP formalism.
Furthermore, in the discussion following Eq.(\ref{scalar:perturb}) we have
recognized that secular terms are not present if the system is in equilibrium,
much in the same manner as the case of pinch singularities as discussed
originally by Altherr~\cite{altherr}. Thus our conclusion is that \emph{pinch
singularities are a temporal Fourier transform representation of linear secular
terms}.

The dynamical renormalization group provides a systematic resummation of these
secular terms and a consistent formulation to implement the renormalization of
the distribution functions suggested by Ni\'egawa.~\cite{niegawa:2}. We
emphasize that the dynamical renormalization group approach explains the
physical origin of the pinch singularities in terms of secular terms and
Fermi's golden rule, and provides a consistent and systematic resummation of
these secular terms that lead to the quantum kinetic equation as a
renormalization group equation that determines the time evolution of the
distribution function. This result justifies in a systematic manner the
conclusions and interpretation obtained in Ref.~\cite{carrington} where a
possible regularization of the pinch singularities was achieved by including
the width of the quasiparticle obtained via the resummation of hard thermal
loops.

Furthermore, we stress that the dynamical renormalization group is far more
general in that it allows to treat situations where the long time evolution is
modified by threshold (infrared) singularities in spectral densities, thereby
providing a resolution of infrared singularities in damping rates and a
consistent resummation scheme to extract the asymptotic time evolution of the
distribution function. The infrared singularities in these damping rates is a
reflection of anomalous (nonexponential) relaxation as a result of threshold
effects.

The pinch singularities signal the breakdown of perturbation theory, just as
the secular terms in real time, however, the advantage of working directly in
real time is that the time scale on which perturbation theory breaks down is
recognized clearly from the real-time perturbative expansion and is identified
directly with the kinetic time scale. The dynamical renormalization group
justifies this identification by providing a resummation of the perturbative
series that improves the solution beyond the intermediate asymptotics.

The resolution of pinch singularities via the dynamical renormalization group
is general. As originally pointed out in Ref.~\cite{altherr} the pinch
singularities typically multiply expressions of the form Eq.~(\ref{gammapinch})
which vanish in equilibrium, just as the linear secular terms multiply similar
terms in the real-time perturbative expansion, as highlighted by
Eq.~(\ref{secularpart}). These terms are of the typical form gain minus loss,
in equilibrium they vanish, but their nonvanishing simply indicates that the
distribution functions are evolving in time and it is precisely this time
evolution that is described consistently by the dynamical renormalization
group.

\section{Conclusions}\label{sec:4.6}
The goal of this chapter is to provide a novel method for
obtaining quantum kinetic equations from a field theoretical and
diagrammatic perturbative expansion that is improved via a
dynamical renormalization group resummation \emph{directly in real
time}.

The first step of this method is to use the microscopic equations of motion to
obtain the evolution equation of the quasiparticle distribution function, i.e.,
the expectation value of the quasiparticle number operator in the initial
density matrix. This evolution equation can be solved in a consistent
diagrammatic perturbative expansion and one finds that the solution for the
time evolution of the distribution function features \emph{secular terms},
i.e., terms that grow in time. In perturbation theory the microscopic and
kinetic time scales are widely separated, hence there is a regime of
intermediate asymptotics in time, within which (i) perturbation theory is valid
and (ii) the secular terms dominate the time evolution of the distribution
function.

The second step is to implement the dynamical renormalization group to resum
the secular terms. A renormalization of the distribution function absorbs the
contribution from the secular terms on a given renormalization time scale,
thereby improving the perturbative expansion. The arbitrariness of this
renormalization scale leads to the dynamical renormalization group equation,
which is interpreted as the quantum kinetic equation. In relaxation time
approximation linear one recognizes that secular terms correspond to usual
exponential relaxation, whereas nonlinear secular terms correspond to anomalous
(nonexponential) relaxation.

We first test this new dynamical renormalization group method within the
familiar self-interacting scalar field theory, not only reproducing but also
generalizing the previous results in the literature. We then move on to apply
the this method to study quantum kinetics of a gas of cool pions and sigma
mesons described by the $O(4)$ linear sigma model in the chiral limit. In the
relaxation time approximation the exponential relaxation of the pions and sigma
mesons is described by the corresponding relaxation rates, which are in
agreement with the damping rate found recently for the same
model~\cite{pisarski:coolpion,patkos,csernai}. This particular model also
reveals a crossover behavior in sigma meson relaxation in the large momentum
limit as a result of threshold singularities associated with the emission and
absorption of massless pions. In the relaxation time approximation we find a
crossover between purely exponential and anomalous nonexponential relaxation
with an exponent of the form $t\ln t$. The crossover time scale is found to be
dependant on the momentum of the sigma resonance. The anomalous relaxation is a
novel result and could be of phenomenological relevance in view of recent
suggestions of novel threshold effects of the sigma resonance near the chiral
phase transition~\cite{hatsuda}, this possibility is worthy of a deeper study
and generalization of the dynamical renormalization group method to reach the
critical region is in progress.

Furthermore, we have established a very close relationship between
the usual renormalization group and the dynamical renormalization
group approach to kinetics. We have shown that the dynamical
renormalization group equation is the quantum kinetic equation,
the collision terms are the equivalent of the beta functions in
the Euclidean renormalization group. Fixed points of the dynamical
renormalization group are identified with stationary solutions of
the kinetic equation and the exponents that determine the
stability of the fixed points are identified with the relaxation
rates in the relaxation time approximation. We have also suggested
that in this language coarse-graining is the equivalent to
neglecting irrelevant couplings in the Euclidean renormalization
program. This identification brings a new and rather different
perspective to kinetics and relaxation that will hopefully lead to
new insights.

There are many advantages in the dynamical renormalization group approach to
quantum kinetics:

(i) This method is based on straightforward quantum field theoretical
diagrammatic perturbation, hence it allows a systematic calculation to any
arbitrary order in perturbation theory. It includes nonequilibrium medium
effects through a consistent resummation of the secular terms, thereby
providing nonequilibrium generalizations of the usual hard thermal loop
resummation in quantum kinetic theory. This feature is worked out in detail in
the scalar field theory.

(ii) It allows a detailed understanding of crossover behavior between different
relaxation phenomena directly in real time. This is important in the case of
wide resonances where threshold effects may lead to nonexponential relaxation
on some time scales, and also in the case of relaxation near phase transitions
where soft collective excitations dominate the dynamics.

(iii) It describes nonexponential relaxation directly in real time whenever
threshold effects are important, thus providing a real-time interpretation of
infrared divergent damping rates in gauge field theories. We consider this one
of the most valuable features of the dynamical renormalization group which
makes this approach particularly suited to study relaxation in gauge field
theories in a medium where the emission and absorption of soft gauge fields
typically lead to threshold infrared divergences.

(iv) This method provides a simple and natural resolution of pinch
singularities often found in nonequilibrium field theory when the distribution
functions are out of thermal equilibrium. Pinch singularities are the temporal
Fourier transform manifestation of the secular terms, and their resolution is
achieved via consistent resummation of secular terms implemented by the
dynamical renormalization group.

We envisage several important applications of the dynamical
renormalization group method primarily to study transport
phenomena and relaxation of collective modes in gauge theories
where infrared effects are important, as well as to study
relaxation phenomena near critical points where soft collective
fluctuations dominate the dynamics. An important aspect of this
method is that it does not rely on the quasiparticle approximation
and allows a direct interpretation of infrared phenomena directly
in real time. We will return to this issue in Chap.~\ref{chap:5},
where we study in detail nonequilibrium dynamics of photons and
fermions in a QED plasma at high temperature.

%% file: chap5.tex
\chapter{Nonequilibrium Dynamics in a QED Plasma at High Temperature}
\label{chap:5}
\section{Introduction}\label{sec:5.1}
The study of nonequilibrium dynamics in hot Abelian and non-Abelian plasmas
under extreme conditions is of fundamental importance in the understanding of
the formation and evolution of a novel phase of matter, the quark-gluon plasma
(QGP)~\cite{qgp,book:qgp}, expected to be produced in current ultrarelativistic
heavy ion experiments at BNL Relativistic Heavy Ion Collider (RHIC) and future
programs at CERN Large Hadron Collider (LHC). Estimates based on energy
deposited in the central collision region at RHIC energies for
$\sqrt{s_{NN}}\sim 200$ GeV suggest that the lifetime of the deconfined QGP is
of order $10$ fm/$c$~\cite{wang,geiger:pcm}. At such unprecedentedly short time
scales, an important aspect is an assessment of thermalization time scales and
the potential for nonequilibrium effects associated with the rapid expansion
and finite lifetime of the plasma and their impact on experimental observables.
Lattice QCD is simply unable to deal with these questions because simulations
are restricted to thermodynamic equilibrium quantities, hance a nonequilibrium
field-theoretical approach is needed for an consistent description of the
formation and evolution of the QGP.

An important and pioneering step in this direction was undertaken
by Geiger~\cite{geiger:pcm}, who applied transport methods
combined with perturbative QCD (pQCD) to obtain a quantitative
picture of the evolution of partons in the early stages of
formation and evolution of the plasma. The consistent study of the
evolution of partons in terms of pQCD cross sections that include
screening corrections to avoid the infrared divergences associated
with small angle scattering lead to the conclusion that quarks and
gluons thermalize on time scales about 1
fm/$c$~\cite{biro,geiger:pcm}. The necessity of a deeper
understanding of equilibrium and nonequilibrium aspects of the
quark-gluon plasma motivated an intense study of the Abelian and
non-Abelian plasmas in extreme environments. A major step towards
a consistent description of nonperturbative aspects was taken by
Braaten and Pisarski~\cite{htl}, who introduced a novel
resummation method that reorganizes the perturbative expansion in
terms of the effective degrees of freedom associated with
collective modes rather than bare particles. This hard thermal
loop (HTL) resummation program is now at the heart of most
treatments of equilibrium Abelian and non-Abelian plasmas at high
temperature~\cite{book:lebellac}.

Thermal field theory provides the tools to study the properties of
plasmas in \emph{equilibrium}~\cite{book:kapusta,book:lebellac},
but the consistent study of nonequilibrium phenomena in real time
requires the methods of nonequilibrium field
theory~\cite{geiger:qcd}. The study of the equilibrium and
nonequilibrium properties of Abelian and non-Abelian plasmas at
high temperature as applied to the QGP has as ultimate goal to
obtain a deeper understanding of the potential experimental
signatures of the formation and evolution of the QGP in
ultrarelativistic heavy ion collisions. Amongst these, photons and
dileptons (electron and/or muon pairs) produced during the early
stages of the QGP are considered as some of the most promising
signals~\cite{feinberg,mclerran,baier:dilepton,kapusta,baier:photon,ruuskanen,alam:em}.
Since photons and lepton pairs interact electromagnetically their
mean free paths are longer than the estimated size of the QGP
fireball $\sim 10-20$ fm and unlike hadronic signals they do not
undergo final state interactions. Therefore photons and dileptons
produced during the early stages of QGP carry clean information
from this phase.

In this chapter we aim to provide a comprehensive study of several
relevant aspects of the nonequilibrium dynamics in a QED plasma at
high temperature directly in real time. This study is of
phenomenological importance in ultrarelativistic heavy ion
collisions, as many features of a QCD plasma at high temperature
are similar to those of a QED plasma at high
temperature~\cite{book:lebellac,htl,weldon:1}. In particular, the
leading contributions in the hard thermal loop approximation can
be straightforwardly generalized from QED to QCD, thus relaxation
of quarks and production of photons in the
QGP~\cite{kapusta,baier:photon} can be understood to leading order
from the study of a QED plasma at high temperature. More
specifically, we focus on the following nonequilibrium aspects.

(i) The real-time evolution of gauge mean fields in linear response in the HTL
approximation. The goal here is to study directly in real time the relaxation
of (coherent) gauge field configurations in the linearized approximation to
leading order in the HTL approximation. Whereas a similar study has been
carried out in scalar quantum electrodynamics (SQED)~\cite{boyanhtl} and
confirmed numerically in Ref.~\cite{rajantie}, the most relevant case of
fermionic QED has not yet been studied in detail.

(ii) The quantum kinetic equation that describes the evolution of the
distribution function of photons in the medium, again to leading order in the
HTL approximation. This aspect is relevant to study photon production via
off-shell effects directly in real time. As explained in detail, this quantum
kinetic equation, obtained from a microscopic field theoretical approach based
on the dynamical renormalization group displays novel off-shell effects that
cannot be captured via the usual kinetic description that assumes completed
collisions~\cite{mclerran,kapusta,baier:photon,ruuskanen}.

(iii) The evolution of fermion mean fields at large times features anomalous
relaxation arising from the emission and absorption of magnetic photons
(gluons) which are only dynamically screened by Landau
damping~\cite{blaizot:qed,takashiba}. The fermion propagator was studied
previously in real time in the Bloch-Nordsieck approximation which provides a
resummation of the infrared divergences associated with soft photon (or gluon)
bremsstrahlung in the medium~\cite{blaizot:qed,takashiba}. In this chapter we
implement the dynamical renormalization group to study the evolution of fermion
mean fields providing an alternative to the Bloch-Nordsieck treatment.

(iv) We obtain the quantum kinetic equation for the fermionic distribution
function for hard fermions via the implementation of the dynamical
renormalization group. There has recently been an important effort in trying to
obtain the effective kinetic (Boltzmann) equations for hard charged
(quasi)particles~\cite{blaizot:be,bodeker,litim} but the collision kernel in
this equation features the logarithmic divergences associated with the emission
and absorption of soft magnetic photons (or gluons)~\cite{blaizot:be}. The
dynamical renormalization group leads to a quantum kinetic equation directly in
real time bypassing the assumption of completed collisions and leads to a
\emph{time-dependent} collision kernel free of infrared divergences.

A fundamental issue that must be addressed prior to setting up our study is
that of gauge invariance. In the Abelian case, it is straightforward to reduce
the Hilbert space to the gauge invariant subspace and to define gauge invariant
field operators for the gauge boson and fermion. The description in terms of
gauge invariant states and operators is best achieved within the canonical
formulation which begins with the identification of the canonical fields and
conjugate momenta as well as the primary and secondary (first-class)
constraints associated with gauge invariance. The physical states are those
annihilated by the constraints and physical operators are those commute with
the constraints. Such a gauge invariant formulation has been implemented
explicitly in the case of SQED~\cite{boyangaugeinv} and the fermionic case can
be treated in the same manner with a few minor technical modifications.

The final result of this formulation is that the Hamiltonian written in terms
of gauge invariant fields and acting on the gauge invariant states is exactly
equivalent to that obtained in Coulomb gauge, which is the statement that
Coulomb gauge describes the theory in terms of the physical degrees of freedom.
Furthermore, the instantaneous Coulomb interaction can be traded for a Lagrange
multiplier~\cite{boyanhtl} leading to the following Lagrangian density
\begin{eqnarray}
\mathcal{L}[\mathbf{A}_T,A_0,\Psi,\bar\Psi]&=&\frac{1}{2}\left[\left(\partial_\mu
\mathbf{A}_T\right)^2 + \left(\nabla A_0\right)^2\right]+
\bar{\Psi}\left(i{\not\!{\partial}}-e\gamma^0 A_0+e{\bbox
\gamma}\cdot\mathbf{A}_T-m\right)\Psi\nonumber\\
&&+\,\mathbf{J}_T\cdot\mathbf{A}_T+\bar{\eta}\Psi+\bar{\Psi}\eta,
\end{eqnarray}
where $e$ is the gauge coupling constant, $\Psi$ is charged fermion field,
$\mathbf{A}_T$ is the transverse component of the gauge (photon) field, and
$A_0$ is \emph{not} to be interpreted as the time component of the gauge filed
\emph{but} is a Lagrange multiplier associated with the instantaneous Coulomb
interaction. We emphasize that the fields $\Psi$, $\mathbf{A}$, and $A_0$ are
all \emph{gauge invariant} (see Refs.~\cite{boyanhtl,boyangaugeinv} for
details). In writing the above Lagrangian density, we have included the
external fermionic (Grassmann) source $\eta$ and electromagnetic source
$\mathbf{J}_T$ to provide an initial value problem for studying relaxation of
the mean fields.\footnote{In this chapter we will not discuss relaxation of the
Lagrangian multiplier $A_0$ associated with the instantaneous Coulomb
interaction, hence the corresponding external source is neglected.}

Furthermore, we will consider a charge neutral plasma with zero fermion
chemical potential at a temperature $T\gg m$, where $m$ is the (physical)
fermion mass at zero temperature, hence in what follows we neglect the fermion
mass unless otherwise stated.

The chapter is organized as follows. In Sec.~\ref{sec:5.2} we first study
relaxation of the photon mean field in the hard thermal loop approximation for
soft $k\leq eT$ and semihard $eT\ll k \ll T$ photon momentum and obtain the
quantum kinetic equation for the (hard and semihard) photon distribution
function. In Sec.~\ref{sec:5.3} we study relaxation of fermionic mean fields
for hard momentum and the quantum kinetic equation for the fermion distribution
function. Our conclusions and some further questions are presented in
Sec.~\ref{sec:5.4}.

\section{Photons Out of Equilibrium}\label{sec:5.2}
\subsection{Relaxation of the gauge mean field}
We begin with the relaxation of \emph{soft} photons of momentum $k\sim eT$. As
discussed in Sec.~\ref{sec:2.2}, the equation of motion for the transverse
photon mean field can be derived from the tadpole method~\cite{tadpole} by
decomposing the full quantum fields into $c$-number expectation values and
quantum fluctuations around the expectation values:
\begin{equation}
\mathbf{A}^\pm_T(\mathbf{x},t)=\mathbf{a}_T(\mathbf{x},t)+
\bbox{\mathcal{A}}^\pm_T(\mathbf{x},t),\quad\mbox{with}\quad
\langle\bbox{\mathcal{A}}^\pm_T(\mathbf{x},t)\rangle=0.
\end{equation}

In momentum space, one obtains the following linearized equation of motion
\begin{equation}
\left(\partial_t^2+k^2 \right)\,\mathbf{a}_T(\mathbf{k},t) +
\int^{t}_{-\infty}dt'\,\Pi_T(\mathbf{k},t-t')\,\mathbf{a}_T(\mathbf{k},t')=
\mathbf{J}_T(\mathbf{k},t), \label{maxwelleq1}
\end{equation}
where $\Pi_T(\mathbf{k},t-t')$ is the transverse part of the retarded photon
self-energy and
$$
\mathbf{a}_T(\mathbf{k},t) \equiv \int d^3x\,e^{-i\mathbf{k}\cdot\mathbf{x}} \,
\mathbf{a}_T(\mathbf{x},t).
$$
Using the nonequilibrium Feynman rules and the free real-time fermion
propagators [see Eq.~(\ref{fermionprops})], we find
\begin{equation}
\Pi_T(\mathbf{k},t-t')=\int_{-\infty}^{+\infty}
d\omega\,\rho_T(\omega,\mathbf{k})\sin[\omega(t-t')],\label{Pit}
\end{equation}
where $\rho_T(\omega,\mathbf{k})$ is the spectral function of the transverse
photon self-energy and to one-loop order reads
\begin{eqnarray}
\rho_T(\omega,\mathbf{k})&=&-e^2\int\frac{d^3 q}{(2\pi)^3}
\Big\{[\delta(\omega-p-q)-\delta(\omega+p+q)] [1+(\mathbf{\hat
k}\cdot\mathbf{\hat p})(\mathbf{\hat k}\cdot\mathbf{\hat q})]
[1-n_F(p)-n_F(q)]\nonumber\\
&&+\,[\delta(\omega-p+q)-\delta(\omega+p-q)][n_F(q)-n_F(p)][1-(\mathbf{\hat
k}\cdot\mathbf{\hat p})(\mathbf{\hat k}\cdot\mathbf{\hat
q})]\Big\},\label{rhot}
\end{eqnarray}
with $\mathbf{p}=\mathbf{k}+\mathbf{q}$.

As before, the source $\mathbf{J}_T(\mathbf{x},t)$ is taken to be switched on
adiabatically from $t=-\infty$ and switched off at $t=0$ to provide the initial
conditions [see Eq.~(\ref{icond})]
\begin{eqnarray}
&&\mathbf{a}_T(\mathbf{k},t=0)=\psi(\mathbf{k},0),
\quad\dot{\mathbf{a}}_T(\mathbf{k},t\leq 0)=0.\label{qed:icond}
\end{eqnarray}
Introduce an auxiliary quantity $\pi_T(\mathbf{k},t-t')$ defined by
$$
\Pi_T(\mathbf{k},t-t')=\frac{\partial}{\partial t'}\pi_T(\mathbf{k},t-t'),
$$
or, equivalently,
$$
\pi_T(\mathbf{k},t-t') = \int_{-\infty}^{+\infty} {d\omega\over \omega}
\rho_T(\omega,\mathbf{k})\cos[\omega(t-t')],
$$
and impose $\mathbf{J}_T(\mathbf{k},t>0)=0$, we can rewrite the equation of
motion for $t>0$ as an initial value problem
\begin{equation}
\left(\partial_t^2+k^2\right)\mathbf{a}_T(\mathbf{k},t) -
\int^{t}_{0}dt'\,\pi_T(\mathbf{k},t-t')\,\dot\mathbf{a}_T(\mathbf{k},t')
+\pi_T(\mathbf{k},0)\,\mathbf{a}_T(\mathbf{k},t)=0,\label{maxwelleq2}
\end{equation}
where the initial conditions specified by Eq.~(\ref{qed:icond}).

This equation of motion can be solved by Laplace transform as befits an initial
value problem. In terms the Laplace transformed quantities
\begin{equation}
\tilde\mathbf{a}_T(s,\mathbf{k})\equiv\int_0^{\infty}dt\,
e^{-st}\,\mathbf{a}_T(\mathbf{k},t),\quad
\tilde{\pi}_T(s,\mathbf{k})\equiv\int_0^{\infty}dt\,e^{-st}\,\pi_T(\mathbf{k},t),
\label{laplatrans}
\end{equation}
where $\mathrm{Re}\,s>0$, the Laplace transformed equation of motion is given
by
\begin{equation}
\left[s^2+k^2+\widetilde{\Pi}_T(s,\mathbf{k})\right]\tilde\mathbf{a}_T(s,\mathbf{k})=
\left[s-\tilde{\pi}_T(s,\mathbf{k})\right]\mathbf{a}_T(\mathbf{k},0),
\label{LTmaxwelleq1}
\end{equation}
where $\widetilde{\Pi}_T(s,\mathbf{k})$ is the Laplace transform of
${\Pi}_T(\mathbf{k},t-t')$
$$
\widetilde{\Pi}_T(s,\mathbf{k})=\pi_T(\mathbf{k},0)-s\,
\tilde{\pi}_T(s,\mathbf{k}).
$$
Eq.~(\ref{LTmaxwelleq1}) has the solution
\begin{equation}
\tilde\mathbf{a}_T(s,\mathbf{k})=\frac{1}{s}\left\{1-
\widetilde{D}_T(s,\mathbf{k})[k^2+\pi_T(\mathbf{k},0)]\right\}\,
\mathbf{a}_T(\mathbf{k},0) \label{laplasol},
\end{equation}
where the retarded transverse photon propagator $\widetilde{D}_T(s,\mathbf{k})$
is given by
\begin{equation}
\widetilde{D}_T(s,\mathbf{k})=[s^2+k^2+
\widetilde{\Pi}_T(s,\mathbf{k})]^{-1}\label{deltatil}.
\end{equation}
The real-time evolution of $\mathbf{a}_T(\mathbf{k},t)$ is obtained by
performing the inverse Laplace transform along the Bromwich contour which is to
the right of all singularities of $\tilde\mathbf{a}_T(s,\mathbf{k})$ in the
complex $s$-plane.

We note that this result is rather \emph{different} from that obtained in
Ref.~\cite{boyanhtl} for SQED in the structure of the solution, in particular
the prefactor $1/s$ in Eq.~(\ref{laplasol}). This can be traced back to the
\emph{different} manner in which we have set up the initial value problem, as
compared to the case studied in~\cite{boyanhtl}. The adiabatic source
(\ref{extsource}) (for $t_0=0$) has a Fourier transform that has a simple pole
in the frequency plane, this translates into the $1/s$ factor in
Eq.~(\ref{laplasol}). This particular case was \emph{not} contemplated in those
studied in Ref.~\cite{boyanhtl} since the source (\ref{extsource}) is
\emph{not} a regular function of the frequency. This difference will be seen to
be at the heart of important aspects of relaxation of the mean field condensate
in the soft momentum limit as discussed in detail below.

It is straightforward to see that the residue vanishes at $s=0$ in the solution
(\ref{laplasol}), hence the singularities of $\tilde\mathbf{a}_T(s,\mathbf{k})$
are those arising from the retarded transverse photon propagator
$\widetilde{D}_T(s,\mathbf{k})$.

\subsubsection{Soft photons $\bbox{k\sim eT}$: real-time Landau damping}
For soft photons of momenta $k\sim eT$, the leading ($\sim e^2T^2$)
contribution to the photon self-energy arises from loop momenta of order $T$
and is referred to as the hard thermal loop (HTL) in the literature~\cite{htl}.
In this region of momenta, the contribution of the hard fermion loop is
nonperturbative (see below).

In the HTL approximation ($s\sim k\ll q$), after some algebra we find that the
leading contribution to $\widetilde\Pi_T(s,\mathbf{k})$ arises exclusively from
the terms associated with $\delta(\omega\mp p\pm q)$ in Eq.~(\ref{rhot}) and
reads
\begin{equation}
\widetilde\Pi_T(s,\mathbf{k})= \frac{e^2 T^2}{12}\left[\frac{is}{k}
\left(1+\frac{s^2}{k^2}\right)\ln\frac{is+k}{is-k}-2\,\frac{s^2}{k^2}\right].
\label{HTLPis}
\end{equation}
The delta functions $\delta(\omega\mp p\pm q)$ have support below the light
cone ($\omega^2 < k^2$) and correspond to Landau damping processes~\cite{htl}
in which the soft photon scatters a hard fermion in the plasma.

The analytic continuation of $\widetilde\Pi_T(s,\mathbf{k})$ in
the complex $s$-plane (physical sheet) is defined by
\begin{equation}
\Pi_T(\omega,\mathbf{k})\equiv\widetilde{\Pi}_T
(s=-i\omega+0^+,\mathbf{k})=\mathrm{Re}\Pi_T(\omega,\mathbf{k})+
i\,\mathrm{Im}\Pi_T(\omega,\mathbf{k}), \label{pi:anacont}
\end{equation}
where the real and imaginary parts are related by the usual dispersion
relations. The analytical continuation of $\widetilde{D}_T(s,\mathbf{k})$ and
$\tilde\pi_T(s,\mathbf{k})$ can be defined analogously, and they are related to
$\Pi_T(\omega,\mathbf{k})$ by
\begin{eqnarray}
&&\Delta_T(\omega,\mathbf{k})=
-[\omega^2-k^2-\Pi_T(\omega,\mathbf{k})]^{-1},\nonumber\\
&&\mathrm{Re}\Pi_T(\omega,\mathbf{k})=\pi_T(\mathbf{k},0)-
\omega\,\mathrm{Im}\,\pi_T(\omega,\mathbf{k}),\nonumber\\
&&\mathrm{Im}\Pi_T(\omega,\mathbf{k})=\omega\,\mathrm{Re}\,\pi_T(\omega,\mathbf{k}).
\end{eqnarray}
One finds from Eq.~(\ref{HTLPis}) that the analytical continuation of
$\widetilde\Pi_T(s,\mathbf{k})$ in the HTL approximation is given by
\begin{eqnarray}
&&\mathrm{Re}\Pi_T(\omega,\mathbf{k})=\frac{e^2 T^2}{12}
\left[2\frac{\omega^2}{k^2}+\frac{\omega}{k}
\left(1-\frac{\omega^2}{k^2}\right)
\ln\left|\frac{\omega+k}{\omega-k}\right|\right],\nonumber\\
&&\mathrm{Im}\Pi_T(\omega,\mathbf{k})=-\frac{\pi e^2
T^2}{12}\frac{\omega}{k}\left(1-\frac{\omega^2}{k^2}\right)
\theta(k^2-\omega^2).\label{Pi:htl}
\end{eqnarray}
We note that $\mathrm{Im}\Pi_T(\omega,\mathbf{k})$ only has support below the
light cone and vanishes \emph{linearly} as $\omega\to k$ from below. As can be
seen from Eq.~(\ref{Pi:htl}) for soft momenta $k\lesssim eT$ the HTL photon
self-energy is comparable in magnitude to or larger than the free inverse
propagator, therefore we have to treat it nonperturbatively for soft photons.

Isolated poles of $\Delta_T(\omega,\mathbf{k})$ in the complex $\omega$-plane
correspond to quasiparticles (or collective excitations)~\cite{book:lebellac}.
The transverse photon pole $\omega_T(k)$ is real and determined by\footnote{In
the following discussion, $\omega_T(k)$ is referred to as the positive pole.}
\begin{equation}
\omega^2_T(k)-k^2-\mathrm{Re}\Pi_T(\omega_T(k),\mathbf{k})=0.
\end{equation}
For ultrasoft photons $k\ll eT$, the dispersion relation of the collective
excitations is~\cite{book:lebellac}
\begin{equation}
\omega_T^2(k)=\omega_{P}^2+\frac{6}{5}k^2+\mathcal{O}\left(\frac{k^4}{e^2
\,T^2}\right),
\end{equation}
where $\omega_{P}=eT/3$ is the \emph{plasma frequency}. Consequently, the
retarded transverse photon propagator $\widetilde{D}_T(s,\mathbf{k})$ has two
isolated poles at $s=\pm i\omega_T(k)$ and a branch cut from $s=-ik$ to $s=ik$.

Having analyzed the analytic structure of
$\widetilde{D}_T(s,\mathbf{k})$, we can now perform the integral
along the Bromwich contour to obtain the real-time evolution of
$\mathbf{a}_T(\mathbf{k},t)$. Closing the contour in the
half-plane with $\mathrm{Re}\,s<0$, we find
$\mathbf{a}_T(\mathbf{k},t)$ for $t>0$ to be given by
$$
\mathbf{a}_T(\mathbf{k},t)=\mathbf{a}^{\it pole}_T(\mathbf{k},t)+
\mathbf{a}^{\it cut}_T(\mathbf{k},t),
$$
with
\begin{eqnarray}
\mathbf{a}^{\it pole}_T(\mathbf{k},t)&=& \frac{k^2Z_T(k)}{\omega_T^2(k)}
\cos[\omega_T(k)t]\,\mathbf{a}_T(\mathbf{k},0),
\label{sol:aTpole}\\
\mathbf{a}^{\it cut}_T(\mathbf{k},t)&=&
k^2\int_{-k}^{+k}\frac{d\omega}{\omega}\, \beta_T(\omega,k)\,e^{-i\omega
t}\,\mathbf{a}_T(\mathbf{k},0),\label{sol:aTcut}
\end{eqnarray}
where
\begin{eqnarray}
Z_T(k)&=&\left[1-\frac{\partial\mathrm{Re}\Pi_T(\omega,\mathbf{k})}
{\partial\omega^2}\right]^{-1}_{\omega=\omega_T(k)},\vspace*{0.5ex}\nonumber\\
\beta_T(\omega,k)&=&\frac{1}{\pi}\frac{\mathrm{Im}\Pi_T(\omega,\mathbf{k})}
{\left[\omega^2-k^2-\mathrm{Re}\Pi_T(\omega,\mathbf{k})\right]^2+
\left[\mathrm{Im}\Pi_T(\omega,\mathbf{k})\right]^2}. \label{gorda}
\end{eqnarray}
The solution evaluated at $t=0$ must match the initial condition, this
requirement leads immediately to the  sum rule~\cite{book:lebellac}
$$
\int_{-\infty}^{+\infty} \frac{dq_0}{q_0}\,\tilde{\rho}_T(q_0,q)=\frac{1}{q^2},
$$
where $ \tilde{\rho}_{T}(q_0,q) $ is the HTL-resummed spectral function for the
transverse photon propagator:\footnote{In our notation the spectral function
for the self-energy is denoted as $\rho$ , whereas that for the corresponding
propagator is denoted as $\tilde\rho$.}
\begin{eqnarray}
\tilde{\rho}_{T}(q_0,q)&=&\frac{1}{\pi}\mathrm{Im}\Delta_T(q_0,q)\nonumber\\
&=&\mbox{sgn}(q_0)\,Z_{T}(q)\,\delta[q^2_0-\omega^2_{T}(q)]+
\,\beta_{T}(q_0,q)\, \theta(q^2-q_0^2). \label{tilderhot}
\end{eqnarray}
A noteworthy feature of  the cut contribution (\ref{sol:aTcut}) is the factor
$\omega$ in the denominator. The presence of this factor can be traced back to
the preparation of the initial state via the adiabatically switched-on external
source, which is the switched off at $t=0$. The Fourier transform of this
source is proportional to $1/\omega$ and results in the prefactor of
$\beta_T(\omega,k)$ in Eq.~(\ref{sol:aTcut}).

For $k\sim eT$, Eq.~(\ref{sol:aTcut}) cannot be evaluated in closed form but
its long-time asymptotics can be extracted by writing the integral along the
cut as a contour integral in the complex $\omega$-plane. The integration
contour $C$ is chosen to run clockwise along the segment $-k<\omega< k$ on the
real axis, the line $\omega=k-iz$ with $0\le z<\infty$, then around an arc at
infinity and back to the real axis along the line $\omega=-k-iz$. After some
algebra we obtain the following expression
\begin{eqnarray}
\mathbf{a}^{\it cut}_T(\mathbf{k},t)&=&k^2\bigg[\left(e^{-ikt}\int_0^\infty
dz\,e^{-zt}\,\frac{i\beta_T(k-iz,k)}{k-iz}+\mbox{c.c.}\right)\nonumber\\
&&-\,2\pi i\! \sum_{\hbox{\scriptsize poles}\atop\hbox{\scriptsize inside $C$}}
\mathrm{Res}\!\left(\frac{\beta_T(\omega,k)}{\omega}e^{-i\omega
t}\right)\bigg]\,\mathbf{a}_T(\mathbf{k},0).
\end{eqnarray}
The contribution from the poles inside the contour $C$ is dominated at long
times by the pole closest to the real axis and is exponentially suppressed. On
the other hand, because of the exponential factor $e^{-zt}$ in the integrals,
the dominant contributions to the integral at long times ($t\gg 1/k$) arise
from the end points of the branch cut with $z\ll k$, thereby leading to a
long-time behavior characterized by a power law:
\begin{equation}
\mathbf{a}^{cut}_T(\mathbf{k},t)\buildrel{kt\gg 1}\over= -\frac{12}{e^2
T^2}\frac{\cos kt}{t^2}\,\mathbf{a}_T(\mathbf{k},0)\left[1 +
\mathcal{O}\left({1 \over t} \right)\right]\quad\mbox{for}\quad k\sim
eT,\label{tail}
\end{equation}
which agrees with the results of Ref.~\cite{boyanhtl} in scalar
quantum electrodynamics (SQED) at high temperature.

\begin{figure}[t]
\begin{center}
\includegraphics[width=3.5truein,keepaspectratio=true]{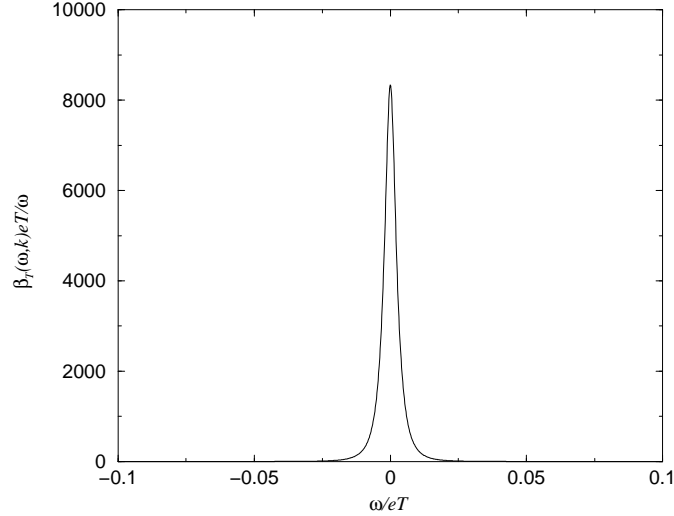}
\caption[The function $\beta_T(\omega,k)/\omega$ plotted as a function of
$\omega$ for ultrasoft photons.]{The function $\beta_T(\omega,k)/\omega$
plotted as a function of $\omega$ for ultrasoft photons
($k/eT=0.1$).}\label{fig:softphoton}
\end{center}
\end{figure}

For very soft momenta $k\ll eT$, the function $\beta_T(\omega,k)/\omega$ is
strongly peaked at $\omega=0$ as depicted in Fig.~\ref{fig:softphoton}.
Furthermore, for $\omega\ll k\ll eT$ the function $\beta_T(\omega,k)/\omega$
features a Breit-Wigner form
\begin{equation}
\frac{1}{\omega}\beta_T(\omega,k)\,\buildrel{\omega\ll k\ll eT}\over=\,
\frac{1}{\pi k^2}\frac{\Gamma(k)}{\omega^2+\Gamma^2(k)}
\quad\mbox{with}\quad\Gamma(k)=\frac{1}{\pi} \frac{12k^3}{e^2 T^2}.
\label{breitwigner}
\end{equation}
Upon using this narrow-width approximation in evaluating Eq.~(\ref{sol:aTcut}),
one obtains
\begin{equation}
\mathbf{a}^{\it cut}_T(\mathbf{k},t) \buildrel{kt\gg 1}\over= \left[
e^{-\Gamma(k)t}-\frac{2\,\Gamma(k)}{\pi k^2 t} \sin kt +
\mathcal{O}\left(\frac{\Gamma(k)}{k^3 t^2} \right)
\right]\,\mathbf{a}_T(\mathbf{k},0)\quad\mbox{for}\quad k\ll eT.
\label{landaudamping}
\end{equation}
The end-point contribution which results in a power law as in Eq.(\ref{tail})
is very small compared with Eq.~(\ref{landaudamping}) for times $t\lesssim
1/\Gamma(k)$. This power law becomes dominant on time scales \emph{much longer}
than $1/\Gamma(k)$, because its amplitude at $t\sim 1/\Gamma(k)$ is of order
$(k/eT)^6\ll 1$ in the soft momentum limit. Thus, upon combining the pole and
cut contributions, we find the following real-time behavior
\begin{equation}
\mathbf{a}_T(\mathbf{k},t)\,\buildrel{{kt\gg 1,}\atop{\Gamma(k)t\lesssim 1}
}\over=\,\mathbf{a}_T(\mathbf{k},0)\left[\frac{k^2 Z_T(k)}{\omega_T^2(k)}
\cos[\omega_T(k)t]+ e^{-\Gamma(k) t}\right]\quad\mbox{for}\quad k\ll eT.
\end{equation}

Two comments here are in order: (i) The exponential decay is not the same as
the usual decay of an unstable particle (or even collisional broadening) for
which the amplitude relaxes to zero at times much longer than the relaxation
time. In this case the asymptotic behavior of the amplitude is completely
determined by the transverse photon pole $\omega_T(k)$, and to this order in
the HTL approximation the collective excitation is stable. The exponential
relaxation does not arise from a resonance at the position of the transverse
photon pole but at zero frequency. Clearly, we expect a true exponential
damping of the collective excitation at higher order as a result of collisional
broadening. (ii) The exponential relaxation given by Eq.~(\ref{landaudamping})
can \emph{only} be probed by adiabatically switching on an external source [see
Eq.~(\ref{extsource})] whose temporal Fourier transform has a simple pole at
$\omega=0$ which is the origin of the prefactor $1/\omega$ in
Eq.~(\ref{sol:aTcut}). That is the adiabatic preparation of the initial state
excites this pole in the Landau damping cut. For external sources whose
temporal Fourier transform is regular at $\omega=0$ no exponential relaxation
arises, in agreement with the results of Ref.~\cite{boyanhtl}. Thus we find
that the exponential relaxation is \emph{not} a generic feature of the
evolution of the mean field, but emerges only for particular (albeit physically
motivated) initial conditions.

Our results confirm those found in SQED at high
temperature~\cite{rajantie}, wherein a numerical analysis of the
relaxation of the mean field was carried out. In that reference
the authors studied the real time evolution in terms of local
Hamiltonian equations obtained by introducing a nonlocal auxiliary
field. The initial conditions chosen there for this nonlocal
auxiliary field correspond precisely to the choice of an
adiabatically switched-on external source leading to an
adiabatically prepared initial value problem.

\subsubsection{Semihard photons $\bbox{eT\ll k \ll T}$: anomalous relaxation}
Most of the studies of the photon self-energy in the hard thermal loop
approximation focused on the soft external momentum region $k \ll eT$ (with
$e\ll 1$)~\cite{weldon:1,htl,book:lebellac} wherein the contribution of the
hard thermal fermion loop must be treated nonperturbatively. However, there are
important reasons that warrant a consideration in the region of \emph{semihard}
photons ($eT\ll k\ll T$). In particular, from a phenomenological standpoint,
hard and semihard photons produced in the QGP phase are important
electromagnetic probes of the deconfined
phase~\cite{mclerran,kapusta,baier:photon,ruuskanen,alam:em}, hence a study of
kinetics, relaxation and production of photons all over the spectrum to explore
potential experimental signatures is warranted.

A more theoretical justification of the relevance of the semihard region is
that whereas the photon self-energy for this region of momentum is still
dominated by the hard thermal fermion loop contribution given by
Eq.~(\ref{HTLPis}), its contribution to the full photon propagator is now
\emph{perturbative} as compared with the free inverse propagator. The validity
of perturbation theory in this regime allows us to study the real-time
evolution with the tools developed in Chap.~\ref{chap:3} that provide a
consistent implementation of the dynamical renormalization group to study
nonequilibrium phenomena directly in real time.

\begin{figure}[t]
\begin{center}
\includegraphics[width=3.5truein,keepaspectratio=true]{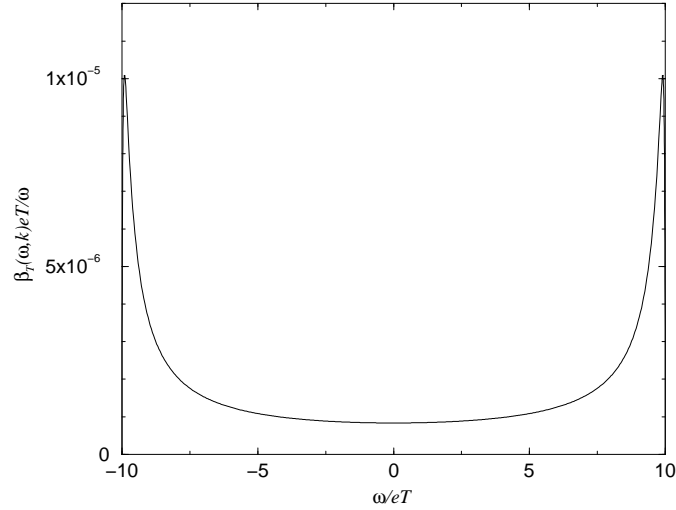}
\caption[The function $\beta_T(\omega,k)/\omega$ plotted as a function of
$\omega$ for semihard photons.]{The function $\beta_T(\omega,k)/\omega$ plotted
as a function of $\omega$ for semihard photons
($k/eT=10$).}\label{fig:semihardphoton}
\end{center}
\end{figure}

As we have shown in Sec.~\ref{sec:4.5}, the dynamical renormalization group
method is particularly suited to study the real-time evolution in the case in
which there are (infrared) threshold singularities in the spectral function. To
understand the potential emergence of anomalous thresholds in the semihard and
hard limit for the photon self-energy, we note that at leading order in the HTL
approximation the Laplace transform of the inverse photon propagator can be
written as [see Eqs.~(\ref{deltatil}) and (\ref{HTLPis})]
\begin{equation}
\widetilde{D}_T^{-1}(s,\mathbf{k})=(s^2+k^2) \left[1+\frac{e^2
T^2}{12}\frac{is}{k^3} \ln\frac{is+k}{is-k}\right] - \frac{e^2
T^2}{6}\frac{s^2}{k^2}.
\end{equation}
The transverse photon poles are at $s=\pm
i[k+e^2T^2/12k+\mathcal{O}(T^4/k^3)]$, which in the limit $k\gg eT$ becomes
$s\to\pm ik$. In this limit the poles merge with the thresholds of the
logarithmic branch cut arising from Landau damping and are no longer isolated
from the continuum, leading to the enhancement of the spectral function near
the threshold as clearly displayed in Fig.~\ref{fig:semihardphoton}. While the
perturbative expansion in the effective coupling $e^2T^2/k^2$ is warranted in
the semihard case under consideration, the wave function renormalization
constant evaluated as the residue at the pole is a nonanalytic function of this
coupling and given by
\begin{equation}
Z_T(k) = 1+ \frac{e^2T^2}{12 k^2}\left[3 +\ln\frac{e^2T^2}{24k^2}+
\mathcal{O}\left( \frac{e^2 T^2}{k^2}\ln \frac{eT}{k}\right)\right].
\end{equation}
This is obviously a consequence of the logarithmic threshold singularities in
the semihard regime. This threshold singularity is reminiscent of those studied
in detail in Ref.\cite{boyanrgir}, where it was shown that the Bloch-Nordsieck
resummation, which is equivalent to a renormalization group improvement of the
space-time Fourier transform of the self-energy, leads to a real-time evolution
that is obtained by the implementation of the dynamical renormalization group
in real time~\cite{boyanrgir}. We now implement the dynamical renormalization
group program to study the relaxation of the photon mean field in the semihard
limit directly in real time.

Perturbation theory in terms of the effective coupling $e^2T^2/k^2$ is in
principle reliable for semihard photons, hence we can try to solve
Eq.~(\ref{maxwelleq2}) by perturbative expansion in powers of $e^2$ (the true
dimensionless coupling is $e^2T^2/k^2$). Writing
\begin{eqnarray}
&&\mathbf{a}_T(\mathbf{k},t)=\mathbf{a}^{(0)}_T(\mathbf{k},t)+
e^2 \,\mathbf{a}^{(1)}_T(\mathbf{k},t)+\mathcal{O}(e^4),\nonumber\\
&&\pi_T(\mathbf{k},t-t')=e^2\,\pi_T^{(1)}(\mathbf{k},t-t')+\mathcal{O}(e^4),
\end{eqnarray}
and expanding Eq.~(\ref{maxwelleq2}) consistently in powers of $e^2$, we obtain
a hierarchy of equations:
\begin{eqnarray}
\left(\partial_t^2+k^2 \right)
\mathbf{a}^{(0)}_T(\mathbf{k},t)&=&0,\nonumber\\
\left(\partial_t^2+k^2 \right) \mathbf{a}^{(1)}_T(\mathbf{k},t)&=&
\int^{t}_{0}dt'\,\pi^{(1)}_T(\mathbf{k},t-t')\,
\dot\mathbf{a}^{(0)}_T(\mathbf{k},t'),\\
\vdots\quad\quad&&\quad\quad\vdots\nonumber
\end{eqnarray}
where we have used the fact that $\pi_T(\mathbf{k},t=0)=0$ in the HTL
approximation. Using the solution to the zeroth-order equation
\begin{equation}\label{solord0}
\mathbf{a}^{(0)}_T(\mathbf{k},t)=\sum_{\lambda=1,2}
\left[A_\lambda(\mathbf{k})\,e^{-ikt}
+A^{\ast}_\lambda(\mathbf{k})\,e^{ikt}\right]
\bbox{\mathcal{E}}_\lambda(\mathbf{k}),
\end{equation}
where $\bbox{\mathcal{E}}_\lambda(\mathbf{k})$ is the polarization vector, and
the retarded Green's function of the unperturbed problem
\begin{equation}
G^{(0)}_R(k,t-t')=\frac{\sin[k(t-t')]}{k}\,\theta(t-t'),
\end{equation}
one finds the solution to the first-order equation reads
\begin{eqnarray}
e^2\mathbf{a}^{(1)}_T(\mathbf{k},t)&=&-\frac{i}{4}\sum_{\lambda=1,2}
A_\lambda(\mathbf{k})\,\bbox{\mathcal{E}}_\lambda(\mathbf{k})
\int_{-\infty}^{+\infty} d\omega\, \frac{\rho_T(\omega,\mathbf{k})}{\omega} \,
\Bigg\{\Bigg[\frac{e^{-ikt}}{\omega-k}
\left(t-\frac{1-e^{-i(\omega-k)t}}{i(\omega-k)}\right)
\nonumber\\&&+\,\frac{e^{ikt}}{\omega+k}
\left(\frac{1-e^{-2ikt}}{2ik}+\frac{1-e^{i(\omega-k)t}}{i(\omega-k)}
\right)\Bigg] +\,(\omega\rightarrow -\omega)\Bigg\}+\mbox{c.c.},\label{aTsol1}
\end{eqnarray}
where $\rho_T(\omega,\mathbf{k})$ is the spectral function of the photon
self-energy in the HTL approximation
\begin{eqnarray}
\rho_T(\omega,\mathbf{k})&=&\frac{1}{\pi}
\mathrm{Im}\Pi_T(\omega,\mathbf{k})\nonumber\\
&=&\frac{e^2T^2}{12}\frac{\omega}{k}\left(1-\frac{\omega^2}{k^2}\right)
\theta(k^2-\omega^2).
\end{eqnarray}

Potential secular terms arise at long times from the regions in $\omega$ where
the denominators in Eq.~(\ref{aTsol1}) are resonant. They can be extracted in
the long-time limit by using the following formulas~\cite{boyanrgir}:
\begin{eqnarray}
\int_0^{\infty} {dy \over y^2} \left(1 -\cos yt\right) p(y)
&\buildrel{t\to\infty}\over=& \frac{\pi}{2} \, t \, p(0) + p'(0)
\,\left[\ln(\mu\,t)+\gamma
\right]\nonumber\\
&& + \int_0^{\infty}\frac{dy}{y^2}\left[p(y)-p(0)-y\,p'(0)\,
\theta(\mu-y)\right]
+ \mathcal{O}\left(t^{-1}\right), \nonumber\\
\int_0^{\infty}{ dy \over y}\left(t-\frac{\sin yt}{y}\right) p(y)&\buildrel{t
\to\infty}\over=&
t \, p(0) \left[ \ln(\mu\,t) + \gamma - 1 \right]\nonumber\\
&&\,+\,t\,\wp\int_0^{\infty} \frac{dy}{y}
\left[p(y)-p(0)\,\theta(\mu -
y)\right]+\mathcal{O}\left(t^{-1}\right),\label{formulas1}
\end{eqnarray}
where $\gamma = 0.577\ldots$ is the Euler-Mascheroni constant and as can be
easily shown the dependence on the arbitrary scale $\mu$ in the above integrals
cancels. For our analysis, the integration variable $y= \omega \pm k$ and $p(y)
=\rho_T(\omega,\mathbf{k})/\omega$. The terms that grow linearly in time are
recognized as those emerging in Fermi's golden rule from elementary
time-dependent perturbation theory in quantum mechanics. We note that $p(0)=0$
and $p'(0)\neq 0$, thus the first integral above gives a contribution to the
\emph{real part} of the mean field (the amplitude) that features a
\emph{logarithmic} secular term, whereas the second integral contributes to the
\emph{imaginary part} (the phase) with a linear secular term, which as will be
recognized below determines a perturbative shift of the oscillation frequency.
Upon substituting $\rho_T(\omega,\mathbf{k})$ into Eq.~(\ref{aTsol1}), one
obtains
\begin{eqnarray}
e^2\mathbf{a}^{(1)}_T(\mathbf{k},t)&=&
-\sum_{\lambda=1,2}A_\lambda(\mathbf{k})\,
\bbox{\mathcal{E}}_\lambda(\mathbf{k})\,e^{-ikt}\nonumber\\
&&\times\,\bigg[\frac{e^2T^2}{12k^2} (\ln 2kt+\gamma-1) + i\,\delta_k\, t+
\mathcal{O}\left(t^{-1}\right)\bigg]+\mbox{c.c.},\label{aTsol1s}
\end{eqnarray}
where
\begin{equation}
\delta_k\equiv\left.\frac{\mbox{Re}\Pi_T(\omega,\mathbf{k})}{2\omega}
\right|_{\omega=k}=\frac{e^2 T^2}{12k}.\label{masshift}
\end{equation}

While the \emph{linear} secular terms have a natural interpretation in terms of
renormalization of the mass (the imaginary part) and a quasiparticle width (the
real part)~\cite{boyanrgir}, the \emph{logarithmic} secular term found above is
akin to those found in Ref.~\cite{boyanrgir} that lead to anomalous relaxation.
Furthermore the origin of these logarithmic secular terms is similar to the
threshold infrared divergences and threshold enhancement of the spectral
function due to the presence of nearby poles studied in Sec.~\ref{sec:4.5}.
Thus we implement the dynamical renormalization group by introducing a
(complex) amplitude renormalization factor in the following manner,
\begin{equation}
A_\lambda(\mathbf{k})=\mathcal{Z}_k(\tau)\,\mathcal{A}_\lambda(\mathbf{k},\tau),\quad
\mathcal{Z}_k(\tau)=1+e^2 z_k(\tau)+\mathcal{O}(e^4),
\end{equation}
where $\mathcal{Z}_k(\tau)$ is a multiplicative renormalization constant,
$\mathcal{A}_\lambda(\mathbf{k},\tau)$ is the renormalized initial value, and
$\tau$ is an arbitrary renormalization scale at which the secular divergences
are cancelled~\cite{boyanrgir}. Choosing
$$
e^2 z_k(\tau)=\frac{e^2T^2}{12k^2}(\ln 2k\tau +\gamma-1)+ i\,\delta_k\, \tau,
$$
we obtain to lowest order in $e^2T^2/k^2$ that the solution of the equation of
motion is given by
\begin{eqnarray}
\mathbf{a}_T(\mathbf{k},t)&=&\sum_{\lambda=1,2}
\mathcal{A}_\lambda(\mathbf{k},\tau)
\,\bbox{\mathcal{E}}_\lambda(\mathbf{k})\,e^{-ikt}
\,\left[1-\frac{e^2T^2}{12k^2}\ln\frac{t}{\tau}
-i\delta_k(t-\tau)\right]\nonumber\\
&&+\,\mbox{nonsecular terms}+\mbox{c.c.},\label{aTren}
\end{eqnarray}
which remains bounded at large times $t$ provided that $\tau$ is chosen
arbitrarily close to $t$. The solution does not depend on the renormalization
scale $\tau$ and this independence leads to the dynamical renormalization group
equation~\cite{boyanrgir}, which to this order is given by
\begin{equation}
\left[\frac{\partial}{\partial\tau}+\frac{e^2T^2}{12k^2\tau}+ i\,
\delta_k\right]\mathcal{A}_\lambda(\mathbf{k},\tau)=0,
\end{equation}
with the solution
\begin{equation}
\mathcal{A}_\lambda(\mathbf{k},\tau)=\mathcal{A}_\lambda(\mathbf{k},\tau_0)\,
e^{-i\delta_k(\tau-\tau_0)}
\left(\frac{\tau}{\tau_0}\right)^{-\frac{e^2T^2}{12k^2}},
\end{equation}
where $\tau_0$ is the time scale such that this intermediate asymptotic
solution is valid and physically corresponds to a microscopic scale, i.e, $
\tau_0\sim 1/k $. Finally, setting $ \tau=t $ in Eq.~(\ref{aTren}) we find that
$\mathbf{a}_T(\mathbf{k},t) $ evolves at intermediate asymptotic times $ t \gg
1/k $  as
\begin{equation}
\mathbf{a}_T(\mathbf{k},t)\simeq\sum_{\lambda=1,2}
\mathcal{A}_\lambda(\mathbf{k},\tau_0)
\,\bbox{\mathcal{E}}_\lambda(\mathbf{k})\,e^{-i(k+\delta_k)(t-\tau_0)}
\left(\frac{t}{\tau_0}\right)^{-\frac{e^2T^2}{12k^2}}+\mbox{c.c.}.
\label{finalsol}
\end{equation}
From Eq.~(\ref{masshift}) we see that $\delta_k$ is consistent with the photon
thermal mass $m^2_{\gamma}= e^2 T^2/6$ for $k^2 \gg
m^2_{\gamma}$~\cite{book:lebellac}.

As discussed in detail in Ref.~\cite{boyanrgir} the dynamical renormalization
group solution (\ref{finalsol}) is also obtained via the Fourier transform of
the renormalization group improved propagator in frequency-momentum space,
hence the above solution corresponds to a renormalization group improved
\emph{resummation} of the self-energy.

This novel anomalous power law relaxation of the photon mean field will be
confirmed below in our study of the kinetics of the photon distribution
function in the linearized approximation. We note that this anomalous power law
relaxation is obviously very slow in the semihard regime in which the HTL
approximation and perturbation theory is valid. At higher orders we expect
exponential relaxation due to collisional processes, which emerges from linear
secular terms~\cite{boyanrgir} in a perturbative solution of the real-time
equations of motion. The power law relaxation will then compete with the
exponential relaxation and we expect a crossover time scale on which relaxation
will change from a power law to an exponential. Clearly an assessment of this
time scale requires a detailed calculation of higher order contributions which
is beyond the scope of this thesis.

A related crossover of behavior will be found below for the evolution of the
photon distribution function and photon production.

\subsection{Quantum kinetics of photons}
As mentioned in Chap.~\ref{chap:4} the first step towards a kinetic description
is to identify the proper degrees of freedom (quasiparticles) and the
corresponding microscopic time scale. For semihard photons of momentum $eT\ll
k\ll T$, it is adequate to choose the free photons as the quasiparticles with
the corresponding microscopic scale $\sim 1/k$. A kinetic description of the
nonequilibrium evolution of the distribution functions assumes a wide
separation between the microscopic and the kinetic time scales, which is
justified in the weak coupling limit under consideration. In the semihard
momentum regime the effective small coupling is $e^2T^2/k^2 \ll 1$ and both the
HTL and the perturbative approximations are valid. Furthermore, we assume that
the fermions are in thermal equilibrium at a temperature $T$ and that there is
no initial photon polarization asymmetry. As before, we consider the case in
which the initial density matrix is diagonal in the basis of the photon
occupation numbers but with nonequilibrium initial photon distribution
functions $n^{\gamma}_\mathbf{k}(t_0)$.

We begin by obtaining the photon number operator from the Heisenberg photon
field operator and its conjugate momentum (in momentum space)
\begin{eqnarray}
\mathbf{A}_T(\mathbf{k},t)&=&\sum_{\lambda=1,2}\,\sqrt{\frac{1}{2k}}
\,\!\Big[a_\lambda(\mathbf{k},t)\,\bbox{\mathcal{E}}_\lambda(\mathbf{k})
+a_\lambda^{\dagger}(-\mathbf{k},t)\,
\bbox{\mathcal{E}}_\lambda(-\mathbf{k})\Big],\nonumber\\
\mathbf{\Pi}_T(\mathbf{k},t)&=&i\sum_{\lambda=1,2}\,\sqrt{\frac{k}{2}}
\,\!\Big[a^{\dagger}_\lambda(-\mathbf{k},t)\,
\bbox{\mathcal{E}}_\lambda(-\mathbf{k})
-a_\lambda(\mathbf{k},t)\,\bbox{\mathcal{E}}_\lambda(\mathbf{k})\Big],
\label{photon:field}
\end{eqnarray}
where $a_\lambda(\mathbf{k},t)$ [$a^\dagger_\lambda(\mathbf{k},t)$] is the
annihilation (creation) operator that destroys (creates) a free photon of
momentum $\mathbf{k}$ and polarization $\lambda$ at time $t$ and
$\bbox{\mathcal{E}}_\lambda(\mathbf{k})$ is polarization vector. The number
operator $N_{\gamma}(\mathbf{k},t)$ that counts the (semihard) photons of
momentum $\mathbf{k}$ is given by
\begin{eqnarray}
N_{\gamma}(\mathbf{k},t)&=&\sum_{\lambda=1,2}a^\dagger_\lambda(\mathbf{k},t)
a_\lambda(\mathbf{k},t)\nonumber\\
&=&\frac{1}{2k}\Big[\mathbf{\Pi}_T(-\mathbf{k},t)\cdot\mathbf{\Pi}_T(\mathbf{k},t)
+\,k^2\mathbf{A}_T(-\mathbf{k},t)\cdot\mathbf{A}_T(\mathbf{k},t)\Big]-\frac{1}{2},
\label{photon:N}
\end{eqnarray}
whose expectation value is interpreted as the number of photons per unit phase
space volume
\begin{equation}
{n}^{\gamma}_\mathbf{k}(t)=\langle {N}_\gamma(\mathbf{k},t)\rangle=(2\pi)^3
\frac{dN}{d^3x d^3 k},\label{photon:n}
\end{equation}
where $N$ is the total number of (semihard) photons in the plasma. Using the
Heisenberg equations of motion, we obtain to leading order in $e$ the following
expression in the CTP formalism
\begin{equation}
\frac{d}{dt}n^\gamma_\mathbf{k}(t)=\lim_{t'\to t}
\frac{e}{4k}\frac{\partial}{\partial t'}
\int\frac{d^3q}{(2\pi)^{3/2}}\big\langle
\bar{\psi}^-(-\mathbf{p},t)\bbox{\gamma}
\cdot\mathbf{A}^+_T(\mathbf{k},t')\psi^-(\mathbf{q},t) \big\rangle +
\mbox{c.c.} ,\label{photon:ndot1}
\end{equation}
where we have separated the time arguments to extract the time derivative from
the expectation value.

The nonequilibrium expectation values can be computed perturbatively in powers
of $e$ using the nonequilibrium Feynman rules and real-time propagators. At
$\mathcal{O}(e)$ the right-hand side of Eq.~(\ref{photon:ndot1}) vanishes
identically. This is a consequence of our choice of initial density matrix
diagonal in the basis of the photon number operator. To $\mathcal{O}(e^2)$, we
obtain
\begin{equation}
\frac{d}{dt}n^\gamma_\mathbf{k}(t)=[1+n^\gamma_\mathbf{k}(t_0)]
\,\Gamma^<_k(t)-n^\gamma_\mathbf{k}(t_0)\,\Gamma^>_k(t),\label{photon:ndot2}
\end{equation}
where the $\Gamma^{<(>)}_\mathbf{k}(t)$ is the \emph{time-dependent} photon
production (absorption) rate
\begin{equation}
\Gamma^\lessgtr_\mathbf{k}(t)=\int_{-\infty}^{+\infty}
d\omega\,\mathcal{R}^\lessgtr_\gamma(\omega,k)\,
\frac{\sin[(\omega-k)(t-t_0)]}{\pi(\omega-k)},
\end{equation}
and
\begin{eqnarray}
\mathcal{R}^<_\gamma(\omega,k)\!\!&=&\!\! \frac{\pi
e^2}{k}\int\frac{d^3q}{(2\pi)^{3}} \bigg[2\,\big[1-(\mathbf{\hat
k}\cdot\mathbf{\hat p})(\mathbf{\hat k}\cdot\mathbf{\hat q})\big]\,
n_F(q)\,[1-n_F(p)]\,\delta(\omega+p-q)\nonumber\\
&&+\,\big[1+(\mathbf{\hat k}\cdot\mathbf{\hat p})(\mathbf{\hat
k}\cdot\mathbf{\hat q})\big]\Big\{n_F(p)n_F(q)\,\delta(\omega-p-q)
+[1-n_F(p)]\nonumber\\
&&\times\,[1-n_F(q)]\,\delta(\omega+p+q)\Big\}\bigg]. \label{Rgamma1}
\end{eqnarray}
In the above expression, $\mathbf{p}=\mathbf{k}+\mathbf{q}$ and
$\mathcal{R}^>_\gamma(\omega,k)$ is obtained from
$\mathcal{R}^<_\gamma(\omega,k)$ through the replacement $n_F\Leftrightarrow
1-n_F$. A comment here is in order. As explained above we are focusing on the
leading HTL approximation, consequently, in obtaining
$\mathcal{R}^\gtrless_\gamma(\omega,k)$ we use the \emph{free} real-time
fermion propagators which correspond to the \emph{hard} part of the fermion
loop momentum. In general there are contributions from the soft fermion loop
momentum region  which will require to use the HTL-resummed fermion
propagators~\cite{htl} for consistency. A detailed study of the contribution
from soft loop momentum is beyond the scope of this thesis, instead we focus
here on the lowest order leading HTL contribution in real time.

Since the fermions are in thermal equilibrium, the Kubo-Martin-Schwinger (KMS)
condition holds:
\begin{equation}
\mathcal{R}^>_\gamma(\omega,k)=e^{\beta\omega} \,
\mathcal{R}^<_\gamma(\omega,k)\label{KMScond},
\end{equation}
where $\beta=1/T$. It is easy to recognize that
$\mathcal{R}^{<(>)}_\gamma(\omega,k)$ has a physical interpretation in terms of
the off-shell (energy-nonconserving) photon production (absorption) processes
in the plasma. The first term in $\mathcal{R}^<_\gamma(\omega,k)$ describe the
process that a fermion (or an antifermion) emits a photon, i.e., bremsstrahlung
from the fermions in the medium, the second term describes annihilation of a
fermion pair into a photon, and the third terms describes creation of a photon
and a fermion pair out of the vacuum. The corresponding terms in
$\mathcal{R}^>_\gamma(\omega,k)$ describe the inverse processes.

As argued above, for semihard photons of momenta $eT\ll k\ll T$, the leading
contribution of $\mathcal{R}^\lessgtr_\gamma(\omega,k)$ arises from the hard
loop momenta $q\sim k$. An analysis of $\mathcal{R}^<_\gamma(\omega,k)$ along
the familiar lines in the hard thermal loop program\cite{htl,book:lebellac}
shows that in the HTL approximation ($q\gg k$)
\begin{equation}
\mathcal{R}^<_\gamma(\omega,k)|_\mathrm{HTL}=\frac{\pi e^2T^3}{12k^2}
\left(1-\frac{\omega^2}{k^2}\right) \,\theta(k^2-\omega^2).\label{Rgamma2}
\end{equation}
Thus we recognize that in the HTL approximation
$\mathcal{R}^{<(>)}_\gamma(\omega,k)$ is completely determined by the off-shell
Landau damping process in which a hard fermion in the plasma emits (absorbs) a
semihard photon, i.e., bremsstrahlung from the fermions in the medium.

\subsubsection{Dynamical renormalization group and the emergence of detailed balance}
We now turn to the kinetics of semihard photons. To obtain a kinetic equation
from Eq.~(\ref{photon:ndot2}), we implement the dynamical renormalization group
resummation as introduced in Chap.~\ref{chap:4}. Direct integration with the
initial condition yields
\begin{eqnarray}
n^\gamma_\mathbf{k}(t)&=&n^\gamma_\mathbf{k}(t_0)+\left[1+n^\gamma_\mathbf{k}(t_0)\right]
\int^t_{t_0}dt'\,\Gamma^<_k(t')
-\,n^\gamma_\mathbf{k}(t_0)\int^t_{t_0}dt'\,\Gamma^>_k(t')\, .\label{ngamma2}
\end{eqnarray}
The integrals that appear in the above expression
\begin{equation}
\int^t_{t_0}dt'\, \Gamma^\gtrless_\mathbf{k}(t') = \int_{-\infty}^{+\infty}
d\omega\,\mathcal{R}^\gtrless_\gamma(\omega,k)\,
\frac{1-\cos[(\omega-k)(t-t_0)]}{\pi(\omega-k)^2}, \label{timedeprates}
\end{equation}
are dominated, in the long-time limit, by the regions of $\omega$ for which the
denominator is resonant, i.e., $\omega\approx k$. The time dependence in the
above integral along with the resonant denominator is the familiar form that
leads to Fermi's golden rule in time-dependent perturbation theory. In the
long-time limit $ t-t_0 \gg 1/k$, Fermi's golden rule approximates the above
integrals by
\begin{equation}
\int^t_{t_0}dt'\, \Gamma^\gtrless_\mathbf{k}(t')\approx \frac{1}{2}\,(t-t_0)\,
\mathcal{R}^\gtrless_\gamma(\omega=k,k)\equiv 0.
\end{equation}
Therefore, the \emph{rate} of photon production, i.e., the coefficient of the
linear time dependence vanishes at this order because of the vanishing of the
imaginary part of the photon self-energy on the photon mass shell. However the
use of Fermi's golden rule, which is the usual approach to extract
(time-independent) rates, misses the important off-shell effects associated
with finite lifetime processes. These can be understood explicitly by using the
first of formulas given in Eq.~(\ref{formulas1}). We find that for $t-t_0\gg
1/k$
\begin{eqnarray}
\int^t_{t_0} dt'\, \Gamma^<_k(t') & \buildrel{k(t-t_0) \gg 1}\over= &
\frac{e^2 T^3}{6k^3}\left\{\ln\left[2k(t-t_0)\right]+
\gamma-1\right\}+\mathcal{O}\left(\frac{1}{k(t-t_0)}\right)\nonumber \\
&=& e^{-\beta k} \int^t_{t_0}dt'\,\Gamma^>_k(t'). \label{detbalance}
\end{eqnarray}
The second equality of this equation displays the condition for \emph{detailed
balance} and holds for time scales $t-t_0 \gg 1/k$. We emphasize that this
condition is a consequence of the fact that the region $\omega \approx k$ (the
resonant denominator) dominates the long-time behavior of the integrals in
Eq.~(\ref{timedeprates}) and the KMS condition (\ref{KMScond}) holds under the
assumption that the fermions are in thermal equilibrium.

Thus we find that the time-dependent rates obey detailed balance which emerges
in the intermediate asymptotic regime $ t-t_0 \gg 1/k$ when the secular terms
dominate the long-time behavior. This is a noteworthy result because it clearly
states that detailed balance emerges on \emph{microscopic} time scales $t-t_0
> 1/k$ at which the secular terms dominate the integrals but for which a
perturbative expansion is still valid. Detailed balance then guarantees the
existence of an asymptotic equilibrium solution which is reached at times of
the order of the kinetic time scale.

The (logarithmic) secular terms in the time-dependent rates (\ref{detbalance})
can now be resummed using the dynamical renormalization group method by
introducing a renormalization of the initial distribution function
\begin{equation}
n^\gamma_\mathbf{k}(t_0)=\mathcal{Z}_k(t_0,\tau)
\,n^\gamma_\mathbf{k}(\tau),\quad \mathcal{Z}_k(t_0,\tau)=1+e^2
z_k(t_0,\tau)+\mathcal{O}(e^4),
\end{equation}
thus rewriting Eq.~(\ref{ngamma2}) consistently to order $e^2$ as
\begin{eqnarray}
n^\gamma_\mathbf{k}(t)&=&n^\gamma_\mathbf{k}(\tau)+ e^2
z_k(t_0,\tau)\,n^\gamma_\mathbf{k}(\tau)+
\left[1+n^\gamma_\mathbf{k}(\tau)\right]\int^t_{t_0}dt'
\,\Gamma^<_k(t')\nonumber\\
&&-\,n^\gamma_\mathbf{k}(\tau) \int^t_{t_0}dt'
\,\Gamma^>_k(t')+\mathcal{O}(e^4).
\end{eqnarray}
The renormalization coefficient $z_k(t_0,\tau)$ is chosen to cancel the secular
divergence on a time scale $t=\tau$. To lowest order in $e^2$ the choice
\begin{equation}
e^2 z_k(t,\tau)\,n^\gamma_\mathbf{k}(\tau)
=-\left[1+n^\gamma_\mathbf{k}(\tau)\right] \int^\tau_{t_0}dt'\,\Gamma^<_k(t')
+\,n^\gamma_\mathbf{k}(\tau)\int^\tau_{t_0}dt'\,\Gamma^>_k(t'),
\end{equation}
leads to an \emph{improved} perturbative solution in terms of the renormalized
occupation number $n^\gamma_\mathbf{k}(\tau)$ as
\begin{eqnarray}
n^\gamma_\mathbf{k}(t)&=&n^\gamma_\mathbf{k}(\tau)+
\left[1+n^\gamma_\mathbf{k}(\tau)\right]\int^t_{\tau}dt'\,\Gamma^<_k(t')
-n^\gamma_\mathbf{k}(\tau) \int^t_{\tau}dt'\,\Gamma^>_k(t')+ \mathcal{O}(e^4),
\end{eqnarray}
which is valid for large times $t \gg t_0$ provided that $\tau$ is chosen
arbitrarily close to $t$. A change in the time scale $\tau$ is compensated by a
change of the $n^\gamma_\mathbf{k}(\tau)$ in such a manner that
$n^\gamma_\mathbf{k}(t)$ does not depend on the arbitrary scale $\tau$. This
independence of $\tau$ leads to the dynamical renormalization group equation
which, consistently to order $e^2$, is given by
\begin{equation}
\frac{d}{d\tau}{n}^\gamma_\mathbf{k}(\tau)=[1+n^\gamma_\mathbf{k}(\tau)]
\,\Gamma^{<}_k(\tau)- n^\gamma_\mathbf{k}(\tau)\,
\Gamma^{>}_k(\tau)+\mathcal{O}(e^4).\label{photon:drg}
\end{equation}
Choosing $\tau$ to coincide with $t$ in Eq.~(\ref{photon:drg}), we obtain the
\emph{quantum kinetic equation} to order $e^2$ to be given by
\begin{equation}
\frac{d}{dt}n^\gamma_\mathbf{k}(t)=[1+n^\gamma_\mathbf{k}(t)]\,\Gamma^{<}_k(t)-
n^\gamma_\mathbf{k}(t)\,\Gamma^{>}_k(t).\label{photon:keq}
\end{equation}

For intermediate asymptotic times $ t-t_0 \gg 1/k $ at which the logarithmic
secular terms dominate the integrals for the time-dependent rates and
\emph{detailed balance} emerges, we find
\begin{equation}
\Gamma^{<}_k(t)\,\buildrel{k(t-t_0) \gg
1}\over=\,\frac{e^2T^3}{6k^3}\frac{1}{t-t_0}
\left[1+\mathcal{O}\left(\frac{1}{k(t-t_0)}\right)\right]=e^{-\beta
k}\,\Gamma^>_k(t),
\end{equation}
where the detailed balance relation is explicitly displayed and the terms being
neglected are oscillatory on time scales $\sim 1/k$ and fall off faster. This
detailed balance relation between the time-dependent rates
$\Gamma^\lessgtr_k(t)$ guarantees the existence of an asymptotic equilibrium
solution of the quantum kinetic equation with the distribution function
$n^\gamma_\mathbf{k}(t=\infty) = 1/(e^{\beta k}-1)$.

The full solution of the quantum kinetic equation (\ref{photon:keq}) is given
by
\begin{equation}
n^\gamma_\mathbf{k}(t)= n^\gamma_\mathbf{k}(t_0)\,e^{-\int^t_{t_0}dt'
\gamma_k(t')}+e^{-\int^t_{t_0}dt'\digamma_k(t')} \int^t_{t_0}
dt'\,\Gamma^{<}_k(t')\,e^{\int^{t'}_{t_0}dt''\digamma_k(t'')},
\label{fullsolkin}
\end{equation}
where $\digamma_k(t)=\Gamma^{>}_k(t)-\Gamma^{<}_k(t)$. However, in order to
understand the relaxation to equilibrium of the distribution functions, we now
focus on the relaxation time approximation, which describes the approach to
equilibrium of a slightly off-equilibrium initial distribution function
$n^\gamma_\mathbf{k}(t_0)=n_B(k)+\delta n^\gamma_\mathbf{k}(t_0)$ while all
other modes are in thermal equilibrium. In the relaxation time approximation,
one obtains
\begin{equation}
\frac{d}{dt}\delta n^\gamma_\mathbf{k}(t)= -\delta n^\gamma_\mathbf{k}(t)
\,(e^{\beta k}-1) \,\Gamma^{<}_k(t).
\end{equation}
For semihard photons of momentum $eT\ll k\ll T$, one can simply replace
$e^{\beta k}-1$ by $k/T$ and upon integration one finds
\begin{equation}
\delta n^\gamma_\mathbf{k}(t)\simeq\delta
n^\gamma_\mathbf{k}(t_0)[k(t-t_0)]^{-e^2 T^2/6k^2}\quad\mbox{for}\quad
k(t-t_0)\gg 1.\label{rta}
\end{equation}
A noteworthy feature of Eq.~(\ref{rta}) is that the relaxation of the
distribution function for semihard photons in the relaxation time approximation
is governed by a power law with an anomalous exponent rather than by the usual
exponential relaxation. This result is similar to that found in scalar quantum
electrodynamics in the Markovian approximation~\cite{boyanhtl}. Comparing
Eq.~(\ref{rta}) and Eq.~(\ref{finalsol}), we clearly see that the anomalous
exponent for the relaxation of the photon distribution function in the
relaxation time approximation is twice that for the linear relaxation of the
photon mean field. This relationship is well known in the case of exponential
relaxation [see Sec.~\ref{sec:3.4}] but our analysis with the dynamical
renormalization group method reveals it to be a more robust feature, applying
just as well to power law relaxation.

The dynamical renormalization group approach to derive quantum kinetic equation
in field theory is different from the one often used in the literature which
involves a Wigner transform, assumption about the separation of fast and slow
variables, and quasiparticle
approximation~\cite{geiger:qcd,heinz,mrowczynski,niegawa:1,danielewicz}. In
particular, this approach reveals clearly the dynamics of off-shell effects
associated with nonexponential relaxation. This aspect will acquire
phenomenological relevance in our study of photon production enhanced by
off-shell effects from a quark-gluon plasma with a finite lifetime in
Chap.~\ref{chap:6}.

\section{Hard Fermions Out of Equilibrium}\label{sec:5.3}
To provide a complete real-time picture of nonequilibrium dynamics
in a QED plasma at high temperature, we now focus on a detailed
study of relaxation of fermionic mean fields (induced by an
adiabatically switched-on Grassmann source) as well as the quantum
kinetics of the fermion distribution function.

In this section we assume that the photons are in thermal equilibrium, and
since we work to leading order in the HTL approximation we can translate the
results \emph{vis \`a vis} to the case of equilibrated gluons. In particular we
seek to study the possibility of anomalous relaxation as a result of the
emission and absorption of magnetic photons. In
Refs.~\cite{blaizot:qed,takashiba} it was found that the relaxation of
fermionic excitations is anomalous and not exponential as a result of the
emission and absorption of magnetic photons that are only dynamically screened
by Landau damping. The study of the real-time relaxation of the fermionic mean
fields in these references was cast in terms of the Bloch-Nordsieck
approximation which replaces the Dirac gamma matrices by the classical velocity
of the fermion. In Ref.~\cite{boyanrgir} the relaxation of a charged scalar
mean field as well as the quantum kinetics of the distribution function of
charged scalars in scalar electrodynamics were studied using the dynamical
renormalization group, both the charged scalar mean field and the distribution
function of charged particles reveal anomalous nonexponential relaxation as a
consequence of emission and absorption of soft magnetic photons. While electric
photons (plasmons) are screened by a Debye mass which cuts off their infrared
contribution, magnetic photons are only dynamically screened by Landau damping
and their emission and absorption dominates the infrared behavior of the
fermion propagator.

While the dynamical renormalization group has been implemented in scalar
theories it has not yet been applied to fermionic theories. Thus the purpose of
this section is twofold: (i) to implement the dynamical renormalization group
to study the relaxation and kinetics of fermions with a detailed discussion of
the technical differences with the bosonic case and (ii) to focus on the
real-time manifestation of the infrared singularities associated with soft
magnetic photons.

\subsection{Relaxation of the fermion mean field}
The equation of motion for a fermion mean field is obtained by following the
strategy described in section II. We begin by writing the fermion field as
\begin{equation}
\Psi^\pm(\mathbf{x},t)=\psi(\mathbf{x},t)+\chi^\pm(\mathbf{x},t),
\quad\mbox{with}\quad\langle\chi^\pm(\mathbf{x},t)\rangle=0.
\end{equation}
Then using the tadpole method~\cite{tadpole} with the external Grassmann source
$\eta(\mathbf{x},t)$ that is adiabatically switched-on from $t=-\infty$ and
switched-off at $t=0$ [see Eq.~(\ref{icond})], we find the Dirac equation of
the fermion mean field for $t>0$ to be given by (in momentum space)
\begin{equation}
\left(i\gamma^0\partial_t-
\bbox{\gamma}\cdot\mathbf{k}\right)\psi(\mathbf{k},t) - \int^{t}_{-\infty}dt'\,
\Sigma(\mathbf{k},t-t')\,\psi(\mathbf{k},t')=0,\label{diraceq1}
\end{equation}
where $\Sigma(\mathbf{k},t-t')$ is the retarded fermion self-energy. As noted
above the relaxation of hard fermions is dominated by the soft photon
contributions, thus in a consistent perturbative expansion one needs to use the
HTL-resummed effective photon propagators to account for the screening effects
in the medium~\cite{book:lebellac}. The HTL-resummed effective photon
propagators can be conveniently written in terms of the resummed spectral
functions:

(i) Transverse components:
\begin{eqnarray}
D_{T}^{++}({\bf q};t,t')&=&  D_{T}^{>}({\bf q};
t,t')\theta(t-t')+D_{T}^{<}({\bf q};t,t')\theta(t'-t),\nonumber \\
D_{L}^{--}({\bf q};t,t')&=& D_{T}^{>}({\bf q};
t,t')\theta(t'-t)+D_{T}^{<}({\bf q};t,t')\theta(t-t'),\nonumber \\
D_{T}^{-+}({\bf q};t,t')&=& D_{T}^>({\bf q};t,t'),
\quad D_{T}^{+-}({\bf q};t,t')= D_{T}^<({\bf q};t,t'),\nonumber\\
D_{T}^{>,ij}({\bf q};t,t')&=&i\,{\cal P}_T^{ij}({\bf\hat
q})\int_{-\infty}^{+\infty} dq_0
\,\tilde{\rho}_T(q_0,q)\,[1+n_B(q_0)]\,e^{-iq_0(t-t')},\nonumber\\
D_{T}^{<,ij}({\bf q};t,t')&=&i\,{\cal P}_T^{ij}({\bf\hat
q})\int_{-\infty}^{+\infty} dq_0\, \tilde{\rho}_T(q_0,q)\,
n_B(q_0)\,e^{-iq_0(t-t')}, \label{wightmanT}
\end{eqnarray}

(ii) Longitudinal components:
\begin{eqnarray}
D_{L}^{++}({\bf q};t,t')&=& -\frac{1}{q^2}\delta(t-t')+ D_{L}^{>}({\bf q};
t,t')\theta(t-t')+D_{L}^{<}({\bf q};t,t')\theta(t'-t),\nonumber \\
D_{L}^{--}({\bf q};t,t')&=& \frac{1}{q^2}\delta(t-t')+ D_{L}^{>}({\bf q};
t,t')\theta(t'-t)+D_{L}^{<}({\bf q};t,t')\theta(t-t'),\nonumber \\
D_{L}^{-+}({\bf q};t,t')&=& D_{L}^>({\bf q};t,t'),\quad
D_{L}^{+-}({\bf q};t,t')= D_{L}^<({\bf q};t,t'),\nonumber\\
D_{L}^{>}({\bf q};t,t')&=& i\int_{-\infty}^{+\infty}
dq_0\,\tilde{\rho}_L(q_0,q)\,
[1+n_B(q_0)]\,e^{-iq_0(t-t^\prime)},\nonumber\\
D_{L}^{<}({\bf q};t,t')&=& i\int_{-\infty}^{+\infty}
dq_0\,\tilde{\rho}_L(q_0,q)\,n_B(q_0)\, e^{-iq_0(t-t^\prime)}.
\label{wightmanL}
\end{eqnarray}
Here $\tilde{\rho}_T(q_0,q)$ is the HTL-resummed spectral function for
transverse photon propagator defined in Eq.~(\ref{tilderhot}) and
$\tilde{\rho}_L(q_0,q)$ is the HTL-resummed spectral function for longitudinal
photon propagator~\cite{book:lebellac}
\begin{eqnarray}
\tilde{\rho}_{L}(q_0,q)&=&\mbox{sgn}(q_0)\,Z_{L}(q)\,
\delta[q^2_0-\omega^2_{L}(q)]
+\,\beta_{L}(q_0,q)\,\theta(q^2-q_0^2), \nonumber\\
\beta_{L}(q_0,q)&=& \frac{\frac{e^2T^2}{6}\frac{q_0}{q}}{\left[q^2+
\frac{e^2T^2}{6}\left(2-\frac{q_0}{q}\ln\frac{q+q_0}{q-q_0}\right)\right]^2+
\big[\frac{\pi e^2T^2}{6}\frac{q_0}{q}\big]^2},\label{rholong}
\end{eqnarray}
where $\omega_{L}(q)$ is the plasmon (longitudinal photon) pole and $Z_{L}(q)$
is the corresponding residue. We note that $\rho_{1(2)}(\omega,\mathbf{k})$ is
an even (odd) function of $\omega$, a property that will be useful in the
following analysis.

Using these HTL-resummed effective photon propagators, one finds to one-loop
order $\Sigma(\mathbf{k},t-t')$ reads
\begin{equation}
\Sigma(\mathbf{k},t-t')=\int_{-\infty}^{+\infty} d\omega \left[
-i\,\gamma^0\,\rho_1(\omega,\mathbf{k})\cos[\omega(t-t')]
+\,\bbox{\gamma}\cdot\mathbf{\hat k}\,\rho_2(\omega,\mathbf{k})
\sin[\omega(t-t')]\right],
\end{equation}
where the spectral functions
\begin{eqnarray}
\rho_1(\omega,\mathbf{k})&=&e^2\int\frac{d^3 q}{(2\pi)^3}
\int_{-\infty}^{+\infty}dq_0
\Big[\tilde{\rho}_T(q_0,q)+\frac{1}{2}\,\tilde{\rho}_L(q_0,q)\Big]
\,[1+n_B(q_0)-n_F(p)]\nonumber\\
&&\times\,[\delta(\omega-p-q_0)+ \delta(\omega+p+q_0)],\label{rho1}\\
\rho_2(\omega,\mathbf{k})&=&e^2\int\frac{d^3 q}{(2\pi)^3}
\int_{-\infty}^{+\infty} dq_0\Big[(\mathbf{\hat k}\cdot\mathbf{\hat q})
(\mathbf{\hat p}\cdot\mathbf{\hat q})\,\tilde{\rho}_T(q_0,q)
-\frac{\mathbf{\hat k}\cdot\mathbf{\hat p}}{2}\,
\tilde{\rho}_L(q_0,q)\Big] \nonumber\\
&&\times\,[1+n_B(q_0)-n_F(p)]\,[\delta(\omega-p-q_0)-
\delta(\omega+p+q_0)],\label{rho2}
\end{eqnarray}
with $\mathbf{p}=\mathbf{k}-\mathbf{q}$. As before, it proves
convenient to introduce the Laplace transform of the retarded
self-energy $\widetilde{\Sigma}(s,\mathbf{k})$, whose analytic
continuation in the complex $s$-plane (physical sheet)
$\Sigma(\omega,\mathbf{k})$ is obtained through the replacement
$s\to -i\omega+0^+$, or more explicitly,
\begin{equation}
\Sigma(\omega,\mathbf{k}) = \int^{+\infty}_{-\infty}
\frac{dk_0}{k_0-\omega-i0^+}
\left[-\gamma^0\rho_1(k_0,\mathbf{k})+\bbox{\gamma}\cdot\mathbf{\hat{k}}
\,\rho_2(k_0,\mathbf{k})\right].
\end{equation}
A comment here is in order. To facilitate the study and maintain
notational simplicity, in obtaining the $\Sigma(\mathbf{k},t-t')$
we have neglect the contribution arising from the
\emph{instantaneous} Coulomb interaction, which is irrelevant to
the relaxation of the fermion mean field and only results in a
perturbative frequency shift.

Following the same strategy in the study of the photon mean field,
by introducing an auxiliary quantity $\sigma(\mathbf{k},t-t')$
defined as
$$
\Sigma(\mathbf{k},t-t')=\partial_{t'}\sigma(\mathbf{k},t-t'),
$$
one can rewrite Eq.~(\ref{diraceq1}) as an initial value problem
\begin{equation}
\left(i\gamma^0 \partial_t -
\bbox{\gamma}\cdot\mathbf{k}\right)\psi(\mathbf{k},t)+ \int^{t}_{0}dt'\,
\sigma(\mathbf{k},t-t')\, \dot\psi(\mathbf{k},t') -\sigma(\mathbf{k},0)\,
\psi(\mathbf{k},t)=0,
\end{equation}
with the initial conditions $\psi(\mathbf{k},0)=\psi_0(\mathbf{k})$ and
$\dot\psi(\mathbf{k},t< 0)=0$.

We are now ready to solve the equation of motion by perturbative expansion in
powers of $e^2$ just as in the case of the gauge mean field. Let us begin by
writing
\begin{eqnarray}
&&\psi(\mathbf{k},t)=\psi^{(0)}(\mathbf{k},t) + e^2\,  \psi^{(1)}(\mathbf{k},t) +\mathcal{O}(e^4)\, , \nonumber\\
&&\sigma(\mathbf{k},t-t')=e^2\, \sigma^{(1)}(\mathbf{k},t-t')+\mathcal{O}(e^4),
\nonumber
\end{eqnarray}
we obtain a hierarchy of equations:
\begin{eqnarray}
\left(i\gamma^0\partial_t
- \bbox{\gamma}\cdot\mathbf{k}\right)\psi^{(0)}(\mathbf{k},t)&=&0,\nonumber\\
\left(i\gamma^0\partial_t -
\bbox{\gamma}\cdot\mathbf{k}\right)\psi^{(1)}(\mathbf{k},t)&=&
\sigma^{(1)}(\mathbf{k},0)\, \psi^{(0)}(\mathbf{k},t) -\int^{t}_{0}dt'\,
\sigma^{(1)}(\mathbf{k},t-t')\, \dot\psi^{(0)}(\mathbf{k},t'),\\
\vdots\quad\quad&&\quad\quad\vdots\nonumber
\end{eqnarray}
These equations can be solved iteratively by using the zeroth-order solution
\begin{equation}
\psi^{(0)}(\mathbf{k},t)=\sum_{s=1,2}\left[ A_s(\mathbf{k})
u_s(\mathbf{k})\,e^{-ikt} +B_s(\mathbf{k}) v_s(-\mathbf{k})\,e^{ikt}\right],
\end{equation}
and the retarded Green's function of the unperturbed problem
\begin{equation}
\mathcal{G}^{(0)}_R(\mathbf{k},t-t')=-i\left[h_+(\mathbf{\hat
k})\,e^{-ik(t-t')}+ h_-(\mathbf{\hat k})\,e^{ik(t-t')}\right]\,\theta(t-t').
\end{equation}
where $h_\pm(\mathbf{\hat k})=(\gamma^0\mp\bbox{\gamma}\cdot\mathbf{\hat k})/2$
and $u_s(\mathbf{k})$ and $v_s(\mathbf{k})$ are the free Dirac spinors that
satisfy
\begin{equation}
h_+(\mathbf{\hat k})\,u_s(\mathbf{k})=0,\quad h_-(\mathbf{\hat k})\,
v_s(\mathbf{k})=0.
\end{equation}

The solution to the first-order equation is found to be given by
$$
\psi^{(1)}(\mathbf{k},t)=\psi^{(1,a)}(\mathbf{k},t)+\psi^{(1,b)}(\mathbf{k},t),
$$
where
\begin{eqnarray}
e^2\psi^{(1,a)}(\mathbf{k},t)\!\!&=&\!\!
-i\,\gamma^{a}(\mathbf{k})\,t\,\sum_{s=1,2}
\big[A_s(\mathbf{k})\,u_s(\mathbf{k})\,e^{-ikt}
-B_s(\mathbf{k})\,v_s(-\mathbf{k})\,e^{ikt}\big],\label{sol1a}\\
e^2\psi^{(1,b)}(\mathbf{k},t)\!\!&=&\!\!\frac{i}{\pi}\sum_s
\int_{-\infty}^{+\infty}
\frac{d\omega}{\omega-k}\,\gamma^{b}(\omega,\mathbf{k})
\Bigg\{A_s(\mathbf{k})\, u_s(\mathbf{k})\,e^{-ikt}
\left[t-\frac{1-e^{-i(\omega-k)t}}{i(\omega-k)}\right]\nonumber\\
&&-\,B_s(\mathbf{k})\, v_s(-\mathbf{k})\,e^{ikt}
\left[t+\frac{1-e^{i(\omega-k)t}}{i(\omega-k)}\right]\Bigg\}, \label{sol1b}
\end{eqnarray}
with
\begin{eqnarray}
&&\gamma^{a}(\mathbf{k})= \int^{+\infty}_{-\infty}\frac{d\omega}{\omega}\,
\rho_2(\omega,\mathbf{k}) =\frac{1}{2}\, \mathrm{Tr}\!\left[h_+(\mathbf{\hat
k})\,\mathrm{Re}\Sigma(0,\mathbf{k})\right],\nonumber\\
&&\gamma^{b}(\omega,\mathbf{k})=\frac{\pi k}{\omega}
\,[\rho_1(\omega,\mathbf{k})-\rho_2(\omega,\mathbf{k})]
=-\frac{k}{2\omega}\,\mbox{Tr}\! \left[h_+(\mathbf{\hat k})\,
\mathrm{Im}\Sigma(\omega,\mathbf{k})\right].\label{gammab}
\end{eqnarray}
We note that in Eq.~(\ref{sol1a}) secular terms are explicitly linear in time
and are purely imaginary, whereas in Eq.~(\ref{sol1b}) potential secular terms
may arise at long times from the resonant denominators. From the expression of
$\gamma^{b}(\omega,\mathbf{k})$ given in Eq.~(\ref{gammab}), we recognize that
$\gamma^{b}(\omega,\mathbf{k})$ evaluated at the fermion mass shell $\omega=k$
is the fermion damping rate computed~\cite{book:lebellac}. However, it has been
shown in the literature that due to the emission and absorption of soft
quasi-static transverse photons which are only dynamically screened by Landau
damping, the fermion damping rate exhibits infrared divergences near the mass
shell in perturbation theory.

The infrared divergences is easily understood from the following analysis. For
soft photons with $q_0,\,q\ll T$, we can replace
\begin{equation}
1+n_B(q_0)-n_F(p)\simeq T/q_0,\quad p\simeq k-q\cos\theta,
\end{equation}
where $\cos\theta=\mathbf{\hat k}\cdot\mathbf{\hat q}$, thus writing
\begin{eqnarray}
\gamma^{b}(\omega,\mathbf{k})&=&\frac{\pi e^2 Tk}{\omega} \int \frac{d^3
q}{(2\pi)^3}\int_{-q}^{+q} \frac{dq_0}{q_0}
\,\Big[(1-\cos^2\!\theta)\,\beta_T(q_0,q)+\beta_L(q_0,q)\Big]\nonumber\\
&&\times\,\delta(\omega-k+q\cos\theta-q_0),\label{gammab1}
\end{eqnarray}
where the subleading pole contributions which corresponding to emission and
absorption of on-shell photons have been neglected~\cite{blaizot:qed}. Recall
that for very soft $q\ll eT$ the function $\beta_T(q_0,q)/q_0$ is strongly
peaked at $q_0=0$ [see Eq.~(\ref{breitwigner}) and Fig.~\ref{fig:softphoton}],
and as $q\to 0$ it can be approximated by
\begin{equation}
\frac{1}{q_0}\,\beta_T(q_0\ll q,q) \to\frac{\delta(q_0)}{q^2}\quad\mbox{as}
\quad q\to 0.\label{singular}
\end{equation}
whose physical origin is the absence of a magnetic photon mass in the plasma.
The infrared divergences near the fermion mass shell become manifest after
substituting Eq.~(\ref{singular}) into Eq.~(\ref{gammab1}).

In order to isolate the singular behavior of $\gamma^b(\omega,\mathbf{k})$, we
follow the steps in Ref.~\cite{blaizot:qed} and write
\begin{equation}
\frac{1}{q_0} \,\beta_T(q_0,q)=\delta(q_0)\left(\frac{1}{q^2}-
\frac{1}{q^2+\omega_{P}^2}\right)+
\frac{1}{q_0}\,\nu_T(q_0,q),\label{betatdecomp}
\end{equation}
where $\omega_{P}=eT/3$ is the plasma frequency and $\nu_T(q_0,q)$ denotes the
regular part of the transverse photon spectral function. Substituting
Eq.~(\ref{betatdecomp}) into Eq.~(\ref{gammab1}), we can then separate
$\gamma^b(\omega,\mathbf{k})$ into an infrared singular part which is
logarithmically divergent near the fermion mass shell
\begin{equation}
\gamma^{b}_{\rm sing}(\omega,\mathbf{k})=\frac{\pi e^2 Tk}{\omega}
\int\frac{d^3 q}{(2\pi)^3}
\left(\frac{1}{q^2}-\frac{1}{q^2+\omega_{P}^2}\right)
\,\delta(\omega-k+q\cos\theta),\label{singpart}
\end{equation}
and a regular part that remains finite near the fermion mass shell
\begin{eqnarray}
\gamma^{b}_{\rm reg}(\omega,\mathbf{k})\!\!&=&\!\!\frac{\pi e^2 Tk}
{\omega}\int\frac{d^3q}{(2\pi)^3} \int_{-q}^{+q}\frac{dq_0}{q_0}
\,\big[\beta_L(q_0,q)-\cos^2\!\theta \,\beta_T(q_0,q)\nonumber\\
&&+\,\nu_T(q_0,q)\big]\,\delta(\omega-k+q\cos\theta-q_0).
\end{eqnarray}
Using the delta function $\delta(q\cos\theta-q_0)$ to perform the angular
integrations, one obtains
\begin{equation}
\gamma^{b}(\omega,\mathbf{k})\buildrel{\omega\to k}\over= \alpha
T\left(\ln\frac{\omega_{P}}{|\omega-k|}
+\mathcal{I}\right)+\mathcal{O}(\omega-k),\label{gammabexpanded}
\end{equation}
where
\begin{equation}
\mathcal{I}=\int_0^\infty dq\, q\int_{-q}^{+q} \frac{dq_0}{q_0}\, \bigg[
\beta_L(q_0,q)-\frac{q_0^2}{q^2}\, \beta_T(q_0,q)
+\,\nu_T(q_0,q)\bigg].\label{I}
\end{equation}
The above double integral has been computed analytically in
Ref.~\cite{blaizot:qed} with the result $\mathcal{I}=\ln 3/2$.

We are now in position to find the secular terms in $\psi^{(1)}(\mathbf{k},t)$
that emerge in the intermediate asymptotic regime. The imaginary part of the
secular terms in $e^2\psi^{(1,a)}(\mathbf{k},t)$ and
$e^2\psi^{(1,b)}(\mathbf{k},t)$ can be combined into the imaginary part of the
secular term
\begin{equation}
\mathrm{Im}S_k(t)=\mp i\,\alpha\,\delta_k\,t
\end{equation}
for the positive (upper sign) and negative (lower sign) energy spinors, where
\begin{equation} \alpha\,\delta_k=\frac{1}{2}\,
\mbox{Tr}\!\left[h_+(\mathbf{\hat k})\,
\mathrm{Re}\Sigma(\omega,\mathbf{k})\right]_{\omega=k}.
\end{equation}
There are no further secular terms arise from the higher order expansion around
the fermion mass shell in Eq.~(\ref{gammabexpanded}). This (linear) imaginary
secular term is thus identified with a perturbative shift of the oscillation
frequency of the fermion mean field~\cite{boyanrgir}. It is a finite quantity
determined by a dispersive integral of the spectral functions
$\rho_{1,2}(\omega,\mathbf{k})$.

The real secular terms are more involved. The finite contribution to
$\gamma^b(\omega,\mathbf{k})$ as $\omega \to k$ leads to a linear secular term,
whereas for the logarithmically divergent contribution as $\omega\to k$ the
following asymptotic result becomes useful~\cite{boyanrgir}:
\begin{eqnarray}
&&\int_{-\infty}^{+\infty} \frac{dy}{y^2}(1-\cos yt)\ln|y| \buildrel{t \to
\infty}\over= \pi\, t\,(1-\gamma-\ln t) + \mathcal{O}\left(t^{-1}\right),
\end{eqnarray}
where we have neglected terms that fall off at long times. Thus, one obtains
the real part of the secular terms to be given by
\begin{equation}
\mathrm{Re}S_k(t)= -\alpha\,T\,t\left(\ln\omega_{P}t +\gamma -1+\frac{\ln
3}{2}\right).
\end{equation}
Gathering the above results, at large times $t\gg 1/\omega_{P}$ the
perturbative solution reads
\begin{eqnarray}
\psi(\mathbf{k},t)\!\!&=&\!\!\sum_{s=1,2}\Big\{[1+S_k(t)]\,A_s(\mathbf{k})\,u_s(\mathbf{k})\,e^{-ikt}
+\,[1+S_k^{\ast}(t)]\,B_s(\mathbf{k})\, v_s(-\mathbf{k})\,e^{ikt}\Big\}
\nonumber\\&&+\,\mbox{nonsecular terms},\label{psiper}
\end{eqnarray}
with
\begin{equation}
S_k(t) = -\alpha\,t\left[\left(\ln\omega_{P}t +\gamma -1+\frac{\ln
3}{2}\right)T+i\,\delta_k\right].
\end{equation}

Obviously, this perturbative solution breaks down on a time scale
$\simeq[\alpha T \ln(1/\alpha)]^{-1}$. To obtain a uniformly valid solution for
large times, we now implement a resummation of the secular terms in the
perturbative series via the dynamical renormalization group. As before,
introducing (complex) renormalization constants for the amplitude
\begin{equation}
A_s(\mathbf{k})=\mathcal{Z}_k(\tau)\,\mathcal{A}_s(\mathbf{k},\tau),\quad
B_s(\mathbf{k})=\mathcal{Z}_k^{\ast}(\tau)\,\mathcal{B}_s(\mathbf{k},\tau),\quad
\mathcal{Z}_k(\tau)=1+\alpha\,z_k(\tau)+\mathcal{O}(e^4)
\end{equation}
and choosing $\alpha\,z_k(\tau)=-S_k(\tau)$ to cancel the secular divergence at
time $\tau$, one finds to order $\mathcal{O}(\alpha)$ the dynamical
renormalization group equations to be given by (after setting $\tau=t$)
\begin{equation}
\left[\frac{d}{dt}-\frac{d S_k(t)}{dt}\right]
\mathcal{A}_s(\mathbf{k},t)=0,\quad \left[\frac{d}{dt}- \frac{d
S_k^{\ast}(t)}{dt} \right]\mathcal{B}_s(\mathbf{k},t)=0.
\end{equation}
Solving the above equations, we find that at asymptotic times ($\omega_{P} t\gg
1$) the fermion mean field relaxes as
\begin{eqnarray}
\psi(\mathbf{k},t)&=&\sum_{s=1,2}\Big[\mathcal{A}_s(\mathbf{k},\tau_0)\,
u_s(\mathbf{k})\,e^{-i\bar{\omega}(k)t} + \mathcal{B}_s(\mathbf{k},\tau_0)\,
v_s(-\mathbf{k}) \,e^{i\bar{\omega}(k)t}\Big]\nonumber\\
&&\times\, e^{-\alpha Tt \left[\ln\omega_P
t+0.126\ldots\right]},\label{fermion:drgsol1}
\end{eqnarray}
where we have replaced $\gamma -1+\ln 3/2 = 0.126\ldots$, $\tau_0\sim
1/\omega_{P}$ is the time scale such that this intermediate asymptotic solution
is valid, and $\bar{\omega}(k)=k+\delta_k$ is the position of the fermion pole
(in hard momentum limit) shifted by one-loop corrections.
Eq.~(\ref{fermion:drgsol1}) reveals a time scale for the relaxation of the
fermion mean field $t_\mathrm{rel}\sim 1/\alpha T \ln(1/\alpha)$, which
coincides with the time scale on which the perturbative solution given by
Eq.~(\ref{psiper}) breaks down. This highlights clearly the nonperturbative
nature of relaxation phenomena.

Our result coincides with that found in Refs.~\cite{blaizot:qed,takashiba} via
the Bloch-Nordsieck approximation and in scalar quantum
electrodynamics~\cite{boyanrgir} using the dynamical renormalization group
method. Furthermore, it provides another important and relevant example of the
reliability and consistency of this novel renormalization group method in study
real-time nonequilibrium dynamics.

\subsection{Quantum kinetics of fermion quasiparticles}
We now study the quantum kinetic equation for the distribution
function of hard fermion quasiparticles. There has recently been
an intense activity to obtain a Boltzmann equation for
quasiparticles in gauge theories~\cite{blaizot:be,bodeker,litim}
motivated in part by the necessity to obtain a consistent
description for baryogenesis in non-Abelian theories. Boltzmann
equations with a diagrammatic interpretation were obtained
in~\cite{bodeker,blaizot:be} in which a collision-type kernel
describes the scattering of hard quasiparticles. In these
approaches this collision kernel reveals the infrared divergences
associated with the emission and absorption of magnetic photons
(or gluons) and must be cutoff by introducing a magnetic mass
$m_\mathrm{mag}\sim\alpha T$ to leading logarithmic
accuracy~\cite{blaizot:be}.

In a derivation of quantum kinetic equations for charged quasiparticles in
SQED~\cite{boyanrgir} using the dynamical renormalization group method, it was
understood that the origin of these infrared divergences is the implementation
of Fermi's golden rule that assumes completed collisions and takes the infinite
time limit in the collision kernels. The dynamical renormalization group
approach leads to quantum kinetic equations in real time in terms of
\emph{time-dependent} scattering kernels without any infrared ambiguity.

In this section we implement this program in fermionic QED to
derive the quantum kinetic equation for hard fermion
quasiparticles. There are several important features of our study
that must be emphasized: (i) as discussed in Sec.~\ref{sec:5.1},
gauge invariance is automatically taken into account by working
directly with gauge invariant operators, thus the operator that
describes the number of fermion quasiparticles is gauge invariant,
(ii) a kinetic description relies on a separation between the
microscopic and the relaxation time scales, this is warranted in a
strict perturbative regime and applies to hard fermion
quasiparticles, (iii) the dynamical renormalization group leads to
a quantum kinetic equation in real time without infrared
divergences since time acts as an infrared cutoff.

As in the case of the photon, we begin by expanding the Heisenberg fermion
field in terms of creation and annihilation operators as (in momentum space)
\begin{eqnarray}
\psi(\mathbf{k},t)&=&\sqrt{\frac{m}{\omega_\mathbf{k}}} \sum_{s}
\left[b_s(\mathbf{k},t)u_s(\mathbf{k})+d^\dagger_s(-\mathbf{k},t)
v_s(-\mathbf{k})\right],\nonumber\\
\psi^\dagger(\mathbf{k},t)&=&\sqrt{\frac{m}{\omega_\mathbf{k}}} \sum_{s}
\left[b^\dagger_s(-\mathbf{k},t)u^\dagger_s(-\mathbf{k})+d_s(\mathbf{k},t)
v^\dagger_s(\mathbf{k})\right],
\end{eqnarray}
where $\omega_\mathbf{k}=\sqrt{\mathbf{k}^2+m^2}$ and $b_s(\mathbf{k},t)$
[$b^\dagger_s(\mathbf{k},t)$] is the annihilation (creation) operator that
destroys (creates) a free fermion of momentum $\mathbf{k}$ and spin $s$ at time
$t$. Here, we have retained the fermion mass to avoid the subtleties associated
with the normalization of massless spinors and the massless limit will be taken
below.

In the hard momentum limit $k\sim T\gg m$, the number operator for
fermion quasiparticles with momentum $\mathbf{k}$ is then given by
\begin{eqnarray}
N_f(\mathbf{k},t)&=&\sum_{s=1,2} b^\dagger_s(\mathbf{k},t) \,
b_s(\mathbf{k},t)\nonumber\\
&=&\psi^\dagger(-\mathbf{k},t)\,h_+(\mathbf{\hat k})\,\gamma^0\,
\psi(\mathbf{k},t).
\end{eqnarray}
Taking time derivative of $N_f(\mathbf{k},t)$ and using the Heisenberg
equations of motion, we find
\begin{eqnarray}
\frac{d}{dt}n^f_\mathbf{k}(t)&\equiv&
\langle\dot{N}_f(\mathbf{k},t)\rangle\nonumber\\
&=&\lim_{t'\to t}\,ie\int\frac{d^3q}{(2\pi)^{3/2}}
\big\langle\bar{\psi}^-(-\mathbf{p},t')
\big[\gamma^0\,A^-_0(-\mathbf{q},t')-
\bbox{\gamma}\cdot\mathbf{A}^{-}_T(-\mathbf{q},t')\big]\nonumber\\
&&\,\times\,h_+(\mathbf{\hat k})\gamma^0 \,
\psi^+(\mathbf{k},t)\big\rangle+\mbox{c.c.}.
\end{eqnarray}
We obtain to $\mathcal{O}(e^2)$
\begin{eqnarray}
\frac{d}{dt} n^f_\mathbf{k}(t)\!\!&=&\!\!
e^2\int\frac{d^3q}{(2\pi)^3}\int_{-\infty}^{+\infty}dq_0\int^{t}_{t_0}dt''
\Big\{\mathcal{N}_1(t_0)\cos[(k-p-q_0)(t''-t_0)]
\big[\tilde{\rho}_L(q_0,q)\nonumber\\
&&\times\, \mathcal{K}^+_{1}(\mathbf{k},\mathbf{q})+2\,\tilde{\rho}_T(q_0,q)
\,\mathcal{K}^+_{2}(\mathbf{k},\mathbf{q})\big]
+\,\mathcal{N}_2(t_0)\cos[(k+p-q_0)(t''-t_0)]\nonumber\\
&&\times\,\big[\tilde{\rho}_L(q_0,q) \,
\mathcal{K}^-_{1}(\mathbf{k},\mathbf{q})+2\,\tilde{\rho}_T(q_0,q) \,
\mathcal{K}^-_{2}(\mathbf{k},\mathbf{q})\big]\Big\},
\end{eqnarray}
where $\mathcal{N}$ and $\mathcal{K}$ denote respectively the following
statistical and kinematic factors
\begin{eqnarray}
&&\mathcal{N}_{1}(t)= [1-n^f_\mathbf{k}(t)]\,n^f_\mathbf{p}(t)
\,n_B(q_0)-n^f_\mathbf{k}(t)\,[1-n^f_\mathbf{p}(t)][1+n_B(q_0)],\nonumber\\
&&\mathcal{N}_{2}(t)=[1-n^f_\mathbf{k}(t)][1-n^f_\mathbf{p}(t)]\,n_B(q_0)
-n^f_\mathbf{k}(t)\,n^f_\mathbf{p}(t)\,[1+n_B(q_0)],\nonumber\\
&&\mathcal{K}^\pm_{1}(\mathbf{k},\mathbf{q})=1\pm \mathbf{\hat
k}\cdot\mathbf{\hat p},\quad \mathcal{K}^\pm_{2}(\mathbf{k},\mathbf{q})= 1\mp
\mathbf{\hat k}\cdot\mathbf{\hat p}\pm \frac{1-(\mathbf{\hat
k}\cdot\mathbf{\hat q})^2}{1-\frac{q}{k}(\mathbf{\hat k}\cdot\mathbf{\hat q})},
\end{eqnarray}
and $\tilde{\rho}_T(q_0,q)$ and $\tilde{\rho}_L(q_0,q)$ are the HTL-resummed
spectral functions for the transverse and longitudinal photons given by
Eqs.(\ref{tilderhot}) and (\ref{rholong}), respectively.

Consistent with the use of the HTL-resummed photon spectral
functions, we work in the relaxation time approximation in which
only the initial distribution function for fermion quasiparticles
of momentum $\mathbf{k}$ is slightly off-equilibrium such that
$n^f_\mathbf{k}(t_0)=n_F(k)+\delta n^f_\mathbf{k}(t_0)$ while all
other (fermion and photon) modes are in thermal equilibrium. Then
upon integrating over $t''$, one obtains
\begin{equation}
\frac{d}{dt}\delta n^f_\mathbf{k}(t)=-\delta n^f_\mathbf{k}(t_0)
\int_{-\infty}^{+\infty} d\omega \, \mathcal{R}_f(\omega,\mathbf{k}) \,
\frac{\sin[(\omega-k)(t-t_0)]}{\pi(\omega-k)},\label{ndot}
\end{equation}
where
\begin{eqnarray}
\mathcal{R}_f(\omega,\mathbf{k})&=&\pi
e^2\int\frac{d^3q}{(2\pi)^3}\int_{-\infty}^{+\infty}dq_0
\left[1+n_B(q_0)-n_F(p)\right]\Big\{\big[\tilde{\rho}_L(q_0,q) \,
\mathcal{K}^+_{1}(\mathbf{k},\mathbf{q})\nonumber\\
&&+\,2\,\tilde{\rho}_T(q_0,q)\,
\mathcal{K}^+_{2}(\mathbf{k},\mathbf{q})\big]
\,\delta(\omega-p-q_0)+\big[\tilde{\rho}_L(q_0,q)\,
\mathcal{K}^-_{1}(\mathbf{k},\mathbf{q})\nonumber\\
&&+\,2\,\tilde{\rho}_T(q_0,q)\,
\mathcal{K}^-_{2}(\mathbf{k},\mathbf{q})\big]\,
\delta(\omega+p+q_0)\Big\}.
\end{eqnarray}
Eq.~(\ref{ndot}) can be integrated directly to yield
\begin{equation}
\delta n^f_\mathbf{k}(t)=\delta n^f_\mathbf{k}(t_0)\left[
1-\int_{-\infty}^{+\infty} d\omega\,\mathcal{R}_f(\omega,\mathbf{k})
\,\frac{1-\cos[(\omega-k)(t-t_0)]}{\pi(\omega-k)^2}\right].\label{n1}
\end{equation}
The time-dependent contribution above is now familiar from the previous
discussions, potential secular terms will emerge at long times from the regions
in which the resonant denominator vanishes. This is the region near the fermion
mass shell $\omega\approx k$, where $\mathcal{R}_f(\omega,\mathbf{k})$ is
dominated by the regions of small $q$ and $q_0$, which physically corresponds
to emission and absorption of soft photons. As before for soft photons with
$q,\,q_0\ll T$, we can replace
\begin{equation}
1+n_B(q_0)-n_F(p)\simeq T/q_0, \quad p\simeq k-q\cos\theta,\quad
\mathcal{K}^+_1\simeq 2,\quad \mathcal{K}^+_2\simeq 1-\cos^2\!\theta,
\end{equation}
thus writing $\mathcal{R}_f(\omega,\mathbf{k})$ at $\omega\approx k$ as
\begin{eqnarray}
\mathcal{R}_f(\omega,\mathbf{k})&=&2\pi e^2 T \int \frac{d^3
q}{(2\pi)^3}\int_{-q}^q \frac{dq_0}{q_0}
\,\big[(1-\cos^2\!\theta)\beta_T(q_0,q)+\beta_L(q_0,q)\big]\nonumber\\
&&\times\,\delta(\omega-k+q\cos\theta-q_0).\label{R1}
\end{eqnarray}
Note that the double integral in Eq.~(\ref{R1}) is exactly the same as that in
Eq.~(\ref{I}). Thus $\mathcal{R}_f(\omega,\mathbf{k})$ features an infrared
divergence near the fermion mass shell as shown in the previous subsection.
Following the analysis carried out in the preceding subsection we obtain
\begin{equation}
\mathcal{R}_f(\omega,\mathbf{k})\buildrel{\omega\to k}\over= -2\alpha
T\left[\ln\frac{|\omega-k|}{\omega_{P}}- \frac{\ln
3}{2}\right]+\mathcal{O}[(\omega-k)^2].\label{R2}
\end{equation}
Substituting  Eq.~(\ref{R2}) into  Eq.~(\ref{n1}), we find the
number of hard fermion quasiparticles at intermediate asymptotic
times $ t-t_0 \gg 1/\omega_{P}$ to be given by
\begin{equation}
\delta n^f_\mathbf{k}(t)=\delta n^f_\mathbf{k}(t_0)\Big\{1-2\alpha T(t-t_0)
[\ln\omega_{P}(t-t_0) + 0.126\ldots]\Big\}\nonumber\\
+\,\mbox{nonsecular terms}.
\end{equation}
As in the case of the fermion mean field relaxation [see Eq.~(\ref{psiper})],
the perturbative solution contains a secular term of the form $t\ln t$.
Obviously, the secular term will invalidate the perturbative solution on time
scales $t_\mathrm{rel}\sim 1/[2\alpha T\ln(1/\alpha)]$ (which will be
identified as the relaxation time scale below). In the intermediate asymptotic
regime $1/k \ll t-t_0 \ll t_\mathrm{rel}$, the perturbative expansion can be
improved by absorbing the contribution of the secular term on a time scale
$\tau$ into a ``renormalization'' of the distribution function. Hence we apply
the dynamical renormalization group method through a renormalization of the
distribution function much in the same manner as the renormalization of the
amplitude in the mean field discussed above,
\begin{equation}
\delta n^f_\mathbf{k}(t_0)=\mathcal{Z}_k(\tau,t_0)\, \delta
n^f_\mathbf{k}(\tau),\quad\mathcal{Z}_k(\tau,t_0)=1+\alpha\,
z_k(\tau,t_0)+\mathcal{O}(\alpha^2).
\end{equation}
The independence of the solution on the time scale $\tau$ leads to the
dynamical renormalization group equation which to lowest order in $\alpha$ is
given by
\begin{equation}
\left\{\frac{d}{d\tau}+2\alpha T[1+\ln\omega_{P}(\tau-t_0)+0.126\ldots
]\right\} \delta n^f_\mathbf{k}(\tau)=0,\label{fermion:drg}
\end{equation}
which is the \emph{quantum kinetic equation} in the relaxation time
approximation, but with a \emph{time-dependent} relaxation rate.

Solving Eq.~(\ref{fermion:drg}) and choosing $\tau$ to coincide with $t$, we
obtain the evolution of the fermion distribution function at asymptotic times $
t-t_0 \gg 1/k$ to be given by
\begin{equation}
\delta n^f_\mathbf{k}(t)\simeq\delta n^f_\mathbf{k}(t_0) \, \exp\{-2\alpha
T(t-t_0)[\ln\omega_{P}(t-t_0)+0.126\ldots]\}. \label{fermion:drgsol2}
\end{equation}
Comparing Eq.~(\ref{fermion:drgsol2}) with Eq.~(\ref{fermion:drgsol1}) we find
that the anomalous exponent that describes the  relaxation of the fermion
distribution function in the linear approximation is twice that for the linear
relaxation of the mean field. A similar relation is obtained between the
damping rate for the single quasiparticle relaxation and the relaxation rate of
the distribution function in the case of time-independent rates and true
exponential relaxation~\cite{book:lebellac}. The dynamical renormalization
group reveals this to be a generic feature even with time-dependent rates.

\section{Conclusions}\label{sec:5.4}
The goals of this chapter are the study of nonequilibrium dynamics
in QED plasmas at high temperature directly in real time. The
focus is a systematic study of relaxation of mean fields as well
as the distribution functions for photons and fermion
quasiparticles. In particular the application of the dynamical
renormalization group method to study anomalous relaxation as a
consequence of the exchange of soft photons.

To begin with, we have cast our study solely in terms of gauge invariant
quantities, this can be done in an Abelian gauge field theory in a
straightforward manner, avoiding potential ambiguities associated with gauge
invariance. The relaxation of photon mean fields revealed important features:
For soft momentum $k\lesssim eT$ mean fields that are prepared by adiabatically
switching-on an external source there is exponential relaxation towards the
oscillatory behavior dominated by the transverse photon pole $\omega_T(k)$. The
source that induces the mean field in this case has a Fourier transform that is
singular at zero frequency and excites the resonance in the Landau damping cut
near zero frequency for the soft photon mean field. Sources that have a regular
Fourier transform would not lead to the exponential relaxation. For semihard
momentum $eT\ll k \ll T$ in principle both the HTL and the perturbative
approximations are valid, however the spectral function for photons becomes
sharply peaked at the edge of the Landau damping continuum consistent with the
fact that the photon pole becomes perturbatively close to the Landau damping
cut. This enhancement of the spectral function near the bare photon mass shell
results in the breakdown of perturbation theory at large times $kt \gg 1$. The
dynamical renormalization group provides a consistent resummation of the photon
self-energy in real time and leads to anomalous power law relaxation of the
mean field. Clearly, higher order terms beyond the HTL approximation will
include collisional contributions leading perhaps to exponential relaxation. We
then expect a crossover between the anomalous power law obtained in lowest
order and the exponential relaxation from higher order collision processes, the
crossover time scale will depend on the details of the different contributions
and requires a study beyond that presented in this thesis.

The dynamical renormalization group provides a consistent and systematic
framework to obtain quantum kinetic equations directly in real time from the
underlying microscopic field theory~\cite{boyanrgir}. This method allows to
extract information that is \emph{not} available in the usual kinetic
description in terms of time-independent collision kernels obtained under the
assumption of completed collisions which only include on-shell
(energy-conserving) processes. The dynamical renormalization group approach to
quantum kinetics consistently includes \emph{off-shell} (energy-nonconserving)
processes and accounts for \emph{time-dependent} collisional kernels. This is
important and potentially phenomenologically relevant in the case of the
quark-gluon plasma which has a finite lifetime. In the case that fermions are
in thermal equilibrium, we have implemented this approach to obtain the lowest
order quantum kinetic equation for the distribution function of semihard
photons in the HTL approximation. This equation features time-dependent rates
and we established that detailed balance, a consequence of fermions being in
thermal equilibrium, emerges on microscopic time scales. The linearization of
the kinetic equation describes relaxation towards equilibrium with an anomalous
exponent, twice as large as that of the photon mean field in the semihard case.

The relaxation of the fermion mean field for hard momentum is
studied with the dynamical renormalization group method. This
method reveals clearly in real time the emergence of the
relaxation time scale $t_\mathrm{rel}\sim 1/[\alpha T
\ln(1/\alpha)]$ at which perturbation theory breaks down. We find
an anomalous exponential relaxation at large times of the form
$\sim\exp[-\alpha T t(\ln\omega_P t+0.126\ldots)]$ which confirms
the results of Refs.~\cite{blaizot:qed,takashiba}, where the
Bloch-Nordsieck approximation was used. We then obtain the quantum
kinetic equation for the distribution function of hard fermion
quasiparticles in the relaxation time approximation. The
collisional kernel is \emph{time-dependent} and \emph{infrared
finite} as the inverse of the time acts like an infrared cutoff.
The linearized kinetic equation describes approach to equilibrium
with an anomalous exponential relaxation, which is twice that of
the fermion mean field. An important payoff of this approach to
quantum kinetics is that it bypasses the assumption of completed
collisions which leads to collisional kernels obtained by Fermi's
golden rule and only describes on-shell (energy-conserving)
processes as in the usual kinetic approach, which in the case
under consideration leads to infrared divergent collisional
kernels~\cite{blaizot:be}.

Perhaps the most phenomenologically pressing aspects that requires
further and deeper study is the photon production by off-shell
(energy-nonconserving) processes in a thermalized quark-gluon
plasma with finite lifetime. This is important in view of an
assessment of experimental electromagnetic signatures of the QGP
which is expected soon to be produced in ultrarelativistic heavy
ion experiments at RHIC. The first step in the study of
nonequilibrium aspects of photon production from a thermalized QGP
must be to include the finite QGP lifetime and to obtain the
momentum distribution of the photons produced as a function of the
temperature and lifetime of the QGP directly in real time via a
quantum kinetic description. Furthermore, to compare with
experimental data, an important next step is to include the
hydrodynamical expansion of the QGP by coupling the quantum
kinetic equation for photons to the hydrodynamic equations for the
space-time evolution of the locally thermalized QGP and to obtain
the momentum spectrum of the photon yield as a function of the
initial temperature of the QGP. These aspects will be studied in
detail in the succeeding chapter.

%% file: chap6.tex
\chapter{Direct Photon Production from the Quark-Gluon Plasma}\label{chap:6}
\section{Introduction}
The first observation of direct photon production in ultrarelativistic heavy
ion collisions has been reported recently by the CERN WA98 collaboration in
${}^{208}$Pb$+{}^{208}$Pb collisions at $\sqrt{s_{NN}}=158$ GeV at the Super
Proton Synchrotron (SPS)~\cite{WA98}. Most interestingly, a clear excess of
direct photons above the background photons predicted from hadronic decays is
observed in the range of transverse momentum $p_T > 1.5$ GeV/$c$ in central
collisions. As compared to proton-induced results at similar incident energy,
the transverse momentum distribution of direct photons shows excess direct
photon production in central collisions beyond that expected from
proton-induced reactions. These findings indicate not only the experimental
feasibility of using direct photons as a signature of the long-sought
quark-gluon plasma (QGP)~\cite{qgp,book:qgp,book:hwa} but also a deeper
conceptual understanding of direct photon production in ultrarelativistic heavy
ion collisions.

Unlike many other new phases of matter created in the laboratory, the formation
and evolution of the QGP in ultrarelativistic heavy ion collisions is
inherently a nonequilibrium phenomena. Currently, it is theoretically accepted
that hard parton scatterings thermalize quarks and gluons on a time scale of
about 1 fm/$c$, after which the plasma undergoes hydrodynamic expansion and
cools adiabatically down to the quark-hadron phase transition. If the
transition is first order, quarks, gluons, and hadrons coexist in a mixed
phase, which after hadronization evolves until freeze-out. Estimates based on
energy deposited in the central collision region at the BNL Relativistic Heavy
Ion Collider (RHIC) energies $\sqrt{s_{NN}}\sim 200$ GeV suggest that the
lifetime of the deconfined QGP phase is of order 10 fm/$c$ with an overall
freeze-out time of order 100 fm/$c$. Different types of signatures are proposed
for each different phase~\cite{signals}. As mentioned in Chap.~\ref{chap:5},
direct photons and dileptons emitted by the QGP (electromagnetic probes) are
free of final state interactions and can provide a clean signature of the early
stages of a thermalized plasma of quarks and gluons. Therefore a substantial
effort has been devoted to a theoretical assessment of the spectra of direct
photons and dileptons emitted from the
QGP~\cite{mclerran,baier:dilepton,kapusta,baier:photon,ruuskanen,aurenche}.

Conventionally, the theoretical framework for studying direct photon production
from a thermalized QGP begins with an equilibrium calculation of the photon
emission rate at finite
temperature~\cite{mclerran,kapusta,baier:photon,ruuskanen,aurenche}. This
static rate is then combined with the hydrodynamical description of QGP to
obtain the total yield of direct photons produced during the evolution of the
QGP~\cite{mclerran,kapusta,ruuskanen,sollfrank,srivastava,alam}. Whereas this
approach is physically intuitive and widely used in the literature, it is
solely based on semiclassical Boltzmann kinetics, which completely neglects
transient effects arising from the finite QGP lifetime. The latter is of
particular importance in the study of experimental signatures of the QGP
because the formation and evolution of the QGP at currently accessible energy
scales is by itself a transient phenomenon.

In this chapter we focus in particular on experimental signatures
associated with processes that would be forbidden by energy
conservation in a QGP of infinite lifetime. We first argue that
the finite lifetime of a transient QGP raises a conceptual
inconsistency in the calculation of direct photon production via
an equilibrium rate. We then introduce a real-time kinetic
description which naturally accounts for the finite-lifetime and
nonequilibrium aspects of the QGP and includes
energy-nonconserving effects. This real-time kinetic approach can
be consistently incorporated with the widely used Bjorken's
hydrodynamics~\cite{bjorken,blaizot:hydro} to obtain the direct
photon yield for an expanding QGP that is expected to be created
in central collisions at RHIC and Large Hadron Collider (LHC)
energies. It is \emph{not} our goal to assess photon production
from the hadronic phase \emph{but} to compare the real-time
kinetic predictions for the QGP phase to those obtained from the
equilibrium calculations.

Recently, Alam \emph{et al}.~\cite{alam} have evaluated the photon
yield for Pb$+$Pb collisions at SPS energies from the initial
state up to freeze-out by using the \emph{equilibrium} photon
production rate for the QGP and hadronic phases within the
framework of (3+1)-dimensional hydrodynamic expansion. These
authors also provide an estimate of the photon yield for Au$+$Au
collisions at RHIC energies. Their results show that the photon
yield from the QGP phase and that from the hadronic phase are of
comparable order. Hence, our goal here is to compare the
nonequilibrium yield from the QGP to that obtained from the
equilibrium formulation with the QGP phase of the plasma. As we
will see below, a substantial enhancement arising from
nonequilibrium effects associated with the transient QGP lifetime
indicates that the nonequilibrium yield from the QGP phase will
stand out over the equilibrium yield from the hadronic phase.

This chapter is organized as follows. In Sec.~\ref{sec:6.2}, we
first review the usual $S$-matrix approach to direct photon
production from a QGP. We then argue that this approach in
obtaining an equilibrium rate has shortcomings and is conceptually
and physically incompatible with photon production from an
expanding QGP with a finite lifetime. In Sec.~\ref{sec:6.3},
Bjorken's hydrodynamical model combined with the $S$-matrix rate
calculation of photon production from the QGP is briefly
summarized and the incompatibility of these two approaches is
highlighted. In Sec.~\ref{sec:6.4}, we introduce the real-time
kinetic approach to photon production from a longitudinally
expanding QGP which is consistently combined with Bjorken's
hydrodynamics. This section contains our main results and we
compare these to those obtained from the usual approach. Our
conclusions are presented in Sec.~\ref{sec:6.5}.

\section[$S$-matrix Approach and its Shortcomings]
{$\bbox{S}$-matrix Approach and its Shortcomings}\label{sec:6.2}
Production of direct photons from a QGP in thermal equilibrium has
been studied
extensively~\cite{feinberg,mclerran,kapusta,baier:photon,aurenche}
because of its relevance as a clean probe of the QGP. We begin
with highlighting the main assumptions that are explicit in
\emph{all} previous calculations of photon production from a
thermalized quark gluon plasma and that are explicitly displayed
by the derivation above. First, the initial state at $t_i$ (which
in the usual calculation is taken to $-\infty$) is taken to be a
thermal equilibrium ensemble of quarks and gluons but the vacuum
state for the physical transverse photons. Furthermore, the
buildup of the photon population is neglected under the assumption
that the mean free path of the photons is larger than the size of
the plasma and the photons escape without rescattering. This
assumption thus neglects the prompt photons produced during the
pre-equilibrium stage. Indeed, Srivastava and
Geiger~\cite{srivageiger} have studied direct photons from a
pre-equilibrium stage via a parton cascade model that includes
pQCD parton cross sections and electromagnetic branching
processes. The usual computation of the prompt photon yield during
the stage of a \emph{thermalized} QGP assumes that these photons
have left the system and the computation is therefore carried out
to lowest order in $\alpha$ with an initial photon vacuum state.
Obviously keeping the pre-equilibrium photon population results in
higher order corrections in $\alpha$. In taking the final time
$t_f$ to infinity in the $S$-matrix element the assumption is that
the thermalized state is \emph{stationary}, while in neglecting
the buildup of the population the assumption is that the photons
leave the system without rescattering and the photon population
never builds up. These assumptions lead to considering photon
production only the lowest order in $\alpha$, since the buildup of
the photon population will necessarily imply higher order
corrections. Although these main assumptions are seldom spelled
out in detail, they underlie all previous calculations of the
photon production from a thermalized quark-gluon plasma.

We now review some important aspects of the usual $S$-matrix
calculation so as to highlight its shortcomings and establish
contact with the real-time kinetic approach to photon production
introduced in Sec.~\ref{sec:6.4}. It is convenient to write the
total Hamiltonian in the form
\begin{equation}
H = H_0+H_{\rm int} \label{totalH}, \quad H_0=H_{\rm QCD}+H_{\gamma}
\label{H0}, \quad H_{\rm int} =e\int d^3x\,J^{\mu} A_{\mu},
\end{equation}
where $H_{\rm QCD}$ is the full QCD Hamiltonian, $H_{\gamma}$ is the free
photon Hamiltonian, and $H_{\rm int}$ is the interaction Hamiltonian between
quarks and photons with $J^{\mu}$ the quark electromagnetic current, $A^{\mu}$
the photon field, and $e$ the electromagnetic coupling constant.

Consider that at some initial time $t_i$ the state $|i\rangle$ is an eigenstate
of $H_0$ with no photons. The transition amplitude at time $t_f$ to a final
state $|f,\gamma_{\lambda}(\mathbf{p})\rangle\equiv
|f\rangle\otimes|\gamma_{\lambda}(\mathbf{p})\rangle$, again an eigenstate of
$H_0$ but with one photon of momentum $\mathbf{p}$ and polarization $\lambda$,
is up to an overall phase given by
\begin{equation}
 S(t_f,t_i) = \langle f,\gamma_{\lambda}(\mathbf{p})|U(t_f,t_i)|i\rangle,\label{transamp}
\end{equation}
where $U(t_f,t_i)$ is the time evolution operator in the interaction
representation
\begin{eqnarray}
U(t_f,t_i) &=& \mathcal{T}\exp\left[-i\int_{t_i}^{t_f}H_{{\rm int},I}(t)dt\right]\nonumber\\
&\simeq& 1-ie\int_{t_i}^{t_f}dt\int d^3x J^{\mu}_{I}(\mathbf{x},t)
A_{\mu,I}(\mathbf{x},t)+{\cal O}(e^2),\label{umatx}
\end{eqnarray}
where the subscript $I$ stands for the interaction representation in terms of
$H_0$. In the above expression we have approximated $U(t_f,t_i)$ to first order
in $e$, since we are interested in obtaining the probability of photon
production to lowest order in the electromagnetic interaction. The usual
$S$-matrix element for the transition is obtained from the transition amplitude
$S(t_f,t_i)$ above in the limits $t_i \rightarrow -\infty$ and $t_f \rightarrow
\infty$
\begin{eqnarray}
S_{fi}&\equiv&S(+\infty,-\infty)\nonumber\\
&=&-\frac{ie}{\sqrt{2E}}\int d^3x \int^{+\infty}_{-\infty}dt\,e^{iP^\mu
x_\mu}\,\varepsilon^{\lambda}_\mu\,\langle f|J^\mu(x)|i\rangle,\label{S}
\end{eqnarray}
where $E=|\mathbf{p}|$ and $P^\mu=(E,\mathbf{p})$ are the energy
and four-momentum of the photon, respectively, and
$\varepsilon^{\lambda}_\mu$ is its polarization four-vector. Since
the states $|i\rangle$ and $ |f\rangle$ are eigenstates of the
\emph{full} QCD Hamiltonian $H_{\rm QCD}$, the above $S$-matrix
element is obtained \emph{to lowest order} in the electromagnetic
interaction, but in principle \emph{to all orders} in the strong
interaction.

The rate of photon production per unit volume from a QGP in thermal equilibrium
at temperature $T$ is obtained by squaring the $S$-matrix element, summing over
the final states, and averaging over the initial states with the thermal weight
$e^{-\beta E_i}/Z(\beta)$, where $\beta=1/T$, $E_i$ is the eigenvalue of $H_0$
corresponding to the eigenstate $|i\rangle$, and $Z(\beta)=\sum_i e^{-\beta
E_i}$ is the partition function. Using the resolution of identity $1=\sum_f
|f\rangle\langle f|$, the sum of final states leads to the electromagnetic
current correlation function. Upon using the translational invariance of this
correlation function, the two space-time integrals lead to energy-momentum
conservation multiplied by the space-time volume $\Omega=V(t_f-t_i)$ from the
product of Dirac delta functions. One can recognize that the limit
$t_f-t_i\to\infty$ is the usual interpretation of $2\pi\delta(0)$ in the square
of the energy conservation delta function.

These steps lead to the following result for the photon production rate in
thermal equilibrium~\cite{mclerran,ruuskanen}
\begin{eqnarray}
\frac{dN}{d^4x}&=&\frac{1}{\Omega}\frac{1}{Z(\beta)}\frac{d^3p}{(2\pi)^3}
\sum_{i,f,\lambda}e^{-\beta E_i}|S_{fi}|^2\nonumber\\ &=& -e^2\,
g^{\mu\nu}\,W^<_{\mu\nu}\frac{d^3p}{2E(2\pi)^3},\label{A}
\end{eqnarray}
where $W^<_{\mu\nu}$ is the Fourier transform of the thermal expectation value
of the current correlation function defined by
\begin{equation}
W^<_{\mu\nu}=\int d^4x\,e^{iP\cdot x}\,\langle J_\mu(0)J_\nu(x)\rangle_\beta.
\end{equation}
In the expression above $\langle\,\cdot\,\rangle_\beta$ denotes the thermal
expectation value. To lowest order in $e^2$ but to all orders in the strong
interactions, $W^<_{\mu\nu}$ is related to the retarded photon self-energy
$\Pi^R_{\mu\nu}$ by~\cite{ruuskanen}
\begin{equation}
e^2\,W^<_{\mu\nu}=\frac{2}{e^{E/T}-1}\,{\rm Im}\Pi^R_{\mu\nu}. \label{genfor}
\end{equation}
Thus, one obtains the (Lorentz) invariant photon production rate
\begin{equation}
E\frac{dN}{d^3p\,d^4x}=-\frac{g^{\mu\nu}}{(2\pi)^3}\,{\rm
Im}\Pi^R_{\mu\nu}\frac{1}{e^{E/T}-1}. \label{invarate}
\end{equation}

Using the hard thermal loop (HTL) resummed effective perturbation theory
developed by Braaten and Pisarski~\cite{htl}, Kapusta {\it et
al}.~\cite{kapusta} and Baier {\it et al}.~\cite{baier:photon} showed that at
one-loop order (in effective perturbation theory) the processes that contribute
to photon production are the gluon-to-photon Compton scattering off (anti)quark
$q(\bar{q})g\rightarrow q(\bar{q})\gamma$ and quark-antiquark annihilation to
photon and gluon $q\bar{q}\rightarrow g\gamma $. The resultant rate of
energetic ($E\gg T$) photon emission for two light quark flavors ($u$ and $d$
quarks) is given by~\cite{kapusta}
\begin{equation}
E\frac{dN}{d^3p\,d^4x}\bigg|_{\mbox{\scriptsize one-loop}}
=\frac{5}{9}\frac{\alpha\alpha_s}{2\pi^2} T^2 e^{-E/T} \ln\left(\frac{0.23
E}{\alpha_s T}\right),\label{rate:kapusta}
\end{equation}
where $\alpha$ is the fine-structure constant and $\alpha_s=g_s^2/4\pi$ with
$g_s$ being the strong coupling constant. In a recent development Aurenche {\it
et al.}~\cite{aurenche} have found that the two-loop contributions to the
photon production rate arising from (anti)quark bremsstrahlung
$qq(g)\rightarrow qq(g)\gamma$ and quark-antiquark annihilation with scattering
$q\bar{q}q(g)\rightarrow q(g)\gamma$ are of the same order as those evaluated
at one loop. The two-loop contributions to the photon production rate
read~\cite{aurenche}
\begin{equation}
E\frac{dN}{d^3p\,d^4x}\bigg|_{\mbox{\scriptsize two-loop}}
=\frac{40}{9}\frac{\alpha\alpha_s}{\pi^5}T^2 e^{-E/T}(J_T-J_L)\left[\ln
2+\frac{E}{3T}\right],\label{rate:aurenche}
\end{equation}
where $J_T\approx 1.11$ and $J_L\approx -1.07$ for two light quark
flavors~\cite{steffen}. Most importantly, they showed that the
two-loop contributions completely dominate the photon emission
rate at high photon energies~\cite{aurenche}. We emphasize that
the thermal photon production rates (\ref{rate:kapusta}) and
(\ref{rate:aurenche}) [or the general result (\ref{invarate})] has
two noteworthy features: (i) the thermal rate is a static,
time-independent quantity as a result of the equilibrium
calculation and (ii) emission of high energy photons is
exponentially suppressed by the Boltzmann factor $e^{-E/T}$.

We have reproduced the steps leading to Eq.~(\ref{invarate}),
which is the expression for the photon production rate used in all
calculations available in the literature, to highlight the
important steps in its derivation in order to compare and contrast
to the real-time analysis discussed below. In particular, in the
above derivation the transition amplitude is obtained via the time
evolution operator $U(t_f,t_i)$ evolved from the initial time
$t_i$ up to a time $t_f$. In the usual $S$-matrix calculation, one
takes the limits $t_i\rightarrow -\infty$ and $t_f \rightarrow
\infty$. Taking the infinite time limit and squaring the
transition amplitude lead to \emph{energy conservation} and a
\emph{time-independent transition rate}, which in turn imply the
validity of semiclassical Boltzmann kinetics.

The main reason that we delve on the main assumptions and specific
steps of the usual computation is to emphasize that there is a
conceptual limitation of this approach when applied to an
expanding QGP of \emph{finite lifetime}. The current theoretical
understanding suggests that a thermalized QGP results from a
pre-equilibrium partonic stage on a time scale of order 1 fm/$c$
after the collision, hence for consistency one must choose $t_i
\sim 1$ fm/$c$. Furthermore, within the framework of hydrodynamic
expansion, studied in detail below, the QGP expands and cools
during a time scale of about 10 fm/$c$. Hence for consistency to
study photons produced by a quark gluon plasma in local thermal
equilibrium one must set $t_f \sim 10$ fm/$c$. Hydrodynamic
evolution is an \emph{initial value problem} indeed the state of
the system is specified at an initial (proper) time surface (to be
local thermodynamic equilibrium at a given initial temperature)
and the hydrodynamic equations are evolved in time to either the
hadronization or freeze out surfaces. The calculation based on the
$S$-matrix approach takes the time interval to infinity, extracts
a \emph{time-independent rate} and inputs this rate, assumed to be
valid for every cell in the comoving fluid, in the hydrodynamic
evolution during a finite lifetime.

As stated in the Introduction, however, the QGP produced in
ultrarelativistic heavy ion collisions is intrinsically a
transient and nonequilibrium state. It is therefore of
phenomenological importance to study nonequilibrium effects on
direct photon production from an expanding QGP with a finite
lifetime with the goal of establishing potential experimental
signatures.

Our main observation is that the usual computations based on
$S$-matrix theory extract a time independent rate after taking the
infinite time interval, which is then used in a calculation of the
photon yield during a \emph{finite time} hydrodynamic evolution.
While we do not question the general validity of the results
obtained via the $S$-matrix approach, we here focus on the
signature of processes available during the \emph{finite} lifetime
of the QGP and that would be forbidden by energy conservation in
the infinite time limit.

\section{Bjorken's Hydrodynamical Model}\label{sec:6.3}
The current understanding of the QGP formation, equilibration, and subsequent
evolution through the quark-hadron (and chiral) phase transitions is summarized
as follows. A pre-equilibrium stage dominated by parton-parton interactions and
strong colored fields which gives rise to quark and gluon production on time
scales $\lesssim 1$ fm/$c$~\cite{geiger:pcm,wang}. The produced quarks and
gluons thermalize via elastic collisions on time scales $\sim 1$ fm/$c$.
Hydrodynamics is probably the most frequently used model to describe the
evolution of the next stage when quarks and gluons are in local thermal
equilibrium (although perhaps not in chemical
equilibrium)~\cite{bjorken,blaizot:hydro}. The hydrodynamical picture assumes
local thermal equilibrium (LTE), a fluid form of the energy-momentum tensor and
the existence of an equation of state for the QGP. The subsequent evolution of
the QGP is uniquely determined by the hydrodynamical equations, which are
formulated as an \emph{initial value problem} with the initial conditions
specified at the moment when the QGP reaches local thermal equilibrium, i.e.,
at an initial time $t_i \sim 1$ fm/$c$. The (adiabatic) expansion and cooling
of the QGP is then followed to the transition temperature at which the equation
of state is matched to that describing the mixed and hadronic
phases~\cite{ruuskanen,srivastava,alam}.

In order to highlight the conceptual limitation of the $S$-matrix
calculation for direct photon production for a transient QGP, we
now review the essential features of the hydrodynamic
description~\cite{bjorken,blaizot:hydro}.  For computational
simplicity we work within Bjorken's hydrodynamical model of a
longitudinally expanding QGP~\cite{bjorken}. The main assumption
in Bjorken's model is longitudinal Lorentz boost invariance in the
central rapidity region of the QGP. This is motivated by the
observation that the particle spectra for the secondaries produced
in $p+$N and N$+$N collisions exhibit a central plateau in the
rapidity space near midrapidity. For a longitudinally expanding
QGP, it is convenient to introduce the proper time $\tau$ and
space-time rapidity $\eta$ variables defined by
\begin{equation}
\tau=\sqrt{t^2-z^2},\quad\eta=\frac{1}{2}\ln\frac{t+z}{t-z},
\end{equation}
where $t$ and $z$, respectively, are the time and spatial coordinate along the
collision axis in the center-of-mass (CM) frame. The transverse spatial
coordinates will be denoted as $\mathbf{x}_T$, hence the space-time integration
measure expressed in terms of $\tau $, $\eta$ and $x_T$ is given by
$d^4x=\tau\,d\tau\,d\eta\,d^2 x_T.$ Invariance under (local) longitudinal
Lorentz boost implies that thermodynamic quantities are functions of $\tau$
only and do not depend on $\eta$.

In Bjorken's scenario~\cite{bjorken,blaizot:hydro} the QGP reaches local
thermal equilibrium at a temperature $T_i$ at a proper time of order
$\tau_i\sim 1\,{\rm fm}/c$ after the maximum overlap of the colliding nuclei.
The initial conditions for hydrodynamical equations are therefore specified on
a hypersurface of constant proper time $\tau_i$. The equation of state for the
locally thermalized QGP is taken to be that of the ultrarelativistic perfect
radiation fluid (corresponding to massless quarks and gluons). The longitudinal
expansion is described by the scaling ansatz $v^z=z/t$, where $v^z$ is the
collective fluid velocity of the hydrodynamical flow and describes free
streaming of the fluid. Hence the space-time rapidity equals to the fluid
rapidity. In terms of $\tau$ and $\eta$ the scaling ansatz implies that the
four-velocity of a given fluid cell in the CM frame is given by
$u^{\mu}=(\cosh\eta,0,0,\sinh\eta)$ with $u^{\mu}u_{\mu}=1$. The conservation
of total entropy leads to adiabatic expansion and cooling of the QGP according
to the cooling law~\cite{bjorken,blaizot:hydro}
\begin{equation}
T(\tau)=T_i\left(\frac{\tau_i}{\tau}\right)^{1/3}.\label{coolinglaw}
\end{equation}
Hence, the QGP phase ends at a proper time $\tau_f=\tau_i(T_i/T_c)^3$, where
$T_c\sim 160$ MeV is the quark-hadron transition temperature. At RHIC energies
the initial thermalization temperature is estimated to be $T_i\sim 200-300$
MeV, which entails that the lifetime of the QGP phase is of order $\lesssim
10\,{\rm fm}/c$.

Within a hydrodynamical model the usual $S$-matrix calculation of direct photon
production from a expanding QGP proceeds as
follows~\cite{kapusta,ruuskanen,sollfrank,srivastava,alam}.
\begin{enumerate}
\item First the rate of direct photon production is calculated within the
$S$-matrix framework described in the previous section, leading to
Eq.~(\ref{invarate}) for the invariant rate. This expression for
the rate describes the photon production rate in the \emph{local
rest} (LR) frame of a fluid cell in which the temperature is a
function of the proper time of the fluid cell. The rate in the CM
frame is obtained by a local Lorentz boost $E \rightarrow
P^{\mu}u_{\mu}$ and the replacement $T(t)\rightarrow T(\tau)$:
\begin{equation}
\frac{dN}{d^2p_T\,dy\,\tau\,d\tau\,d\eta\,d^2x_{T}}\bigg|_{\rm CM}=
E\frac{dN}{d^3p\,d^4x}\bigg|_{\rm LR} [P^{\mu}u_{\mu},T(\tau)],\label{CMrate}
\end{equation}
where $p_T$ and $y$, respectively, are the transverse momentum and rapidity of
the photon.
\item The direct photon yield is now obtained by integrating the
rate over the space-time history of the QGP, from the initial hypersurface of
constant proper time $\tau_i$ to the final hypersurface of constant proper time
$\tau_f$ at which the phase transition occurs. This leads to the following form
of the total direct photon yield in the CM frame for central collisions:
\begin{equation}
\frac{dN}{d^2p_T\,dy}\bigg|_{\rm CM}=\pi R_A^2
\int_{\tau_i}^{\tau_f}d\tau\,\tau
\int_{-\eta_{\rm cen}}^{\eta_{\rm cen}}d\eta 
\,E\frac{dN}{d^3p\,d^4x}\bigg|_{\rm LR}
[P^{\mu}u_{\mu},T(\tau)],\label{yieldSmtx}
\end{equation}
where $R_A$ is the radius of the nuclei and $-\eta_{\rm cen} < \eta < \eta_{\rm
cen}$ denotes the central rapidity region in which Bjorken's hydrodynamical
description holds. The fact that the $S$-matrix calculation for the rate
results in a time-independent rate determines that the only dependence of the
rate in the LR frame on the proper time is through the temperature which is
completely determined by the hydrodynamic expansion.
\end{enumerate}

It is at this stage that the conceptual incompatibility between
the $S$-matrix calculation of the photon production rate and its
use in the evaluation of the total photon yield from an expanding
QGP of \emph{finite lifetime} becomes manifest. The hydrodynamic
evolution is treated as an initial value problem with a
distribution of quarks and gluons in \emph{local thermal
equilibrium} on the initial hypersurface of constant proper time
$\tau_i \sim 1$ fm/$c$. The subsequent evolution determines that
the QGP is a transient state with a lifetime of order $\lesssim
10$ fm/$c$. The direct photon yield is obtained by integrating the
rate over this \emph{finite lifetime}. The $S$-matrix calculation
of the rate for a QGP in thermal equilibrium, on the other hand,
implicitly assumes that $\tau_i \rightarrow -\infty$ and $\tau_f
\rightarrow \infty$ as discussed above in detail. Therefore while
the rate has been calculated by taking the time interval to
infinity, it is integrated during a finite time interval to obtain
the total yield.

The question that we now address, which is the focus of this chapter, is the
following: \emph{Is this conceptual incompatibility of physical relevance and
if so what are the experimental observables?} In order to answer this question
and to assess the potential experimental signatures from nonequilibrium
effects, we must depart from the $S$-matrix formulation and provide a
\emph{real-time} calculation of the direct photon production rate based on
nonequilibrium quantum field theory.

\section{Real-Time Kinetic Approach}\label{sec:6.4}
Having highlighted the conceptual shortcomings of the usual $S$-matrix approach
to photon production from the QGP, in this section we provide an alternative
approach which bypasses the above mentioned conceptual shortcomings. This
approach, based on the dynamical renormalization group method to quantum
kinetics discussed in details in previous chapters, treats the production of
photons as an initial value problem directly in real time without invoking
Fermi's golden rule. Compared to the usual approach, this real-time kinetic
approach has the following noteworthy advantages: (i) It is capable of
capturing energy-nonconserving effects arising from the finite lifetime of the
plasma, as completed collisions are not assumed \textit{a priori}. (ii) Since
both the real-time kinetic approach and hydrodynamics are formulated as initial
value problems, they can be incorporated consistently on the same footing. This
last point proves very important in the hydrodynamic description of photon
production.

\subsection{Nonexpanding QGP}
We begin our discussion with the calculation of the invariant photon production
rate from a nonexpanding QGP. This calculation is relevant because the result
is interpreted as the invariant photon production rate in the local rest frame
of a fluid cell. The corresponding rate in the CM frame is obtained simply by a
local Lorentz boost and the direct photon yield is obtained by integrating the
rate over the space-time history of the QGP as explained in the previous
section.

Because of the Abelian nature of the electromagnetic interaction, we will work
in a gauge invariant formulation in which physical observables (in the
electromagnetic sector) are manifestly gauge invariant and the physical photon
field is transverse~\cite{boyangaugeinv}.

The real-time kinetic approach begins with the initially prepared
density matrix and the time evolution of the expectation value of
the photon number operator (see Chap.\ref{chap:4} for details).
Consistent with the hydrodynamical initial value problem, we
consider that at the initial time $t_i$ quarks and gluons are
thermalized such that the initial state of the QGP is described by
a thermal density matrix at a given initial temperature $T_i$.
Furthermore, we consider that photons are not present at the
initial time. Although this last assumption can be relaxed
allowing an initial photon distribution, the usual approach is to
assume that the photons produced during the \emph{pre-equilibrium}
stage had left the plasma without building up a population.
Therefore the initial density matrix $\rho$ is of the form
\begin{equation}
\rho= \rho_{\rm QCD}\otimes |0_{\gamma}\rangle\langle 0_{\gamma}|,\quad
\rho_{\rm QCD}=e^{-H_{\rm QCD}/T_i}.\label{densmatx}
\end{equation}

We remark that the assumption that the initial density matrix is
that of a thermalized system (after the strong interactions
thermalize quarks and gluons on a time scale $\sim 1~\mbox{fm}/c$)
underlies the program that studies the \emph{equilibrium}
properties of the QGP. This is our \emph{only} assumption, i.e.,
that of a thermalized QGP at an initial time scale $t_i \approx 1~
\mbox{fm}/c$ and is consistent with the general assumptions behind
the equilibrium program. This assumption is elevated to that of
LTE, again consistent with the hydrodynamical description of an
expanding QGP.

As usual we will focus on \emph{hard} photon production, since hard photons
suffer less from the background contamination and has been measured in recent
experiments at SPS~\cite{WA98}. By assuming photons escape directly from the
QGP without further interaction, one can treat them as free asymptotic
particles. Consequently, the number operator for photon can be obtained from
the Heisenberg photon field (and the conjugate momentum) in the same manner as
we did for the hot QED plasma [see Eqs.~(\ref{photon:field}) and
(\ref{photon:N})]. The number of photons per unit phase space volume at time
$t$ is then given by Eq.~(\ref{photon:n}), where the expectation value is now
taken with the initial density matrix $\rho$ specified above in
Eq.~(\ref{densmatx}). The invariant photon production rate
$E\,dN(t)/d^3p\,d^4x$ is obtained by using the Heisenberg equations of motion
and can be written in the CTP formalism as
\begin{equation}
E\frac{dN(t)}{d^3p\,d^4x}=\lim_{t'\to t}\frac{\partial}{\partial
t'} \sum^{N_f}_{f=1}\frac{e\,e_f}{2(2\pi)^3}
\int\frac{d^3q}{(2\pi)^3}
\big\langle\bar{\psi}_f^-(-\mathbf{k},t){\bbox\gamma}
\cdot\mathbf{A}^+_T(\mathbf{p},t')\psi_f^-(\mathbf{q},t)\big\rangle
+\mbox{c.c.},\label{a:rate1}
\end{equation}
where $\mathbf{k}=\mathbf{p}+\mathbf{q}$. In above expression, $N_f$ is the
number of quark flavors, $e_f$ is the quark charge in units of the
electromagnetic coupling constant $e$, $\mathbf{A}_T$ is the transverse
component of the photon field, $\psi_f$ is the (Abelian) gauge invariant quark
fields (with color index suppressed).

\begin{figure}[t]
\begin{center}
\includegraphics[width=3.5truein,keepaspectratio=true]{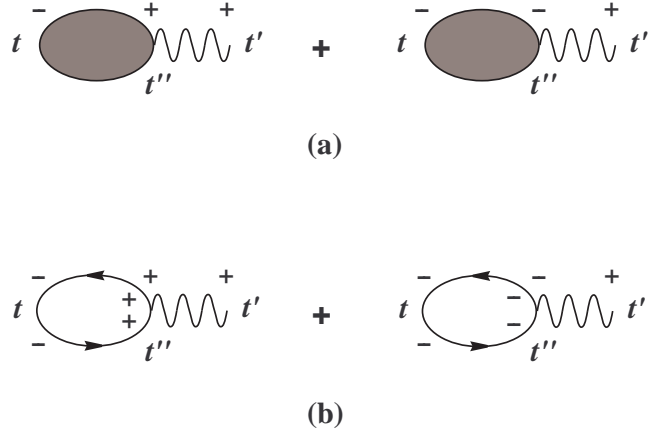}
\caption[The Feynman diagrams that contribute to the invariant
photon production rate from the QGP.]{The Feynman diagrams that
contribute to the invariant photon production rate from the QGP.
Figure~(a) shows the contribution to order $\alpha$ and to all
orders in $\alpha_s$ with the blob denoting the full quark
electromagnetic current correlation function to all orders in
$\alpha_s$. Figure~(b) is the lowest order contribution, of order
$\alpha$.} \label{fig:loop}
\end{center}
\end{figure}

We assume the weak coupling limit $\alpha \ll \alpha_s \ll 1$. Whereas the
first limit is justified and is essential for the interpretation of
electromagnetic signatures as clean probes of the QGP, the second limit can
only be justified for very high temperatures, and its validity in the regime of
interest can only be assumed so as to lead to a controlled perturbative
expansion. As mentioned above since there are no photons initially and those
produced escape from the plasma without building up their population, therefore
the QGP is effectively treated as the vacuum for photons. Consequently, the
nonequilibrium expectation values on the right-hand side of Eq.~(\ref{a:rate1})
are computed perturbatively to order $\alpha$ and in principle to all orders in
$\alpha_s$ by using nonequilibrium Feynman rules and real-time propagators (but
with vacuum photon propagators). The corresponding Feynman diagrams are
depicted in Fig.~\ref{fig:loop}.

We focus here on the \emph{lowest order} (one-loop) contribution, which is
missed by all the previous investigations using the $S$-matrix approach
implicitly assuming an infinite QGP lifetime. To lowest order in perturbation
theory, neglecting the vacuum contribution one obtains the invariant photon
production rate for a QGP of two light quark flavors ($u$ and $d$) to be given
by
\begin{eqnarray}
E\frac{dN(t)}{d^3p\,d^4x}&=&\frac{2}{(2\pi)^3} \int_{t_i}^{t}dt''
\int_{-\infty}^{+\infty}\frac{d\omega}{\pi}\,
\mathcal{R}(\omega)\,\cos[(\omega-E)(t-t'')],\label{a:noneqratei}
\end{eqnarray}
where
\begin{eqnarray}
\mathcal{R}(\omega)&=&\frac{20\,\pi^2\alpha}{3}\int\frac{d^3q}
{(2\pi)^{3}}\Big[2\big[1-(\mathbf{\hat p}\cdot\mathbf{\hat k})(\mathbf{\hat
p}\cdot\mathbf{\hat q})\big] n(q) [1-n(k)]\delta(\omega+k-q)\nonumber\\&&
+\big[1+(\mathbf{\hat p}\cdot\mathbf{\hat k})(\mathbf{\hat p}\cdot\mathbf{\hat
q})\big] n(q)n(k)\delta(\omega-k-q)\Big],\label{Ri}
\end{eqnarray}
with $n(q)=1/[e^{q/T_i}+1]$ the quark distribution function at the
\emph{initial time} $t_i$. We note that in obtaining the above
expressions, we have used the free quark propagators which
correspond to hard quark loop momentum ($q\gtrsim T_i$). Soft
quark lines ($q\ll T_i$) require using the HTL-resummed effective
quark propagator~\cite{htl}, thereby leading to higher order
corrections. Indeed, the one-loop diagram with soft quark loop
momentum is part of the higher order contribution of order
$\alpha\alpha_s$ that has been calculated in
Refs.~\cite{kapusta,baier:photon}.

We note that apart from color and flavor factors $\mathcal{R}(\omega)$ given by
Eq.~(\ref{Ri}) has the same structure as the first \emph{two} terms in
$\mathcal{R}^<_\gamma(\omega,k)$ given by Eq.~(\ref{Rgamma1}). From the
discussion there, one finds that the first delta function $\delta(\omega+k-q)$
has support below the light cone ($\omega^2<E^2$) corresponding to the Landau
damping cut whereas the second delta function $\delta(\omega-k-q)$ has support
above the light cone corresponding to the usual two-particle cut. Furthermore,
$\mathcal{R}(\omega)$ has a clear physical interpretation in terms of the
following \emph{energy-nonconserving} photon production processes: the first
term describes (anti)quark bremsstrahlung $q(\bar{q})\rightarrow
q(\bar{q})\gamma$, and the second term describes quark-antiquark annihilation
to photon $q\bar{q}\rightarrow\gamma$.

The dependence of $\mathcal{R}(\omega)$ on the quark distribution at the
initial time $t_i$ is a consequence of the fact that in nonequilibrium quantum
field theory the important ingredient is the \emph{initial density matrix},
which consistent with hydrodynamics is taken to be thermal for quarks and
gluons. Since the calculation is carried out consistently in perturbation
theory, the quark (and gluon) propagators are in terms of the equilibrium free
particle distribution functions. In the case of a nonexpanding and thermalized
QGP, the integral over $t''$ in Eq.~(\ref{a:noneqratei}) can be carried out
directly leading to
\begin{equation}
E\frac{dN(t)}{d^3p\,d^4x}=\frac{2}{(2\pi)^3}
\int_{-\infty}^{+\infty}d\omega\,\mathcal{R}(\omega)
\,\frac{\sin[(\omega-E)(t-t_i)]}{\pi(\omega-E)},\label{a:noneqratei2}
\end{equation}

At this stage we can make contact with the $S$-matrix calculation and highlight
the importance of the finite-time, nonequilibrium analysis. If, as is implicit
in the $S$-matrix calculation, the QGP is assumed in thermal equilibrium and
with an infinite lifetime (entailing that $t_i\rightarrow -\infty$ and $t_f
\rightarrow \infty$), then we can take the infinite time limit $t_i\rightarrow
-\infty$ in the argument of the sine function in Eq.~(\ref{a:noneqratei2}) and
use the approximation
\begin{equation}
\frac{\sin[(\omega-E)t]}{\pi(\omega-E)}\buildrel{t\rightarrow
\infty}\over{\approx}\delta(\omega-E).\label{goldenrule}
\end{equation}
This is the assumption of completed collisions that is invoked in
time-dependent perturbation theory leading to Fermi's golden rule
and energy conservation. The delta function $\delta(\omega-E)$ is
a manifestation of energy conservation for each \emph{completed}
collision. Under this assumption one finds a
\emph{time-independent} photon production rate proportional to
$\mathcal{R}(E)$, provided that the latter is finite. In the
present situation, however, the delta functions in
$\mathcal{R}(\omega)$ cannot be satisfied on the photon mass
shell. Therefore, under the assumption of completed collisions the
\emph{lowest order} energy-nonconserving contribution to the
photon production rate simply vanishes due to kinematics.
Therefore this lowest order contribution is absent (by energy
conservation) in the $S$-matrix calculation, but is present at any
finite time.

The relevant question to ask is that \emph{how this finite-time contribution of
order $\alpha$ compares to the higher order $S$-matrix contribution to the
photon yield}.

For finite QGP lifetime the time-dependent rate given in
Eq.~(\ref{a:noneqratei2}) is finite and nonvanishing, thus leading to a
nontrivial contribution to direct photon production. The photon yield (per unit
volume) is obtained by integrating the rate over the lifetime of the QGP. Using
Eq.~(\ref{a:noneqratei2}), one obtains
\begin{equation}
E\frac{dN}{d^3p~d^3x}=\frac{2}{(2\pi)^3} \int_{-\infty}^{+\infty}
d\omega\,\mathcal{R}(\omega)\,
\frac{1-\cos[(\omega-E)(t_f-t_i)]}{\pi(\omega-E)^2},\label{a:yield}
\end{equation}
where $t_f-t_i\sim10\,{\rm fm}/c$ is the lifetime of the QGP.

Before proceeding further, we give an analytic estimate of the behavior of the
photon yield in the HTL approximation. In this approximation the leading
contribution of $\mathcal{R}(\omega)$ is dominated by the Landau damping cut,
which corresponds to off-shell (anti)quark bremsstrahlung, and is found to be
given by
\begin{equation}
{\cal R}_\mathrm{HTL}(\omega)=\frac{20}{3}\frac{\pi^2 \alpha T^2}{12}
\frac{\omega}{E}\left(1-\frac{\omega^2}{E^2}\right)n_B(\omega)\,
\theta(E^2-\omega^2),\label{RHTL}
\end{equation}
where use has been made of the KMS condition, valid for quarks in thermal
equilibrium.

\begin{figure}[t]
\begin{center}
\includegraphics[width=3.5truein,keepaspectratio=true]{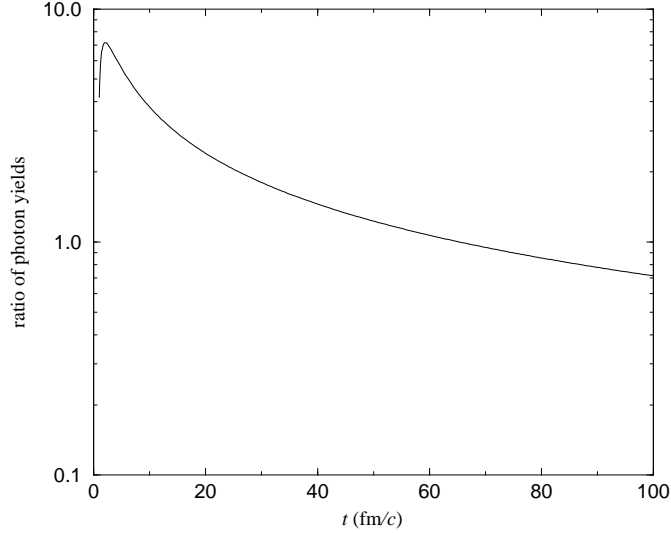}
\caption[The ratio of the nonequilibrium and equilibrium hard photon yields
plotted as a function of time for $\alpha=1/137$, $\alpha_s=0.4$ and $T=200$
MeV.]{The ratio of the nonequilibrium and equilibrium hard ($E\sim T$) photon
yield plotted as a function of time for $\alpha=1/137$, $\alpha_s=0.3$ and
$T=200$ MeV (see text for details).} \label{fig:photonratio}
\end{center}
\end{figure}

For $E\ll T$, ${\cal R}_\mathrm{HTL}(\omega)$ can be further simplified as [see
Eq.~(\ref{Rgamma2})]
\begin{equation}
{\cal R}_{\rm HTL}(\omega)\buildrel{E\ll T}\over\simeq \frac{20}{3}\frac{\pi^2
\alpha T^3}{12E} \left(1-\frac{\omega^2}{E^2}\right)
\theta(E^2-\omega^2).\label{RHTL2}
\end{equation}
The dominant contribution of the $\omega$ integral in Eq.~(\ref{a:yield}) for
$E\ll T$ arises from the region where the resonant denominator vanishes
($\omega\approx E$). Using Eq.~(\ref{RHTL2}), we obtain
\begin{equation}
E\frac{dN(t)}{d^3p\,d^3x}\buildrel{E\ll T}\over=
\frac{5}{9}\frac{\alpha}{2\pi^2}\frac{T^3}{E^2} \left[\ln
2Et+\gamma-1\right]+{\cal O}\left(\frac{1}{t}\right), \label{yieldhtl}
\end{equation}
where $\gamma=0.577\ldots$ is the Euler-Mascheroni constant.
Figure~\ref{fig:photonratio} displays the ratio of the hard ($E\sim T$) photon
yield extrapolated from Eq.~(\ref{yieldhtl}) and that obtained from the
equilibrium rate of Kapusta \textit{et al}.\ given by Eq.~(\ref{rate:kapusta}).
We see clearly that the lowest order $\mathcal{O}(\alpha)$ nonequilibrium
contribution to the photon yield is \emph{comparable at early times} to the
higher order $\mathcal{O}(\alpha\alpha_s)$ equilibrium contribution.

\begin{figure}[t]
\begin{center}
\includegraphics[width=3.5truein,keepaspectratio=true]{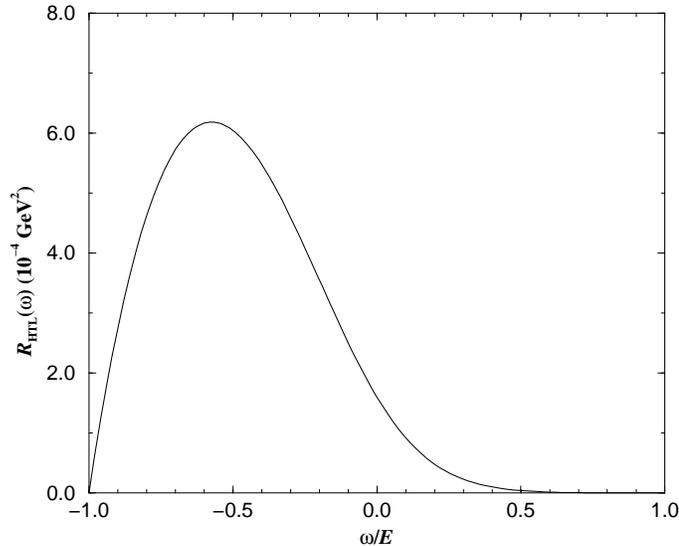}
\caption[The function ${\cal R}_{\rm HTL}(\omega)$ plotted in the limit $E\gg
T$.]{The function ${\cal R}_{\rm HTL}(\omega)$ plotted in the limit $E\gg T$.
Here we take $E=2$ GeV and $T=200$ MeV.} \label{fig:RHTL}
\end{center}
\end{figure}

For $E\gg T$, as depicted in Fig.~\ref{fig:RHTL}, ${\cal
R}_\mathrm{HTL}(\omega)$ is exponentially suppressed in the region
$T<\omega<E$. From this observation we emphasize the following
important features of the nonequilibrium yield given by
Eq.~(\ref{a:yield}). (i) Because for $E\gg T$ the threshold
contribution near the photon mass shell $\omega\approx E$ is
exponentially suppressed, the $\omega$ integral in
Eq.~(\ref{a:yield}) is now dominated by the interval
$-E<\omega<T$, which corresponds to highly off-shell (anti)quark
bremsstrahlung. (ii) As the integrand in Eq.~(\ref{a:yield}) is
positive-definite and the function $1-\cos[(\omega-E)t]$ averages
to 1 for large $t$ in the region $-E<\omega<T$, for \emph{fixed}
$E\gg T$ the yield approaches a constant at large times. (iii) In
contrast to those obtained from equilibrium rates, the yield for
$E\gg T$ is \emph{not} suppressed by the Boltzmann factor
$e^{-E/T}$.

\begin{figure}[t]
\begin{center}
\includegraphics[width=3.5truein,keepaspectratio=true]{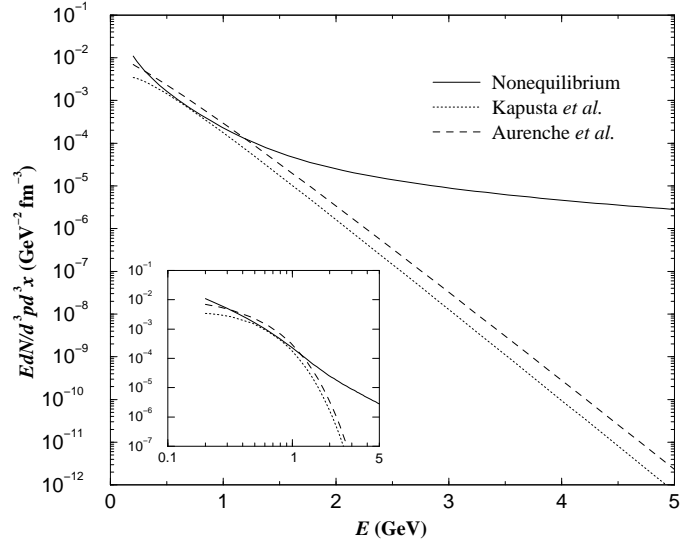}
\caption[Comparison of various contributions to the direct photon yield from a
QGP of temperature $T=200$ MeV and lifetime $t=10$ fm/$c$.]{Comparison of
various contributions to the direct photon yield from a QGP of temperature
$T=200$ MeV and lifetime $t=10$ fm/$c$. The inset shows the figure on a log-log
plot.} \label{fig:spectrum}
\end{center}
\end{figure}

Figure~\ref{fig:spectrum} shows a comparison of the nonequilibrium
and equilibrium contributions to the hard photon yield in the
range of energy $T < E < 5$ GeV from a QGP of temperature $T=200$
MeV and lifetime $t=10$ fm/$c$. Whereas the two-loop contribution
dominates the direct photon yield for smaller values of $E$, a
significant enhancement of the direct photon yield due to the
nonequilibrium contribution is seen at $E>2$ GeV as a consequence
of its power law falloff for $E\gg T$. Numerical evidence shows
that the nonequilibrium yield $E dN(t)/d^3p\,d^3x$ falls off with
a power law $E^{-\nu}$ with $\nu\approx 2.14$ for $T\ll E<5$ GeV.
In particular, the nonequilibrium contribution is larger than the
equilibrium contributions by several orders of magnitude for $E>3$
GeV. As the nonequilibrium contribution for fixed $E\gg T$
approaches a constant at large times, the equilibrium
contributions, which grow linearly in time, will eventually
dominate the yield if the QGP has a very long lifetime. However,
the linear growth in time of the equilibrium contributions has to
compensate the Boltzmann suppression for $E\gg T$. Therefore we
emphasize that for $E\gg T$ the equilibrium contributions
\emph{could} dominate the yield \emph{only if} the QGP lifetime is
of order $10^2-10^3$ fm/$c$ or larger, which nevertheless is very
unrealistic at RHIC energies.

While these results in the nonexpanding case revealed the importance of the
nonequilibrium and finite-lifetime aspects, the most experimentally relevant
case to study is that of an expanding QGP. This experimentally relevant case
can only be investigated by a detailed numerical study.

\subsection{Longitudinally expanding QGP}
As a prelude to photon production from an expanding QGP, we first focus on the
invariant photon production rate from each individual fluid cell of the QGP.
Since the proper time equals to the local time in the local rest frame of any
fluid cell, we can follow the same real-time nonequilibrium analysis presented
in the proceeding subsection to calculate the nonequilibrium invariant photon
production rate in the local rest frame of each fluid cell. We obtain
\begin{equation}
E\frac{dN(\tau)}{d^3p\,d^4x}\bigg|_{\rm
LR}=\lim_{\tau'\rightarrow\tau} \frac{\partial}{\partial \tau'}
\sum^{N_f}_{f=1}\frac{e\,e_f}{2(2\pi)^3}
\int\frac{d^3q}{(2\pi)^3}\big\langle\bar{\psi}_f^-(-{\bf
k},\tau){\bbox\gamma} \cdot{\bf A}^+_T({\bf p},\tau')\psi_f^-({\bf
q},\tau)\big\rangle+\mbox{c.c.}.\label{rate1}
\end{equation}

As before, the nonequilibrium expectation values on the right-hand
side of Eq.~(\ref{rate1}) is computed perturbatively to order
$\alpha$ and in principle to all orders in $\alpha_s$ by using
real-time Feynman rules and propagators. To \emph{lowest order} in
perturbation theory, the result for two light quark flavors ($u$
and $d$) reads
\begin{equation}
E\frac{dN(\tau)}{d^3p\,d^4x}\bigg|_{\rm LR}=\frac{2}{(2\pi)^3}
\int_{\tau_i}^{\tau}d\tau'' \int_{-\infty}^{+\infty}
\frac{d\omega}{\pi}\,\mathcal{R}(\omega)
\,\cos[(\omega-E)(\tau-\tau'')], \label{noneqratei}
\end{equation}
where $\mathcal{R}(\omega)$ is the same as that given in
Eq.~(\ref{Ri}), but now with $n(q)=1/[e^{q/T_i}+1]$ being the
quark distribution function \emph{at initial proper time}
$\tau_i$, at which the QGP reaches local thermal equilibrium.

In principle in the case of an expanding QGP under consideration, the photon
production rate given by Eq.~(\ref{noneqratei}) has to be supplemented by
kinetic equations that describe the evolution of the quark and gluon
distribution functions so as to setup a closed set of coupled equations.
However the assumption of the validity of (ideal) hydrodynamics entails that
the quarks and gluons form a perfectly coupled fluid. This in turn implies that
the mean free paths of the quarks and gluons are much shorter than the typical
wavelengths and the relaxation time scales are much shorter than the typical
time scales, i.e., the quark and gluon distribution functions adjust to local
thermal equilibrium instantaneously. Thus the assumption of the validity of
(ideal) hydrodynamics bypasses the necessity of the coupled kinetic equations:
quarks and gluons are in local thermal equilibrium at all times. Therefore,
within the framework of hydrodynamics we obtain the invariant photon production
rate by directly replacing the initial quark distribution function $n(q)$ in
Eq.~(\ref{Ri}) by the ``updated'' distribution function
$n[q,T(\tau'')]=1/[e^{q/T(\tau'')}+1]$ at proper time $\tau''>\tau_i$, where
$T(\tau'')$ is determined by the cooling law Eq.~(\ref{coolinglaw}). The
nonequilibrium invariant photon production rate that is consistent with the
underlying hydrodynamics is then given by
\begin{eqnarray}
&&E\frac{dN}{d^3p\,d^4x}\bigg|_{\rm LR}[\tau,E,T(\tau)]=\frac{2}{(2\pi)^3}
\int_{\tau_i}^{\tau}d\tau'' \int_{-\infty}^{+\infty}\frac{d\omega}{\pi}\,
\mathcal{R}[\omega,T(\tau'')]
\,\cos[(\omega-E)(\tau-\tau'')], \label{noneqrate}
\end{eqnarray}
with
\begin{eqnarray}
\mathcal{R}[\omega,T(\tau)]&=&\frac{20\,\pi^2\alpha}{3} \int\frac{d^3q}
{(2\pi)^{3}}\Big[2\big[1-({\bf\hat{p}}\cdot{\bf\hat{k}})
({\bf\hat{p}}\cdot{\bf\hat{q}})\big]
n[q,T(\tau)]\bar{n}[k,T(\tau)]\delta(\omega+k-q) \nonumber\\
&&+\,\big[1+({\bf\hat{p}}\cdot{\bf\hat{k}})
({\bf\hat{p}}\cdot{\bf\hat{q}})\big]n[q,T(\tau)]
n[k,T(\tau)]\delta(\omega-k-q)\Big],\label{R}
\end{eqnarray}
where $n[q,T(\tau)]=1/[e^{q/T(\tau)}+1]$ and $\bar{n}=1-n$. The momentum $q$
integrals in Eq.~(\ref{R}) are calculated in the LR frame and hence are
equivalent to those of the nonexpanding case above.

Before proceeding further, we emphasize two noteworthy features of
the nonequilibrium invariant photon production rate: (i) The
photon production processes do \emph{not} conserve energy. (ii)
The rate depends on (proper) time not only implicitly through the
local temperature but also explicitly. Furthermore, this explicit
(proper) time dependence is non-Markovian as clearly displayed in
Eq.~(\ref{noneqrate}). Whereas these two features seem rather
uncommon in transport phenomena in heavy ion collisions (see,
however, Refs.~\cite{rau,reinhard}), they are not unusual in
nonrelativistic many-body quantum kinetics under extreme
conditions~\cite{book:haug}. In particular, energy-nonconserving
transitions and memory effects which cannot be explained by usual
(semiclassical) Boltzmann kinetics have been observed recently in
ultrafast spectroscopy of semiconductors with femtosecond laser
pulses~\cite{ultrafast}.

The real-time kinetic approach when incorporated consistently with the
hydrodynamical evolution of the QGP reveals clearly that photon production is
inherently a nonequilibrium quantum effect associated with the expansion and
finite lifetime of the QGP.

At RHIC and LHC energies the quark distribution function
$n[q,T(\tau)]$ depends on the proper time $\tau$ very weakly
through the temperature $T(\tau)$ within the lifetime of the QGP
phase, hence a Markovian approximation (MA) in which the
temperature in $\mathcal{R}[\omega,T(\tau'')]$ is taken at the
upper limit of the integral is reasonable and hence the memory
kernel may be simplified. In this Markovian approximation,
$\mathcal{R}[\omega,T(\tau'')]$ in Eq.~(\ref{noneqrate}) is
replaced by $\mathcal{R}[\omega,T(\tau)]$ and taken outside of the
$\tau''$-integral. Thus  Eq.~(\ref{noneqrate}) becomes
\begin{equation}
E\frac{dN}{d^3p\,d^4x}\bigg|^{\rm MA}_{\rm
LR}[\tau,E,T(\tau)]=\frac{2}{(2\pi)^3}
\int_{-\infty}^{+\infty}d\omega\,\mathcal{R}[\omega,T(\tau)]
\,\frac{\sin[(\omega-E)(\tau-\tau_i)]}{\pi(\omega-E)}. \label{noneqrate2}
\end{equation}
A computational advantage of this Markovian nonequilibrium production rate is
that it provides the ``updated'' quark distribution functions locally in time.
Physically, the motivation for this approximation is that the most important
aspect of the nonequilibrium effect is the nonconservation of energy originated
in the finite lifetime of the QGP, a feature that is missed by the usual
$S$-matrix calculation, while the proper time variation of the temperature is a
secondary effect and accounted for in the $S$-matrix approach.

It is worth noting that a connection with the Boltzmann
approximation can be obtained by assuming completed collisions,
i.e., taking the limit $\tau_i\rightarrow -\infty$ in the argument
of the sine function in Eq.~(\ref{noneqrate2}) and using the
approximation given by Eq.~(\ref{goldenrule}). Consequently, the
\emph{lowest order} nonequilibrium photon production rate vanishes
in the Boltzmann approximation due to kinematics. This highlights,
once again,  that the usual approach to photon production outlined
in Sec.~\ref{sec:6.2} and used in the literature corresponds to
the Boltzmann approximation and therefore fails to capture the
energy-nonconserving and memory effects that occur during the
transient stage of evolution of the QGP.

We are now in a position to calculate the direct photon yield from a
longitudinally expanding QGP. In the CM frame the invariant production rate for
photons of four-momentum $P^{\mu}$ from a fluid cell with four-velocity $u^\mu$
can be obtained from Eq.~(\ref{noneqrate2}) through the replacement
$E\rightarrow P^{\mu}u_{\mu}$. In terms of the photon transverse momentum $p_T$
and rapidity $y$, one finds $P^{\mu}u_{\mu}=p_T\cosh(y-\eta)$. The photon yield
is obtained by integrating the nonequilibrium rate over the space-time
evolution of the expanding QGP. Assuming  a central collision of identical
nuclei, in the Markovian approximation we find the invariant nonequilibrium
photon yield to lowest order in perturbation theory to be given by
\begin{equation}
\frac{dN}{d^2p_T\,dy}\bigg|^{\rm MA}_{\rm CM}=\pi R_A^2
\int_{\tau_i}^{\tau_f}d\tau\,\tau \int_{-\eta_{\rm cen}}^{\eta_{\rm cen}}d\eta
\,E\frac{dN}{d^3p\,d^4x}\bigg|^{\rm MA}_{\rm LR}
[\tau,P^{\mu}u_{\mu},T(\tau)],\label{yield}
\end{equation}
where $R_A$ is the radius of the nuclei and $-\eta_{\rm cen}<\eta<\eta_{\rm
cen}$ denotes the central rapidity region within which Bjorken's hydrodynamical
model is valid.

As remarked above, at RHIC and LHC energies the quark distribution
function $n[q,T(\tau)]$ depends on the proper time $\tau$ very
weakly through the temperature $T(\tau)$ within the lifetime of
the QGP phase, therefore the resultant photon yield is expected to
qualitatively resemble the nonequilibrium photon yield from a
nonexpanding QGP studied in above. This will be numerically
verified below. We note that the expanding and nonexpanding cases
differ mainly by the Jacobian $\tau$ in the $\tau$ integral in
Eq.~(\ref{yield}) that accounts for the longitudinal expansion of
the QGP, and by the replacement $E\rightarrow P^{\mu}u_{\mu}$ in
the argument of the invariant photon production rate that accounts
for the shift of the photon energy under local Lorentz boost .

\subsection{Numerical analysis: central collisions at RHIC and LHC energies}
We now perform a numerical analysis of the nonequilibrium photon
yield and compare the results to the equilibrium one obtained from
higher order equilibrium rate calculations given by
Eqs.~(\ref{rate:kapusta}) and (\ref{rate:aurenche}). The
nonequilibrium photon yield in the Markovian approximation given
by Eq.~(\ref{yield}) contains a four-dimensional integral that is
performed numerically for the values of parameters of relevance at
RHIC and LHC energies.

\begin{figure}[t]
\begin{center}
\includegraphics[width=3.5in,keepaspectratio=true]{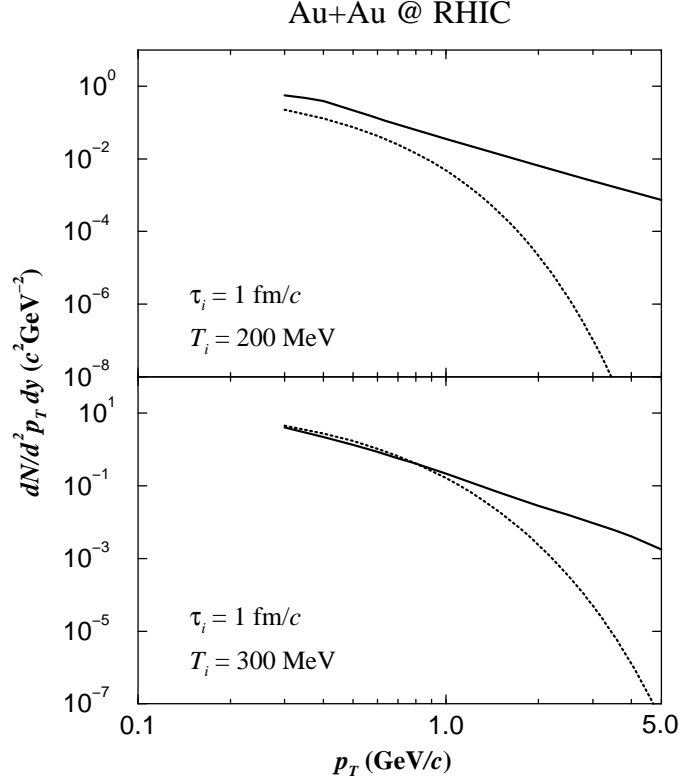}
\caption[Comparison of nonequilibrium and equilibrium photon
yields at midrapidity from a longitudinally expanding QGP at RHIC
energies.]{Comparison of nonequilibrium (solid) and equilibrium
(dotted) photon yields at midrapidity ($y=0$) from a
longitudinally expanding QGP at RHIC energies with initial
conditions given by $\tau_i=1$ fm/$c$ and $T_i=200$ (top), 300
(bottom) MeV.} \label{fig:rhic}
\end{center}
\end{figure}

For central ${}^{197}$Au$+{}^{197}$Au collisions at RHIC energies
$\sqrt{s_{NN}}\sim 200$ GeV, we take $R_A\simeq 1.2\,A^{1/3}$ fm
$\approx 7$ fm~\cite{book:lebellac} and $\eta_{\rm cen}=2$. The
initial thermalization time is taken to be $\tau_i=1$
fm/$c$~\cite{alam,bjorken}, the final proper time $\tau_f$ is
determined when the critical temperature for the quark-hadron
transition is reached at $T(\tau_f)\simeq 160$ MeV and is obtained
from the cooling law given by Eq.~(\ref{coolinglaw}) for a given
initial temperature $T_i$ at proper time $\tau_i$.

The nonequilibrium photon yield at midrapidity ($y=0$) in the
range of transverse momentum $0.3 < p_T < 5$ GeV/$c$ is shown on a
log-log plot in Fig.~\ref{fig:rhic} for initial temperatures
$T_i=200$ and 300 MeV. For comparison we also plot the
corresponding equilibrium yield obtained by integrating
Eqs.~(\ref{rate:kapusta}) and (\ref{rate:aurenche}) with the
transformation to the CM frame as specified by
Eq.~(\ref{yieldSmtx}), and using the value of
$\alpha_s$~\cite{srivastava,alam,karsch}
\begin{equation}
\alpha_s[T(\tau)]=\frac{6\pi}{(33-2N_f)\ln[8\,T(\tau)/T_c]},
\end{equation}
with $N_f=2$. Furthermore, Fig.~\ref{fig:rapidity} shows the
rapidity distribution of the nonequilibrium photon yield at
different values of $p_T$ for $T_i=200$ and $300$ MeV. Several
noteworthy features are gleaned from these figures:

\begin{figure}[t]
\begin{center}
\includegraphics[width=3.5in,keepaspectratio=true]{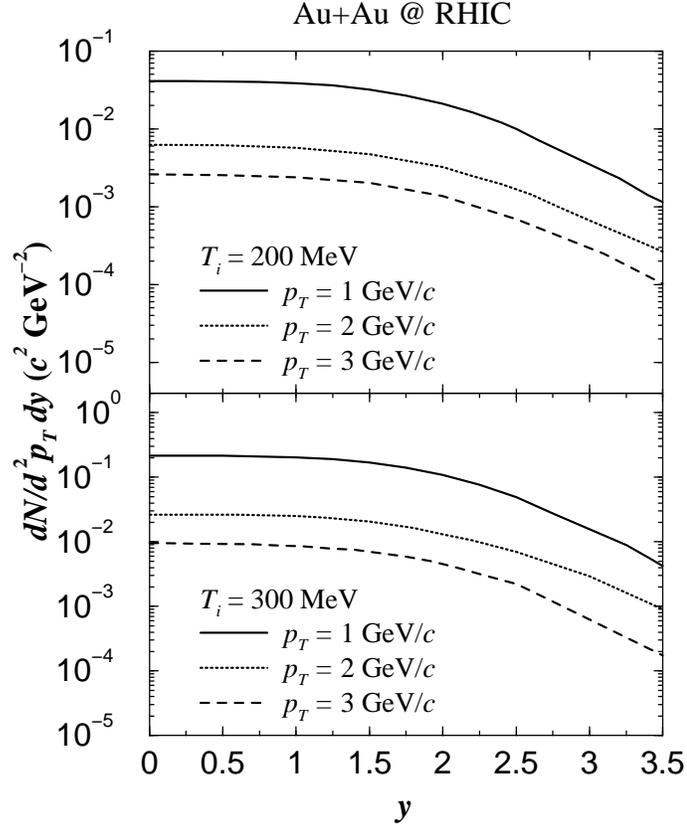}
\caption[Rapidity distribution of the nonequilibrium photon yield
at $p_T=1$, 2, and 3 GeV for a longitudinal expanding QGP at RHIC
energies.]{Rapidity distribution of the nonequilibrium photon
yield at $p_T=1$ (solid), 2 (dotted), and 3 (dashed) GeV for a
longitudinal expanding QGP at RHIC energies with initial
conditions given by $\tau_i=1\,{\rm fm}/c$ and $T_i=200$ (top),
300 (bottom) MeV. The distribution is symmetric at $y=0$.}
\label{fig:rapidity}
\end{center}
\end{figure}

\begin{enumerate}
\item{Whereas the equilibrium yield dominates the total yield for small $p_T$,
the nonequilibrium yield becomes significantly dominant in the
range $p_T>1.0-1.5$ GeV/$c$. Perhaps coincidentally, this is the
range in which the CERN WA98 data for central Pb$+$Pb collisions
at SPS energies shows a distinct excess~\cite{WA98}.}

\item{While the equilibrium yield leads to a transverse momentum distribution
that falls off approximately with the Boltzmann factor
$e^{-p_T/T_i}$, the nonequilibrium yield is \emph{not} Boltzmann
suppressed \emph{and} falls off algebraically. It is found
numerically that for $p_T\gg T_i$ and within the region $1
\lesssim p_T \lesssim 5$ GeV/$c$ the nonequilibrium yield at
midrapidity falls off with a power law $p_T^{-\nu}$ with
$\nu\simeq 2.47$ and $2.77$ for $T_i=200$ and 300 MeV,
respectively. This is a remarkable consequence of the fact that
bremsstrahlung is the dominant process. As discussed in detail
above at large energies the dominant process is bremsstrahlung
corresponding to the contribution from the term $n(q)[1-n(k)]$
with $k = |\mathbf{p}+\mathbf{q}|$ in Eq.~(\ref{Ri}). For large
photon energy $p$ the important contribution, which is not
exponentially suppressed  arises from the small $q$ region. }

\item{A central plateau in the range of rapidity $y\lesssim 2$ is seen
clearly. The rapidity distribution begins to bend down when
$y\simeq\eta_{\rm cen}$, i.e., when the photon rapidity probes the
fragmentation region.}

\item{The numerical analysis of the expanding case reveals
features very similar to those found in the nonexpanding case,
where we estimate that for $p_T \geq 1.5$ GeV/$c$ the higher order
equilibrium contribution to the direct photon yield (which grows
linearly in time) becomes of the same order as the lowest order
nonequilibrium contribution (which grows at most logarithmically
in time) only if the lifetime of the QGP is of the order longer
than 1000 fm/$c$. Therefore, we conclude that these nonequilibrium
effects dominate during the lifetime of the QGP in a realistic
ultrarelativistic heavy ion collision experiment.}
\end{enumerate}

\begin{figure}[t]
\begin{center}
\includegraphics[width=3.5in,keepaspectratio=true]{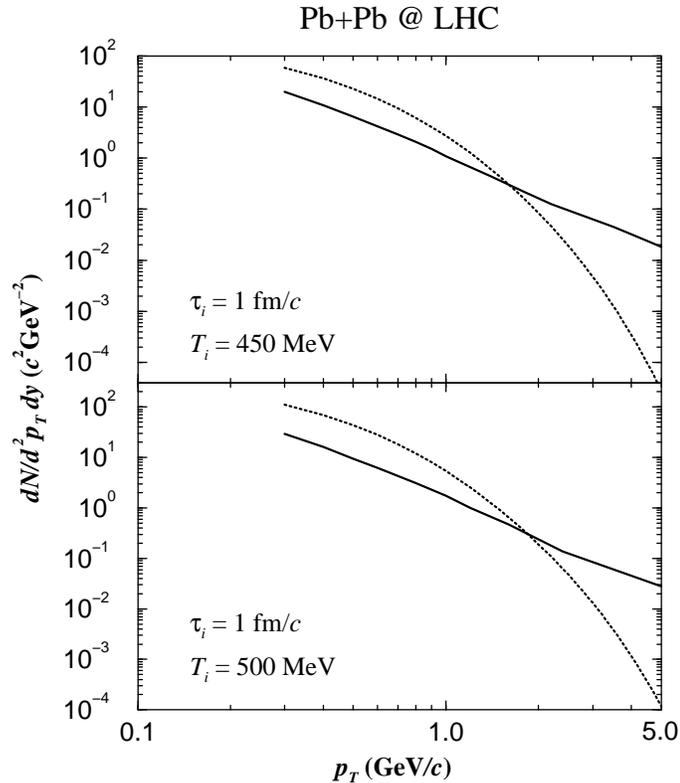}
\caption[Comparison of nonequilibrium and equilibrium photon
yields at midrapidity from a longitudinally expanding QGP at LHC
energies.]{Comparison of nonequilibrium (solid) and equilibrium
(dotted) photon yields at midrapidity ($y=0$) from a
longitudinally expanding QGP at LHC energies with initial
conditions given by $\tau_i=1$ fm/$c$ and $T_i=450$ (top), 500
(bottom) MeV.} \label{fig:lhc}
\end{center}
\end{figure}

Similar numerical analysis can be performed for central
${}^{208}$Pb$+{}^{208}$Pb collisions at higher LHC energies
$\sqrt{s_{NN}}\sim 5500$ GeV, for which we take $R_A\approx 7$ fm,
$\eta_{\rm cen}=5$ and $\tau_i=1$ fm/$c$~\cite{alam,bjorken}. The
comparison of nonequilibrium and equilibrium photon yields at
midrapidity is displayed in Fig.~\ref{fig:lhc} for $T_i=450$ and
500 MeV. The dominance of the nonequilibrium yield at high $p_T$
remains but now at higher transverse momentum $p_T\gtrsim 2$ GeV.
This can be understood as a consequence of the longer QGP lifetime
resulting from higher initial temperature at LHC energies.
Furthermore, it is found numerically that for $1 \lesssim p_T
\lesssim 5$ GeV/$c$ the nonequilibrium yield at midrapidity falls
off with a power law $p_T^{-\nu}$ with $\nu\simeq 2.52$ and $2.56$
for $T_i=450$ and 500 MeV, respectively.

These results indicate a clear manifestation of the nonequilibrium
aspects of direct photon production associated with a transient
QGP of finite lifetime. The most experimentally accessible signal
of the nonequilibrium effects revealed by this analysis is the
\emph{power law} falloff of the transverse momentum distribution
for direct photons in the range $1 \lesssim p_T \lesssim$ 5
GeV/$c$ with an exponent $2.5 \lesssim \nu \lesssim 3$ for
temperatures expected at RHIC and LHC energies. Our numerical
studies reveal that this exponent increases with initial
temperature and therefore with the initial energy density of the
QGP and the total multiplicity rapidity distribution
$dN_\pi/dy$~\cite{bjorken}. This could be a clean experimental
\emph{nonequilibrium signature} of a transient QGP since the
photon distribution from the hadronic gas is expected to feature a
Boltzmann type exponential suppression for $p_T \gg
T_i$~\cite{kapusta,sollfrank,srivastava,alam}.

\section{Conclusions}\label{sec:6.5}
In this chapter we study an important phenomenological application
of the real-time quantum kinetic techniques developed in previous
chapters. Our goal is to search for clear experimental signatures
of direct photon production associated with nonequilibrium aspects
of the transient QGP created in ultrarelativistic heavy ion
collisions at RHIC and LHC energies.

We first argue that the usual $S$-matrix approach to direct photon
production from an expanding nonequilibrium QGP has conceptual
limitations. Instead we introduce a real-time kinetic approach
that allows a consistent treatment of photon production from a
transient nonequilibrium state of \emph{finite lifetime}.

We focus on obtaining the direct photon yield from a thermalized
QGP undergoing Bjorken's hydrodynamical expansion in central
Au$+$Au collisions at RHIC ($\sqrt{s_{NN}}\sim 200$ GeV) and LHC
($\sqrt{s_{NN}}\sim 5500$ GeV) energies. The lifetime of a QGP for
these collisions is of order $10-30$ fm/$c$. We find that
energy-nonconserving (anti)quark bremsstrahlung
$q(\bar{q})\rightarrow q(\bar{q})\gamma$ and quark-antiquark
annihilation $q\bar{q}\rightarrow\gamma$, \emph{both of lowest
order in $\alpha$}, dominate during such short time scales with
bremsstrahlung dominating for $p_T \gg T_i$. The contribution from
these processes is a consequence of the transient nature and the
finite lifetime of the QGP. As compared to the equilibrium rate
calculations, the energy-nonconserving processes lead to a
substantial enhancement in direct photon production for $1-2
\lesssim p_T \lesssim 5$ GeV/$c$ near midrapidity. In striking
contrast with the equilibrium calculation that predicts an
exponential suppression of the transverse momentum distribution
for $p_T \gg T_i$ ($T_i$ is the initial temperature), the
nonequilibrium processes lead to a \emph{power law} behavior
instead. We find that at RHIC and LHC energies the direct photon
transverse momentum distribution near midrapidity is of the form
$p^{-\nu}_T$ with $2.5 \lesssim \nu \lesssim 3$ for $1 \lesssim
p_T \lesssim 5$ GeV/$c$ and that photon rapidity distribution (for
fixed $p_T$) is almost flat in the interval $|y| \lesssim
\eta_{\rm cen}$, where $|\eta|<\eta_{\rm cen}$ denotes the central
rapidity region of the QGP. The exponent $\nu$ is numerically
found to increase with the initial temperature, hence increases
with the total multiplicity rapidity distribution $dN_\pi/dy$,
which is an experimental observable.

Thus, as the main conclusion, we propose that direct photons are
distinct experimental signatures of the transient nonequilibrium
QGP created at RHIC and LHC energies, both in the form of a large
enhancement at $1-2 \lesssim p_T \lesssim$ 5 GeV/$c$ as well as a
power law transverse momentum distribution $p^{-\nu}_T$ with an
exponent $\nu$ that is within the range $2.5 \lesssim \nu \lesssim
3$ and increases with total multiplicity rapidity distribution
$dN_\pi/dy$.

Our study of direct photon production focuses solely on the QGP
phase and neglects contributions from pre-equilibrium stage as
well as the mixed and hadronic phases, because we focus on a
comparison between the usual equilibrium approach and the kinetic
real-time approach that allows for following the time evolution of
the density matrix as an initial value problem consistent with
hydrodynamics. Furthermore, as discussed elsewhere in the
literature, since most of the high-$p_T$ photons originate from
the very early, hot stage in the QGP
phase~\cite{mclerran,kapusta,ruuskanen} and photons produced in
the mixed phase as well as the subsequent hadronic phase are
mainly in the lower-$p_T$ region~\cite{alam}, the nonequilibrium
yield from the QGP phase contributes dominantly to the total
high-$p_T$ photons.

We have provided a systematic real-time description compatible
with the initial value problem associated with hydrodynamic
evolution, however more needs to be understood for a complete
description of \emph{all} the different stages. The parton cascade
approach to describe the early pre-equilibrium stage after the
collision is an important first step in a full microscopic
description, but perhaps a more consistent description of the
initial condition for QGP formation must be based on the recent
notions of a color glass condensate~\cite{glass,raju}. Hence, a
complete treatment of the direct photon yield must, in principle,
begin from the initial stage, possibly a color glass condensate,
and obtain the real-time evolution of photon production. Clearly
there must be many more advances in this field before such a
program becomes feasible. The finite-temperature equilibrium
calculations \emph{assume} that the equilibrium thermal state
always prevailed, thus ignores completely not only the initial
stages but also the time evolution. Our approach while
incorporating the time evolution consistently, also neglects the
initial stage. We have focused on direct photons from a transient
QGP for a direct comparison with equilibrium calculations, but
obviously photons will continue to be produced during the mixed
and hadronic phases. A detailed understanding of the transition as
well as the hadronic photon production matrix elements is
necessary for a more reliable estimate of the potential
nonequilibrium effects \emph{after} hadronization.

Although we do not claim to have provided a completely detailed
understanding of potential nonequilibrium effects due to the lack
of knowledge of initial conditions in heavy ion collisions, we do
claim that our approach provides a more systematic description of
the finite-lifetime effects associated with a transient QGP. These
effects lead to a distinct experimental prediction: a power law
falloff of the distribution $dN/d^2p_T dy$ near the central
rapidity region which is distinguishable from a thermal tail for
$1-2 \lesssim p_T \lesssim 5$ GeV/$c$. We note that the
parton-cascade results of photon production from the
pre-equilibrium stage~\cite{srivageiger} also seem to reveal a
power law falloff of the photon distribution in this range of
transverse momentum with comparable order of magnitude and
exponent, as can be gleaned from Figs.~2 and 3 of
Ref.~\cite{srivageiger}. Perhaps these two effects, namely the
prompt photons from the pre-equilibrium stage and the direct
photons from the hydrodynamically expanding QGP with a finite
lifetime cannot be distinguished experimentally. Nevertheless, we
emphasize that a power law departure from a thermal tail in the
direct photon spectrum may very well be explained by
nonequilibrium effects, either by a finite QGP lifetime as
advocated in this chapter or by prompt photons from the
pre-equilibrium stage.

An important and very relevant question is that why not treat high
energy particle collisions in the same manner, i.e., with the
real-time evolution rather than with the $S$-matrix approach. The
answer to this question hinges on the issue of time scales. In a
typical high energy particle collider experiment, the colliding
``beams'' are actually bunches or packets with a typical spatial
extent of order $1-10$ cm and hence the typical time interval for
the collision is of order $10^{-9}$ sec. This time scale is many
orders of magnitude longer than the typical hadronic interaction
time scale $\sim 10^{-23}$ sec, thus taking the infinite time
limit is amply justified. This of course is the basis for using
the $S$-matrix calculation in terms of asymptotic {\sl in} and
{\sl out} states (at $t=\mp\infty$): the total interaction time is
much, much longer than the typical hadronic time scale. With
regard to the heavy ion collision, the consensus is that after an
initial pre-equilibrium stage, a locally thermalized QGP is
formed. It then evolves hydrodynamically, hadronizes and
eventually the freeze-out of hadronic gas ensues. Current
theoretical estimates indicate a total time between formation and
freeze-out of order $50-100$ fm/$c$, with a QGP phase lasting for
about $10$ fm/$c$. These time scales are \emph{not} several orders
of magnitude larger than the hadronic interaction time scale, thus
the infinite time limit taken in the $S$-matrix calculation is at
best questionable. Therefore while in typical particle collider
experiments the $S$-matrix approach is valid (as has been
demonstrated by half a century of experiments), the transient QGP
with a finite lifetime of the order of the hadronic time scales
merits a \emph{different} treatment.

As many recent investigations~\cite{shuryak} have suggested that
the QGP produced at RHIC and LHC energies is not expected to be in
local chemical equilibrium, i.e., the distribution functions of
quarks and gluons will probably be undersaturated, an important
extension of our work  will be  the study of nonequilibrium
effects on direct photon production from a chemically
nonequilibrated QGP. Another important aspect that requires
further investigation is the finite size of the QGP. Much in the
same way as the finite lifetime introduces nonequilibrium effects
associated with energy nonconserving transitions, we expect that
the finite size $\sim 7$ fm of the QGP will introduce uncertainty
in the momentum of the emitted particles. This is clearly an
important topic that deserves further and deeper study but is
beyond the scope of this thesis. However, at this stage we
speculate that if such effects are present they could bear an
imprint in transverse flow~\cite{flow}.

%% file: chap7.tex
\chapter{Summary}\label{chap:7}
The central theme of this thesis is the study of real-time
nonequilibrium quantum dynamics with specific focus on the
real-time relaxation of the mean fields and quantum kinetics in a
variety of physically relevant quantum field theories with
applications to cosmology and ultrarelativistic heavy ion
collisions. This study could be generalized to nonequilibrium
problems in condensed matter physics, quantum optics and atomic
physics. To conclude this thesis, we summarize our work and
highlight the main results.

\textbf{Fermion damping in a fermion-scalar plasma}. We study the
real-time relaxation of the fermion mean field in a fermion-scalar
plasma which is relevant to models of baryogenesis at the
electroweak scale. We begin by obtaining directly in real time the
effective in-medium Dirac equation for the fermion mean field
induced by an adiabatically switched-on external source. The
in-medium Dirac equation is fully renormalized, retarded and
causal, thereby leading to an initial value problem for the
fermion relaxation.

For the light scalar which cannot decay into fermion pairs, we
find that to lowest order in the Yukawa coupling the only medium
effect on fermions is dispersive and in-medium propagation of
fermion excitations is undamped. On the other hand, for the heavy
scalar such that its decay into fermion pairs is kinematically
allowed, we find a novel dissipative medium effect arises from
scalar decay in the medium and leads to \emph{damping of fermion
excitations}. That is, the fermions acquire a width due to the
decay of the heavy scalar in the medium and become quasiparticles
with a finite lifetime. Solving the effective Dirac equation by
Laplace transform, we obtained the time evolution of the fermion
mean field which allows a clear identification of the damping
rate. We calculate the damping rate in the narrow width
approximation to one loop order for arbitrary values of the
fermion and scalar masses (provided the scalar is heavy enough to
decay), temperature and fermion momentum. An all-order expression
for the damping rate in terms of the exact quasiparticle wave
functions is established by analyzing the structure of the fermion
self-energy.

Finally, a Boltzmann-type kinetic equation for the fermion distribution
function in the relaxation time approximation confirms the damping of fermion
excitations as a consequence of the decay of heavy scalars in the medium, and
displays the relationship between the fermion damping rate and interaction rate
to lowest order in the Yukawa coupling directly in real time.

\textbf{Dynamical renormalization group approach to quantum
kinetics}. We provide an alternative first-principle derivation of
quantum kinetic equations in quantum field theory by implementing
a diagrammatic perturbative expansion improved by a resummation
via the dynamical renormalization group directly in real time.

This method begins by obtaining the equation of motion for the quasiparticle
distribution function in perturbation theory. The solution of this equation of
motion at large times reveals secular terms that grow with time, as a
consequence of the fact that the perturbative analysis breaks down at large
times. We implement the dynamical renormalization group to resum these secular
terms directly in real time, the corresponding dynamical renormalization group
equation, which describes the time evolution of the quasiparticle distribution
function insensitive to the physics on microscopic time scales, is interpreted
as the quantum kinetic equation. A remarkable feature of this method is that it
allows to include consistently medium effects via resummation akin to the hard
thermal loops but away from equilibrium.

We establish a close relationship between the dynamical
renormalization group approach and the renormalization group in
Euclidean field theory. In particular, coarse graining, stationary
solutions, relaxation time approximation and relaxation rates have
a natural parallel as irrelevant operators, fixed points,
linearization near the fixed point and stability exponents,
respectively, in the usual Euclidean renormalization group. This
identification brings a new and rather different perspective to
kinetics and relaxation that will hopefully lead to a new
insights.

We then apply this novel method to study quantum kinetics of pions and sigma
mesons in the $O(4)$ chiral linear sigma model. We derive the quantum kinetic
equations that describes the time evolution of the pion and sigma meson
distribution functions. The pion and sigma meson relaxation rates calculated in
relaxation time approximation and agree with the corresponding damping rates
found in the literature. We also find that in large momentum limit the emission
and absorption of massless pions result in threshold infrared divergence in
sigma meson relaxation rate and lead to a crossover behavior in relaxation that
is unambiguously captured by the dynamical renormalization group analysis.

Furthermore by establishing a direct correspondence between pinch singularities
and secular terms, we show that the dynamical renormalization group approach to
quantum kinetics provides a natural framework to interpret and resolve the
issue of pinch singularities in nonequilibrium quantum field theory.

\textbf{Nonequilibrium dynamics in a QED plasma at high
temperature}. We study real-time nonequilibrium dynamics in a QED
plasma at high temperature within the hard thermal loop (HTL)
approximation by implementing the dynamical renormalization group.
The goal is to understand the relaxation of photon and fermion
mean fields directly in real time and quantum kinetics of the
photon and fermion distribution functions beyond the usual
Boltzmann description of kinetics. As many features of a hot QCD
plasma are similar to those of a hot QED plasma, in addition to
its direct impacts on astrophysics and cosmology, this study is
also of phenomenological importance in ultrarelativistic heavy ion
collisions. In particular, the leading contributions in the hard
thermal loop approximation can be generalized easily from hot QED
to hot QCD, thus relaxation of quarks and production of photons in
the QGP can be understood to leading order from the study of a hot
QED plasma.

For semihard photons of momentum $eT\ll k \ll T$, we find to
leading order in the HTL approximation that the photon mean field
relaxes with a power law $\sim t^{-e^2T^2/12k^2}$ at large times
$t\gg 1/k$, as a result of infrared enhancement of the photon
spectral density near the Landau damping threshold. The quantum
kinetic equation for the distribution function of semihard
photons, which includes off-shell effects missed by the usual
Boltzmann description, is derived directly in real time from first
principles in the relaxation time approximation by using the
dynamical renormalization group method. We find that the
distribution function also relaxes with a power law, with an
exponent which is twice that for the photon mean field. The
dynamical renormalization group analysis reveals the emergence of
detailed balance on microscopic time scales larger than $1/k$
while the rates are still varying with time.

We show directly in real time that as a consequence of the
emission and absorption of soft, quasi-static magnetic photons,
hard fermion mean field with momentum $k\sim T$ exhibits a
nonexponential relaxation at large times as $\sim e^{-\alpha
Tt(\ln\omega_P t+0.126\ldots)}$, where $\omega_P=eT/3$ is the
plasma frequency and $\alpha=e^2/4\pi$. A quantum kinetic equation
for the distribution function of hard fermions in the relaxation
time approximation is obtained directly in real time by
implementing the dynamical renormalization group resummation.
Unlike that of the usual Boltzmann kinetic equation derived in the
quasiparticle approximation, the collision term of this quantum
kinetic equation is \emph{time-dependent} and \emph{infrared
finite}. The distribution function is found to relax
nonexponentially with an anomalous exponent twice as large as that
for the hard fermion mean field.

Our real-time analysis reveals clearly the limitation of the usual Boltzmann
description of kinetics and the quasiparticle approximation in the study of
nonequilibrium quantum dynamics in gauge field theory and has important impacts
on direct photon production from the quark-gluon plasma.

\textbf{Direct photon production from the quark-gluon plasma}. As
an important phenomenological application of real-time
nonequilibrium dynamics in quantum field theory, we study direct
photon production from a longitudinally expanding QGP at RHIC and
LHC energies. Instead of focusing on obtaining an equilibrium
photon production rate using the usual $S$-matrix formalism, we
study photon production as an initial value problem with the
real-time quantum kinetic description developed in this work.

This real-time approach allows us to incorporate consistently the
kinetic description of photons with the hydrodynamical description
of the QGP. Within Bjorken's hydrodynamical model, we show that
the lowest order energy-nonconserving (anti)quark bremsstrahlung
$q(\bar{q})\rightarrow q(\bar{q})\gamma$ and quark-antiquark
annihilation $q\bar{q}\rightarrow \gamma$ contribute dominantly to
photon production during the transient lifetime of the QGP. For
central collisions at RHIC ($\sqrt{s_{NN}}\sim 200$ GeV) and LHC
($\sqrt{s_{NN}}\sim 5500$ GeV) energies, we find a significant
excess of direct photons in the range of transverse momentum $1-2
\lesssim p_T \lesssim 5$ GeV/$c$ as compared to higher order
equilibrium estimates. The rapidity distribution of photons is
fairly flat near the midrapidity region. The remarkable result is
that the transverse momentum distribution of photons at
midrapidity falls off with a power law $p^{-\nu}_T$ with $2.5
\lesssim \nu \lesssim 3$ as a consequence of the
energy-nonconserving processes taking place during the transient
lifetime of the QGP, providing a distinct nonequilibrium
experimental signature of the QGP formation.

As summarized above, in this thesis we have advanced several
aspects of real-time nonequilibrium dynamics in quantum field
theory. In particular, we have established a theoretical framework
for studying nonequilibrium dynamics of quantum multiparticle
systems directly in real time from first principles. This
framework is used to study a variety of physically relevant
quantum field theories. It not only reproduces the usual results
in the literature when the same approximations and assumptions are
invoked, but also leads us to a new territory which has not been
explored before. Most importantly, it is capable of making
concrete predictions that can be tested experimentally at BNL RHIC
and CERN LHC. We envisage our work will also have direct impacts
on the understanding of nonequilibrium phenomena in condensed
matter physics, quantum optics and cosmology.

%% file: appendix.tex
%
\appendix
\chapter{Renormalization Group and Asymptotic Analysis}
\label{chap:app}

In this appendix we present a pedagogical introduction to the
renormalization group (RG) method in studying asymptotic analysis
of ordinary differential equations. In order to introduce the
basic idea, we consider a simple but illuminating example: a
weakly damped harmonic oscillator. The reader is referred to
Refs.~\cite{goldenfeld,kunihiro} for generalization to the much
more complicated problems.

The differential equation governing a weakly damped harmonic
oscillator is given by
\begin{equation}
\ddot{y}+y=-\epsilon \dot{y},\quad\epsilon\ll 1\label{damposc}
\end{equation}
with initial conditions $y(t_0)$ and $\dot{y}(t_0)$ specified at
time $t_0$, where the overdot denotes the derivative with respect
to $t$. Note that Eq.~(\ref{damposc}) can be solved exactly with
the exact solution given by
\begin{equation}
y(t)=R_0\,e^{-\epsilon(t-t_0)/2}\,\cos\left[\sqrt{1-\frac{\epsilon^2}{4}}\,
(t-\Theta_0)\right],\label{exact}
\end{equation}
where $R_0$ and $\Theta_0$ are constants determined by the initial
conditions. However, in order to demonstrate the renormalization
group method, we attempt to solve Eq.~(\ref{damposc}) in a
perturbative expansion in powers of $\epsilon$
\begin{equation}
y=y_0+\epsilon y_1 +\epsilon^2 y_2 + \cdots.
\end{equation}
Upon substituting the perturbative expansion into
Eq.~(\ref{damposc}), one obtains a hierarchy of equations:
\begin{eqnarray}
\ddot{y}_0+y_0 &=& 0,\nonumber\\
\ddot{y}_1+y_1 &=& -\dot{y}_0,\\
\ddot{y}_2+y_2 &=& -\dot{y}_1,\nonumber\\
\vdots&\quad &\vdots\nonumber
\end{eqnarray}
Starting from the solution to the zeroth-order equation
\begin{equation}
y_0(t)=R_0 \cos(t-\Theta_0),
\end{equation}
and the zeroth-order retarded Green's function
\begin{equation}
G_0^R(t-t')= \sin(t-t')\,\theta(t-t'),
\end{equation}
one obtains
\begin{eqnarray}
y_1(t) &=& -\int^t_{t_0} dt' \sin(t-t')\,\dot{y}_0(t')\nonumber\\
&=& -\frac{R_0}{2}[(t-t_0)\cos(t-\Theta_0)-
\cos(t_0-\Theta_0)\sin(t-t_0)],\nonumber\\
y_2(t) &=& -\int^t_{t_0} dt' \sin(t-t')\,\dot{y}_1(t')\nonumber\\
&=&\frac{R_0}{8} \left[\left((t-t_0)^2 + \frac{1}{2}\right) \cos
(t-\Theta_0) - (t-t_0)\sin(t-2t_0+\Theta_0)-\cos(t-2t_0+\Theta_0)
\right].\nonumber\\
\end{eqnarray}
Hence, to order $\mathcal{O}(\epsilon^2)$ the solution is given by
\begin{eqnarray}
y(t)&=&R_0\Biggl\{\left[1-\frac{\epsilon}{2}
 (t-t_0)+\frac{\epsilon^2}{8}(t-t_0)^2\right]\cos(t-\Theta_0)
-\frac{\epsilon^2}{8}
(t-t_0)\sin(t-2t_0+\Theta_0)\Biggr.\nonumber\\
&&+\,\Biggl.\frac{\epsilon}{2}\cos(t_0-\Theta_0)\sin(t-t_0)
+\frac{\epsilon^2}{8}\left[ \frac{1}{2}\cos(t-\Theta_0)
-\cos(t-2t_0+\Theta_0)\right]\Biggr\} +
\mathcal{O}(\epsilon^3).\nonumber\\\label{dosol}
\end{eqnarray}

It is noted that this solution contains secular terms that grow in
$t$. This perturbative expansion breaks down for $\epsilon
(t-t_0)\gtrsim 1$ because of these secular terms. The presence of
secular terms in a perturbative expansion indicates that
perturbation method is only suitable to study short time behavior
of a nonequilibrium system. To extract information of the long
time behavior, one needs to go beyond the perturbation method. A
close inspection of the secular terms in Eq.~(\ref{dosol})
suggests the prefactor multiplying $\cos(t-\Theta_0)$ resembles a
time-dependent amplitude $R(t)$ and that multiplying
$\sin(t-2t_0+\Theta_0)$ resembles a time-dependent phase
$\Theta(t)$.

The way to solve this difficulty of the perturbation method is to
take account of its limitation and hence apply it only to describe
short time behavior. The perturbative solution is evolved in time
from $t_0$ \emph{only} up to a time, say $t_1$, such that
$\epsilon(t_1-t_0)\ll 1$ and hence the perturbative solution is
still reliable. Then the amplitude $R(t_1)$ and phase
$\Theta(t_1)$ is used as the \emph{updated} initial conditions at
$t_1$ to iterate the perturbative solution forward in time to
$t_2$ say $t_1$, such that $\epsilon(t_2-t_1)\ll 1$. Repeating
this procedure, one obtains an improved perturbative solution up
to an arbitrary time $\tau$. The idea of renormalization group
comes in when one wants to find out how the amplitude $R(\tau)$
and phase $\Theta(\tau)$ change as $\tau$ changes.

Let us assume that $R(\tau)$ and $\Theta(\tau)$ are related to
$R_0$ and $\Theta_0$, respectively, by a multiplicative
\emph{amplitude renormalization} $Z_R(\tau)$ and an additive
\emph{phase renormalization} $Z_\Theta(\tau)$ such that
\begin{equation}
R_0=Z_R(\tau)R(\tau),\quad\Theta_0=\Theta(\tau)+
Z_\Theta(\tau),\label{renconst}
\end{equation}
with $Z_R(\tau)$ and $Z_\Theta(\tau)$ being expanded in power of
$\epsilon$ as
\begin{equation}
Z_R(\tau)=1+ \epsilon a_1(\tau) + \epsilon^2 a_2(\tau) +
\cdots,\quad Z_{\Theta}(\tau)= \epsilon b_1(\tau) + \epsilon^2
b_2(\tau)+\cdots.\label{phaserenexp}
\end{equation}
Inserting the above expressions into Eq.~(\ref{dosol}), one
obtains to order $\mathcal{O}(\epsilon)$
\begin{eqnarray}
y(t)&=& R\,\biggl\{\cos(t-\Theta) - \epsilon\left[\frac{1}{2}
(t-t_0)-a_1(\tau)\right] \cos(t-\Theta)\nonumber\\
&&+\,\frac{\epsilon}{2}\left[\cos(\Theta-t_0) \sin(t-t_0)+2
b_1(\tau) \sin(t-\Theta)\right]\biggr\} + \mathcal{O}(\epsilon^2).
\label{1stsol:res1}
\end{eqnarray}
Now we want to remove the term that grows with $t_0$. Because for
a fixed (eventually large) $t$, this is a secular term. Choosing
\begin{equation}
a_1(\tau)=\frac{1}{2} (\tau-t_0),\quad b_1(\tau)= 0,
\end{equation}
we can rewrite Eq.~(\ref{1stsol:res1}) as
\begin{equation}
y(t,\tau)=R(\tau)\left\{\left[1-\frac{\epsilon}{2}
(t-\tau)\right]\cos[t-\Theta(\tau)] +\frac{\epsilon}{2}
\cos[\Theta(\tau)-t_0]\sin(t-t_0)\right\} +
\mathcal{O}(\epsilon^2). \label{1stsol:res2}
\end{equation}

A remarkable feature of Eq.~(\ref{1stsol:res2}) is that it is
valid for large times as long as (i) $\tau$ is chosen such that
$\epsilon(t-\tau)\ll 1$, and (ii) $R(\tau)$ and $\Theta(\tau)$ are
uniquely determined by the initial conditions. The first condition
is trivial as the variable $\tau$ introduced through
Eq.~(\ref{renconst}) is complete arbitrary. To determine $R(\tau)$
and $\Theta(\tau)$, we notice that since $\tau$ done not appear in
the original problem, the solution $y(t,\tau)$ \emph{cannot}
depend on the arbitrary variable $\tau$. This leads to
\begin{equation}
\frac{\partial y}{\partial \tau}=\left[\frac{dR}{d\tau}
+\frac{\epsilon}{2}R\right]\cos[t-\Theta(\tau)]
+R\frac{d\Theta}{d\tau} \sin[t-\Theta(\tau)]=0.\label{CS1}
\end{equation}
Using the fact that $\cos[t-\Theta(\tau)]$ and
$\sin[t-\Theta(\tau)]$ are linearly independent functions, to this
order in $\epsilon$, Eq.~(\ref{CS1}) reduces to
\begin{equation}
\frac{d R}{d\tau}=-\frac{\epsilon}{2}R(\tau) +
\mathcal{O}(\epsilon^2),\quad \frac{d
\Theta}{d\tau}=0+\mathcal{O}(\epsilon^2).\label{RG1}
\end{equation}
This set of equations are analogous to the renormalization group
equations in Euclidean field theory in the sense that they
describe the \emph{flow} of the amplitude $R(\tau)$ and phase
$\Theta(\tau)$ under the change of $\tau$. The time scale $\tau$
can be thought of as an arbitrary \emph{renormalization scale},
and the RG equation determine how the ``renormalized'' amplitude
and phase depend on the renormalization scale in such a way that
the change in the renormalization scale is compensated by the
change in the renormalized amplitude and phase. It is easy to see
that the configuration $R(\tau)=0$ and $\Theta(\tau)=\mbox{const}$
corresponds to the \emph{fixed point} of the RG equations.

Solving the above RG equations and setting $\tau=t$, one obtains
\begin{equation}
R(t)=R(0) e^{-\epsilon(t-t_0)/2}+\mathcal{O}(\epsilon^2 t), \quad
\Theta(t)=\Theta(0)+\mathcal{O}(\epsilon^2 t),\label{RGEsol}
\end{equation}
where $R(0)$ and $\theta(0)$ are constants determined by initial
conditions at $t_0$. Upon setting $\tau=t$ in
Eq.~(\ref{1stsol:res2}), we obtain an ``improved'' perturbative
solution $y(t)$ which to this order does not contain any secular
terms and hence is valid for large times. With the initial
conditions $y(t_0)=1$ and $\dot{y}(t_0)=0$, one finds $R(0)=1$,
$\Theta(0)=t_0$ and the RG-improved solution reads
\begin{equation}
y(t)=e^{-\epsilon(t-t_0)/2} \left[\cos(t-t_0)
+\frac{\epsilon}{2}\sin(t-t_0)\right] + \mathcal{O}(\epsilon^2).
\end{equation}

This renormalization group method can be extended to higher order
in $\epsilon$. Specifically, to order $\mathcal{O}(\epsilon^2)$
the choice of renormalization constants
\begin{equation}
a_2(\tau)=\frac{1}{8} (\tau-t_0)^2,\quad b_2(\tau)=-\frac{1}{8}
(\tau-t_0),
\end{equation}
can be shown to completely remove all of the secular terms to this
order. This is a sign of the renormalizability to all orders in
perturbation theory. After amplitude and phase renormalization the
solution is found to be given by
\begin{eqnarray}
y(t,\tau)&=&R(\tau)\left\{\left[1-\frac{\epsilon}{2} (t-\tau) +
\frac{\epsilon^2}{8}\left((t-\tau)^2+\frac{1}{2}\right)\right]
\cos[t-\Theta(\tau)] +\frac{\epsilon}{2}
  \cos[\Theta(\tau)-t_0] \sin(t-t_0)\right.\nonumber\\
&& -
\left.\frac{\epsilon^2}{8}\left[(t-\tau)\sin[t-2t_0+\Theta(\tau)]
+\frac{1}{2}\cos[t-2t_0+\Theta(\tau)]\right]\right\} +
\mathcal{O}(\epsilon^3). \label{2ndsol_res1}
\end{eqnarray}
The fact that $y(t,\tau)$ is independent of $\tau$ leads to
\begin{equation}
\frac{\partial y}{\partial \tau}=\left[\frac{dR}{d\tau}
+\frac{\epsilon}{2}R\right]\cos[t-\Theta(\tau)]
+\left[R\frac{d\Theta}{d\tau}+\frac{\epsilon}{4}\frac{dR}{d\tau}\right]
\sin[t-\Theta(\tau)]+\mathcal{O}(\epsilon^3)=0.\label{CS2}
\end{equation}
Hence, the RG equations to order $\mathcal{O}(\epsilon^2)$ read
\begin{equation}
\frac{dR}{d\tau}=-\frac{\epsilon}{2}R(\tau)+\mathcal{O}(\epsilon^3),\quad
\frac{d\Theta}{d\tau}=\frac{\epsilon^2}{8}+\mathcal{O}(\epsilon^3).\label{RG2}
\end{equation}

The solution of the above RG equations (\ref{RG2}) is given by
\begin{equation}
R(t)=R(0)\,e^{-\epsilon(t-t_0)/2}+\mathcal{O}(\epsilon^3 t), \quad
\Theta(t)=\Theta(0)+ \frac{\epsilon^2}{8} t +
\mathcal{O}(\epsilon^3 t),\label{RGEsol2}
\end{equation}
where $R(0)$ and $\theta(0)$ are constants determined by initial
conditions. Using the initial conditions specified above, to order
$\epsilon^2$, one finds the RG-improved solution is given by
\begin{eqnarray}
y(t)&=& e^{-\epsilon(t-t_0)/2}\,
\bigg\{\left(1+\frac{\epsilon^2}{16}\right)
\cos\left[\left(1-\frac{\epsilon^2}{8}\right)(t-t_0)\right]
+\frac{\epsilon}{2}\cos\left[\frac{\epsilon^2}{8}(t-t_0)\right]
\sin(t-t_0)\nonumber\\
&&-\,\frac{\epsilon^2}{16}\cos\left[\left
(1+\frac{\epsilon^2}{8}\right)(t-t_0)\right]\bigg\} +
\mathcal{O}(\epsilon^3). \label{2ndsol:res3}
\end{eqnarray}

The interpretation of the renormalization group method is very
clear in this simple example: the perturbative expansion is
carried out to a time scale $\ll 1/\epsilon$ within which
perturbation theory is valid. The correction is recognized as a
change in the amplitude and phase, so at this time scale the
correction is absorbed into the renormalization of the amplitude
and phase. The perturbative expansion is carried out to a later
time but in terms of the amplitude and phase \emph{at the
renormalization time scale}. The dynamical renormalization group
equation is the differential form of this procedure of evolving in
time, absorbing the corrections into the amplitude and phase, and
continuing the evolution in terms of the renormalized amplitudes
and phases.
\newpage

%% file: references.tex
%